%% file: thesis.tex
\documentclass[12pt,a4paper,oneside]{report}
\usepackage{fullpage}
\usepackage{graphicx}
\usepackage{graphics}
\usepackage{epsfig}
\usepackage{subfigure}
\usepackage{multirow}
\usepackage{epsfig}
\usepackage{subfigure}
\usepackage{multirow}
\usepackage{graphicx}
\usepackage{amsmath}
\usepackage{amssymb}
\usepackage{amsfonts}

\usepackage{rotating}
\usepackage[usenames]{color}
\usepackage[space]{cite}
\usepackage{fancyhdr}

\renewcommand{\baselinestretch}{1.5}

\nocite{*}

\begin{document}
\newcommand{\pz}{\phantom{0}}
\newcommand{\Zg}{Z/\gamma^*}
\newcommand{\BR}{\mbox{Br}}
\newcommand{\Br}{\mbox{Br}}
\newcommand{\metcal}{\mbox{ $\slash\hspace*{-1.5ex}E_{T}^{cal}$ }}
\newcommand{\metcalx}{\mbox{ $\slash\hspace*{-1.5ex}E_{x}^{cal}$ }}
\newcommand{\metcaly}{\mbox{ $\slash\hspace*{-1.5ex}E_{y}^{cal}$ }}
\newcommand{\met}{\mbox{$\slash\hspace*{-1.5ex}E_T$}}
\newcommand{\metx}{\mbox{$\slash\hspace*{-1.5ex}E_x$}}
\newcommand{\mety}{\mbox{$\slash\hspace*{-1.5ex}E_y$}}
\newcommand{\ppbar}{p\overline{p}}
\newcommand{\bbbar}{$b\overline{b}$ }
\newcommand{\wplus}{$W^{+}$}
\newcommand{\wminus}{$W^{-}$}
\newcommand{\nbb}{$\nu\beta\beta$}
\newcommand{\pa}{$\rm ^{234m}Pa$}
\newcommand{\nd}{$\rm ^{150}Nd$}
\newcommand{\tl}{$\rm ^{208}Tl$}
\newcommand{\Mmm}{$M_{\mu \mu}$}
\def\picwidth{13.5cm}
\pagenumbering{arabic}
\input{Preamble.tex}

\pagenumbering{arabic}
\renewcommand{\baselinestretch}{1}
\normalsize
\newpage
\pagestyle{plain}

\tableofcontents
\input{abstract.tex}
\input{Declaration.tex}
\input{author.tex}
\input{Acknoledgments.tex}

\newpage
\pagestyle{plain}
\input{introduction.tex}
\input{theory.tex}

\input{detector.tex}

\input{technical.tex}
\input{background.tex}
\input{twoneutrino.tex}

\input{zeroneutrino.tex}

\input{conclusion.tex}
\appendix

\newpage
\bibliography{thesis.bib}









\end{document}

%% file: Preamble.tex
\vspace{2in}
\pagestyle{empty}
\renewcommand{\baselinestretch}{2}
\normalsize
\vspace{0.7in}
\begin{center}
\begin{large}
{\bf Measurement of the Double Beta Decay Half-life of \boldmath$^{150}$Nd and Search for Neutrinoless Decay~Modes \\with the NEMO--3~Detector }
\end{large}

\vspace{0.5in}
\renewcommand{\baselinestretch}{1}
\normalsize
A thesis submitted to the University of Manchester for the degree of Doctor of Philosophy  in the Faculty of Engineering and Physical Sciences

\vspace{0.5in}
\begin{large}
{\bf 2009}
\end{large}

\vspace{1.2in}
\begin{large}
{\bf Nasim Fatemi-Ghomi}
\end{large}
\vspace{0.2in}

\begin{large}
{\bf Particle Physics Group}\\
{\bf School of Physics and Astronomy}\\
\end{large}

\end{center}

%% file: abstract.tex
\newpage
\renewcommand{\baselinestretch}{1.6}
\normalsize
\pagestyle{plain}
\begin{center}
\begin{huge}
\bf{Abstract}
\end{huge}
\end{center}
The half-life for two-neutrino double beta decay of $^{150}$Nd has been measured  with data taken by the NEMO~3 experiment at the Modane Underground Laboratory. Using $924.7$ days of data recorded with $36.55$~g of $^{150}$Nd the half-life  of this process is measured to be
\begin{equation}
T^{2\nu}_{1/2}=(9.11^{+0.25}_{-0.22}~{\rm(stat.)}\pm0.62~{\rm(syst.)})\times 10^{18}~{\rm y}.   \nonumber
\end{equation}
This result is the world's most accurate half-life measurement of this isotope. 
A search for neutrinoless double beta decay of this isotope is presented using the same data taking period. No significant excess of events above the background expectation is observed and the limit on the half-life of this process is set to be
\begin{equation}
T_{1/2}^{0\nu}>1.8\times10^{22}~{\rm y}~(90\%{\rm~CL}).      \nonumber
\end{equation}
This result has significantly improved the previous limit on neutrinoless double beta decay of this isotope. Limits are also set on several other neutrinoless double decay modes.

%% file: Declaration.tex
\newpage
\begin{center}
\begin{huge}
\bf{Declaration}
\end{huge}
\end{center}
\renewcommand{\baselinestretch}{1.6}
\normalsize
No portion of the work referred to in this dissertation has been
 submitted in support of an application for another degree or 
qualification of this or any other university or other institute of learning.

The author of this thesis (including any appendices and/or schedules to this 
thesis) owns any copyright in it (the ``Copyright'') and she has given The University of Manchester the right to use such Copyright for any administrative, 
promotional, educational and/or teaching purposes. 

Copies of this thesis, either in full or in extracts, may be made only in accordance with the regulations of the John Rylands University Library of Manchester. 
Details of these regulations may be obtained from the Librarian. This page must 
form part of any such copies made. 

The ownership of any patents, designs, trade marks and any and all other 
intellectual property rights except for the Copyright (the ``Intellectual Property Rights'') and any reproductions of copyright works, for example graphs and tables (``Reproduction''), which may be described in this thesis, may not be owned 
by the author and may be owned by third parties. Such Intellectual Property 
Rights and Reproductions cannot and must not be made available for use without 
the prior written permission of the owner(s) of the relevant Intellectual Property 
Rights and/or Reproductions. 

Further information on the conditions which disclosure, publication and exploitation of this thesis, the Copyright and any Intellectual Property Rights and 
or Reproductions described in it may take place is available from the Head of 
School of Physics and Astronomy. 

%% file: author.tex
\newpage
\begin{center}
\begin{huge}
\bf{The Author}
\end{huge}
\end{center}
\renewcommand{\baselinestretch}{1.6}
\normalsize
The author gained a first class BSc degree in Physics at the Science and Research Branch of Tehran Azad University, before obtaining a MSc degree from the School of Physics and Astronomy at the University of Manchester in 2005. The work presented here was undertaken in Manchester between 2005 and 2009.

%% file: Acknoledgments.tex
\newpage
\pagestyle{plain}
\begin{center}
\begin{huge}
\bf{Acknowledgements}
\end{huge}
\end{center}
First and foremost I would like to thank my supervisor, Stefan 
S\"oldner-Rembold, for his guidance and advice and most importantly encouragement through the past three years. I am grateful to  Vladimir Vasiliev of UCL  for getting me started with NEMO~3 data analysis and striving to answer any and all of the questions I had over the last three years.  Thanks to Fred Loebinger for being so friendly and always willing to help and Sabah Salih for his support. My thanks goes to all  members of the Manchester HEP group who made my time  here enjoyable.

Thanks to all the members of the  NEMO~3 analysis group for all the interesting discussions and useful meetings related to this analysis, especially Ruben Saakyan, Xavier Sarazin, Alexander Barabash, Victor Tretyak and Vera Kovalenko. I am also grateful  to Wade Fisher at                                                Fermilab and Mark Owen who provided help and guidance on running the limit setting program used in this thesis.   
I would also like  to thank Irina Nasteva, Chris Jackson,  Steve Snow and Lisa Alexander for proof reading my thesis.

I am lucky for having  a wonderful family and I would like to thank them all. Thank you Navid, Natalie, Parisa, Darius and Leila for being there for me whenever I needed a break.  Thank you Mark for being supportive and patient when I was stressed out with work and for making me feel happy.
Thank you mum for being so great, for bringing me up to be a feminist independent woman  and for being my best friend through all of my life. 

I would like to dedicate this thesis to my father Taghi (baba)  and my brother Nader (dadash Nana) who I lost during my teenage years. You have been  always with me and I will always remember you.

%% file: introduction.tex
\renewcommand{\baselinestretch}{1.6}
\normalsize
\chapter{Introduction}
Neutrinoless double beta decay  is a lepton number violating process which provides unique evidence that neutrinos are Majorana particles, i.e.~their own anti-particle. Neutrinoless double beta decay is also sensitive to the neutrino mass scale. The experimental signature of $0\nu\beta\beta$ decay is the observation of two electrons, for which the total energy sum is equal to the nuclear transition energy.

This thesis presents a measurement of the half-life of  neodymium-150 ($^{150}$Nd) two-neutrino double beta decay ($2\nu\beta\beta$) and a search for different modes of neutrinoless double beta decay ($0\nu\beta\beta$). The data   used for this thesis were collected by NEMO~3  between 2003 and 2006, corresponding to $924.7$ days of data taking. 

The $^{150}$Nd isotope  has a nuclear transition energy of $3.367$~MeV, which is higher than for most of the natural radioactive sources of background, and has  a  large phase space factor. These properties have made $^{150}$Nd a strong candidate for next generation double beta decay experiments, such as SuperNEMO~\cite{supernemo} and SNO+~\cite{snop}.

 In NEMO~3 the backgrounds to $0\nu\beta\beta$ are divided into   $2\nu\beta\beta$  decay and  radioactive backgrounds.  The $2\nu\beta\beta$ background is irreducible as it has the same event topology as $0\nu\beta\beta$ decay. The precise half-life measurement of $2\nu\beta\beta$ decay is therefore important for  $0\nu\beta\beta$ searches. It also helps to improve the understanding of nuclear matrix elements (NME), which are the major source of uncertainty in the derivation of the neutrino mass from the $0\nu\beta\beta$ half-life. The radioactive backgrounds are reduced by applying two-electron event selection criteria. The measurement of their activities  is necessary for estimating the number of remaining events due to these backgrounds. This is achieved by studying control  channels with final states different from the signal.



The  thesis is set out as follows.
The theoretical background to the work presented in this thesis and  the current status of double beta decay experiments are described in Chapter~\ref{chap-theory}. Chapter~\ref{chap-detector} describes the NEMO~3 experiment. The analysis techniques used are discussed in Chapter~\ref{chap-technique}. The radioactive backgrounds to double beta decay of $^{150}$Nd  are introduced and their activity measurements are described  in  Chapter~\ref{chap-bgr}. This chapter also shows that  the background estimation can describe data well in several different analysis channels.

Chapter~\ref{chap-2nbb} is dedicated to a measurement of the $^{150}$Nd $2\nu\beta\beta$ half-life. The systematic uncertainty on the  measurement is estimated  and the  result is compared to other measurements of the $^{150}$Nd half-life. Chapter~\ref{chap-0nbb} presents the limits on half-lives of different neutrinoless double beta modes and compares these results with other searches for new physics in double beta decay experiments. Finally, Chapter~\ref{chap-summary} summarises the work described and gives a conclusion.

%% file: theory.tex
\renewcommand{\baselinestretch}{1.6}
\normalsize
\chapter{Theoretical background}
\label{chap-theory}
\section{Introduction}
The neutrino was first proposed by Wolfgang Pauli in 1930~\cite{pauli}  as a light neutral particle  to solve the observed non-conservation of energy in beta decay.   Since then much work 
has been done to establish a theory that can describe the fundamental particles and the observed interactions between them. As a result, neutrinos have become one of the building blocks of  
  the Standard Model (SM) of  particle physics, but several of their properties  such as mass and their Dirac or Majorana nature are not known.

This chapter presents the theoretical background for the measurements and searches presented in this thesis. It begins with a brief introduction to the Standard Model. Sections~\ref{sec-neutrinomass} to \ref{sec-pmns}  review the properties of neutrinos with emphasis on the properties  which are not known.   These sections also  explain  how $0\nu\beta\beta$ decay can answer some of the questions regarding the nature and masses of the neutrinos. Section~\ref{sec-dbdth}   gives details of the different double beta decay theories and modes. Section~\ref{sec-nme} introduces the nuclear matrix elemenet of the double beta decay.
\section{The  Standard Model}
In the SM there are two general classes of fundamental particles: fermions, which have non-integer spin; and bosons, which have integer spin. The twelve  types of fermions are subdivided into two groups, leptons and quarks. Leptons have three flavours and consist of the charged electron, muon and tau, together with three corresponding charge-neutral neutrinos: electron neutrino, muon neutrino and tau neutrino.
 The interactions of particles in the SM are mediated by
 the exchange of gauge bosons. There are three types of interaction in the SM: electromagnetic, weak and strong interactions. The properties of fermions and gauge bosons in the SM are given in Tables~\ref{table:fermions} and Table~\ref{table:bosons}, respectively. In the  SM  fermions and gauge bosons obtain masses through the Higgs mechanism~\cite{higgs} which introduces an additional field with an associated particle, the Higgs boson. 
\begin{table}[htp!]
\begin{center}
\begin{tabular}{|r||c||c||c|}
\hline                                               
&fermions & charge &  mass  \\
\hline
leptons & electron $(e)$&-1 & 0.51~MeV\\
       & electron-neutrino ($\nu_{e}$) &0& $<2$~eV\\
\cline{2-3}

        &muon $(\mu )$ &-1 & 105.6~MeV\\
       & muon-neutrino ($\nu_{\mu}$) & 0 & $<2$~eV\\
\cline{2-3}
       &tau $(\tau)$ & -1 & 1777~MeV\\
       &tau-neutrino ($\nu_{\tau}$) & 0 & $<2$~eV\\
\hline
 quarks & up (u) & $+2/3$& 1.5--4~MeV\\
     & down (d) & $-1/3$ & 4--8~MeV\\
\cline{2-3}
       & charm (c) & $+2/3$& 1.15--1.35~GeV\\
       & strange (s) & $-1/3$ & 80--130~MeV\\
\cline{2-3}
       & top (t) & $+2/3$ & $174 \pm 5$~GeV\\
       & bottom   (b)  & $-1/3$ & 4.1--4.4~GeV\\
\hline
\end{tabular}
\caption[The three generations of fermions in the SM,
their electric charge and mass]{The three generations of fermions in the SM,
their electric charge and mass\cite{pdg}.}
\label{table:fermions}
\hspace{10cm}
\begin{tabular}{|r||c|c|}
\hline
 gauge bosons & interaction & mass (GeV)\\
\hline
$\gamma$ (photon) & electromagnetic & 0\\
 $Z$ boson & weak & 91.188$\pm$0.002  \\
 $W^{\pm}$ boson & weak & 80.425 $\pm$ 0.038\\
g (gluon) & strong & 0 \\
\hline
\end{tabular}
\end{center}
\caption[The gauge bosons of the SM and their masses]{The gauge bosons of the SM and their masses \cite{pdg}.}
\label{table:bosons}
\end{table}

\subsubsection{Definition of lepton number in SM}
Each generation of leptons has associated with it  a quantum  number. There are three lepton numbers: electron number ($L_{e}$), muon number ($L_{\mu}$) and tau number ($L_{\tau}$). The electron and electron neutrino have $L_{e}=1$ and the positron and electron anti-neutrino have $L_{e}=-1$. For all other leptons $L_{e}=0$. Similarly, $L_{\mu}=0$ and $L_{\tau}=0$ for leptons not in the muon and tau generations, respectively.  In the SM, the sums of lepton numbers are conserved in all known interactions.

\section{The nature of  massive neutrinos}
\label{sec-neutrinomass}
Despite  the success of the SM, it  is not able to account for  massive neutrinos.
In the SM, neutrinos are considered to be left-handed and this precludes the possibility that neutrinos have mass, as  helicity is conserved only for a massless particle.  However, the observation of the oscillation between the  different flavours of neutrinos~\cite{oscillation,oscillation2} has shown that the neutrino mass eigenstates ($\nu_{1}, \nu_{2}$ and $\nu_{3}$) and the flavour eigenstates ($\nu_{e}$, $\nu_{\mu}$ and $\nu_{\tau}$) are different, and thus neutrinos have mass.  This leads to two fundamental questions: what is the nature of the neutrino mass and what is the mass scale of the neutrino? The latter is discussed in Section~\ref{sec-mass}. This section focuses on the question of the nature of the neutrino. Neutrinos are either Dirac particles, which means that they are distinct from their own anti-particles, or they are Majorana particles, meaning they are  their own anti-particles.
\subsection{Dirac neutrino}
The Dirac mass term of the neutrino Lagrangian is written as \cite{diracmass}
\begin{equation}
\label{eq-md}
\mathcal{L}_{m_{D}}=m_{D}\left(\overline{\nu_{R}^{0}}\nu_{L}^{0}+\overline{\nu_{L}^{0}}\nu_{R}^{0}\right),
\end{equation}
where $m_{D}$ is the non-diagonal  Dirac mass matrix, and  $\nu_{R}^{0}$ and $\nu_{L}^{0}$ are the chirally right-handed and left-handed flavour neutrino fields  defined as
\begin{equation}
\nu_{L}^{0}=\left(\begin{array}{c}
\nu_{eL}^{0}\\
\nu_{\mu L}^{0}\\
\nu_{\tau L}^{0}\end{array}\right),
\nu_{R}^{0}=\left(\begin{array}{c}
\nu_{eR}^{0}\\
\nu_{\mu R}^{0}\\
\nu_{\tau R}^{0}\end{array}\right).
\end{equation}
The second term in Equation~\ref{eq-md} is the Hermitian conjugate ($h.c.$) of the first term.
As right-handed neutrinos do not couple through the weak interaction, the Dirac model predicts three right-handed sterile  neutrinos which do not interact and can not be detected. 
Dirac mass terms require four independent components ($\nu_{L}^{0}$,$\nu_{R}^{0}$, $\overline\nu_{L}^{0}$, $\overline\nu_{R}^{0}$) and thus suggest that neutrinos and anti-neutrinos are distinct and lepton number is conserved. The Dirac model leads to  small coupling of the neutrino to the Higgs field (in comparison  with the coupling of other leptons).  The Dirac model  can not explain why the neutrino mass is much lower than that of other leptons.

\subsection{Majorana neutrino}
The Majorana mass term may be constructed out of solely  the left-handed or right-handed neutrino. However, the left-handed mass is not invariant under the  electroweak gauge group. Thus, it is more common to see the mass term in the right-handed construction.  The   neutrino mass  Lagrangian can be written as~\cite{diracmass}
\begin{equation}
\mathcal{L}=\frac{1}{2}m_{RM}\overline{(\nu_{R}^{0})^{c}}\nu_{R}^{0}+h.c. ,
\label{eq-maj}
\end{equation}
where $m_{RM}$ is the right-handed symmetric Majorana mass matrix, and $\overline{(\nu_{R}^{0})^{c}}$ is the charge-conjugate field of $\overline{\nu_{R}^{0}}$, which satisfies the Majorana condition  $(\nu_{R}^{0})^{c}=C\overline{\nu^{0}}^{T}_{R}$, where $C$ is the charge conjugation matrix, and $T$ denotes transposition. 
This field also has the property that $(\nu_{R}^{0})^{c}=P_{L}(\nu^{0})^{c}$, where $P_{L}$ is the left-handed projection operator, that is $(\nu_{R}^{0})^{c}$ is left-handed. This can be considered in terms of the picture of moving into a reference frame where the helicity of the massive neutrino flips.

The Majorana neutrino field is self-conjugate and therefore Majorana neutrinos are their own antiparticles. Among all fermions only neutrinos can be Majorana particles as they do not carry charge. Since the Majorana field has only two independent components ($\overline{\nu_{R}^{0}}$, $\nu_{R}^{0}$) compared to the four in the Dirac field, the Majorana theory is simpler and more natural than the Dirac theory.
As there is no reason for the Majorana term to be zero,  neutrinos are assumed  to be Majorana particles in most theories  beyond the SM.

\subsubsection{The see-saw mechanism}
The see-saw mechanism~\cite{seesaw} is a Dirac-Majorana model which is  developed in order to justify the smallness of the mass of the neutrinos in comparison with other fermions in the SM\@. 
If the neutrino has a Dirac mass as other  fermions in the SM  do and also a right-handed Majorana mass defined through Equation~\ref{eq-maj}, 
 then its total Dirac-Majorana neutrino mass term $\mathcal{L}_{mass}$ is constructed as 
\begin{eqnarray}
\mathcal{L}_{mass}=\mathcal{L}_{D}+\mathcal{L}_{RM}=&\left(m_{D}\overline{\nu_{R}}\nu_{L}+\frac{1}{2}m_{RM}\overline{(\nu_{R}})^{c}\nu_{R}\right)+h.c.&\nonumber\\
=&\frac{1}{2}\left((\overline{\nu_{L}})^{c}~\overline\nu_{R}\right)\left(\begin{array}{cc}
0 &m_{D}\\
m_{D} & m_{RM}
\end{array}\right)\left(\begin{array}{c}
\nu_{L}\\
\nu_{R}^{c}
\end{array}\right)&
\label{eq-see}
\end{eqnarray}
The matrix $\mathcal{M}_{\nu}$=$\left(\begin{array}{cc}
0 &m_{D}\\
m_{D} & m_{RM}
\end{array}\right)$ is referred to as the neutrino mass matrix. 
Equation~\ref{eq-see} can be diagonalised to give the mass eigenstates with eigenvalues
\begin{equation}
M_{1,2}=\frac{1}{2}\left(m_{RM}\pm\sqrt{(m_{RM}^{2}+4m_{D}^{2})}\right).
\end{equation}
If it is assumed that the Dirac mass, $m_{D}$ of the neutrino is of the same order of magnitude as the Dirac mass of other fermions in the SM and that the right-handed Majorana mass term is much heavier than the Dirac mass term ($m_{RM}\gg m_{D}$),  the first approximation of these eigenvalues gives $M_{1}=\frac{m_{D}^{2}}{m_{RM}}$ and $M_{2}=m_{RM}$.  Thus this model predicts   two physical neutrinos: the light left-handed neutrino and a very heavy right-handed Majorana neutrino. If the heavy neutrino mass is at the GUT (Grand Unified Theory) scale ($\sim10^{14}$~GeV), then the  mass of the light neutrino is order of few eV\@. Therefore this model can explain why the mass of the SM neutrino is much lower than the mass of the other fermions. As the Majorana mass term appears in both neutrino mass eigenvalues, this mechanism predicts that neutrinos are Majorana particles.



\section{PMNS  matrix }
\label{sec-pmns}
In the full three neutrino mixing framework the weak eigenstates can be expressed as superpositions of three neutrino mass eigenstates ($m_{1}$, $m_{2}$, $m_{3}$) linked via a unitary matrix $U$:
\begin{equation}
\left( \begin{array}{c}
\nu_{e}  \\
\nu_{\mu} \\
\nu_{\tau}  \end{array} \right)=
\left(\begin{array}{ccc}
 U_{e1} & U_{e2} &U_{e3} \\
U_{\mu1} & U_{\mu2} & U_{\mu3} \\
 U_{\tau1} & U_{\tau2} & U_{\tau3}\end{array}\right)
\left(\begin{array}{c}
\nu_{1}  \\
\nu_{2} \\
\nu_{3}  \end{array} \right).
\end{equation}
This can be rewritten as 
\begin{equation}
|\nu_{\alpha}\rangle=\sum_{i}{U_{\alpha i}}|\nu_{i}\rangle.
\end{equation}
This unitary mixing matrix is known as the Pontecorvo-Maki-Nakagawa-Sakata~(PMNS) matrix~\cite{pmns,pmns2}.  The unitary matrix $U$ can be parametrised in the following form
\begin{equation}
U=\left(\begin{array}{ccc}
 c_{12}c_{13} & s_{12}c_{13} & s_{13}e^{-i\delta} \\
-s_{12}c_{23}-c_{12}s_{23}s_{13}e^{i\delta}& c_{12}c_{23}-s_{12}s_{23}s_{13}e^{i\delta} & s_{23}c_{13} \\
s_{12}s_{23}-c_{12}s_{23}s_{13}e^{i\delta} & -c_{12}s_{23}-s_{12}c_{23}s_{13}e^{i\delta}& c_{23}c_{13}
\end{array}\right)D,
\end{equation}
$D$ is the diagonal matrix defined as 
\begin{equation}
D=\left(\begin{array}{ccc}
1 & 0 &0\\
0 &e^{\phi_{1}} & 0 \\
0 & 0 & e^{\phi_{2}}
\end{array}\right),
\end{equation} 
where $s_{ij}$, $c_{ij}$ are the sine and cosine of the mixing angle $\theta_{ij}$,  $\delta$ is the Dirac phase and $\phi_{1}$ and $\phi_{2}$ are the Majorana phases, which only affect Majorana neutrinos. Thus the matrix $D$ only appears in the PMNS matrix if neutrinos are Majorana particles.
\subsubsection{Dirac and Majorana phases}
\label{sec-phases}
The Dirac and Majorana phases in the PMNS matrix are CP-violating phases. 
The Dirac CP-violating phase, $\delta$, can be measured by oscillation experiments via comparing the 
probability of neutrino and anti-neutrino oscillations from a certain flavour to another. However, the oscillation experiments are not able to measure the Majorana CP-violation phases $\phi_{1}$ and $\phi_{2}$ as they only appear on the leading diagonal of the PMNS matrix and therefore cancel in all measurable quantities in oscillation experiments. Neutrinoless double beta decay $(0\nu\beta\beta)$ experiments may be able to provide a possible constraint on CP-violation phases, as $0\nu\beta\beta$  is sensitive to the value of the effective neutrino mass:
\begin{equation}
\langle m_{\nu}\rangle=\arrowvert\sum_{i}{U_{\alpha i}^{2}m_{i}}\arrowvert=\arrowvert c_{12}^{2}c_{13}^{2}m_{1}+s_{12}^{2}c_{13}^{2}m_{2}e^{-2i\phi_{1}}+s_{13}^{2}m_{3}e^{-2i(\phi_{2}-\delta)}\arrowvert,
\label{eq-hi}
\end{equation}
where $m_{i}=\nu_{i}$.
More details about neutrinoless double beta decay are given in Section~\ref{sec-0nbbth}.
\subsection{Neutrino mass hierarchy problem}
\begin{figure}
\centering
\includegraphics[width=12.0cm]{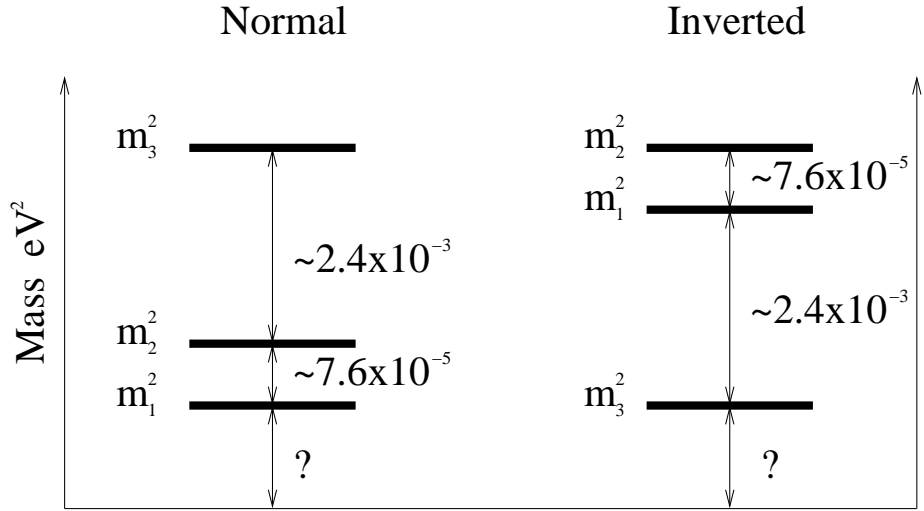}
\caption[A schematic view of the hierarchies models consistent with data from oscillation experiments.] {A schematic view of the hierarchy models consistent with data from oscillation experiments. The question mark illustrates that the absolute mass scale of neutrinos is not known.}
\label{fig-normal}
\end{figure}

\begin{figure}
\centering
\includegraphics[width=10cm]{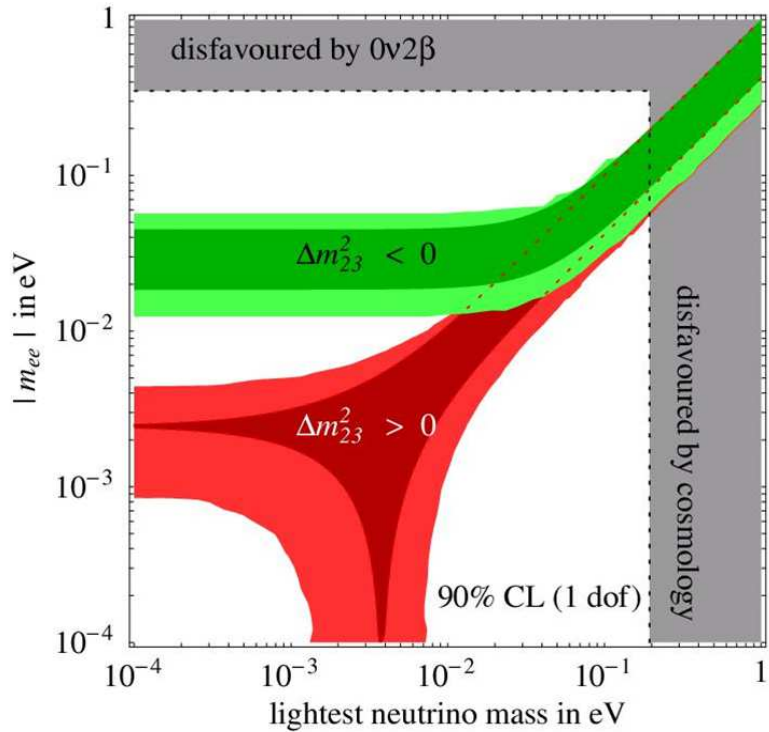}
\caption[Bounds on the effective neutrino mass with respect to the lightest neutrino mass at 90\% confidence level]{Bounds on the effective neutrino mass with respect to the lightest neutrino mass at 90\% confidence level~\cite{massscale}. The  normal hierarchy is shown in red and the inverted hierarchy is shown in green.  The degenerate region is where the green overlaps with the red. The darker regions of the plot show how the bounds on the effective neutrino mass would be constrained if the current  oscillation parameters were calculated with negligible error. Also shown in grey are the regions disfavoured by current $0\nu\beta\beta$-decay limits and from cosmology.}
\label{fig-himass}
\end{figure}

The probability for a neutrino to change from one flavour to another   is related to the difference  between the masses squared of the mass eigenstates, $\Delta m_{ij}^{2}=m_{i}^{2}-m_{j}^{2}$.
 Based on the observations from neutrino oscillations, various mass models have been proposed~\cite{massscale}. These can be categorised as the  normal hierarchy in which $m_{1}$ has the lowest mass among the three, and the inverted hierarchy in which $m_{3}$ has the lowest mass. In both scenarios $m_{1}$ and $m_{2}$ have similar masses. The third model is degenerate, where the three mass eigenstates have  similar masses. Figure~\ref{fig-normal}  shows a schematic view of  the normal and inverted hierarchy models. The approximate mass squared difference between the  mass eigenstates found by oscillation experiments is shown~\cite{dmsol,dmatm}. 
%
Figure~\ref{fig-himass} shows the distribution of $\langle m_{\nu}\rangle$ with respect to the oscillation parameters for normal (red) and inverted hierarchy (green).   The next generation double beta decay  experiments will be sensitive to the full degenerate and inverted  hierarchy regions. If no neutrinoless double beta decay event is observed by these experiments, then the inverted hierarchy can be excluded. 
\subsection{Absolute mass scale of neutrino}
\label{sec-mass}
Oscillation experiments are trying to measure the difference between mass eigenstates of the neutrinos, but they are not sensitive to the absolute mass scale of the neutrino. Experimentally there are  three possible ways to constrain or measure the absolute mass scale. One way is to search for the neutrino rest mass by exploring the endpoint energy of the electron spectrum in tritium beta decay and thus find the mass scale of the neutrino ($\langle m_{\nu}\rangle$) directly from the kinematics of the decay.~Currently a limit for the electron neutrino mass of less than $2.2$~eV has been achieved~\cite{tl-beta1,tl-beta2}. The advantage of this experimental approach is that it is independent of the nature of the neutrino (Dirac or Majorana); however, the measurement of the mass from the decay spectrum is limited by the energy resolution of the experiments and background considerations. This makes it experimentally difficult to improve the sensitivity of the experiment to an energy scale of  meV.

The second way to constrain the neutrino mass is from analysing cosmological data. The astrophysical experiments  are able to set limits on the sum of the neutrino masses ($\sum_{i}{m_{i}}$). These limits are based on the cosmological models which give the neutrino contribution to the density of the universe. 
The most recent limits for  $\sum_{i}{m_{i}}$ range from $0.7$ to $2$~eV at $90\%$ confidence level~\cite{cosmology}. This range can constrain  the lightest neutrino mass ($m_{1}$ or $m_{3}$). The disfavoured region of lightest neutrino mass is shown by a  vertical gray bound in Figure~\ref{fig-himass}. As the cosmological models are based on several assumptions, it would be  challenging to improve these results.

The third approach to constrain the absolute mass scale of neutrinos is searching for neutrinoless double beta decay $(0\nu\beta\beta)$.  As mentioned previously in Section~\ref{sec-phases}, neutrinoless double beta decay is sensitive to the effective neutrino mass $\langle m_{\nu}\rangle$ if the neutrino is a Majorana particle. 
\section{Theory of  double beta decay} 
\label{sec-dbdth}
\subsection{Beta decay}
Beta ($\beta$) decay is a type of radioactive decay in which  the weak interaction converts a down quark of the neutron into an up quark of a proton while emitting an electron and an antineutrino. Thus, this process changes the atomic number, $Z$, of a nucleus by one unit, while the atomic mass $A$ remains the same:
\begin{equation}
(A,Z)\rightarrow(A,Z+1)+e^{-}+\bar\nu_{e},\\
\end{equation}
There are also two other possible  modes of this process which are known as $\beta^{+}$ and electron capture (EC) and  defined as:
\begin{eqnarray}
&(A,Z)&\rightarrow(A,Z-1)+e^{+}+\nu_{e}~~~(\beta^{+}~{\rm decay}),\\
&(A,Z)&+~e^{-}\rightarrow(A,Z-1)+\nu_{e}~~~({\rm EC}).
\end{eqnarray}
\subsection{Double beta decay}
\label{sec-2nbbth}
\begin{figure}[tb]
\centering
\includegraphics [width=10.0cm]{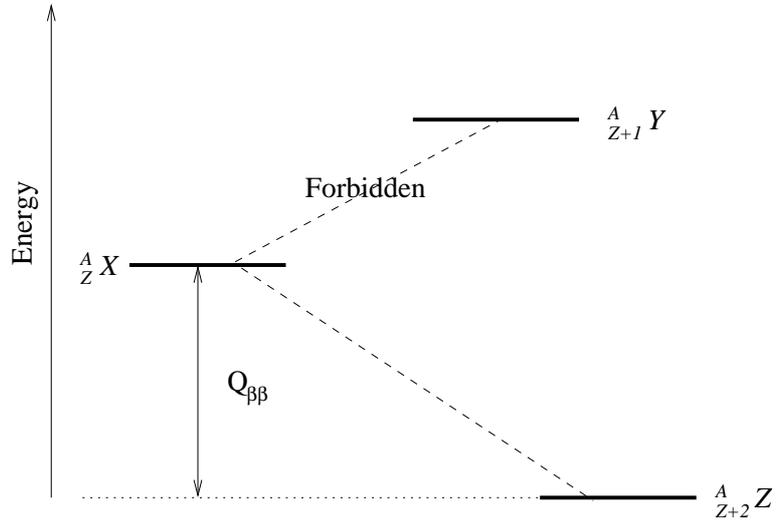}
\caption[A  diagram of the energy requirements for double beta decay]{A schematic diagram of the energy requirements for double beta decay. A parent nucleus $^{A}_{Z}X$ decays to a daughter nucleus $^{A}_{Z+2}Z$ because the intermediate single beta decay to $^{A}_{Z+1}Y$ is energetically forbidden.}
\label{fig:dbd_schematic}
\end{figure}

\begin{figure}[tb]
\centering
\includegraphics[width=15.30cm]{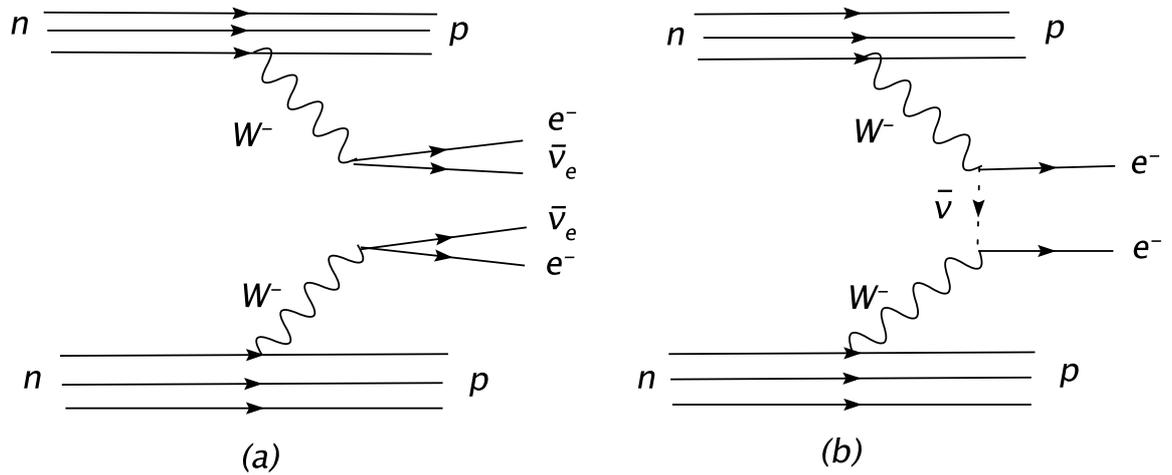}
\caption{Two-neutrino  double beta decay and  neutrinoless double
beta decay diagrams.}
\label{fig:dbdf}
\end{figure}

Double beta decay, $2\nu\beta\beta$~\cite{dbdt}, is a process in which two beta decays occur simultaneously with emission of two electrons and two antineutrinos:
\begin{equation}
(A,Z)=(A,Z+2)+2e^{-}+2\bar\nu.
\end{equation} 
Double beta decay  can only occur in nuclei with even atomic number and atomic mass (even-even). In these nuclei the single beta decay can be  either energetically forbidden or strongly suppressed.  The nuclear transition energy, $Q_{\beta\beta}$, for this process is defined as:
\begin{equation}
Q_{\beta\beta}=m(A,Z)-m(A,Z+2)-2m_{e},
\end{equation}
where $m(A,Z)$ and $m(A,Z+2)$ are, respectively, the masses of the initial and final  nuclei and $m_{e}$ is the mass of the electron. 
Figure~\ref{fig:dbd_schematic} shows a  diagram of the energy requirements for double beta decay. This process is the rarest known kind of radioactive decay, and it is predicted in only $36$ isotopes.  Double beta decay occurs within the SM and involves a second order weak interaction as shown in Figure~\ref{fig:dbdf}a. The  rate of the process is characterised by its very long lifetime (more than $10^{18}$ years).

The half-life ($T^{2\nu}_{1/2}$) of the double beta decay process is expressed by the equation:
\begin{equation}
(T^{2\nu}_{1/2})^{-1}=G^{2\nu}(Q,Z)|M^{2\nu}|^{2},
\end{equation}
where $G^{2\nu}(Q_{\beta\beta},Z)$ is an analytically calculable phase space integral which scales with  $Q_{\beta\beta}^{11}$ and the atomic number, $Z$~\cite{bookkai}. The term $M^{2\nu}$ is the nuclear matrix element (NME) (Section~\ref{sec-nme}) for double beta decay which gives the probability of the process and can be calculated theoretically.  To test the methods used to calculate NMEs, it is important to determine the half-life of $2\nu\beta\beta$ experimentally. Two neutrino double beta decay is  an irreducible background for searches for   beyond the  Standard Model (BSM) double beta decay processes for detectors with weaker energy resolutions.

\subsection{Neutrinoless double beta decay}
\label{sec-0nbbth}
Neutrinoless double beta decay ($0\nu\beta\beta$)~\cite{0nbbt} involves a transition of two neutrons into two protons with the emission of two electrons and no neutrinos (Figure~\ref{fig:dbdf}b):
\begin{equation}
(A,Z)=(A,Z+2)+2e^{-}.
\end{equation}
In this process a right-handed antineutrino emitted at one vertex undergoes a helicity flip and is absorbed into the other vertex as a left-handed neutrino. This can only happen if   neutrinos and antineutrinos are identical (Majorana particles) and have mass. This process violates total lepton number by two units and is forbidden in  the  SM\@. Observation of  neutrinoless double beta decay would prove that neutrinos are Majorana particles and  would answer the most fundamental question about the nature of the neutrinos.

\begin{figure}
\centering
\includegraphics[width=14.0cm]{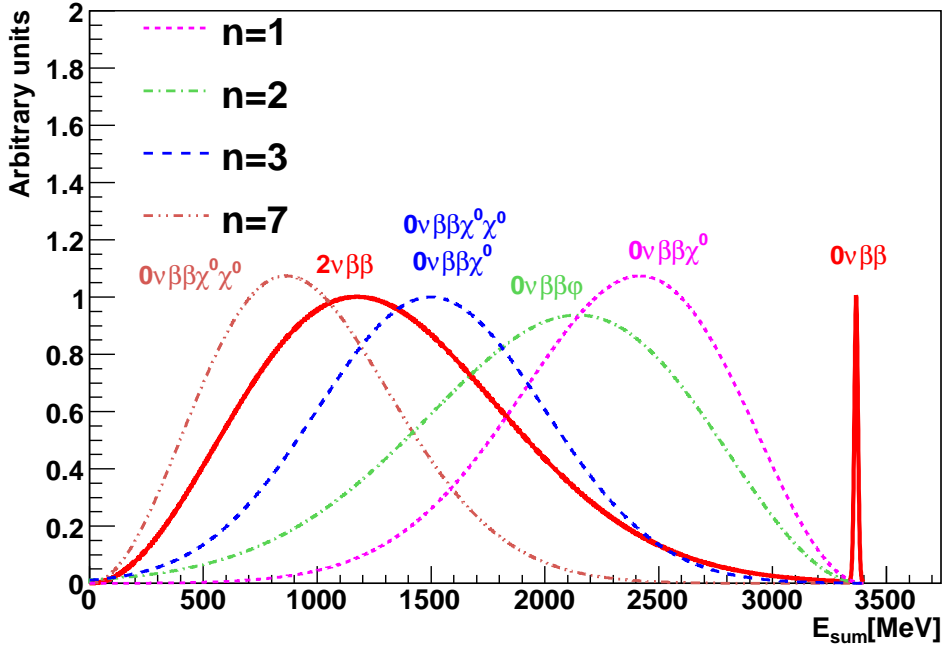}
\caption{Energy sum of the two electrons for $2\nu\beta\beta$, $0\nu\beta\beta$,  Majoron mode one (n=1), Majoron mode two (n=2), Majoron mode three (n=3) and Majoron mode seven (n=7) for an arbitrary isotope.  }
\label{fig-2nbb-0nbb}
\end{figure}

As there are no neutrinos in the final state, the experimental signature of $0\nu\beta\beta$ is two electrons, with energies summing to the nuclear transition energy ($Q_{\beta\beta}$). For $2\nu\beta\beta$ decay the energy sum  of the two electrons forms a continuous spectrum (Figure~\ref{fig-2nbb-0nbb}). 

The half-life of $0\nu\beta\beta$ is expressed by:
\begin{equation}
(T^{0\nu}_{1/2})^{-1}=G^{0\nu}(Q_{\beta\beta},Z)|M^{0\nu}|^{2}\left(\frac{\langle m_{\nu_{e}}\rangle}{m_{e}}\right)^{2},
\label{eq-0nbbhalflife}
\end{equation}
where $\langle m_{\nu_{e}}\rangle$ is the effective Majorana neutrino mass and $m_{e}$ is the electron rest mass. Thus the experimentally measured half-life, or in case of non-observation of the process its lower limit, can provide the effective neutrino mass or an  upper limit on it.
The phase space integral, $G^{0\nu}$, is proportional to $Q_{\beta\beta}^{5}$ and the  atomic number, $Z$~\cite{g0n}.  Thus the neutrinoless double beta decay rate is higher for isotopes with high $Q_{\beta\beta}$ value.
\subsubsection{Right-handed current  {\boldmath $0\nu\beta\beta$}}
The $0\nu\beta\beta$ mechanism that involves the helicity flip of the neutrino due to its mass is known as the mass mechanism. This mechanism is not the only possibility  that  leads to emission of two electrons in the final state.
There are theories beyond the Standard Model which predict the existence of pure right-handed $W$ bosons ($W_{R}$) or  $W$ bosons with mixtures of left-handed and right-handed $W$ bosons ($W=W_{R}+W_{L}$)~\cite{bookkai}. If $W_{R}$ exists and the neutrino is a Majorana particle, then a right-handed neutrino can interact at the other vertex without need for a helicity flip. This mechanism is known as the right-handed current mechanism of neutrinoless double beta decay. In this thesis a limit is set only on the half-life of pure right-handed neutrinoless double beta decay.
The half-life of this process is inversely related to  the phase space factor of this process, the NME ($M^{0\nu}_{\lambda}$) and $\lambda$, the coupling constant of the right-handed neutrino with $W_{R}$,
\begin{equation}
(T_{1/2}^{0\nu})^{-1}=G^{0\nu}_{\lambda}|M^{0\nu}_{\lambda}|^{2}\langle\lambda\rangle^{2}.
\end{equation}

\subsection{Double beta decay with emission of Majorons}
\label{sec-majoronth}
\begin{figure}
\centering      
\includegraphics[width=6.5cm]{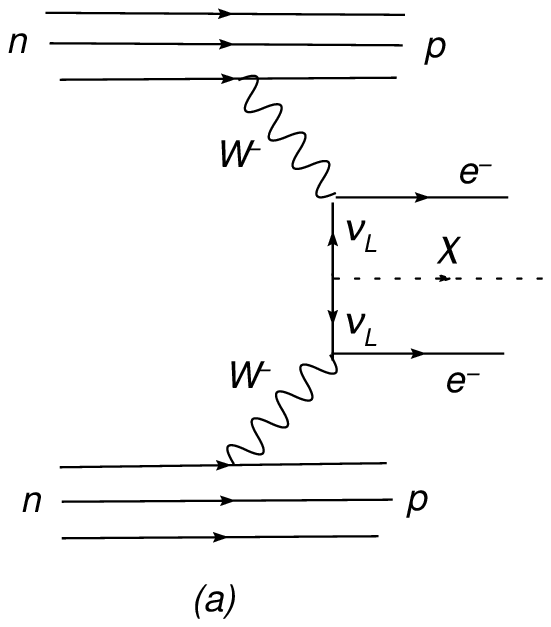}
\includegraphics[width=6.5cm]{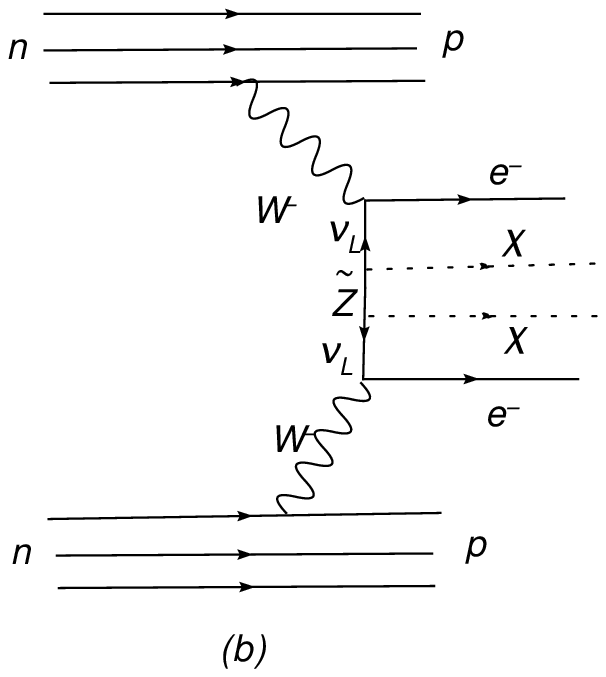}
\caption{Diagram of double beta decay with emission  of one Majoron in the   singlet Majoron model     and of two Majorons from the supersymmetric Majoron model.}
\label{fig-majoron}
\end{figure}

 Several beyond the SM models exist  in which global B-L (Baryon-Lepton) symmetry is broken spontaneously. These  lead to the prediction  of a massless Goldstone boson, called the Majoron~\cite{majoron} which can couple to the neutrino. There are three types of Majoron models: singlet Majoron, doublet Majoron and triplet Majoron~\cite{bookmajoron}.  The measurement of the $Z$ boson width at LEP~\cite{zlep} has ruled out the doublet and triplet Majoron models, but  singlet and dominantly singlet Majoron models are still possible. These models predict that the two neutrinos in double beta decay couple with  a Majoron, $\chi^{0}$:
\begin{equation}
(A,Z)\rightarrow (A,Z+2)+2e+\chi^{0},
\end{equation}
  Figure~\ref{fig-majoron}a shows a diagram of such an emission. 
 The drawback of the singlet Majoron models is that in these models the Majoron couples to the neutrino with a coupling strength of $g=(m_{\nu_{L}}/M_{B-L})$~\cite{majoronfinetuning} where $m_{\nu_{L}}$ is the mass of the light neutrino and $M_{B-L}$ is the symmetry-breaking scale (which is higher than the electroweak scale $\sim$ $90$~GeV). In order to have an observable rate for Majoron emitting double beta decay ($0\nu\beta\beta\chi^{0}$), the singlet Majoron model requires severe fine tuning.
To avoid the fine tuning problem, several  new Majoron models have been constructed. Here the term Majoron means light or massless boson with couplings to neutrinos. In this definition the  Majoron is not constrained to be a Goldstone boson.

There is a doublet Majoron model in supersymmetry which predicts neutrinoless double beta decay with emission of two Majorons~\cite{bookmajoron}:
\begin{equation}
(A,Z)\rightarrow (A,Z+2)+2e+2\chi^{0},
\end{equation}
Figure~\ref{fig-majoron}b shows the neutrinoless double beta decay diagram with the emission of two Majorons. In this diagram the Zino, $\tilde{Z}$, is  the  fermionic  supersymmetric partner of the $ Z$ boson. 
There are also other Majoron models which predict double beta decay with emission  of one or two Majorons. Majorons in these  models are  predicted to be vector     bosons and carry lepton charge and have  mass~\cite{mvector1,mvector2,vectormajoron}. 

A model for neutrino masses is proposed in the context of large extra dimensions. In this model the global B-L symmetry is broken spontaneously by a gauge singlet Higgs field in a bulk~\cite{majoronbulk}.  This leads to a bulk singlet Majoron which is observable in neutrinoless double beta decay.

There are in total  $10$ Majoron models (one bulk Majoron and $9$ light or massless bosonic Majorons) that are  of  interest in double beta decay experiments. The  energy of the final state electrons   from these ten  models can  form  four distinguishable distributions~\cite{mvector2}.  The possible two electron energy spectra ($E_{sum}$) for different  Majoron modes of $^{150}$Nd are shown in Figure~\ref{fig-2nbb-0nbb}. In the figure, the index $n$ defines the shape of the spectrum by modifying the distribution with a factor $(Q_{\beta\beta}-E_{sum})^{n}$.  Majoron mode one $(n=1)$ denotes the theories which lead to emission of one Majoron, Majoron mode two $(n=2)$ denotes the bulk Majoron emission, Majoron mode three $(n=3)$ denotes the emission of one or two massless lepton number carrying  Majorons and Majoron mode seven $(n=7)$ denotes the emission of two light Majorons. Note that electrons from the   $n=7$ mode  have on average lower energies   than $2\nu\beta\beta$ electrons;  thus, it is more difficult  to extract experimentally, since most backgrounds tend to dominantly produce low-energy electrons. 
The half-life of Majoron-emitting double beta decay is expressed by
\begin{equation}
(T^{0\nu\beta\beta\chi^{0}(\chi^{0})}_{1/2})^{-1}=G^{\beta\beta}_{\alpha}(Q_{\beta\beta},Z)|M^{\beta\beta}_{\alpha}|^{2}\langle g_{\alpha}\rangle^{2}.
 \end{equation}
The index $\alpha$ indicates the Majoron mode, $\langle g_{\alpha}\rangle$ is the effective coupling constant of Majoron mode $\alpha$, $G^{\beta\beta}_{\alpha}$ is the phase space factor and $M^{\beta\beta}_{\alpha}$ is the nuclear matrix element of each Majoron mode.
\subsection{Nuclear Matrix elements}
\label{sec-nme}
If the half-life of neutrinoless double beta decay is measured, the effective neutrino mass can be calculated from Equation~\ref{eq-0nbbhalflife}. Accurate calculations of the NME are necessary in order to measure the effective neutrino mass with minimum uncertainty.

Two main theoretical methods have been widely used to calculate the nuclear matrix elements  for neutrinoless double beta decay: the nuclear shell model (NSM)~ and the quasi-particle random phase approximation (QRPA). 
The NSM~\cite{shell2nbb} is useful for calculating  single particle states close to the Fermi level  and is only reliable for light nuclei  such as $^{48}$Ca, $^{76}$Ge  and $^{82}$Se. For medium to heavy double beta decay isotopes the shell  model calculations are difficult to carry out.

 For heavier nuclei  the proton-neutron QRPA (pn-QRPA)~\cite{qrpa} has been found to be a powerful model. The QRPA can handle a great number of intermediate states.  In double beta decay the initial nucleus decays to the final nucleus through virtual excitations of all states of the intermediate nucleus. The intermediate excited states are obtained by solving the QRPA equations.  
The QRPA calculations contain two two-body interaction matrix elements:  particle-particle (pp)  which is correlated to the  proton-proton interaction and particle-hole (ph) which is correlated to the proton-neutron interaction.  Both matrices contain independent interaction constants $g_{ph}$ and $g_{pp}$~\cite{shiva81}.

The ph matrix elements mainly  affect the giant Gamow-Teller resonance~\cite{ggt}, which is reproduced accurately by the QRPA and as such $g_{ph}$ is generally fixed.    The $g_{pp}$ constant, which  has a large effect on NME and double beta decay,  is a free parameter. This parameter causes uncertainties in NME calculations and thus much of the work into  QRPA has been focused on the $g_{pp}$ problem. One method in the early stage of development is to fix the value of $g_{pp}$ by fitting it to the available $2\nu\beta\beta$ and $\beta$ decay data.  Another important issue in QRPA is that the nuclei of experimental interest   are assumed to have spherical symmetry. As many heavy nuclei are deformed~\cite{deformation} this approximation does not give a realistic calculation for these nuclei (such as $^{150}$Nd). 
Table~\ref{tab-2nbb} presents the  most recent calculated $0\nu\beta\beta$ NMEs ($M^{0\nu}$) for several double beta decay isotopes. 
\begin{table}
\centering
\begin{tabular}{|c|c|}
\hline
isotope &  $M^{0\nu}$ \\
\hline
$^{76}$Ge   & 3.33--4.58\\
$^{82}$Se   & 2.01--4.17\\
$^{96}$Zr &   1.01--1.28            \\
$^{100}$Mo  &  2.22 --3.53 \\
$^{130}$Tl  &  2.27--3.77\\
$^{150}$Nd  & 3.14--4.04\\
\hline
\end{tabular}
\caption[$0\nu\beta\beta$ NME  ($M^{0\nu}$)  for several double beta decay isotopes, using the QRPA method]{$0\nu\beta\beta$ NME  ($M^{0\nu}$)  for several double beta decay isotopes, using the QRPA  method of calculation~\cite{rodin}.}
\label{tab-2nbb}
\end{table}

\chapter{Double beta decay experiments}
The challenging task in double beta decay experiments is to search for  the $0\nu\beta\beta$ peak  in the electron energy spectrum within the continuum of  $2\nu\beta\beta$ and natural radioactive background. This chapter presents the experimental criteria and status of double beta decay experiments.

\section{Experimental criteria}
 In order to search for $0\nu\beta\beta$ signal and minimise backgrounds several factors need to be considered in $0\nu\beta\beta$ experiments~\cite{eliot}:
\begin{itemize}
\item The double beta decay isotopes and the detector components must have high purity. This goal can be achieved via purification of double beta decay sources and constructing  smaller and more  granular detector components. 

\item To suppress cosmic ray backgrounds the  detectors are  situated underground. 
\item The isotopes with large $Q_{\beta\beta}$ are favoured as  the energy region of interest is above potential  backgrounds (more details are given in Section~\ref{sec-isotopeth}).
\item  A good energy resolution is required to prevent the tail of the $2\nu\beta\beta$ spectrum from extending into the $0\nu\beta\beta$ region of interest.  This  produces a good signal over background ratio.
\item To reject background, event reconstruction and good particle identification is required.

\item The NME is understood for some isotopes more than  others. The interpretation of limits or, in case of discovery, $0\nu\beta\beta$ signals requires measurements for a range of isotopes. 
\item Several  $0\nu\beta\beta$ modes can be studied through energy and angular distributions of the electrons in the final states.   
\end{itemize}
So far, no experiment searching for $0\nu\beta\beta$ has managed to include all the above criteria for its $0\nu\beta\beta$ searches. Section~\ref{sec-experiments} gives details of the current experimental status of double beta decay detectors.
\subsection{Choice of   isotopes }
\label{sec-isotopeth}
As mentioned in Section~\ref{sec-2nbbth}, $2\nu\beta\beta$ occurs in $36$ isotopes. However, it is not experimentally favourable (and interesting) to detect the decay in all  these isotopes. Most of these isotopes have a low $Q_{\beta\beta}$ value  which is similar to  the $Q$ values of other natural radioactive decays. As double beta decay is rare,   it  will be difficult to  detect it above background.
 Currently  there are nine  double beta decay isotopes which are experimentally considered for neutrinoless double beta decay searches: $^{48}$Ca, $^{76}$Ge,  $^{82}$Se, $^{96}$Zr, $^{100}$Mo,  $^{116}$Cd, $^{130}$Te, $^{136}$Xe and  $^{150}$Nd. Table~\ref{tab-qvalue} gives the $Q_{\beta\beta}$ value and natural abundance of each of these isotopes.
\begin{table}[htp]
\centering
\begin{tabular}{|c||c|c|}
\hline
Transition & $Q_{\beta\beta}$(keV) & Natural Abundance (\%)\\
\hline
\hline
$^{76}\rm Ge \rightarrow ^ {76}$Se &   2039      & 7.8 \\
\hline
$^{136}\rm Xe \rightarrow ^ {136}$Ba &   2479      & 8.9 \\
\hline
$^{130}\rm Te \rightarrow ^ {130}$Xe & 2533 & 34.5\\
\hline
$^{116}\rm Cd \rightarrow ^{116}$Sn & 2802 & 7.5\\
\hline
$^{82} \rm Se \rightarrow ^{82}$Kr & 2995 & 9.2\\
\hline
$^{100}\rm Mo \rightarrow ^{100}$Ru & 3034 &9.6\\
\hline
$^{96}\rm Zr \rightarrow ^{96}$Mo & 3350 & 2.8\\
\hline
$^{150} \rm Nd \rightarrow ^{150}$Sm & 3367 & 5.6\\
\hline
$^{48} \rm Ca \rightarrow ^{48}$Ti & 4271 & 0.187\\
\hline
\end{tabular}
\caption[Isotopes used for double beta decay studies in experiments]{Isotopes used for double beta decay studies in experiments. The nuclear transition energy ($Q_{\beta\beta}$) and the natural abundance of each isotope are shown~\cite{bookkai}.}
\label{tab-qvalue}
\end{table}
This thesis is about the double beta decay study of $^{150}$Nd, thus the rest of this section describes the properties of this isotope.
\subsection{Neodymium-150}
\begin{figure}
\centering
\includegraphics[width=10cm]{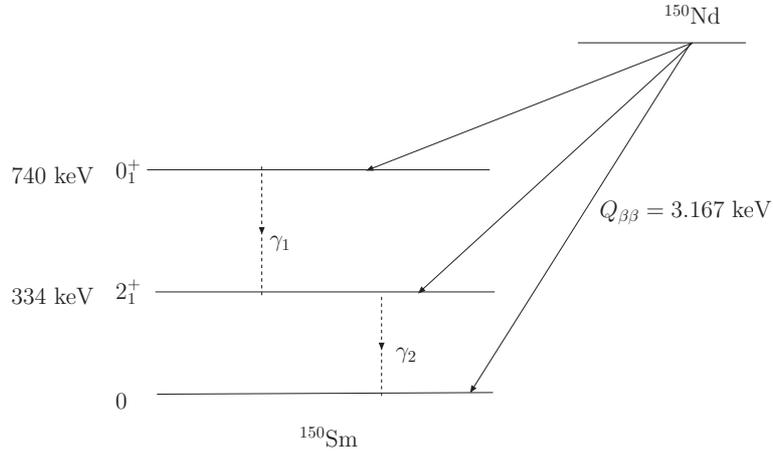}
\caption{A decay scheme of $^{150}$Nd to ground state of $^{150}$Sm.}
\label{fig-nddiagram}
\end{figure}

The double beta decay of  $^{150}$Nd to the ground state of $^{150}$Sm is accompanied by the simultaneous emission of two electrons. The  $2\nu\beta\beta$ decay to the  ground state of $^{150}$Sm  has already been observed  by several experiments. Figure~\ref{fig-nddiagram} shows a  decay  scheme of this isotope.
In addition,  $^{150}$Nd can decay to $^{150}$Sm $0^{+}_{1}$ and $2^{+}_{1}$ excited states and then de-excite by emitting photons. In the case of a decay to the  $0^{+}_{1}$ excited state, the two electrons in the final state are accompanied by two photons with energies $410$~keV and $334$~keV, and in the case of decaying to the $2^{+}_{1}$ state, electrons are emitted with one $334$~keV  photon.  The $2\nu\beta\beta$  half-life of these processes is  predicted to be greater than $10^{20}$~y~\cite{barabashnd}. 
A precise measurement of the $2\nu\beta\beta$ decay of this isotope to the ground state is important since it is  an irreducible background to $0\nu\beta\beta$ and also can be used to improve current NME calculations.

\begin{figure}[h]
\centering
\includegraphics[width=11cm]{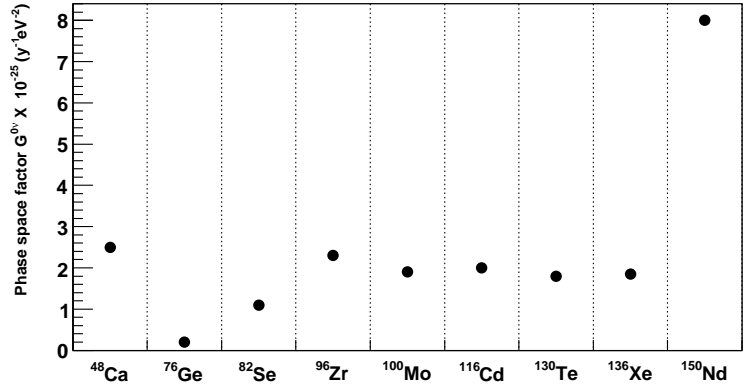}
\caption [The approximate phase space factor of several double beta decay isotopes]{The approximate phase space factor of several double beta decay isotopes~\cite{Gnend}. }
\label{fig-gnd}
\end{figure}

The Neodymium-150 ($^{150}$Nd) isotope  has the  second  highest $Q_{\beta\beta}$ value of all the double beta decay isotopes (after $^{48}$Ca). This, combined  with a high atomic number, makes the neutrinoless double beta decay  phase space factor $G^{0\nu}(Q_{\beta\beta},Z)$) the highest of all isotopes. Figure~\ref{fig-gnd} shows approximate   $G^{0\nu}(Q_{\beta\beta},Z)$ values for several double beta decay isotopes.
  From Equation~\ref{eq-0nbbhalflife} it is shown that the half-life of $0\nu\beta\beta$ process is inversely related to    $G^{0\nu}(Q_{\beta\beta},Z)$, thus the event rate for $^{150}$Nd is expected to be higher than all other isotopes, assuming identical NMEs. This  feature makes $^{150}$Nd a strong candidate for next generation  double beta decay experiments such as SNO+~\cite{snop} and SuperNEMO~\cite{supernemo} (Section~\ref{sec-experiments})\@.
~Currently, the only drawback of using $^{150}$Nd for $0\nu\beta\beta$ searches is the high uncertainty on the NME calculation  due to its mass and deformation~\cite{nddeform}. 
\section{Experimental status}
\label{sec-experiments}
The observation of neutrinoless double beta decay can answer important questions in neutrino physics. In the past ten years several experiments have been therefore constructed and    many more are currently in the  R\&D stage.
~The double beta detectors are divided into two main categories: homogeneous   (source equal to detector) experiments in which  the double beta decay source is part of the active detector and  heterogeneous  (source not equal to detector) experiments in which  the source  is separated from the detector. The  homogeneous detection technique gives  good energy  resolution.
~The  heterogeneous experiments contain  tracking detectors which give them a good particle identification.
\subsection{Experiments following the homogeneous  system }
\begin{figure}
\centering
\includegraphics[width=9.5cm]{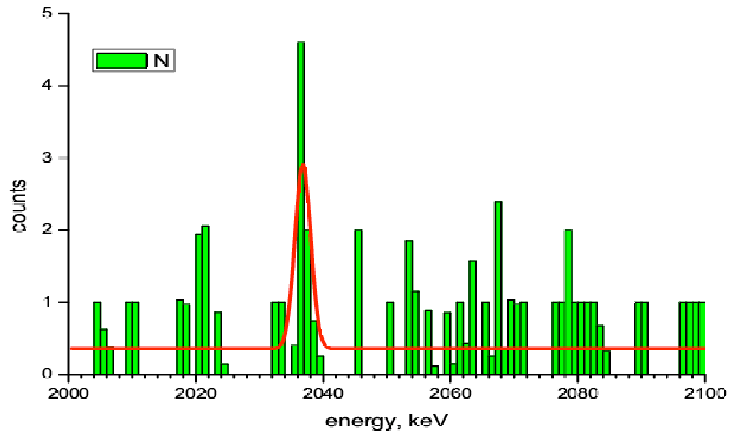}
\caption [Energy sum of the two electrons in $0\nu\beta\beta$ region for $^{76}$Ge  isotope of Heidelberg-Moscow experiment]{Energy sum of the two electrons in $0\nu\beta\beta$ region for $^{76}$Ge  isotope of Heidelberg-Moscow experiment.~A  peak is observed around the  $Q_{\beta\beta}$ value of  $^{76}$Ge~($2.039$~MeV)~\cite{moscowclaim}.}
\label{fig-moscow}
\end{figure}

Semiconductor germanium (Ge) detectors  are among the most popular double beta decay experiments. This is because  of  the cheap enrichment process for $^{76}$Ge. The  NME of this isotope is also relatively  well known.
There have been   two previous semiconductor $^{76}$Ge detectors  which   produced important $0\nu\beta\beta$ search results: Heidelberg-Moscow and IGEX\@.
 Heidelberg-Moscow was  a semi-conductor germanium detector, enriched to $86\%$ in $^{76}$Ge. The final design of the detector consisted  of five individual sub-detectors  with a total mass of $11.5$~kg. The experiment ran between $1990$ and $2003$.  
A claim for discovery of $0\nu\beta\beta$ was made in $2001$ by a sub-group of the Heidelberg-Moscow collaboration~\cite{moscowclaim}. The half-life value of $0\nu\beta\beta$ was obtained to be $T_{1/2}^{0\nu}=1.19\pm^{2.99}_{0.50}\times 10^{25}$~years, with $\langle m_{\nu}\rangle=0.2-0.6$~eV. Figure~\ref{fig-moscow} shows the result  of this experiment. However, this result has received  criticism~\cite{nodiscovery}. It is believed that the background and the systematic uncertainties of this experiment are underestimated.
The IGEX experiment used a similar experimental technique to the Heidelberg-Moscow experiment and produced a limit on the half-life  of $T_{1/2}^{0\nu}>1.57\times 10^{25}$~years  at $90\%$ confidence level (CL)~\cite{igexlimit}.

To confirm or disprove the  Heidelberg-Moscow results, two experiments are being designed to study the region highlighted by the Heidelberg-Moscow experiment: GERDA~\cite{gerda} and MAJORANA~\cite{igexlimit}.  Both of these detectors will use Ge enriched to $86\%$ $^{76}$Ge. In the first phase  of its running, GERDA will utilize the existing Heidelberg-Moscow  and IGEX experiments (18~kg of $^{76}$Ge) and will reach a  half-life sensitivity of  $3\times 10^{25}$~years.  This sensitivity can rule out or confirm the Heidelberg-Moscow results. In the second phase of its running another $38$~kg will be added, giving a half-life sensitivity of $\sim$ $1.4\times 10^{26}$ years, which corresponds to  $\langle m_{\nu}\rangle$~$\sim 124$~meV. MAJORANA's eventual half-life sensitivity is hoped to be $5.5\times 10^{26}$~y, corresponding to  $\langle m_{\nu}\rangle~\sim$ 61~meV. The final design of this experiment is still under development.


One running source equal detector experiment is CUORICINO~\cite{courichino}. This experiment uses  bolometers containing double beta decay isotopes, running at extremely low temperatures (order of mK). When a double beta decay event occurs, the electrons of the decay increase the temperature by depositing energy inside a crystal. By measuring the temperature rise the deposited energy is measured. 
Each bolometer is $5\times 5\times 5$~cm$^{3}$ and is fixed into a layered tower structure. The total mass of the crystals is $41$~kg, corresponding to 11~kg of $^{130}$Te. This detector has set a limit on the half-life of neutrinoless double beta decay of $^{130}$Te to be more than $3.0\times 10^{24}$~years, corresponding to $\langle m_{\nu}\rangle <0.42-0.58$~eV~\cite{courichino}.
CUORE~\cite{cuore}, which is the next generation of the CUORICINO experiment, will  have $760$~kg of TeO$_{2}$ crystals with a total of 206~kg $^{130}$Te. It will consist of $35$ towers which have the same size as the CUORICINO tower. It is  expected to reach a half-life sensitivity of $2.5~\times 10^{26}$ years which corresponds to a mass sensitivity of  $\langle m_{\nu}\rangle\sim45-53$~meV, depending on NME.


The COBRA~\cite{cobra} experiment is in the R\&D stage and  is going to use  CdZnTe semiconductor  crystals. CdZnTe contains nine double beta isotopes, with $^{130}$Te and $^{116}$Cd having  the highest abundance in the detector. This experiment is also searching for   $0\nu\beta^{+}\beta^{+}$, $0\nu\beta^{+}$~EC  and $0\nu$EC~EC  processes by studying  $^{106}$Cd, $^{64}$Zn, $^{108}$Cd and $^{120}$Te decays. The finished experiment will hold 64000, $1$~cm$^{3}$ CdZnTe detectors. The  detector will have  $418$~kg mass, in which 183~kg will be Cd enriched to 90\% in $^{116}$Cd. The half-life sensitivity of this experiment will be greater than $10^{26}$~years, corresponding to  $\langle m_{\nu}\rangle\sim38-96$~meV.

The SNO+ experiment  is planning to use most of the SNO experiment infrastructure and shielding (SNO was a successful experiment that confirmed neutrino oscillations in solar neutrinos), replacing the heavy water  with $1000$ tones of Nd-loaded liquid scintillator. A $1\%$ Nd-loading would  correspond to 560~kg of natural Nd~\cite{kaizubersno} or 56~kg of $^{150}$Nd. It is also a possibility that this experiment could load the scintillators with enriched $^{150}$Nd. 
 This experiment  is multi-purpose and apart from double beta decay studies is going to focus on  low energy solar neutrinos as well as geo-neutrinos, reactor and supernova neutrinos~\cite{kaizubersno}.

\subsection{Experiments following the heterogeneous system}
These type of experiments are divided into two categories: the time projection chambers (TPCs), where the double beta decay isotope is a filling gas or liquid, and tracking-calorimeter experiments, where the double beta decay source is a solid foil.  The  EXO~\cite{exo} experiment is a TPC experiment.  Its goal is to use between 1 to 10 tons of Xe enriched to 80\% in $^{136}$Xe. 
 The experiment is  developing a high resolution liquid xenon TPC with good tracking capability.~A high voltage cathode will be located in the middle of the TPC volume.~At each end there will be $114$ horizontal and vertical anode wires. These wires will be for charge collection, allowing energy measurement  and particle trajectory reconstruction. The liquid xenon is also an efficient scintillator. The use of both collected charge and scintillator light improves the energy  resolution. The experiment will have two phases; the first phase is called EXO-200 (200~kg of  $^{136}$Xe is used) and is expected to reach a mass sensitivity of 30~meV. In the second phase the EXO detector is expected to trap and identify the 
$^{136}$Ba ions, the daughter isotope of $^{136}$Xe, by laser spectroscopy. These ions can be identified via atomic spectroscopy by optical pumping with blue and red lasers~\cite{baexo}. The radioactive background events can be   significantly reduced by Ba tagging.   

NEMO~3~\cite{technical} follows the tracker-calorimeter technique and  is currently the only running experiment of this type. More details about this detector are given in Chapter~\ref{chap-detector}.  The SuperNEMO detector is a next generation double beta decay experiment which is going to use the same experimental approach as NEMO~3 but will improve the sensitivity to neutrino mass scale by one order of magnitude ($\langle m_{\nu}\rangle<45-85$~meV). This experiment will comprise $20$ individual modules, each of which has a  rectangular shape with  the source foil in the centre surrounded by the tracker and the calorimeter. This experiment will use $100$~kg--$200$~kg  $^{150}$Nd and/or $^{82}$Se isotope. 
\begin{table}
\centering
\footnotesize
\begin{tabular}{|c|c|c|c|c|}
\hline
Experiment & Isotope & mass (kg) & Expected $\langle m_{\nu}\rangle$(meV) &Time scale\\
\hline
GERDA  Phase~1     &    $^{76}$Ge  &  $18$ &  $<$250-440 & 2010\\
GERDA  Phase~2     &    $^{76}$Ge  &   $56$ &  $<$124 & 2011\\            
CUORICINO & $^{130}$Te & $11$ &  $<$420--580 & current limit \\
CUORE     &  $^{130}$Te & $206$ & $<$45--53 & 2011--2016 \\
SNO+      & $^{150}$Nd &  56 or 560 & under study    & 2012 \\ 
COBRA    & $^{116}$Cd  &   $164$      & $<$38--96   & -- \\
EXO-200       &  $^{136}$Xe &  200     &    $<$30     &   2009 \\
NEMO~3      &  $^{100}$Mo &    7        &       $<$ 630--110         &       current limit     \\ 
SuperNEMO     & $^{150}$Nd/$^{82}$Se & 100--200 &  $<$45--85    & 2012 \\  
MOON     &       $^{100}$Mo & 1000 &    $<$50     &    $>2011$        \\
DCBA phase~1   & natural Nd  & 600 &  $<$120   &     $>2011$        \\  
\hline 
\end{tabular}
\caption{Neutrinoless double beta decay experiments, the  mass of the isotopes they use, their expected sensitivity and their running time scale}
\label{tab-experiments}
\end{table}

There are also two other experiments which are planning to use  heterogeneous techniques: MOON~\cite{moon} and  DCBA~\cite{dcba}. MOON will  use one tonne of $^{100}$Mo and hopes to achieve an effective neutrino mass sensitivity of $\langle m_{\nu}\rangle<50$~meV. DCBA will use natural $^{150}$Nd in its first phase of data taking and then enriched $^{150}$Nd in the second phase.  They hope to reach a sensitivity of $\langle m_{\nu}\rangle <0.12$~eV. Table~\ref{tab-experiments} summarises the current and future double beta decay experiments, the  isotopes they use for $0\nu\beta\beta$ searches, the isotope masses, their expected effective neutrino mass sensitivity and their running time scale.

%% file: detector.tex
\renewcommand{\baselinestretch}{1.6}
\normalsize
\chapter{The NEMO~3 detector}
\label{chap-detector}
The Neutrino Ettore Majorana Observatory 3 (NEMO~3) experiment has been running
in the Fr\'ejus Underground Laboratory (Laboratoire Souterrain de Modane) in France since February 2003. The main objective of the experiment is to search for evidence of neutrinoless double beta decay in  a variety of  isotopes.

The NEMO~3 detector is a  heterogeneous system, in which the radioactive sources do not make up part of the active detector. Particle tracking and energy measurements are also performed independently, giving good particle identification and understanding of radioactive backgrounds. 

\section{General description of the NEMO~3 detector}
NEMO~3 is cylindrical in design and is segmented into 20 sectors. The sectors are numbered  from  0 to  19. Each sector contains tracking chambers and calorimeters in order to directly detect decay particles emanating from the radioactive source foils. The source foils are positioned in the centre of each sector and divide the sectors into two parts: the inner part, which is from the source foil to internal wall of each sector; and the outer part, which is from the source foil to external wall of each sector. The whole detector is surrounded by a solenoid providing a magnetic field of 25~Gauss, external neutron shielding and an anti-radon tent. 

\begin{figure}[h]
\centering
\includegraphics[width=3.5cm]{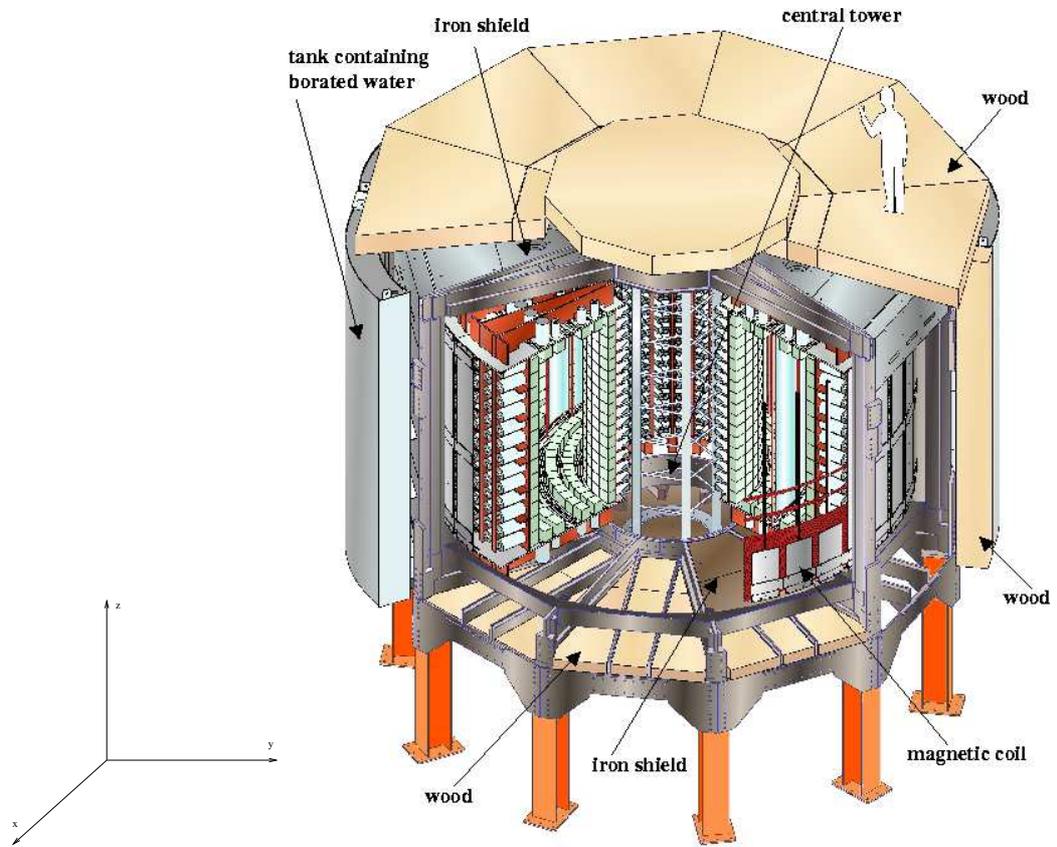}
\includegraphics[width=10.0cm]{pictures/NEMO3vcolor.epsi}
\caption[A cutaway view of the NEMO~3 detector]{A cutaway view of the NEMO~3 detector~\cite{technical}. The magnetic coil, iron $\gamma$-ray shield, and neutron shielding are shown.}
\label{fig:general_layout}
\end{figure}

\begin{figure}[tb]
\centering
\includegraphics[width=11.0cm]{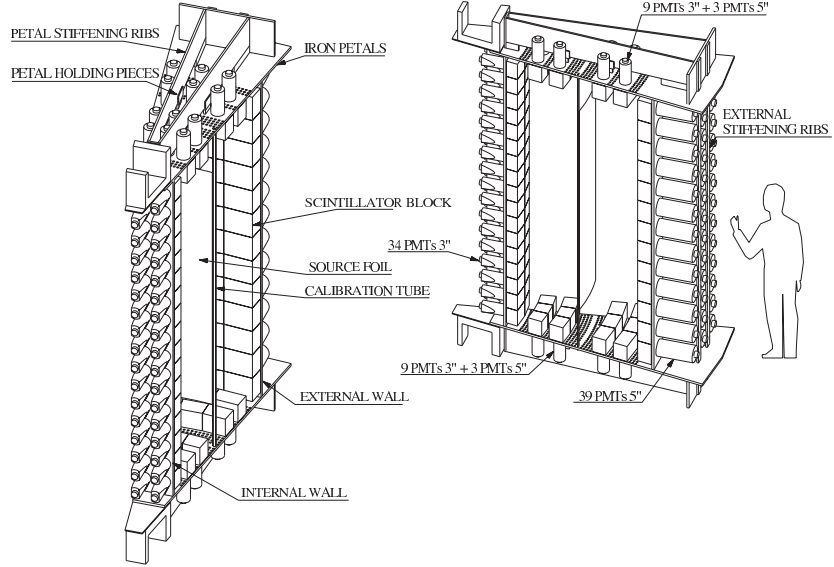}
\caption[A sector of NEMO~3 viewed from two different angles]{A sector of NEMO~3 viewed from two different angles~\cite{technical}. The source foil, scintillator blocks and photomultipliers are shown. The tracker cells are located between the internal and external walls on both sides of the foil. The tracker wires, which run vertically,  are not shown for clarity.   }
\label{f-onefoil}
\end{figure}

The NEMO~3 detector can be defined in both right-handed Cartesian and cylindrical coordinate systems.  In the Cartesian system, the $x$ axis  starts from the centre of the detector and is  along the edge of sector~0. The $y$ axis starts  from the centre of the detector and is along the edge of sector~5. The $z$ axis
starts  from the centre of the detector and points vertically upward. By performing notational conversion, the radius ($R=\sqrt{x^{2}+y^{2}}$), polar angle  ($\phi=\arccos(\frac{x}{R})$)  and $z$ of the cylindrical coordinates are  formed.
 Figures~\ref{fig:general_layout} and \ref{f-onefoil} show a cutaway view of the NEMO~3 detector and a  view of one of the sectors of the  detector, respectively. 

\section {The NEMO~3 sources}
Unlike all other currently running double beta decay experiments where  calorimeters serve as both  double beta decay source and detector, the sources in NEMO~3 are independent from the detector. Inside  NEMO~3 there are seven double beta decay isotopes, thus permitting study and comparison of results from different isotopes.
 Each sector contains seven foils mounted vertically at a radius of 155~cm from the centre within a metallic support frame. This structure allows multiple isotopes to be mounted within one sector.  Each source foil has a height of 2.5~m, a  width of 65~mm and an area density of 30--60~mg/cm$^2$. 
 
The choice of source nuclei for NEMO~3 was  based on  several
factors:  the double beta decay transition energy, $Q_{\beta\beta}$, the phase space factors, $G^{0\nu}$ and $G^{2\nu}$, corresponding to neutrinoless decay and two-neutrino decay; 
the nuclear matrix elements  ($M^{0\nu}$ and $M^{2\nu}$);  the background in the energy region around the $Q_{\beta\beta}$ value;
and  the natural  abundance and enrichment possibilities of the isotope.

\begin{figure}[tb]
\centering
\includegraphics[width=9.0cm]{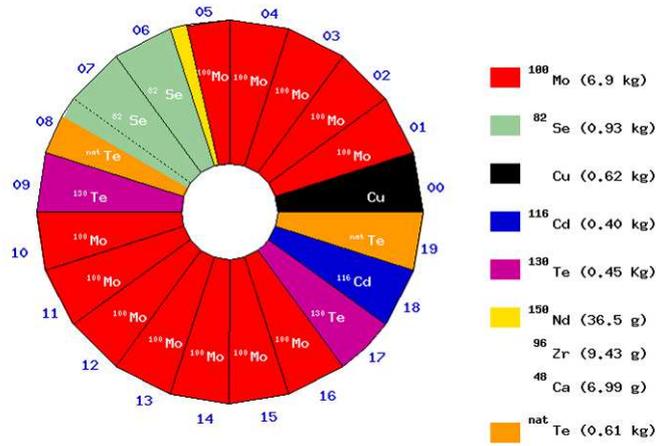}
\caption{The source foil distribution in the 20 sectors of NEMO~3.}
\label{fig-source_foil}
\end{figure}

Nine isotopes in various quantities were chosen for NEMO~3 
and these are placed in the order and quantities displayed in Figure~\ref{fig-source_foil}. 
Double beta decay occurs in seven of them: 
 $^{116}$Cd, $^{82}$Se, $^{100}$Mo, $^{96}$Zr,  $^{150}$Nd,
 $^{48}$Ca and $^{130}$Te.
The two other isotopes, Cu and natural Te,  have  a negligible impurity and are mainly used for measuring the background from external sources. The $^{100}$Mo and $^{82}$Se foils have the highest mass in the experiment, thus much effort has been focused by the NEMO collaboration into neutrinoless double beta decay searches of these isotopes~\cite{mo-ground,mo-excited,momajoron}.
\subsubsection{The neodymium  source foil}
\begin{figure}
\centering
\includegraphics [width=4.5cm]{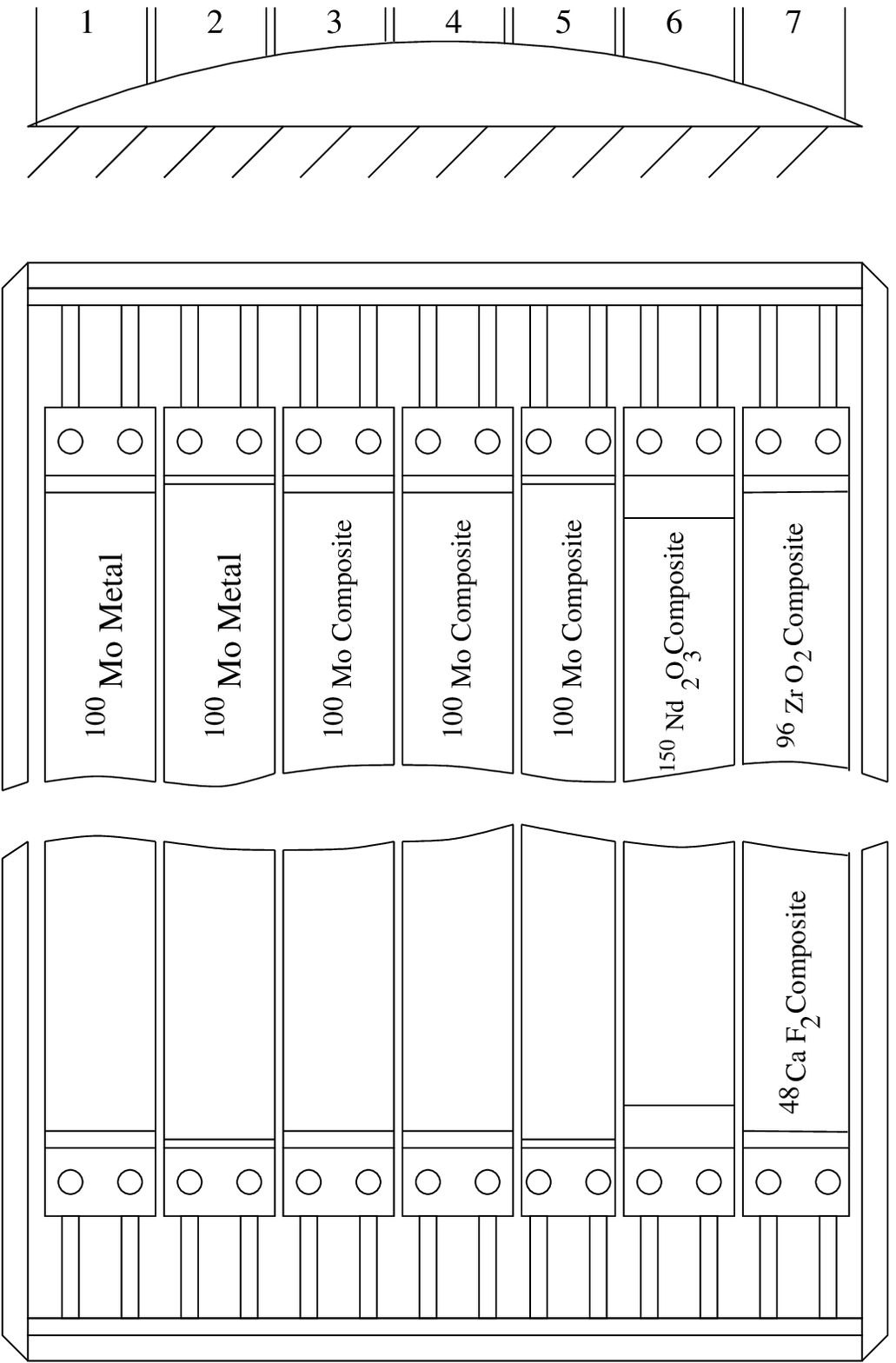}
\caption[A view of the sector~5 of the NEMO~3]{A view of the sector~5 of the NEMO~3. The $^{150}$Nd foil is shown with $^{100}$Mo foil on its left and $^{48}$Ca and  $^{96}$Zr on its right. }
\label{fig-sec5}
\end{figure}

The neodymium-150 composite foil  (Nd$_{2}$O$_{3}$) was enriched (95.0 $\pm$0.5\% of isotope 150) by the Institute for Nuclear Research of Moscow. It  is situated in foil~6 of sector~5 (Figure~\ref{fig-sec5}). The length of the active foil is 234~cm and the   width is $6.5$~cm.
 The total mass of the $^{150}${Nd} foil is 57~g,
 in which $36.5~\pm 0.1$~g  is  $^{150}$Nd,  6.458~g is a Mylar support film and the remainder is composed of foil impurities and polyvinyl alcohol (PVA\@).
The impurities inside the foil were  measured using a high purity germanium (HPGe) detector to be  $^{234m}$Pa, $^{207}$Bi,  $^{154}$Eu, $^{152}$Eu,$^{214}$Bi, $^{214}$Pb, $^{208}$Tl and $^{40}$K\@~\cite{technical}. 
\section{The tracking detector}
\begin{figure}[tb]
\centering
\includegraphics[width=10.0cm]{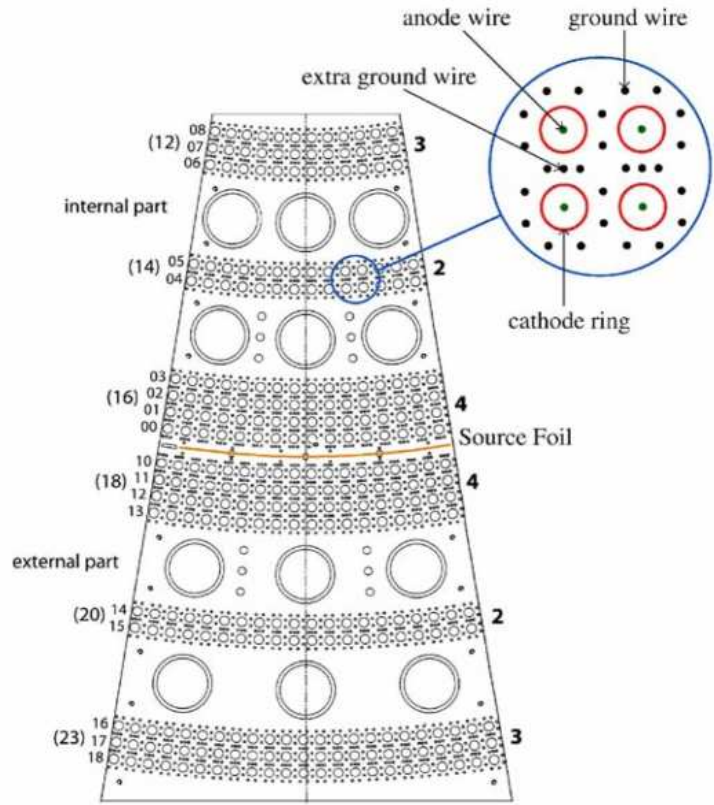}
\caption[A top view of a tracking sector of  NEMO~3]{A top view of a tracking sector of  NEMO~3. The 4-2-3 drift cell layer configurations are shown in both the external and internal parts~\cite{technical}.}
\label{fig-geigertubes}
\end{figure}

The tracker consists of $6180$ octagonal drift cells operating in Geiger mode \cite{geiger}. Each of the 20 sectors  has an internal and external tracking volume, each containing  drift cells  arranged in a 4-2-3 layer configuration as
 shown in Figure~\ref{fig-geigertubes}. 
This gives a total of nine drift cell layers on
 each side of the source foil to reconstruct the particle tracks.

\begin{figure}[tb]
\centering
\includegraphics[width=11.0cm]{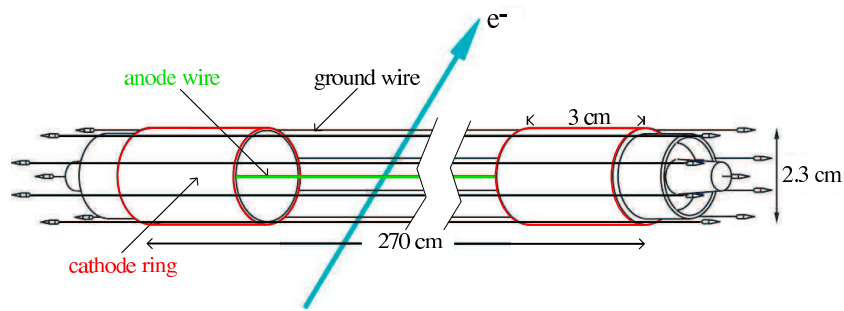}
\caption[A drift cell from NEMO~3]{A drift cell from NEMO~3~\cite{technical}. The anode wire, cathode rings and the ground wires are illustrated.}
\label{fig-onecell}
\end{figure}

Each  cell  has a diameter of 3~cm and a  length of 270~cm,  and contains a central  anode wire surrounded by eight ground wires. The layers of drift cells are separated by an extra ground wire in order to reduce electrostatic cross-talk.
All wires are composed of stainless steel and have a diameter of 50~$\mu$m. The tracking cells are  strung between top and bottom walls. 
At the top and bottom of the wires there are 3~cm long copper cathode rings with a diameter of 2.3~cm. Figure~\ref{fig-onecell}  shows a diagram of a drift cell in NEMO~3.

\subsubsection{ Drift  cell  operation in NEMO~3}
The tracking system is immersed in a gas mixture of helium (95\%), argon (1\%),  ethyl alcohol (4\%) and water (1500~ppm). The Geiger mode operating voltage of the cells is around 1620~V\@. When a charged particle passes through a cell, it causes ionisation of the gas atoms. The result is a track of He$^{+}$ ions and free electrons. 
These electrons drift towards the anode wire and cause further ionization, triggering an avalanche process as the electrons accelerate. 
The avalanche process becomes significant in the region very 
close to the wire (around 1~mm), where the
electric field is high. When the avalanche arrives on the anode, it causes a pulse which  has a typical rise time of 10~ns.  The time to digital convertor (TDC) connected to  the anode gives  the arrival time of the initial avalanche to the anode and thus the transverse position of the particle track. 

The helium ions produced by the charged particle and the resulting avalanche create UV photons which travel further along the cell. These cause further ionisation in the drift cell and  new avalanches. Under the high voltage environment of the drift cell this process is smooth and propagates at a constant velocity toward both ends of the cell. Upon reaching the ends, the  ions are collected by the copper cathode rings.  The arrival time of the plasma at each cathode, which is measured by the TDC of each cathode, is used to determine the vertical position of the particle track passing through the cell.


Helium is a low atomic number gas, thus it minimises the energy loss by a charged particle passing through the gas. 
Ethyl alcohol plays the role of the quencher,  which neutralises helium ions and limits the creation of the UV photons and therefore avalanches. A small amount of  argon~(1\%) and water (1500~ppm) were added later during the commissioning of the detector in order to increase the plasma propagation efficiency and  to reduce noise and  improve the cell stability~\cite{water}.



\section{The calorimeter}
\label{sec-cal}
\begin{figure}
\centering
\includegraphics[width=12.0cm]{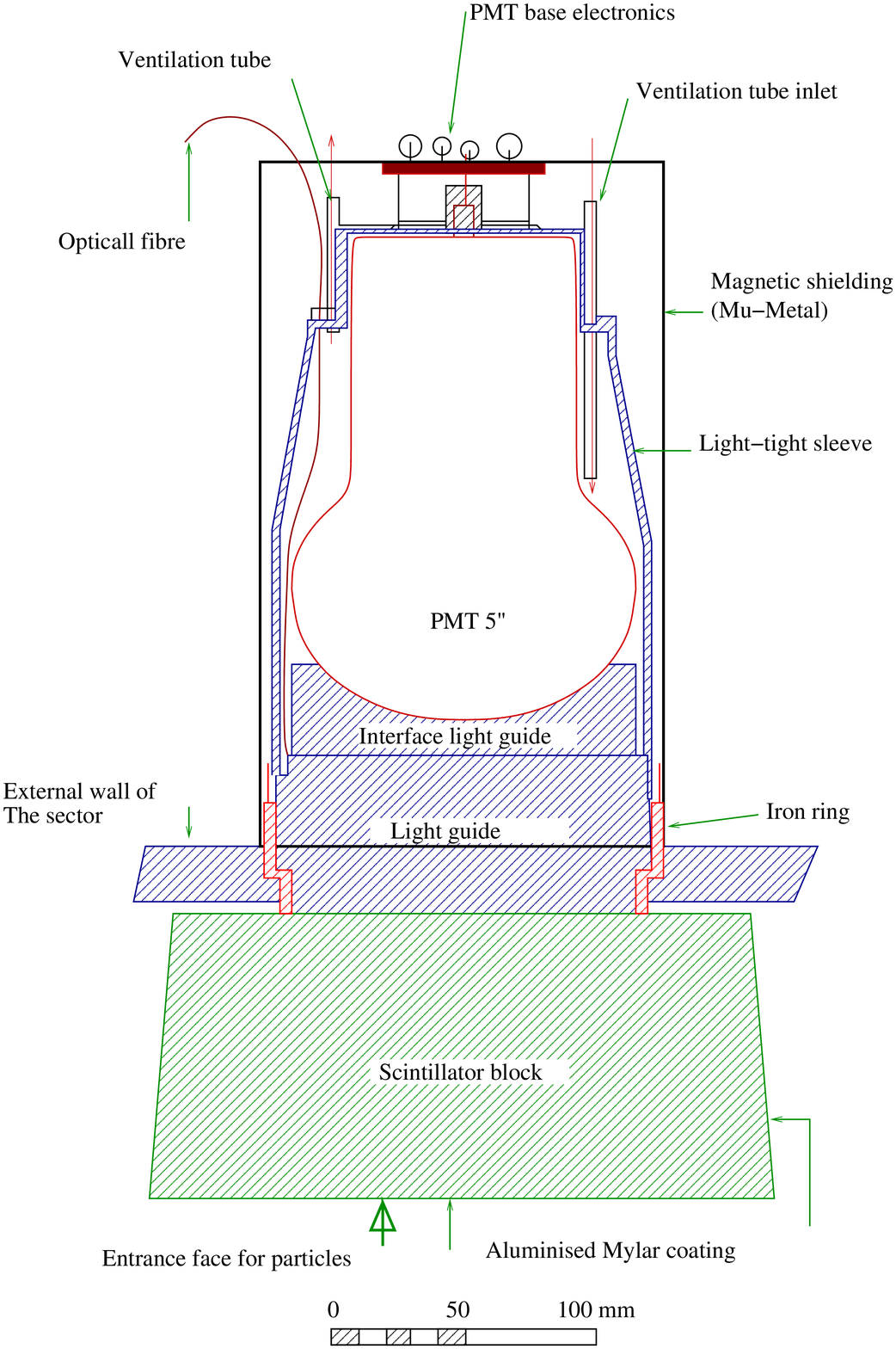}
\caption[A schematic view of a scintillator counter with a  light guide and 5'' PMT  in the NEMO~3 calorimeter]{A schematic view of a scintillator counter with a  light guide and 5'' PMT  in the NEMO~3 calorimeter~\cite{technical}.} 
\label{fig-onepmt}
\end{figure}

The calorimeter in NEMO~3 measures the 
 energy and the time of flight (TOF) of particles. It consists of 1940 large
 scintillator blocks coupled to 3'' and 5''  photomultiplier tubes (PMTs) via light guides.
The scintillator blocks are produced from low radioactive polystyrene which also provides minimum back scattering. The blocks cover the inner and outer cylindrical walls that surround the tracking volume and the top and bottom of each sector. In order to minimise energy loss, the scintillator blocks are mounted inside the helium-alcohol gas mixture of the tracking volume.  Figure~\ref{fig-onepmt} shows a schematic view of one of the scintillation detectors with a 5'' PMT\@. The energy resolution (full width half maximum, FWHM), $\sigma_{E}/E$, of the NEMO~3 calorimeter is 14\% at 1~MeV and 8\% at 3~MeV,  and the time resolution is 250~ps. 
In addition to 1940 PMTs in the calorimeter, there are six reference PMTs situated outside the detector that  are only used during laser calibrations (Section~\ref{sec-laser}). 

\subsection{The scintillators and the light guides}
The scintillators of the calorimeter are made in seven different shapes designed to completely cover the cylindrical  geometry of NEMO~3. The thickness of the scintillators  is approximately~10~cm throughout. The width and length are between 11~cm and 15~cm.  Charged particles lose energy rapidly when passing through polystyrene via  molecule excitation. As the molecules de-excite, photons are produced  at a wavelength transparent to polystyrene. Photons incident on the scintillator lose energy via Compton scattering and  the resulting electrons then lose energy as described previously. The scintillation detectors have a detection efficiency of 50\% at 500~keV.

In order to transmit scintillation photons to the PMTs, polymethyl methacrylate (PMMA) light guides are optically glued to the back face of the scintillator blocks. The light guides are 60~mm thick and are glued to the detector walls to provide a seal against the helium-alcohol environment of the tracker.


\subsection{The photomultiplier tubes}
The photomultiplier tubes (PMTs) convert  optical signals from scintillators to measurable electrical signals. They achieve this via the photoelectric effect and avalanche multiplication.  
There are two sizes of PMTs used in the calorimeter, 3'' and 5''. The 3'' PMTs are mounted on scintillators at the top and bottom regions of each sector, and the 5'' PMTs are mounted on scintillators around the inner and outer cylinder walls.  The PMTs used have been specially produced to have three times lower radioactive contamination than a standard PMT\@. The gain of the PMTs is adjusted to cover a range of energies up to 12~MeV\@.
 In order to protect them from the external light the PMTs are contained in black plastic boxes.
The output signals from PMTs pass through discriminators. These produce logic high signals when the PMT pulse passes a minimum threshold.
The discriminator signals  are used  to start the  time measurement and charge integration which are performed by time to digital convertors (TDCs) and  analogue to digital convertors (ADCs), respectively.

%


\section{The Trigger and readout}
\label{sec-trigger}
The trigger  ensures that  only events of interest are read out. As NEMO~3 is a low count rate experiment a simple hardware trigger is used to achieve this. 
The  trigger is connected to the calorimeter electronics. When it receives a signal from a PMT which has passed a discriminator threshold of 48~mV, it generates the STOP-PMT signal after $20$~ns. This signal stops the charge integration and time measurements performed by an ADC and a TDC, respectively.

In the next stage, the tracking readout system is programmed to search for  activation of drift cells in each half-sector (inner or outer). If enough cells are activated (3 out of 9 layers), the STOP-A signal is generated and sent to the drift cell acquisition boards with a programmable delay set at $6.14$~$\mu$s after the STOP-PMT\@. The STOP-A signal stops the TDC measurement of the drift cell anodes (TDC$_{A}$) and the anode drift times are calculated.

 Finally, the  STOP-$\alpha$ signal is sent to the drift cell acquisition boards with a fixed delay of 710~$\mu$s after the STOP-PMT signal. This signal is to stop  the TDC$_{\alpha}$, which is a TDC independent of TDC$_{A}$, and is used for calculation of the drift time for delayed hits. This is designed  to detect alpha ($\alpha$) particles  from daughter isotopes of radon  and  have a half-life of 164~$\mu$s. The value of 710~$\mu$s is more than four times  this half-life to have high detection efficiency for $\alpha$--particles.   


\section{Magnetic coil and  shielding}
 A solenoid surrounding the entire detector produces a 25 Gauss magnetic
field vertically through the detector. The coil is cylindrical  and is 5220~mm in diameter and 2713~mm in height. This causes charged particle tracks to bend as they pass through the detector allowing the identification of electron and positron tracks.


Iron plates, wood panelling, borated water and an anti-radon tent form 
the external shielding for the detector. The iron plates are 20~cm thick and use low radioactivity iron  to stop photons  coming from outside  the detector. The  water and wood comprise a  thick shield which slows down fast neutrons and captures them. The wood covers the top and bottom of the detector and is  28~cm thick. The borated water tanks, which  are 35~cm thick, cover the cylindrical external walls of the  detector.

\section{The anti-radon facility}
After running the detector for approximately one year, it was discovered  that  the radon level inside the detector was too high (with a total activity of around 0.7~Bq) and  adversely affecting the experiment's sensitivity. This was  caused by radon diffusion through the  glued joints between sectors.  In order to reduce the radon contamination inside the detector  an anti-radon facility was installed in September 2004. It consists of  an airtight tent and a radon trapping facility. 

The detector is fully enclosed in an airtight tent that is  made of two  layers of polyethylene. It allows only radon-free air to pass  through from the radon-trapping facility into the detector and isolates it from laboratory air. The radon trapping system  uses  activated charcoal that has been treated with oxygen to open up  numerous  pores inside the material. When air  passes through the charcoal, radon is trapped in these pores. The trapping time is greater than the radon decay time.  Therefore radon decays before  reaching the tent.  

\begin{figure}
\centering
\includegraphics[width=10.0cm]{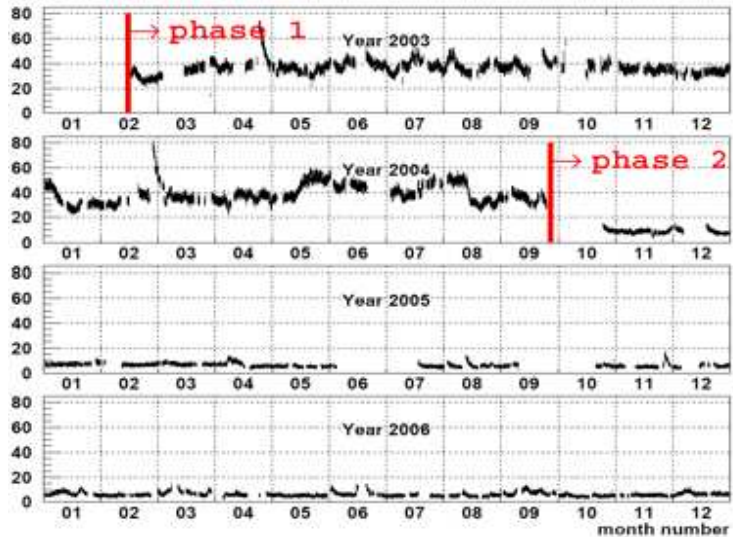}
\caption[Evolution of the total radon activity in the NEMO~3 before~(Phase~1) and after~(Phase~2) the installation of the anti-radon facility]{Evolution of the total radon activity in the NEMO~3 before~(Phase~1) and after~(Phase~2) the installation of the anti-radon facility~\cite{externalbkg}.}
\label{fig:evolutionrn}
\end{figure}

As a result of this system, the radon activity inside the detector has been reduced by a factor of six. Figure~\ref{fig:evolutionrn} shows the level of  radon inside NEMO~3 before and after the installation of the anti-radon system. 

\section{Calibration of the calorimeters}
\label{sec-laser}
In order to convert   the charge output of the PMTs to  energy, a calibration procedure is performed. In NEMO~3 there are dedicated calibration runs which take place once per month and last 24 hours. During this time radioactive sources with well known energy spectra are introduced near the double beta decay sources.
As NEMO~3  is mainly focused on electron
energy measurements, the selected  radioactive sources  emit electrons. 
The $^{207}$Bi source decays by a  conversion process to two electrons~\cite{bi207decay} (see Section~\ref{sec-eg})  with energies of 482~keV and 976~keV, and is suitable for energy calibration up to 1.5~MeV\@.  In order to calibrate the energy up to 3~MeV and higher, $^{90}$Y, which is a pure beta emitter with transition energy  of 2.283~MeV, is used. 

The timing responses of the PMTs are not identical. The time calibration is performed  to make the timing response uniform for all the PMTs in the calorimeter. 
 Currently,  electrons and photons from $^{207}$Bi sources are used for the time calibration.   The particle with the higher energy decays with a delay of 133~$\mu$s. This time is subtracted from the arrival time of the delayed signal. The time difference between the prompt and delayed signals is then used to make the time response uniform.  

Daily studies of the stability of the PMTs (in time and energy measurements) are performed using a laser survey system.
The purpose of this system is to check  the absolute energy and time calibration. It also measures the PMT response linearity between 0 and 12~MeV\@. Laser light with a known  intensity is sent to the PMTs of the calorimeter via optical fibres.
Six reference PMTs  are connected to scintillators with embedded $^{207}$Bi sources and  also to the laser survey system via optical fibres.  They are used  to check the stability of the laser light intensity   which is received by the PMTs in the calorimeter. The laser energy calibration is not applied to the data used in this thesis, as the laser data analysis was not   ready for use with the analysis presented.  The effect of not using this correction has been studied and is included as systematic uncertainty~(Sections~\ref{sec-systematic2n2b}).
\subsubsection{The laser time correction}
\label{sec-ltc}
\begin{figure}
\centering
\includegraphics[width=12.5cm]{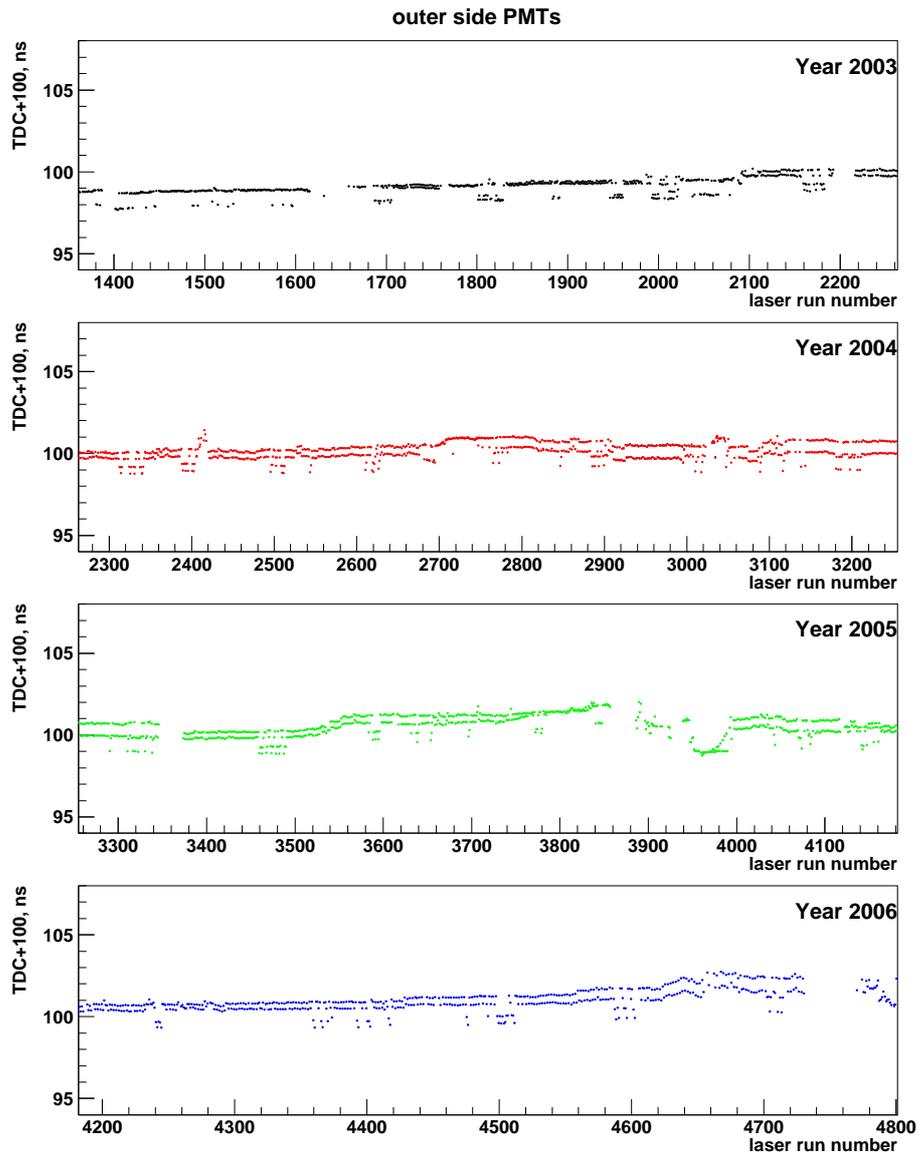}
\caption[The mean  TDC values for PMTs in outer part of the sector 5 versus the laser run number for years 2003, 2004, 2005 and 2006]{The mean  TDC values for PMTs in outer part of the sector 5 versus the laser run number for years 2003, 2004, 2005 and 2006~\cite{vera1}. The TDC jumps are seen from the end of 2003 onwards.}
\label{fig-ltc}
\end{figure}

During the analysis of NEMO~3 data  in September 2006, it was discovered that the timing measurements of the PMTs were not correct. The time measured by the TDC counter of a given PMT was expected to  remain constant for different laser runs; however, a discontinuity was observed in the distribution of the TDC time~(Figure~\ref{fig-ltc}). This effect was seen in the PMTs of all the 20 sectors. To solve the timing  problem a correction was introduced known as laser time correction (LTC)~\cite{vera}.  The measured TDC for each PMT is compared to the reference run as follows:
\begin{equation}
{\rm LTC}^{i}_{j}=tdc^{i}_{j}-tdc^{i}_{1404}
\end{equation}
where $tdc^{i}_{j}$ is the time measured by the PMT $i$ in the laser run $j$, and $1404$   is the reference laser run. In this run  the  maximum number of PMTs were active and  they were known to be relatively stable and the timing measurements performed by the TDCs were known to be correct. The value of LTC is added to the TDC values of the PMTs and calculated in each laser run to correct the deviation in timing.




 


%% file: technical.tex
\renewcommand{\baselinestretch}{1.6}
\fancyhead[RO]{\rightmark\hspace{1cm}\thepage}
\fancyhead[LE]{\thepage\hspace{1cm}\leftmark}
\fancyhead[RE,LO]{}
\fancyhead[CE,CO]{}
\normalsize
 \chapter{Analysis technique}
\label{chap-technique}
This chapter describes techniques used  to prepare data for analysis, including the  reconstruction of raw data and Monte Carlo (MC)  simulation. The $^{150}$Nd foil geometry is defined in Section~\ref{sec-ndg}. Section~\ref{sec-tof} describes the time of flight (TOF) selection  that is used in all analysis channels. The statistical methods used to compare simulated events to data and set limits on several $ 0\nu\beta\beta$  processes  are described in Section~\ref{sec-fit} and Section~\ref{sec-colie}, respectively.
\section{Reconstruction of particle tracks}
The raw  hits in the tracker that are written to data storage tapes are  processed by reconstruction software. 
  According to the tracker anode and cathode time values, the triggered drift cell hits  are classified as ``in time hits'',  ``delayed hits''  or as  cells fired by neighbouring cells (``re-fired cells'' and ``noisy cells'').  ``In time hits'' are usually due to the electrons or positrons and ``delayed hits'' are caused by $\alpha$ particles. 
Hits are only  reconstructed as tracks if they are ``in time hits'' or ``delayed hits''. For ``in time hits'' the anode time values must not be more than a maximum value (currently $1.74$~$\mu$s found from Geiger calibrations). For  ``delayed hits'', the time measured by the anode  is up to 710~$\mu$s. The anode drift time  and the cathode times provide the transverse and longitudinal position of a hit in a drift cell, respectively.


Tracks are reconstructed if there are ``in time hits'' in at least three of the nine drift cell layers, with at least two of the hits occurring  in neighbouring layers.  The pattern recognition is carried out using a cellular automation algorithm~\cite{cellular-automation}. This algorithm defines segments as the lines which connect  two hits  in  neighbouring layers. In order to take into account all inefficiencies, the segments may connect two hits which are not in neighbouring layers if there is no hit in between. The angle between the two segments with a common hit can not be more than  $40$ degrees. 
To connect two cells with a segment there are four possibilities, as each cell has two edges, right and left. The ambiguity is resolved by finding the longest and smoothest pattern of combined segments.

 Once the pattern  is found, an iterative fit is performed assuming the track follows a helix (due to the magnetic field). The curvature of the track indicates  if the particle is an electron or positron. The reconstructed track is extrapolated back to the source foil and projected onto an associated scintillator surface. The   coordinates where the track crosses the foil and the scintillator are calculated.   Assuming that the tracks  originate from the foil, the former gives the vertex of the event.
 For ``delayed hits'', the time which is measured by TDC$_{\alpha}$ is used to  reconstruct the track in  the $(x,y)$ plane. 
\subsubsection{The  vertex resolution}
The resolution of the reconstructed vertex of the two electrons in the transverse plane depends on their energy. By using two electrons  from $^{207}$Bi source foils (Section~\ref{sec-laser}), the vertex resolution is found to be $3$~mm at 0.5~MeV and $2$~mm at 1~MeV in the $(x,y$) plane. The longitudinal vertex resolution depends on both energy and position of the electrons in the $z$ direction. For 1~MeV electrons, the resolution is $5$~mm if the vertex is in the top or bottom of the detector and  $4$~mm if it is in the centre. For 0.5~MeV these values change to $8$~mm and $6$~mm, respectively~\cite{vertex-resolution}.
\section{Particle identification}
Particle identification is possible by combining the information provided by the tracker and the calorimeter. NEMO~3 is able to identify electrons, positrons, photons and alpha particles. 

Due to the magnetic field charged particle trajectories are curved. In NEMO~3 the curvature sign of a trajectory is found with respect to the surface of the source foil. Thus, an electron (positron)  originating from the foil is identified as a track of negative (positive) curvature, which is  associated with only one scintillator hit.  A photon ($\gamma$) is identified as a scintillator hit which is not associated with a track. As the energy losses between the  scintillator blocks are not known, events are rejected if the photons or electrons are detected by two scintillators. The rate of  fake scintillator hits is negligible.

An alpha ($\alpha$) particle  can be identified only if it is  accompanied by at least one electron track. Alpha particles  are mainly produced by $^{214}$Po decay which is  in the  radon ($^{222}$Rn) decay chain. They  are defined by drift cell hits delayed by more than 70~$\mu$s.
 As the range of $\alpha$-particles from $^{214}$Bi decay in the helium gas of the tracking chamber is  25--40~cm~\cite{vera1}, the distance between the reconstructed vertex of  the electron and the delayed drift cell(s) corresponding to an $\alpha$-particle is required to be $|\Delta z|<30$~cm in the vertical direction and $|\Delta r|=\sqrt{(\Delta x)^{2}+(\Delta y)^{2}}<25$~cm in the  $x-y$ plane.
\section{Event simulation in NEMO~3}
Monte Carlo (MC) programs are used as event generators that simulate initial particles from isotope decays. The generated events are passed through simulations of the detector response so that the output can be compared to data. Resolution and detector acceptance effects are taken into consideration in the simulation.
MC simulations are necessary for analysing experimental data. The MC events are used to  measure  detector acceptances and to compare the generated physics signal with that seen in the detector. Selection criteria can be optimised by studying the signal to background ratio for specific processes.

The NEMO~3 simulation program~\cite{genbb} uses GENBB as the generator. GENBB  simulates the initial kinematics of the particles for two-neutrino double beta decay, different theories of neutrinoless double beta decay and  all the possible radioactive backgrounds to the double beta decay processes. GENBB also provides the possibility to generate the kinematics of  Compton scattering of  photons from sources outside the detector (external photons)  and  M$\o$ller scattering of external electrons.

The particles generated with GENBB (photons, electrons, positrons and alphas)   pass through the various regions of the detector. The description of the geometry of the detector and  the simulation  of the detector response is developed in the framework of the  GEANT~3 package~\cite{geant}.  The interactions of the particles with the source foils, tracking wires, the scintillators and other material in the sectors are taken into account in the simulation.   
\subsection{The  reconstruction of the simulated events}
The analysis in this work involves estimating the composition of data samples based on MC simulations of the various   processes. The simulated events were reconstructed in the same way as the experimental data. 
The functionality of the calorimeter and the tracker components in each run period is slightly different. This gives different detection efficiency for each run (for example having  noisy PMTs disconnected from the detector decreases the efficiency).  For the reconstructed MC simulation  to have real detector conditions,
the simulated events  are assigned to the real data run periods. The number of events assigned to each run  depends on the duration of the run. The detector conditions during the particular run are then applied  to the simulated events associated to that run.
\section{The data set}        
\label{sec-dataset}
 The analysis is performed on data taken between $14^{\rm th}$  February 2003 and $31^{\rm st}$ December 2006. This corresponds to runs  1869 to 5468. Each run represents  a period of time in which data acquisition was performed. The length of the runs is typically $12$ hours.  Data are removed from the analysis if any of the following conditions are met:
\begin{itemize}
\item There was a major problem with the electronics of the detector during the run.
\item The run has been  taken less than 24 hours after a general electronics shut down, as the PMTs of the calorimeter need time to stabilise.
\item They are taken less than 24 hours after $^{207}$Bi calibration runs. During such calibrations PMTs are known to have  a high counting rate and require time to stabilise~\cite{christine}.
\item  No laser run has been performed on the day that data were taken. The laser runs are vital as they give information about the behaviour of the PMTs, and also give the value of LTC~(see Section~\ref{sec-ltc}).
\item No appropriate energy or time calibration  of the PMTs is available. 
\end{itemize}
After removing these data, the effective data taking time is calculated to be $924.7$ days. 
The data  are divided into two sets: Phase~1 and Phase~2. Phase~1 is the data taken before installation of the anti-radon facility in September 2004 and Phase~2 is the data taken after the installation. 
The Phase~1 and Phase~2 data are combined throughout this thesis unless stated otherwise.
\section{Definition of the \boldmath$^{150}$Nd source foil boundaries}
\label{sec-ndg}
To define the foil position in the transverse plane, the polar angle,~$\phi$, is   replaced by   a sector number. The sector number is related to the polar angle by:
\begin{equation}
\rm sector~number =\frac{20}{2\pi} \times \phi.
\end{equation}

\begin{figure}[htbp]
\centering
\subfigure{
a)\includegraphics[width=9.0cm]{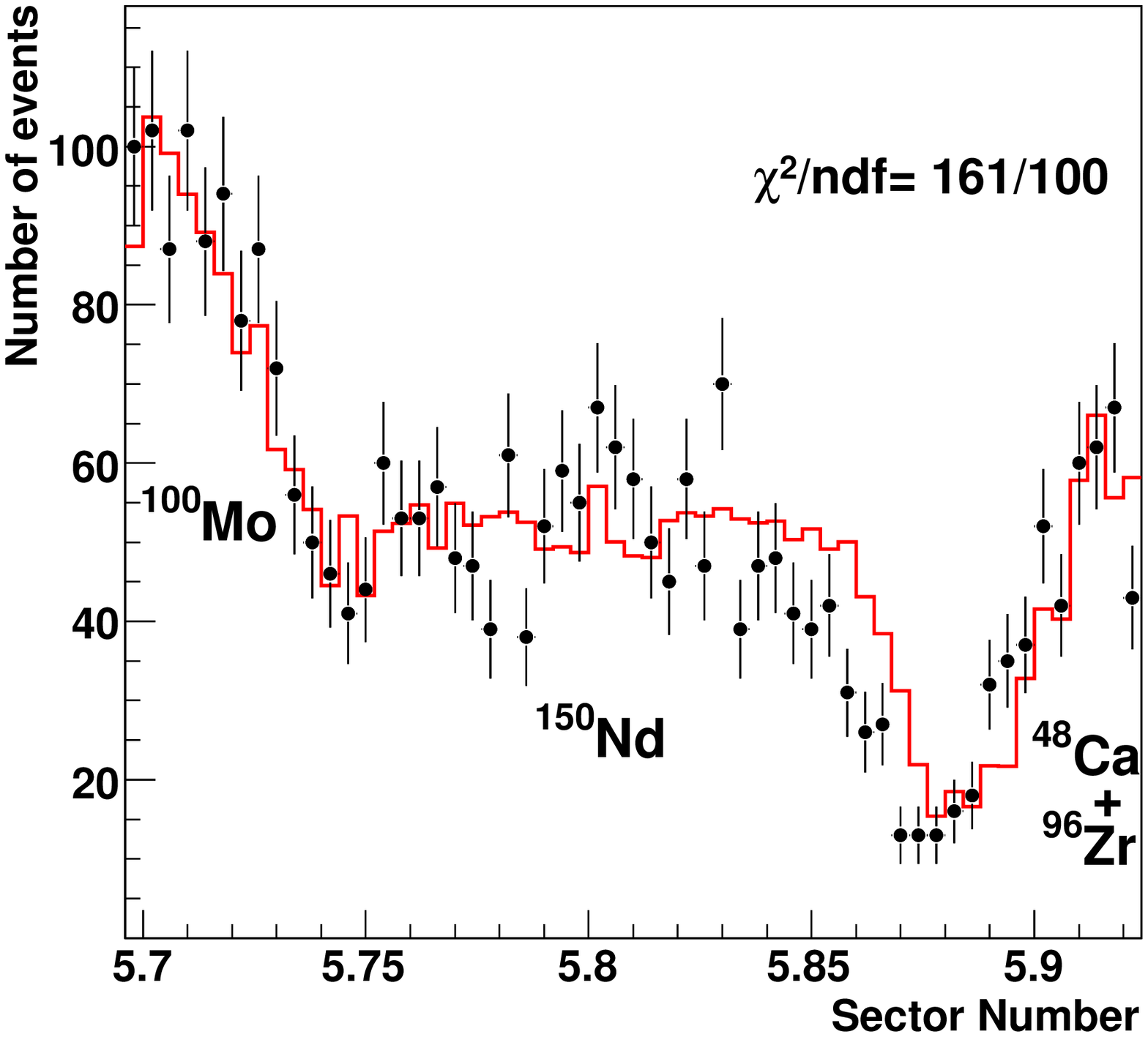}}
\subfigure{
b)\includegraphics[width=9.0cm]{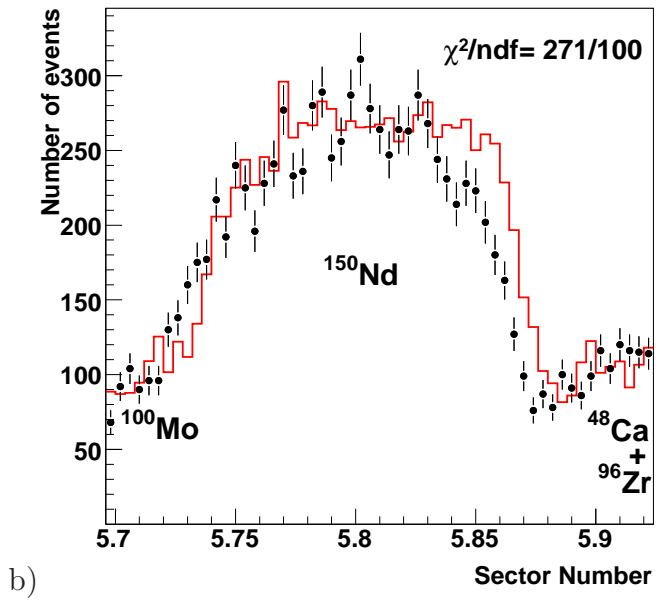}}
\caption[The vertex position of  two-electron  and one electron plus one photon events from $^{150}$Nd  and part of $^{48}$Ca, $^{96}$Zr and $^{100}$Mo foils.]{The vertex position of the a) two electron  and b) one electron plus one photon events from $^{150}$Nd  and part of the  $^{48}$Ca, $^{96}$Zr and $^{100}$Mo foils. The original MC geometry is shown in red and NEMO 3 data are shown as points. The statistical uncertainties on the data points are shown  with  error bars. The shift of data is clearly seen in both plots.  }
\label{fig-sec5}
\end{figure}

\begin{figure}[htbp]
\centering
\subfigure{
a)\includegraphics[width=9.0cm]{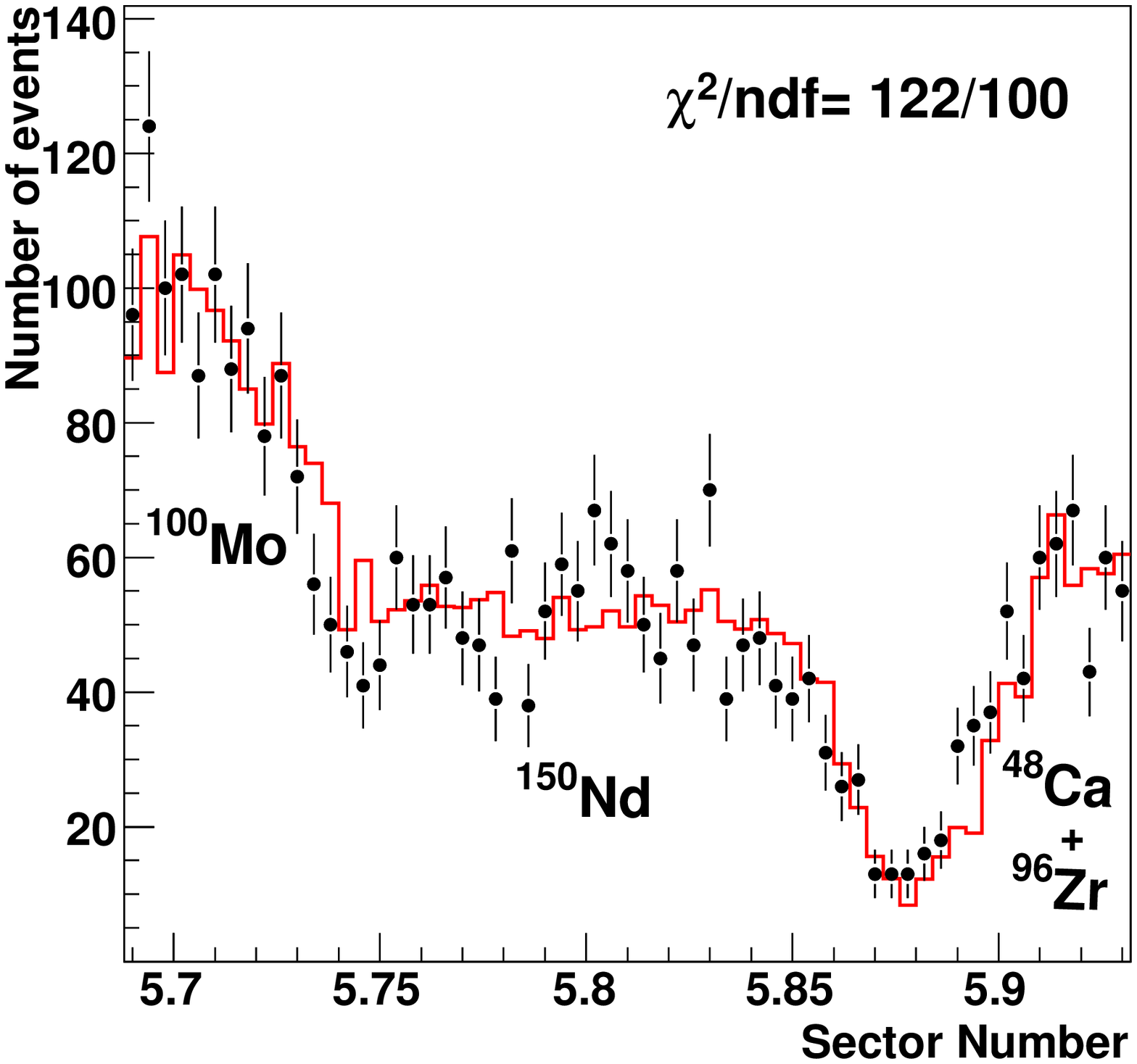}}
\subfigure{
b)\includegraphics[width=9.0cm]{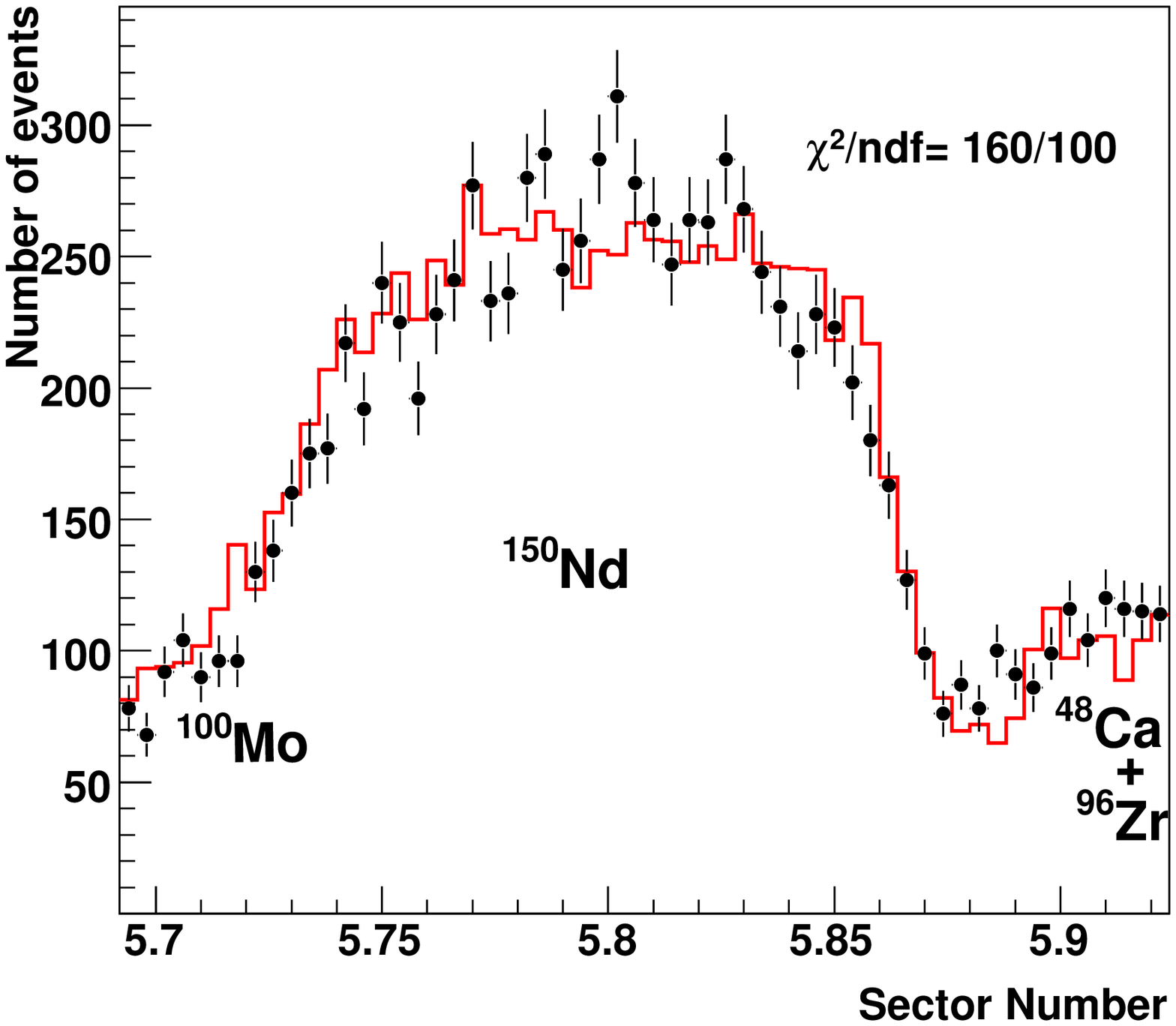}}
\caption[The vertex position of  two-electron  and  one electron plus one photon events from $^{150}$Nd  and part of $^{48}$Ca, $^{96}$Zr and $^{100}$Mo foils]{The vertex position of the a) two electron  and b) one electron plus one photon events from $^{150}$Nd  and part of the $^{48}$Ca, $^{96}$Zr and $^{100}$Mo foils. The new definition of the  MC geometry is shown in red and NEMO 3 data are shown as points. The statistical uncertainties on the data points are shown with  error bars.}
\label{fig-sec5c}
\end{figure}

In the construction design of $^{150}$Nd, this value is between $5.74$ and $5.87$. However,  by studying  the distribution of event vertex positions, it is   observed that data are shifted  with respect to the defined MC geometry.  Figure~\ref{fig-sec5}a shows the distribution of the vertex position for two electron events coming from the foil (higher sector number). The discrepancy between  data  points and  MC simulation  can be observed on the right side of the $^{150}$Nd foil.  From these plots no conclusion can be drawn for the left side of the foil, since  two-electron events from $^{100}$Mo overlap with events from $^{150}$Nd. 
To check if the same effect is also observable on the left side,  the electron-photon   decay channel was analysed. In this channel the event rate for $^{100}$Mo is lower than $^{150}$Nd as this isotope has less contamination than $^{150}$Nd. Figure~\ref{fig-sec5}b shows the distribution of the vertex position for one electron and one photon events coming from the foil~(full details of the  electron-photon selection are given in Section~\ref{sec-eg}). The discrepancy between simulation and data is  observed on both left and right sides of the $^{150}$Nd foil. 
In order to correct this discrepancy, the $^{150}$Nd sector number was shifted by  0.01 of a sector number to lower values, in the  simulation. The effect of this shift is shown in Figures~\ref{fig-sec5c}a and~\ref{fig-sec5c}b. The $\chi^{2}/ndf$ values of the data and MC comparison improve significantly. Therefore the $^{150}$Nd position was  redefined to be between sector number  5.73 and 5.86.
The active height of the $^{150}$Nd foil is 234~cm, thus the boundaries of $^{150}$Nd in the $z$ coordinate are defined to be $|z|<117$~cm.



\section{Time of flight selection criteria}
\label{sec-tof}
It is possible that a  particle from  a source outside the foil  deposits energy in one scintillator block,
 enters the tracking volume,
 interacts with the source foil, scatters off it and is then detected by another scintillator. 
These types of 
events, which are known as external events,  can mimic two electron events originating from the foil (internal events). 
 In order to reduce  the number of these  background events, a time of flight analysis is performed~\cite{tof}. 
In this analysis, the time differences between the two calorimeter hits are calculated for two  hypotheses.
 In the first hypothesis  the two calorimeter hits and tracks are caused by two electrons from the foil
(internal hypothesis). The second hypothesis assumes that the hits are from external sources (external hypothesis).
 The time differences calculated for both scenarios are then compared with the measured time difference of the calorimeter hits.

 For  each hypothesis, $\chi^{2}_{th}$ is defined as:
\begin{equation}
\chi^{2}_{th}=\frac{(\Delta t_{mes}- \Delta t_{th})^{2}}{\sigma_{mes}^{2}+\sigma_{th}^{2}},
\end{equation}
where $\Delta t_{mes}$ is the measured time difference between the two scintillator hits;
 $\Delta t_{th}$ is the calculated time difference  for the internal and external hypothesis; 
 and   $\sigma_{mes}$ and $\sigma_{th}$ are
 the uncertainty on the measured and calculated time differences, respectively.
The value of $\Delta t_{th}$ is related to the length of the particle trajectories, $l_{1}$ and $l_{2}$, and the energies measured by the calorimeter, $E_{1}$ and $E_{2}$. In the case of the internal (external) hypothesis for two-electron events, $\Delta t_{th}$ is   defined as:
\begin{eqnarray}
\Delta t_{th}^{int}=\frac{l_{1}}{\beta_{1}}-\frac{l_{2}}{\beta_{2}},\\
\Delta t_{th}^{ext}=\frac{l_{1}}{\beta_{1}}+\frac{l_{2}}{\beta_{2}},
\end{eqnarray}
 where the relativistic factor, $\beta_{i}$, is related to the energy measured by each calorimeter block ($E_{i}$) and the electron rest mass ($ m_{0}$):
\begin{equation}
\beta_{i}=\frac{\sqrt{(E_{i}+2m_{0})}}{E_{i}+m_{0}}.
\end{equation}
The theoretical uncertainty, $\sigma_{th}$, is found from  differentiation:
\begin{equation}
(\sigma_{th}^{i})^{2}=\sum_{i=1}^{2}\left(\frac{t_{th}^{i}m_{0}}{E_{i}(E_{i}+m_{0})(E_{i}+2m_{0})}\right)^{2}\sigma_{E}^{2}+\sum_{i=1}^{2}\left(\frac{1}{\beta_{i}}\right)^{2}\sigma_{l}^{2},
\end{equation}
where $t_{th}^{i}=\frac{l_{i}}{\beta_{i}}$,  $\sigma_{E}$ 
is the uncertainty on the energy measurement and $\sigma_{l}$ is the uncertainty on the track length.

The scenario for internal and external electron-photon  events is similar except  photons do not leave tracks in the tracker. To measure the length of the photon path, the distance between the scintillator hit and the intersection point of the electron track with the foil (the electron-photon vertex) is found. The length of the photon path, $l_{\gamma}$, and the time calculated for the photon to traverse this length, $t_{th}^{\gamma}$,  are found by:
\begin{equation}
l_{\gamma}=\sqrt{(x_{ve}-x_{scin})^{2}+(y_{ve}^{2}-y_{scin})^{2}+(z_{ve}-z_{scin})^{2}},\textnormal{\hspace{0.7cm}and\hspace{0.7cm}}t_{th}^{\gamma}=\frac{l_{\gamma}}{c},
\end{equation}
where $x_{ve}$, $y_{ve}$ and $z_{ve}$ are the coordinates of the vertex position of the events from the  detector centre, and $x_{scin}$, $y_{scin}$ and $z_{scin}$ are the coordinates of the scintillator hit by the photon. As it is not known where in the scintillator the photon hits, these coordinates correspond to the centre of the face. The uncertainty on $t_{th}^{\gamma}$ is related to the maximum dimensions of the scintillator face and the energy  measured by the scintillator.
\begin{figure} 
\centering
\subfigure{
a)\includegraphics[width=9.0cm]{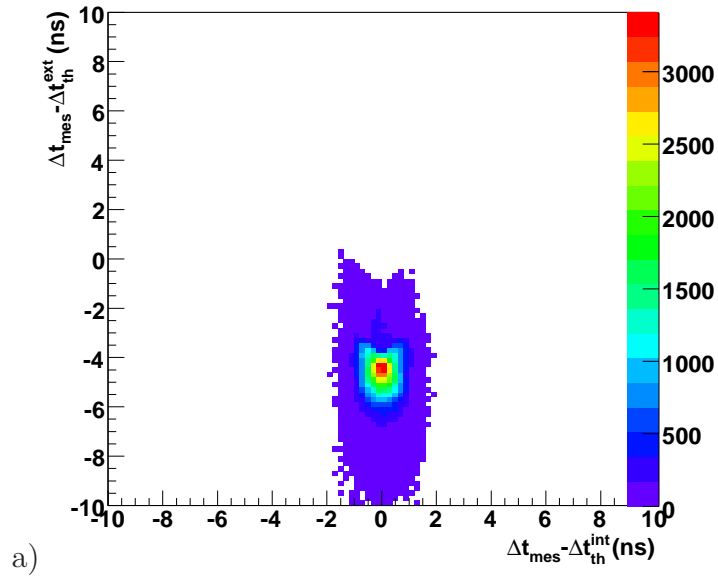}}
\subfigure{
b)\includegraphics[width=9.0cm]{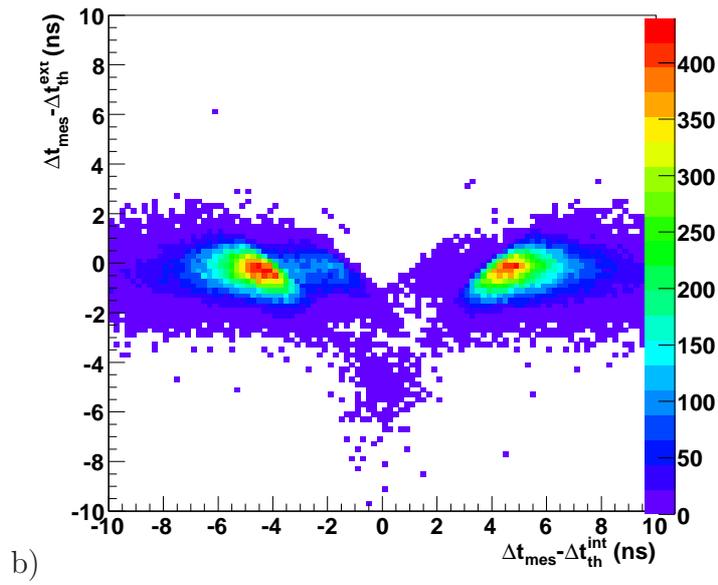}}
\caption[Distribution of $\Delta t_{mes}- \Delta t_{th}^{int}$ versus  $\Delta t_{mes}- \Delta t_{th}^{ext}$ for  two  tracks from $^{150}$Nd  and $^{214}$Bi in the PMTs decay]{Distribution of $\Delta t_{mes}- \Delta t_{th}^{int}$ versus  $\Delta t_{mes}- \Delta t_{th}^{ext}$ for  two  tracks  from  a) $2\nu\beta\beta$ decay of $^{150}$Nd  foil and b) external electrons  originating from $^{214}$Bi contaminants in the PMTs. The simulations are normalised by an arbitrary factor.}
\label{fig-dt}
\end{figure}
\begin{figure}[htbp]
\centering
\subfigure{
a)\includegraphics[width=5.0cm]{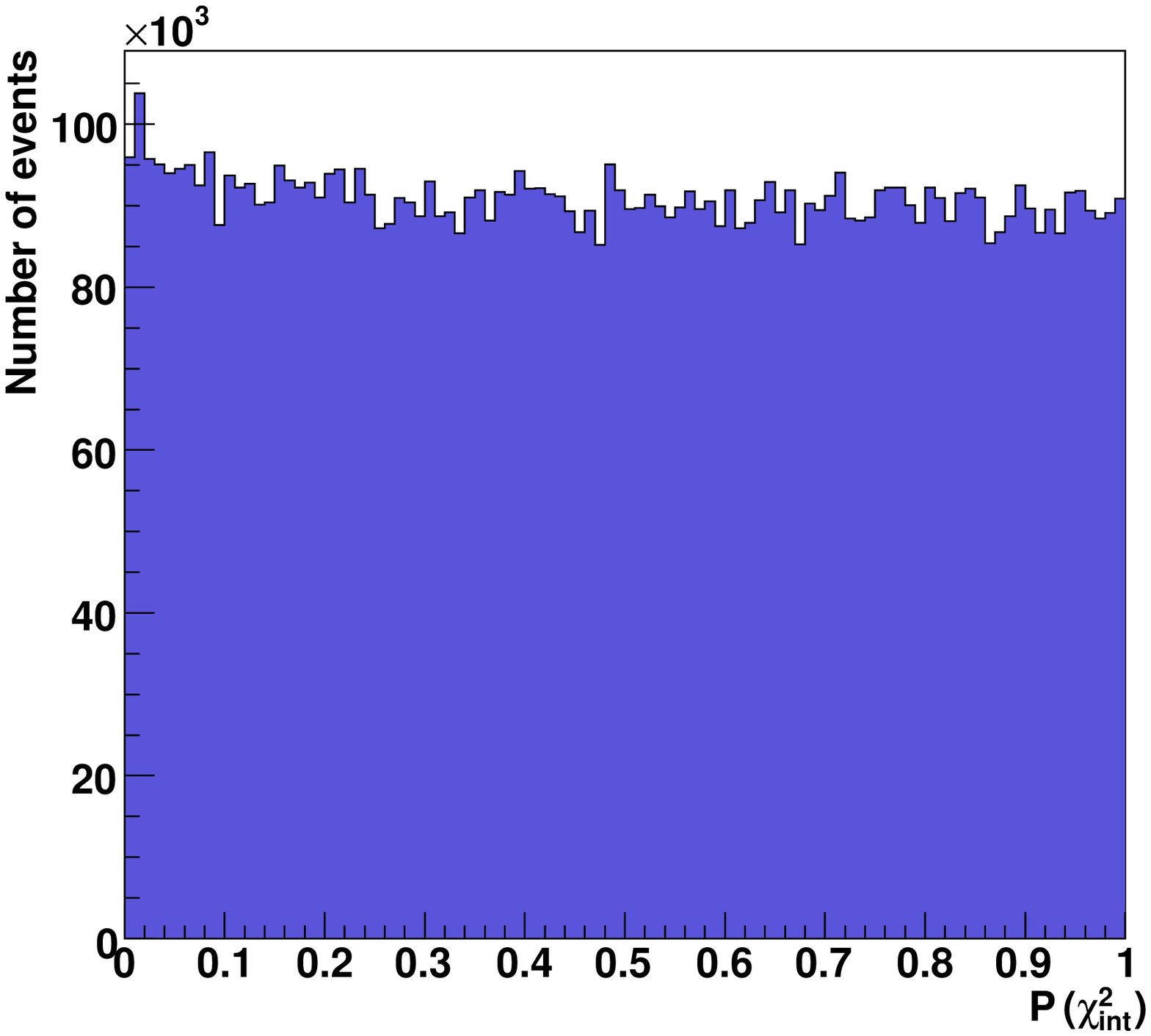}}
\subfigure{
b)\includegraphics[width=5.0cm]{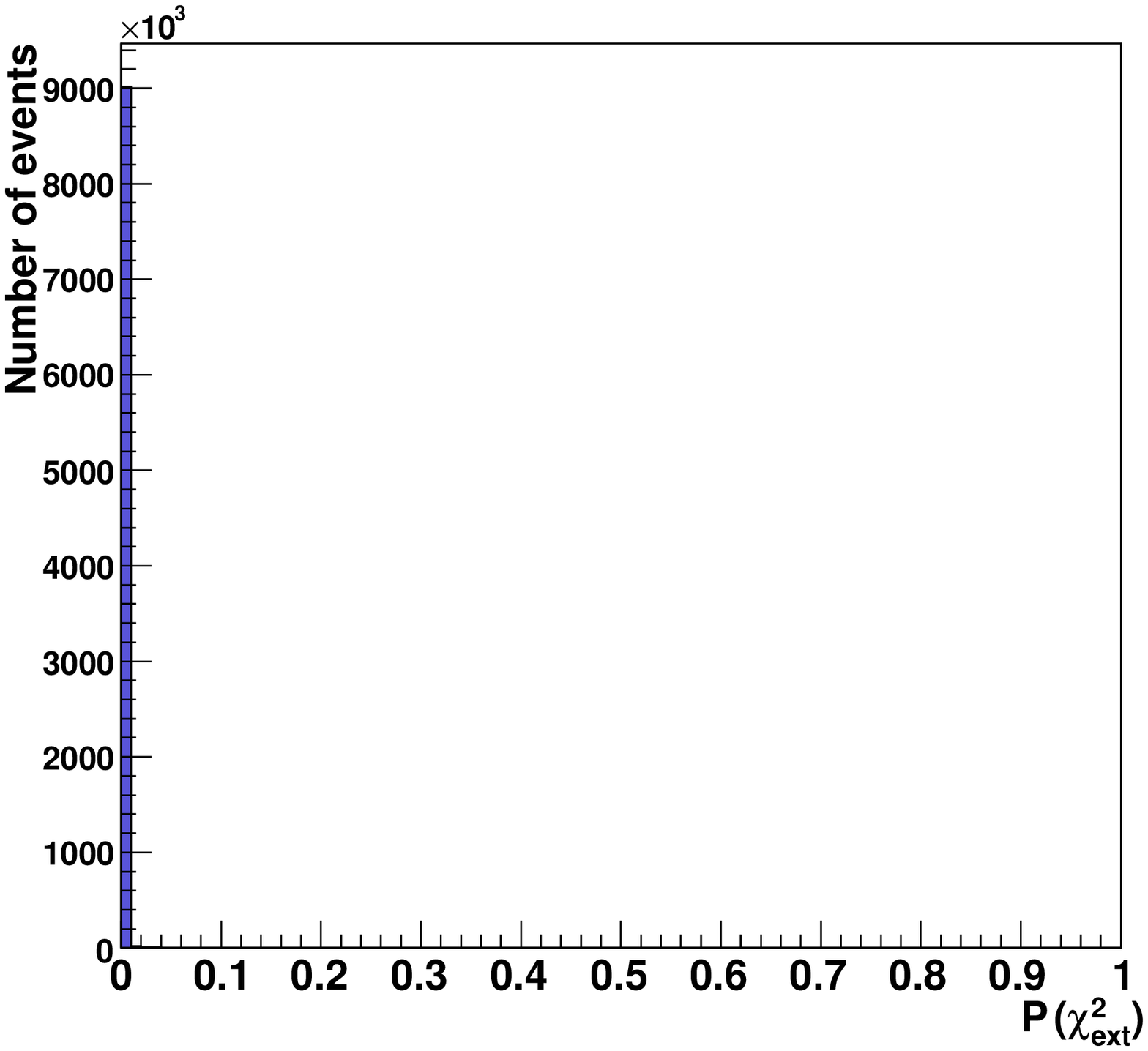}}
\subfigure{
c)\includegraphics[width=5.0cm]{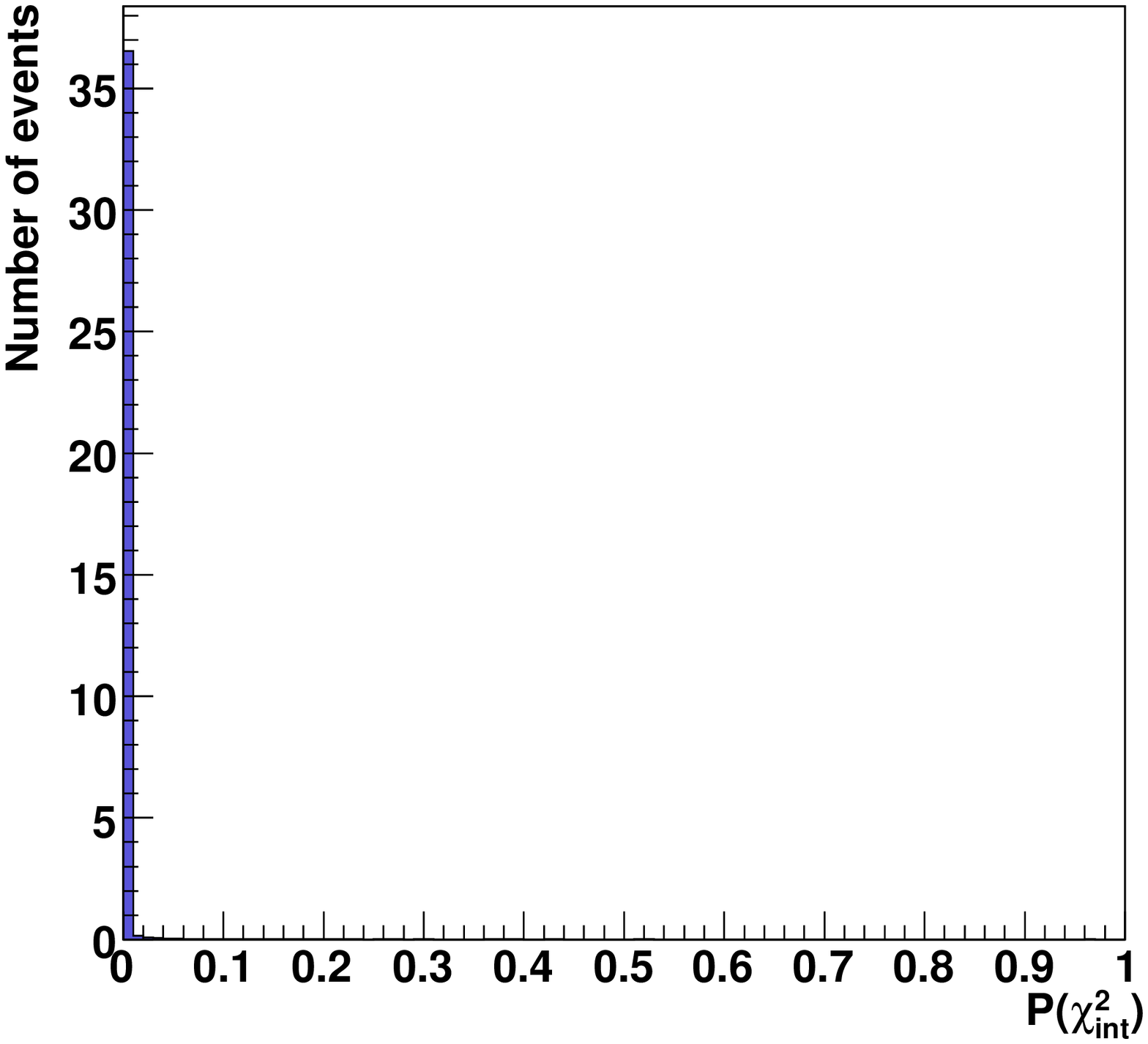}}
\subfigure{
d)\includegraphics[width=5.0cm]{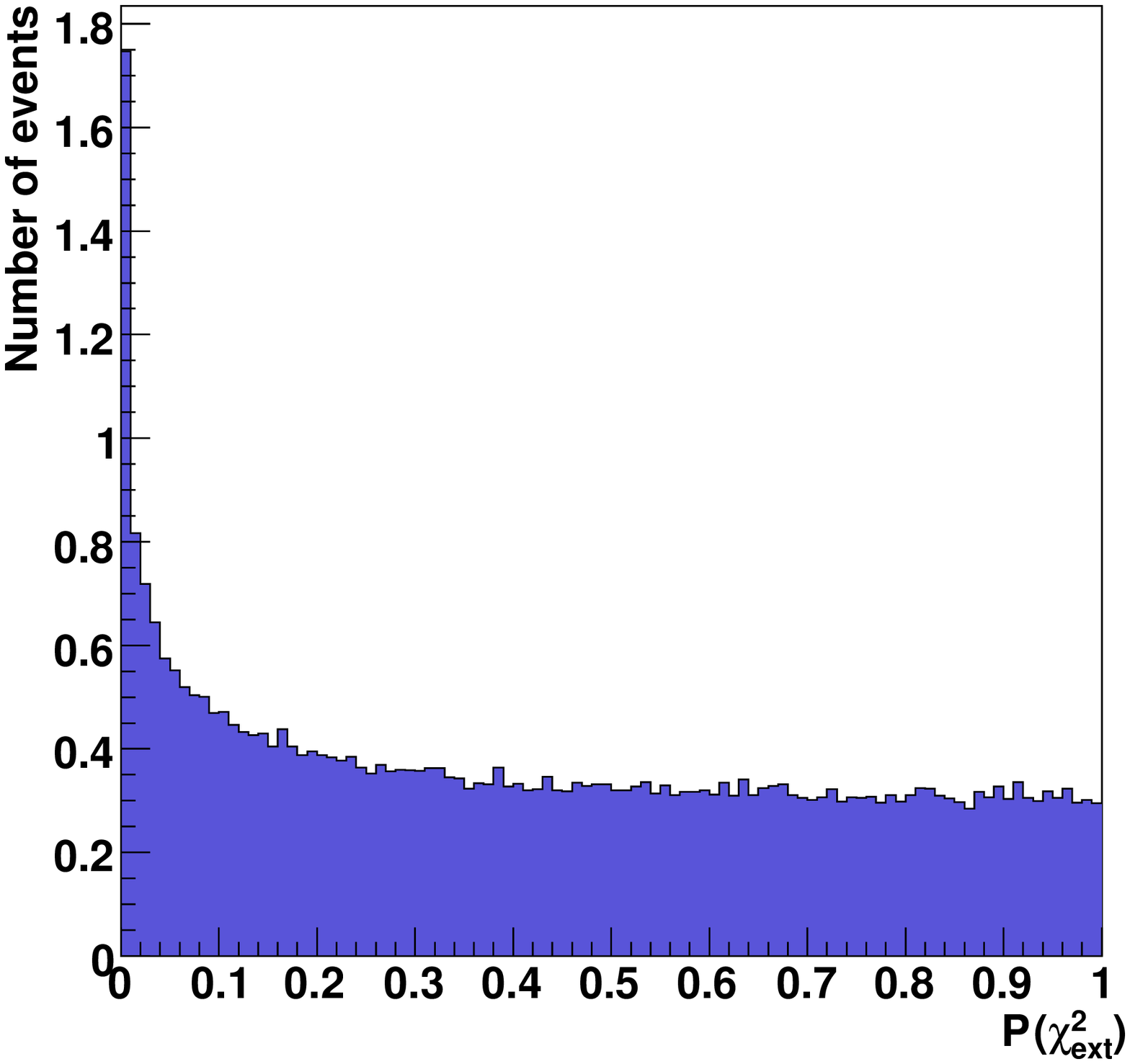}}
\caption[The distribution of the internal and external hypothesis probability  for $2\nu\beta\beta$ events from  $^{150}$Nd foil and for internal $^{214}$Bi]{The distribution of the internal a) and external b) hypotheses probability, $P(\chi^{2}_{int})$ and $P(\chi^{2}_{ext})$, for $2\nu\beta\beta$ events from  $^{150}$Nd foil. c) The $P(\chi^{2}_{int})$ and d) $P(\chi^{2}_{ext})$ distribution  for two electron tracks created by $^{214}$Bi contamination in the PMTs. The simulated events are normalised by  arbitrary factors. }
\label{fig-pchi2-2e}
\end{figure}
For an internal event, $\Delta t_{mes}- \Delta t_{th}^{int}$ is expected to be zero ns with some uncertainty  due to the time resolution, whereas  $\Delta t_{mes}- \Delta t_{th}^{ext}$ has a value of the order -$5$~ns. The opposite  applies
 for external events.
Figures~\ref{fig-dt}a and \ref{fig-dt}b show  $\Delta t_{mes}- \Delta t_{th}^{int}$ versus  $\Delta t_{mes}- \Delta t_{th}^{ext}$ for internal two electron events (simulated  $2\nu\beta\beta$  decay of $^{150}$Nd) and external two electron events (simulated $^{214}$Bi contamination in the PMTs), respectively.

\begin{figure}[htbp]
\centering
\subfigure{
a)\includegraphics[width=5.0cm]{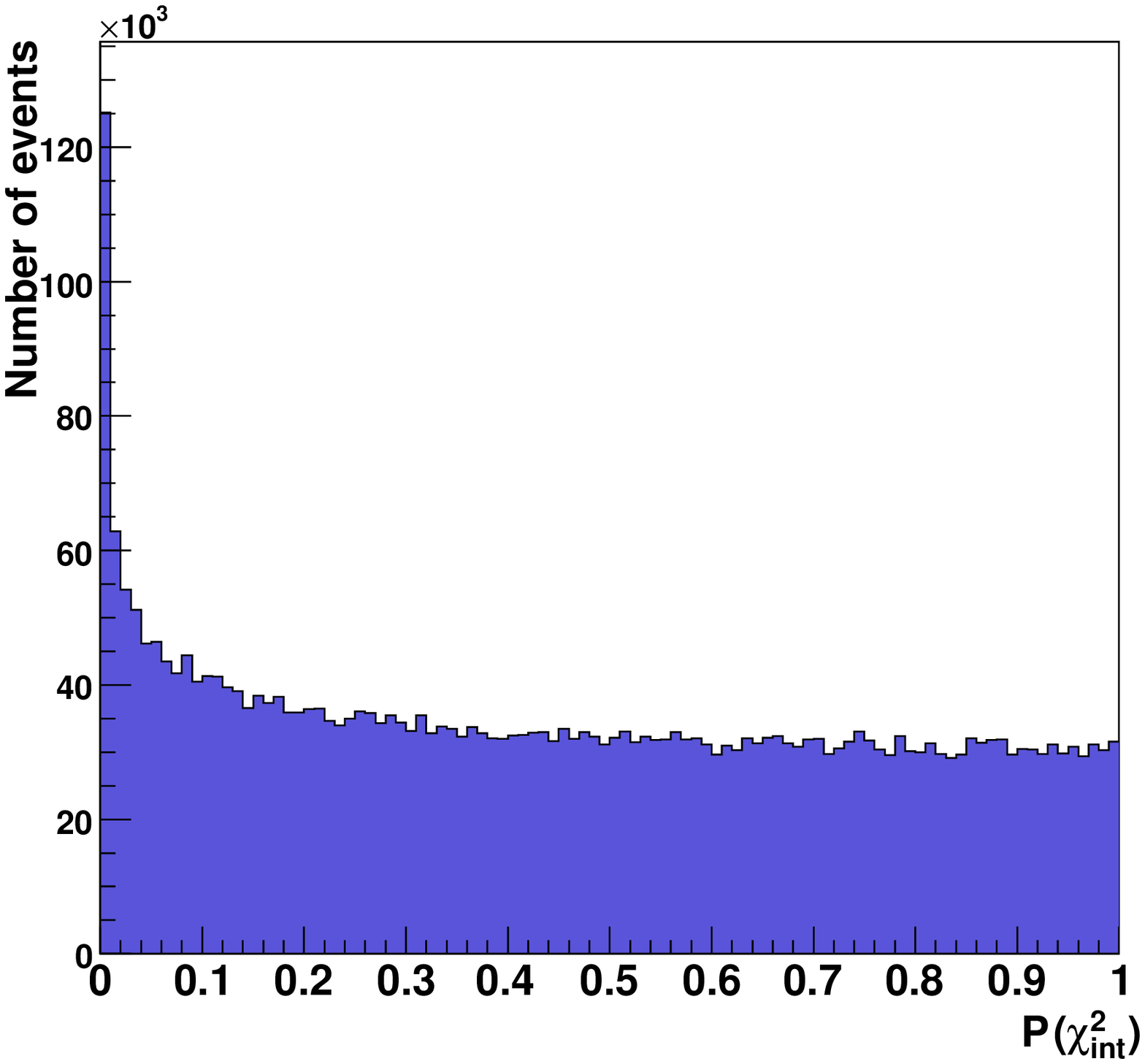}}
\subfigure{
b)\includegraphics[width=5.0cm]{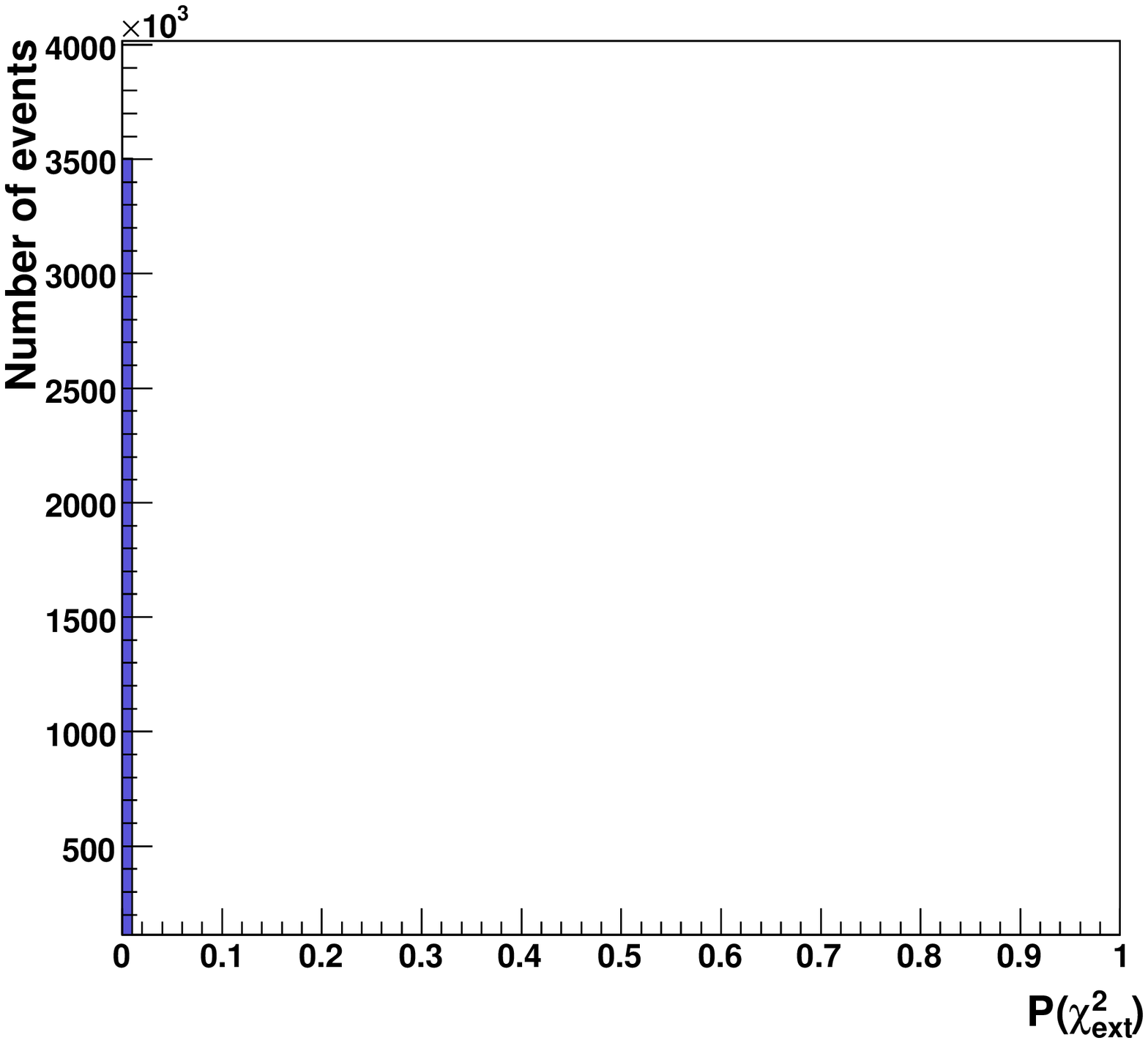}}
\subfigure{
c)\includegraphics[width=5.0cm]{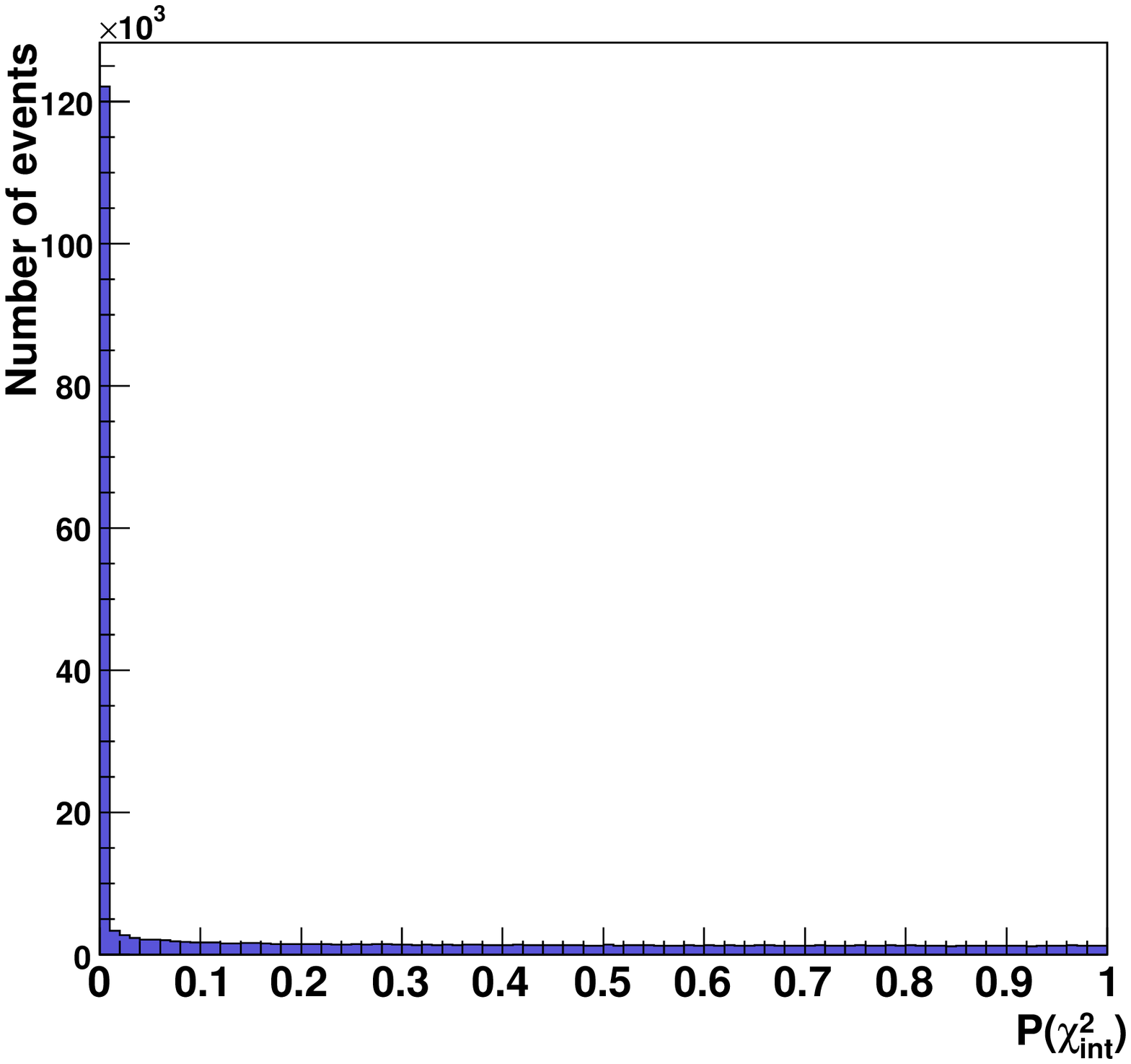}}
\subfigure{
d)\includegraphics[width=5.0cm]{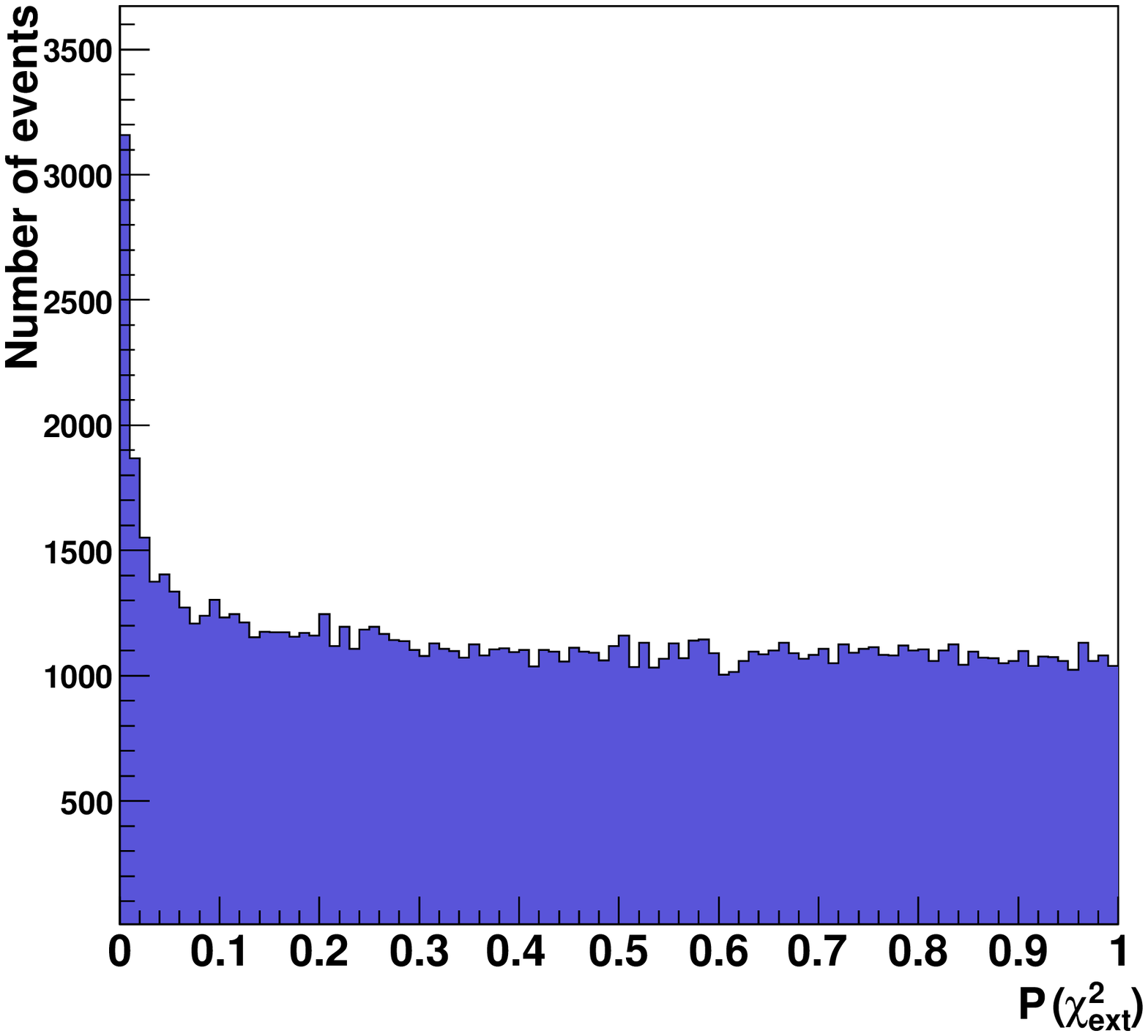}}
\caption[The distribution of the internal and external hypothesis probability for $e\gamma$ events from $^{214}$Bi contaminant inside the $^{150}$Nd foil and  for  $e\gamma$ events created by $^{214}$Bi contamination in the PMTs]{ The distribution of the internal a) and external b) hypothesis probability, $P(\chi^{2}_{int})$ and $P(\chi^{2}_{ext})$, for $e\gamma$ events from $^{214}$Bi contaminant inside the $^{150}$Nd foil, and  for c) and d) $e\gamma$ events created by $^{214}$Bi contamination in the PMTs. The simulated  events are normalized by   arbitrary factors. }
\label{fig-pchi2-1e1g}
\end{figure}

For both hypotheses, the  probabilities, $P(\chi^{2}_{int})$ and $P(\chi^{2}_{ext})$, are calculated. The $P(\chi^{2}$) 
is defined as:
\begin{equation}
P(\chi^{2})=1-\frac{2}{\sqrt\pi}\int_0^{\chi^{2}} e^{x^{2}}dx,
\end{equation}
where $x=\frac{1}{1+\sqrt{2\chi^{2}}}$. 
 Figure~\ref{fig-pchi2-2e}   shows  the $P(\chi^{2}_{int})$ and  $P(\chi^{2}_{ext})$ for simulated internal two electron events ($2\nu\beta\beta$ decay of $^{150}$Nd) and simulated crossing electrons events ($^{214}$Bi contaminant of the PMTs). 
Figure~\ref{fig-pchi2-1e1g} shows $P(\chi^{2}_{int})$ and  $P(\chi^{2}_{ext})$ for simulated internal electron-photon  events ($^{214}$Bi contaminant in $^{150}$Nd foil) and external electron-photon events ($^{214}$Bi contaminant of the PMTs).

By using these MC distributions, data selections are optimised in order to select events originating from the foil or from the detector components~\cite{corrine}. The criteria for the internal events are
\begin{equation}
P(\chi^{2})_{int}>0.04~~~ {\rm and}~~~P(\chi^{2}_{ext})<0.01,
\end{equation}
and for the external events are
\begin{equation}
P(\chi^{2}_{int})<0.01~~~ {\rm and} ~~~P(\chi^{2}_{ext})>0.04.
\end{equation}

\section{Fitting  Monte Carlo samples to data}
\label{sec-fit}
Chapters~\ref{chap-bgr} and~\ref{chap-2nbb} are dedicated to the measurement of the radioactive background activities and $2\nu\beta\beta$ half-life by comparing MC simulations to data.  In the situation where only one MC component is not known, it is sufficient to do a simple fit based on the  total number of data and simulated events. In this method the simulated events are normalised to the same number of events as total data events.

 However, in the situation where more than one MC components should  be estimated, a  binned maximum likelihood fit~\cite{rogermc} is  used. 
In this  technique, to fit $m$ number of MC sources to data, the shapes of the distributions are taken into account.  The predicted (normalised) number of MC events in bin $i$ is given by
\begin{equation}
f_{i}=N_{D}\sum_{j=1}^{m}\frac{P_{j}A_{ji}}{N_{j}},
\end{equation}
where $N_{D}$ is the total number of events in the data sample, $N_{j}$ is the total number in the MC sample for source $j$, $P_{j}$ is the proportion of  source $j$ in data, and $A_{ij}$ is the expected number of MC simulated  events from source $j$ in bin $i$ (which includes the unknown uncertainty in the bin).

  The probabilities of observing a particular number of data events, $d_{i}$, and a particular number of simulated events, $a_{ji}$, follow  Poisson statistics:
\begin{eqnarray}
P(d_{i};f_{i})=e^{-f_{i}}\frac{f_{i}^{d_{i}}}{d_{i}!},\\
P(a_{ji};A_{ji})=e^{-A_{ji}}\frac{A_{ji}^{a_{ji}}}{a_{ji}!}.
\end{eqnarray}
The total logarithm of likelihoods is then defined by:
\begin{equation}
\label{eq-like}
\ln L=\ln P(d_{i};f_{i})+\ln P(a_{ji};A_{ji})=\sum_{i=1}^{n}d_{i}\ln f_{i}-f_{i} +\sum_{i=1}^{n}\sum_{j=1}^{m}a_{ji}\ln A_{ji}-A_{ji}.
\end{equation}
The proportion of each  MC source, $P_{j}$,  is estimated by maximising Equation~\ref{eq-like} using computing packages~\cite{minuit,root}.
\section{Limit setting}
\label{sec-colie}
The search for new physics in a double beta decay experiment involves dealing with  small numbers of expected signal events  (in the case of $0\nu\beta\beta$)
 and  much larger number of  background events than expected signal events (in the case of $0\nu\beta\beta$ with the emission of Majoron(s)). The exclusion of a signal at a particular confidence level (CL)  can be significantly improved relative to a simple counting experiment by using the binned distribution of data, signal and background. In this way each bin is treated as an independent search channel. 

A typical new physics search analysis is described by  a final variable which  is designed to distinguish between two possible hypotheses: the production and detection of the new physics along with the background (S+B hypothesis), or the presence of only the background (B-only hypothesis). 
The consistency between data and the signal and background models in this final variable is  used to set a limit on the  maximum number of signal events consistent with the data. 
The likelihood ratio test statistics ~\cite{cls,cls2} is an optimal choice for searches with small statistics~\cite{wade8},
\begin{equation}
Q=L(S+B)/L(B),
\end{equation}
 where $L(S+B)$ and $L(B)$ are the Poisson likelihood for the S+B and B-only hypotheses. 

The limits in this thesis  are calculated by using a likelihood-fitter~\cite{wade}   developed at  the D$\O$ experiment using a log-likelihood ratio (LLR) test statistic. A  description of this method is given in~\cite{markowen}. The LLR is defined as~\cite{cls}: 
\begin{equation}
{\rm LLR}\equiv \chi=-2\ln{Q}=-2\sum_{i}\left(s_{i}-n_{i}\ln(1+\frac{s_{i}}{b_{i}})\right),
\label{eq-llr}
\end{equation}
where $i$ is the the $i^{th}$ bin in the distribution used, $s_{i}$ is the number of expected signal events in that bin, $b_{i}$ is number of expected background  events and $n_{i}$ is the number of data events observed  in bin $i$. 
The expected LLR distributions for both hypotheses are generated by performing a large number of toy Monte Carlo pseudo-experiments. In  pseudo-experiments, the pseudo-data are the coherent sum of signal and background or background processes and thus their value in each bin is obtained by drawing a random number from a Poisson distribution  where $P(x;p)=p^{x}e^{-p}/x!$ is the probability for obtaining $x$ events, given an expectation of $p$ events.  
The expectation of a number of events in each bin for each pseudo-experiment is varied according to the systematic uncertainties. The systematic uncertainties are  introduced into signal and background expectations, $p^{0}$, via Gaussian distributions,
\begin{equation}
p=p^{0}\left(1+\sum_{j}g(\sigma_{j})\right),
 \end{equation}
where $g$  is a random number taken from a Gaussian distribution with a mean of zero and width of one and $\sigma_{j}$ is  the $j^{th}$ fractional uncertainty on the rate of the $p^{0}$.  The same random Gaussian number is used for correlated uncertainties.
The LLR distributions for S+B and B-only hypotheses, built up from the pseudo-experiments can be compared with the observed value of LLR in the data  (LLR$_{\rm obs}$ or $\chi_{d}$).   The LLR$_{\rm obs}$  are found by  substituting  $n_{i}$ in Equation~\ref{eq-llr} with the number of observed data events ($d_{i}$).  
The confidence level in S+B hypothesis is given by
\begin{equation}
CL_{S+B}=P_{S+B}(\chi>\chi_{d})=\int_{\chi_{d}}^{\infty}\frac{dP_{S+B}}{d\chi}d\chi,
\label{eq-clsb}
\end{equation}
where $P_{S+B}(\chi>\chi_{d})$ is the probability for the S+B hypothesis to produce an outcome which is less signal-like than that observed in the data and is defined by integrating the expected LLR distribution for S+B hypothesis.
Similarly, the calculation of the confidence limit for the background only hypothesis is given by the probability for the B hypothesis to produce an outcome which is less signal-like  than that observed in the data and is found by integrating the expected LLR distribution for B-only hypothesis,
\begin{equation}
CL_{B}=P_{B}(\chi>\chi_{d})=\int_{\chi_{d}}^{\infty}\frac{dP_{B}}{d\chi}d\chi.
\end{equation}
Downward fluctuations of the background   lead to inconsistencies between expected background and observed data and  create  inaccurate exclusion limits using $CL_{S+B}$~\cite{wade8}.  To deal with this problem,  $CL_{S+B}$ is divided by $CL_{B}$ and  the value of the signal confidence level ($CL_{s}$) is defined as~\cite{cls}:
\begin{equation}
CL_{S}=\frac{CL_{S+B}}{CL_{B}}.
\label{eq-cls}
\end{equation}

\begin{figure}
\centering
\subfigure{
a)\includegraphics[width=12cm]{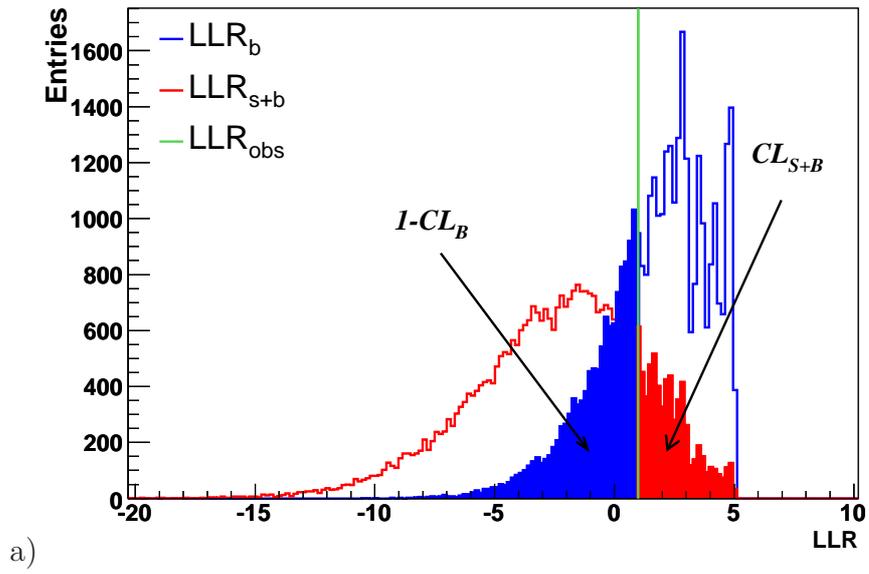}}
\subfigure{
b)\includegraphics[width=12cm]{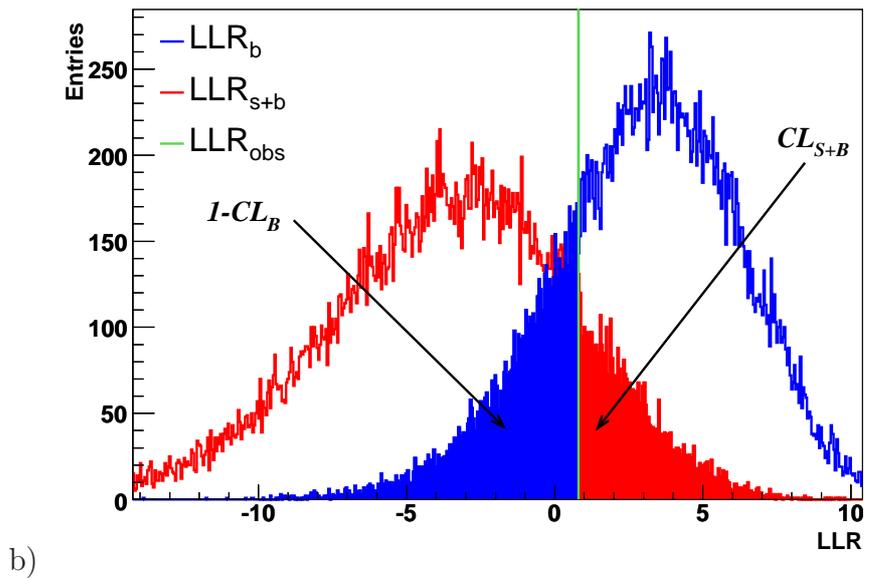}}
\caption[LLR distributions for the expectation in the B-only hypothesis  and in the S+B hypothesis  and the observation in the data for  $0\nu\beta\beta$ and  $0\nu\beta\beta \chi$ search]{LLR distributions for the expectation in the B-only hypothesis (blue line) and in the S+B hypothesis (red line) and the observation in the data (green line) for a)  $0\nu\beta\beta$ and b) $0\nu\beta\beta \chi$ search. The regions integrated to obtain $1-CL_{B}$ and $CL_{S+B}$ are also shown. }
\label{fig-llr}
\end{figure}

Figures~\ref{fig-llr}a and~\ref{fig-llr}b show two examples of LLR distributions. The data originating from $^{150}$Nd foil are used to create these distributions. In both figures the background model consists of $2\nu\beta\beta$ plus all other radioactive background (See Chapter~\ref{chap-2nbb}). In Figures~\ref{fig-llr}a,    the signal is $0\nu\beta\beta$ and  in Figure~\ref{fig-llr}b, it is  $0\nu\beta\beta$ with the emission of a Majoron particle. 

In order to calculate the observed limit on the signal rate, this rate is given a new value  and the expected LLR distributions   are re-generated from another series of  pseudo-experiments and the observed LLR is recalculated. This leads to a new value of $CL_{s}$. This process is repeated through a range of signal rates until $CL_{s}$ reaches the desired confidence level, which  in this thesis, by convention,  is $CL_{s}=0.1$ ($1-CL_{s}=0.9$) or 90\% confidence level. 

\subsection{Definition of observed and expected limit}
There are two distinct limits, the median expected and observed limit. The values of  observed LLR (${\rm LLR_{obs}}$) and median expected LLR (${\rm LLR_{med}}$)  are found by setting $n_{i}=d_{i}$ and $n_{i}=b_{i}$ in Equation~\ref{eq-llr}, respectively. The value of ${\rm LLR_{obs}}$ relative to ${\rm LLR_{med}}$ indicates  whether data  appears to be more background-like or not.  The expected limit on the signal rate gives the experiment sensitivity without any bias from actual observation. Thus, different experiments searching for a particular  process can be compared using the expected limit results. 

\subsection{The profile likelihood ratio}
\label{sec-llrprofile}
Systematic uncertainties are often estimated rather than measured and frequently  correspond to an upper bound on the true value. These systematic uncertainties    widen the expected LLR distributions and  decrease the analysis sensitivity for setting limits. This effect of systematics can be reduced by comparing the predicted shape  of the final variable to that  observed in data, and finding the best fit for systematic uncertainties  to data observation~\cite{wadenote}:
\begin{equation}
\chi^{2}(H)=2\sum\left((p(H)^{\prime}_{i}-n_{i})-n_{i}\ln\frac{p(H)^{\prime}_{i}}{n_{i}}\right)+\sum_{k}{S(H)_{k}^{2}},
\label{eq-chi2collie}
\end{equation}
 where  $p(H)_{i}^{\prime}$ is the systematically varied prediction in bin $i$ for the two hypotheses: 
\begin{equation}
p(H)_{i}^{\prime}=p(H)_{i}\prod_{k=1}^{k}{(1+S(H)_{k})}.
\label{eq-uncertainty}
\end{equation}
In the case of the S+B hypothesis $p_{i}=s_{i}+b_{i}$, and in the case of the B-only hypothesis  $p_{i}=b_{i}$.  The fitted value for systematics $k$   is indicated as $S(H)_{k}$.  
The optimal choice of the test statistic in this case is the profile likelihood ratio~\cite{wade,junk}.  This test statistic is the ratio of the $\chi^{2}$ minimised  for each of the two hypothesis.
~The LLR is defined as:
\begin{equation}
{\rm LLR}=-\ln{\frac{\chi^{2}_{min}(H_{B})}{\chi^{2}_{min}(H_{S+B})}},
\label{eq-llrchi}
\end{equation}
To find the LLR distributions for the S+B and B-only hypotheses,  $n_{i}$ in Equation~\ref{eq-chi2collie} is replaced with the MC pseudo-data, and to find ${\rm LLR_{obs}}$ and ${\rm LLR_{med}}$, $n_{i}$ is replaced with the number of observed data and expected background events. The  $p(H)_{i}^{\prime}$ term in Equation~\ref{eq-chi2collie}  results in narrower LLR distributions than the ones resulting  from Equation~\ref{eq-llr} and therefore increases the analysis sensitivity for setting limits. 
The  Equations~\ref{eq-clsb} to~\ref{eq-cls} are used to set limits on the number of observed and expected events at $1-CL_{s}=0.9$ by varying the signal rate as before.

\subsection{Floating the background  normalisation}
\label{sec-float}
The $\chi^{2}$ function used in the fit model in Equation~\ref{eq-chi2collie}
includes a term ($S(H)_{k}^{2}$) that adds a Gaussian constraint on each systematic uncertainty  . If  the normalisation on the background is not fully known, it can be ``floated'' by removing the Gaussian constraint. This  allows a free normalization of the background in the fit.



%% file: background.tex
\renewcommand{\baselinestretch}{1.6}
\normalsize

 \chapter{Estimation of the  radioactive background}
\label{chap-bgr}
\begin{figure}
\centering
\includegraphics[width=15.5cm]{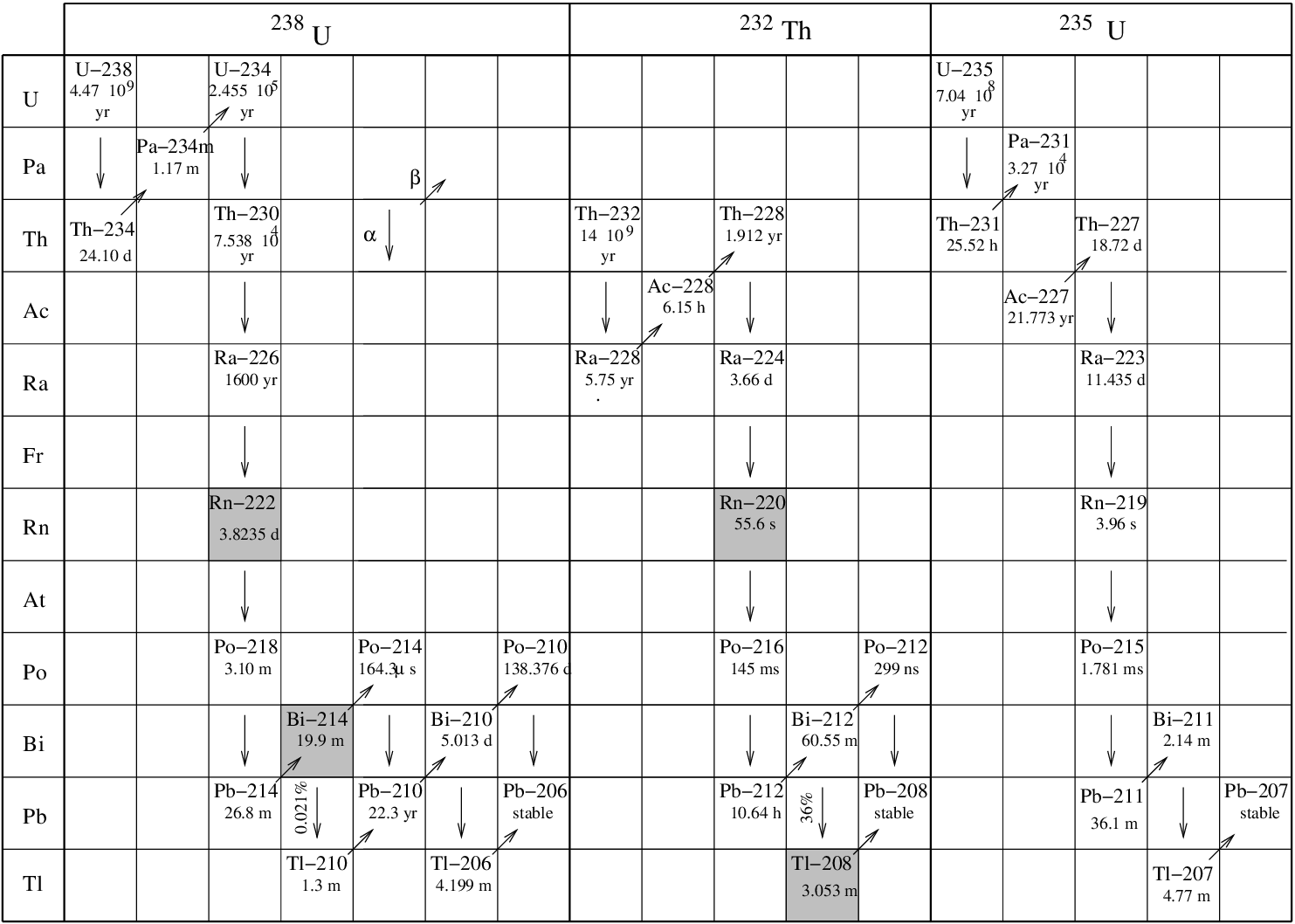}
\caption[The three main decay chains of natural radioactivity (uranium-238, thorium-232 and uranium-235)]{The three main decay chains of natural radioactivity (uranium-238, thorium-232 and uranium-235)~\cite{technical}. The two former are the main source of background to double beta decay in NEMO~3. The isotopes in grey produce the main background to neutrinoless double beta decay.}
\label{fig-decaychain}
\end{figure}

This chapter describes the measurement of the radioactive backgrounds to double beta decay of $^{150}$Nd in NEMO~3. The  dominant decay channels of each background are discussed. 
The backgrounds to $0\nu\beta\beta$ are divided into  $2\nu\beta\beta$ background and radioactive backgrounds that can  mimic double beta decay. The former is discussed in Chapter~\ref{chap-2nbb}.  
The latter, which are also backgrounds to $2\nu\beta\beta$,  are   divided into two categories: internal and external. The internal backgrounds are due to radioactive contamination inside the NEMO~3 source foils, whereas  the external background is defined as events created by the radioactivity located in the detector's components and the material surrounding the detector. Most  contamination  comes from the two  main  natural decay chains of uranium ($^{238}$U) and thorium ($^{232}$Th). These decay chains are shown in  Figure~\ref{fig-decaychain}. The detector components and the source foils are also contaminated with potassium-40~($^{40}$K), which is a naturally occurring radionuclide. 
 
\section{The internal background of  the $^{150}$Nd foil}
\begin{figure}[tb]
\centering
\includegraphics[width=16.0cm]{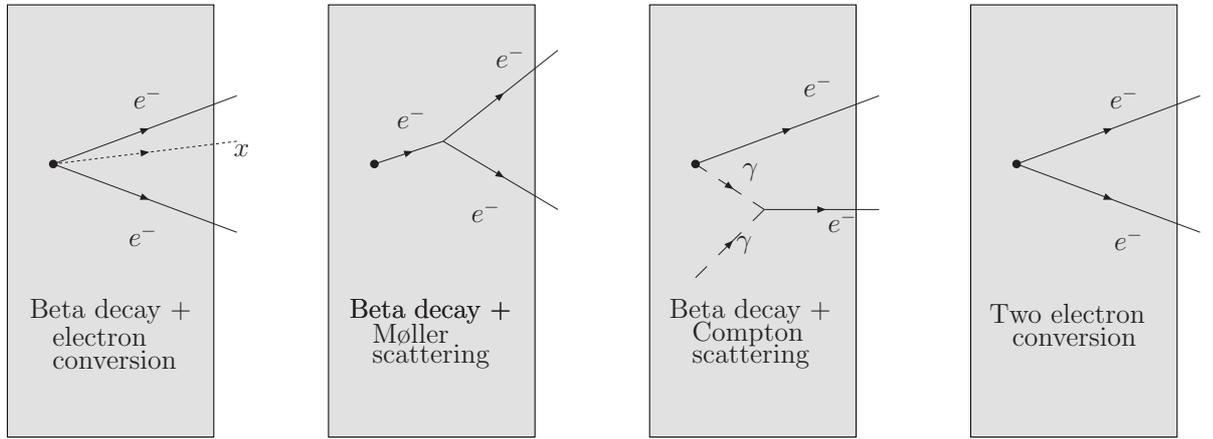}
\caption{ Four ways that the internal contaminants inside $^{150}$Nd source foil can mimic double beta decay. }
\label{fig:internal}
\end{figure}

The internal contaminants can mimic double beta decay of $^{150}$Nd in four ways:
\begin{itemize}
\item A $\beta$ decay accompanied by an electron from an electron conversion process. The latter happens when a photon from an excited daughter nucleus interacts with an electron in one of the inner electron shells, causing the electron to be ejected from the atom.  
\item M$\o$ller scattering of a $\beta$-decay electron in the foil which leads to emission of two electrons.
\item A $\beta$-decay accompanied by a photon. In this case a second electron  is produced via a Compton scattering. 
\item  Two-electron conversion processes due to the de-excitation of the daughter  nuclei of  $^{207}$Bi, $^{152}$Eu  and $^{154}$Eu contaminants in the $^{150}$Nd foil. 
\end{itemize}
Figure~\ref{fig:internal} shows the schematic of 
 each of these four processes that can give two electrons with sufficient energy from a single vertex.

The internal contaminants vary in the different source foils of  NEMO~3. The level of contamination was obtained (after the isotope purification) from measurements using a low-background high purity germanium (HPGe) detector~\cite{technical}.  This HPGe detector was able to detect photons from de-excitation of the daughter isotopes.  The energy and the intensity of the photons indicated the contaminant and its abundance in the foil.

The HPGe radioactivity measurements for the Nd$_{2}$O$_{3}$ source foil are shown in Table~\ref{table-hpge}. The half-life ($T_{1/2}$) and the nuclear transition energy ($Q$) are also included for each element. 
\begin{table}[htb]
\scriptsize
\centering
\begin{tabular}{|p{0.56in}|p{0.41in}|p{0.39in}|p{0.34in}|p{0.37in}|p{0.39in}|p{0.30in}|p{0.65in}|p{0.52in}|p{0.52in}|}
\hline
          & \multicolumn{2}{c|}{$^{238}$U chain} &  \multicolumn{3}{c|}{$^{232}$Th chain}&  $^{235}$U chain& $^{40}$K & $^{152}$Eu &  $^{207}$Bi\\
\cline{2-6}
          &  $^{214}$Pb $^{214}$Bi & $^{234}$Th & $^{228}$Ac & $^{208}$Tl & $^{212}$Pb &  & & & \\ 
\hline
$T_{1/2}$         &  26.8~min 19.9~min & 24.10~d &6.15~h &3.05~min &10.64~h & ~~-- &1.25$\times$10$^{9}$~y & 13.54~y& 32.9~y\\
\hline
$Q$~(MeV)         & 1.02 3.27      & 0.27 & 2.13     &4.99 &0.57 & ~~-- &1.31 & 1.82 & ~~--  \\ 
\hline
 Activity (mBq/kg)& $< 3.0$ & $<66$  & $20 \pm 7$  &$10 \pm 2$ & $30\pm5$ & $<1$ &$<70$ &  $40 \pm 5$ & $130 \pm 5$ \\ 
 \hline         
\end{tabular}
\label{table-hpge}
\caption[Radioactivity measurements of Nd$_{2}$O$_{3}$ source foils performed by HPGe  detector, after  purification]{Radioactivity measurements of Nd$_{2}$O$_{3}$ source foils performed by the HPGe  detector, after  purification~\cite{technical}. }
\end{table}
All of these contaminants are $\beta$ emitters, except $^{207}$Bi.
 The $^{214}$Bi  and $^{208}$Tl with high $Q$ value  are the main radioactive background to $0\nu\beta\beta$ of $^{150}$Nd. The other elements given in Table~\ref{table-hpge} can mimic $2\nu\beta\beta$ events, except $^{234}$Th and $^{212}$Pb whose  $Q$ value is not high enough to mimic  $2\nu\beta\beta$. However, $^{234}$Th decays to $^{234m}$Pa  which  has  a half-life of $1.17$~min and undergoes a $\beta$ decay with $Q$ value of $2.29$~MeV  and  is thus a background to $2\nu\beta\beta$. The same statement is true for $^{212}$Bi which is a daughter isotope of $^{212}$Pb and has $Q$ value of $2.25$~MeV and half-life of $60.5$~min.

The HPGe measurements show that the $^{150}$Nd foil is contaminated with $^{207}$Bi
and $^{152}$Eu atoms. Both of these isotopes are artificial as they are not found in nature. Thus it is likely that the equipment used for $^{150}$Nd enrichment was polluted with small amounts of this isotope.
Europium-152 is created from  natural europium and is usually produced together with $^{154}$Eu (natural europium is a  mixture of $^{151}$Eu and $^{153}$Eu)~\cite{eu_154d,eu152eu154}. 
This means, in addition to $^{152}$Eu atoms, there should be some $^{154}$Eu  contamination, which the HPGe detector was not able to measure.
Europium-154 has  very similar decay schemes and half-life as $^{152}$Eu. ($Q=1.97$~MeV and $T_{1/2}$=8.6~y). Both isotopes can  undergo either  $\beta$ decay (branching ratio of $27\%$) or an electron capture ($73\%$).



The  HPGe detector  measurements are not definitive and are performed for a small sample of $^{150}$Nd  and before installing the foil inside the NEMO~3 detector. 
In order to have a correct estimation of each internal radioactive background,   precise measurements of the activities with NEMO~3 data need to be  performed.

Background contaminants  often decay in multiple channels (and not only the two-electron channel). It is vital to measure the background activities in channels which are independent of the $2\nu\beta\beta$ signal.
The activity of each background is determined using the decay channel with the highest branching ratio. The activity (A) of each contaminant (in Becquerel) is defined by the number of decays per second,
\begin{equation}
A=\frac{(N_{data}-N_{bgr})}{t\epsilon},
\end{equation}
where $N_{data}$ is the number of  data events found for the decay channel, $N_{bgr}$ is the number of background events to a specific contaminant, $t$ is the total data taking period in seconds and $\epsilon$ is the efficiency of the event selection criteria, which is found by the ratio of simulated events that pass the selections to the total number of  generated events. 
\subsection{Electron-photon channel}
\label{sec-eg}
\begin{figure}
\centering
\includegraphics[width=7.0cm]{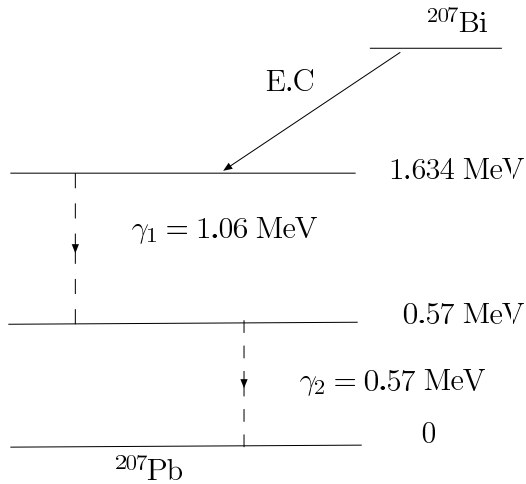}
\caption{A  decay scheme of $^{207}$Bi.}
\label{fig:bi207decay}
\end{figure}

The isotopes $^{207}$Bi, $^{152}$Eu and $^{154}$Eu  decay predominantly to an electron and a photon.  $^{207}$Bi can undergo electron capture  and decay to the excited state of $^{207}$Pb. In this case the photon is produced by the de-excitation of the excited $^{207}$Pb and the electron is produced  by the internal conversion process~\cite{bi207decay}.  Figure~\ref{fig:bi207decay} shows a decay scheme of $^{207}$Bi, in which only the most intense transitions are shown. This decay scheme shows that the electron and photon energy   peaks are at  0.57~MeV and 1.06~MeV. In the NEMO~3 experiment, due to energy loss in the foil, the energies of these peaks  are slightly shifted. 

 $^{152}$Eu  can either undergo a $\beta$ decay  ($27\%$) or electron capture   ($73\%$) to produce several excited states  of $^{152}$Gd or $^{152}$Sm. These daughter isotopes  de-excite to the ground states resulting in the production of  photons of various energies~\cite{eu_152}.  Similarly, $^{154}$Eu  undergoes $\beta$ decay to produce $^{154}$Gd~\cite{eu_154d} .

 $^{208}$Tl decays predominantly into  one electron and two photons; however, it is possible that only one of the photons is detected and thus the activity of this isotope can be checked using the electron-photon channel. Further details of this isotope's decay scheme  are given in   Section~\ref{sec-egg}.

\begin{figure}
\centering
\subfigure{
a)\includegraphics[width=6.9cm]{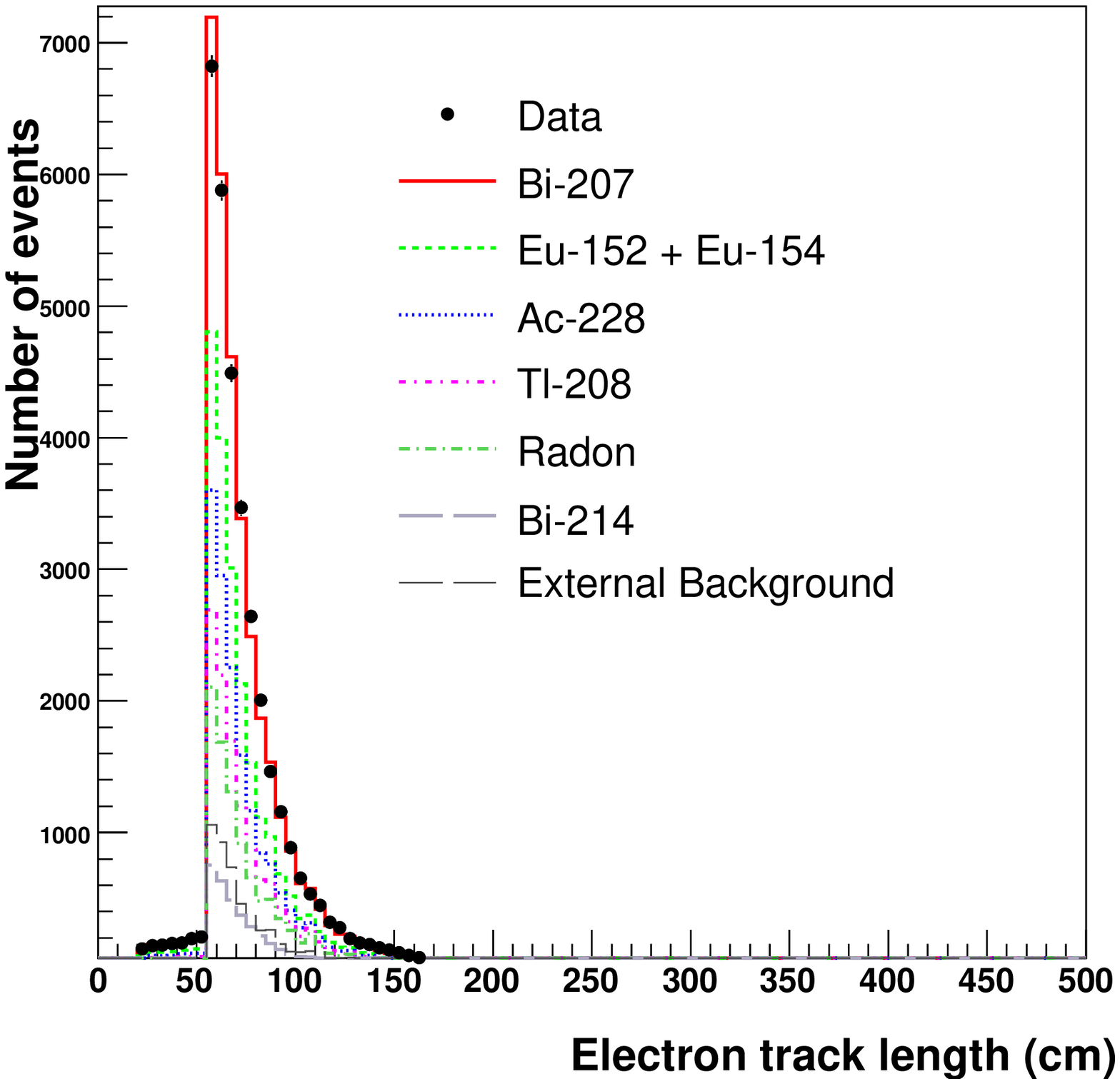}}
\subfigure{
b)\includegraphics[width=6.9cm]{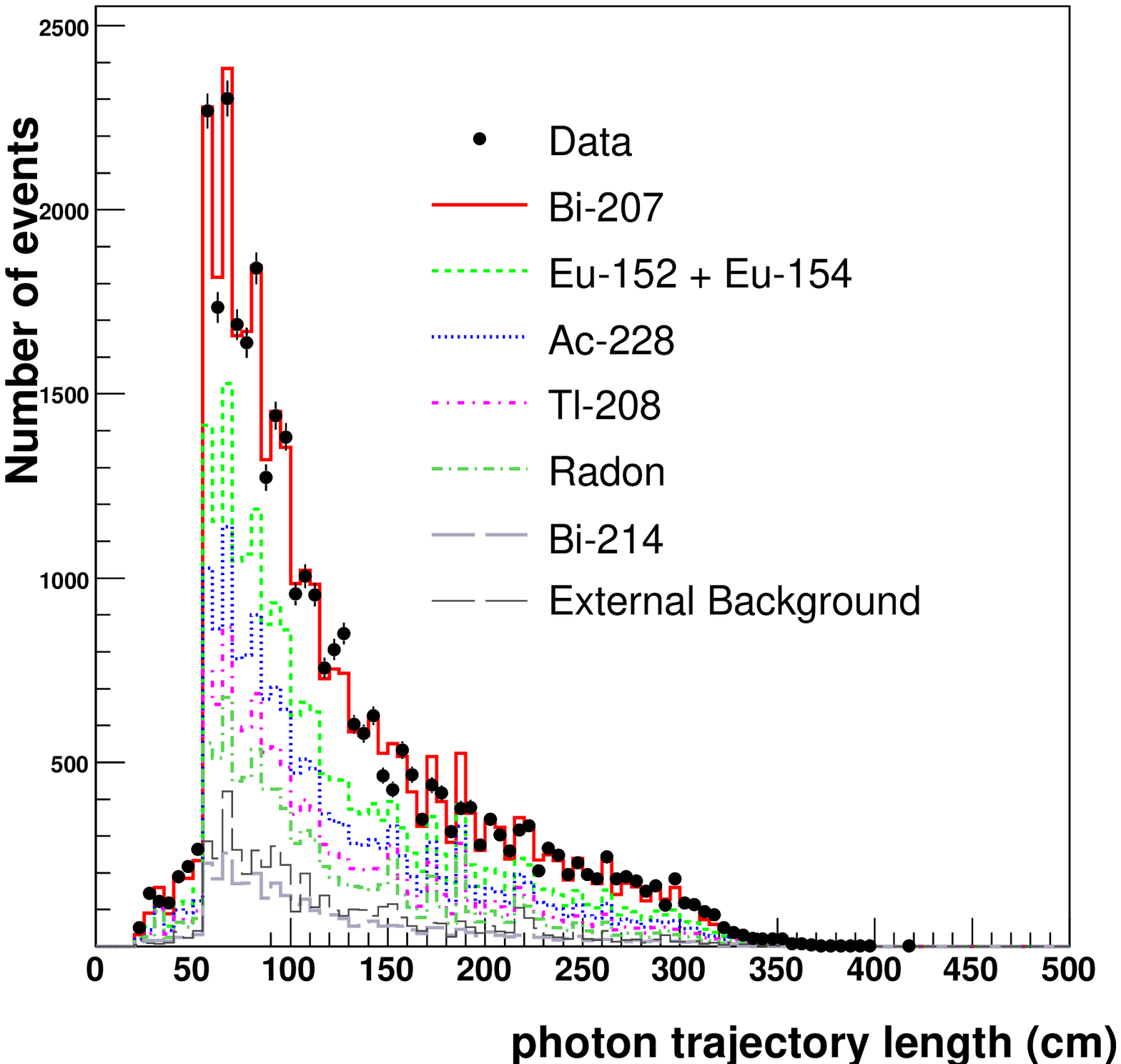}}
\caption[The trajectory length of the electron and the photon for $e\gamma$ events]{a) The track length of the electron  and b) trajectory length of the photon in $e\gamma$ events. The statistical uncertainties on the data points are shown with error bars. }
\label{fig-eglength}
\end{figure}

\begin{figure}
\centering
\includegraphics[width=10.0cm]{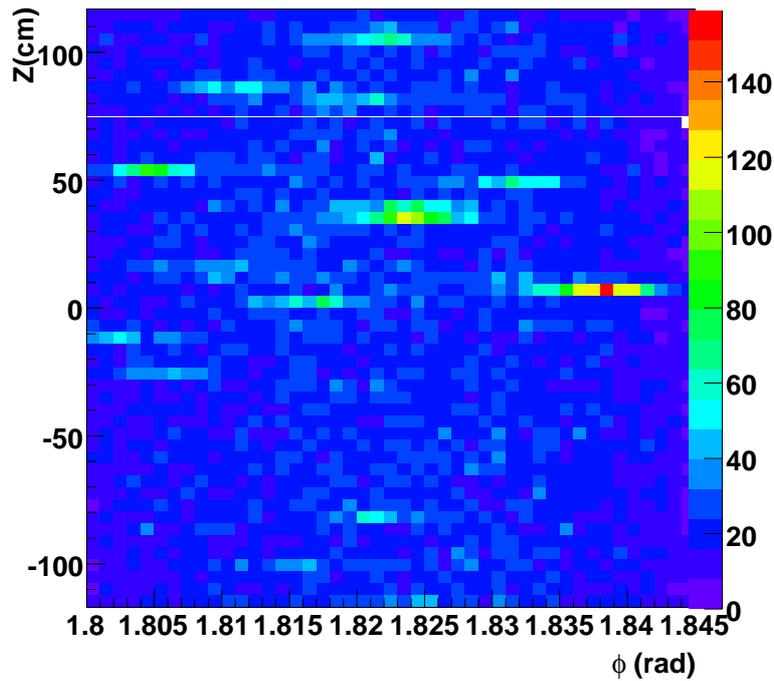}
\caption[Distribution of event vertices in the $^{150}$Nd source foil for $e\gamma$ events]{Distribution of event vertices in the $^{150}$Nd source foil for $e\gamma$ events. Hot-spots are  seen in the plot.}
\label{fig-hsbi207}
\end{figure}

\begin{figure}
\centering
\subfigure{
a)\includegraphics[width=6.9cm]{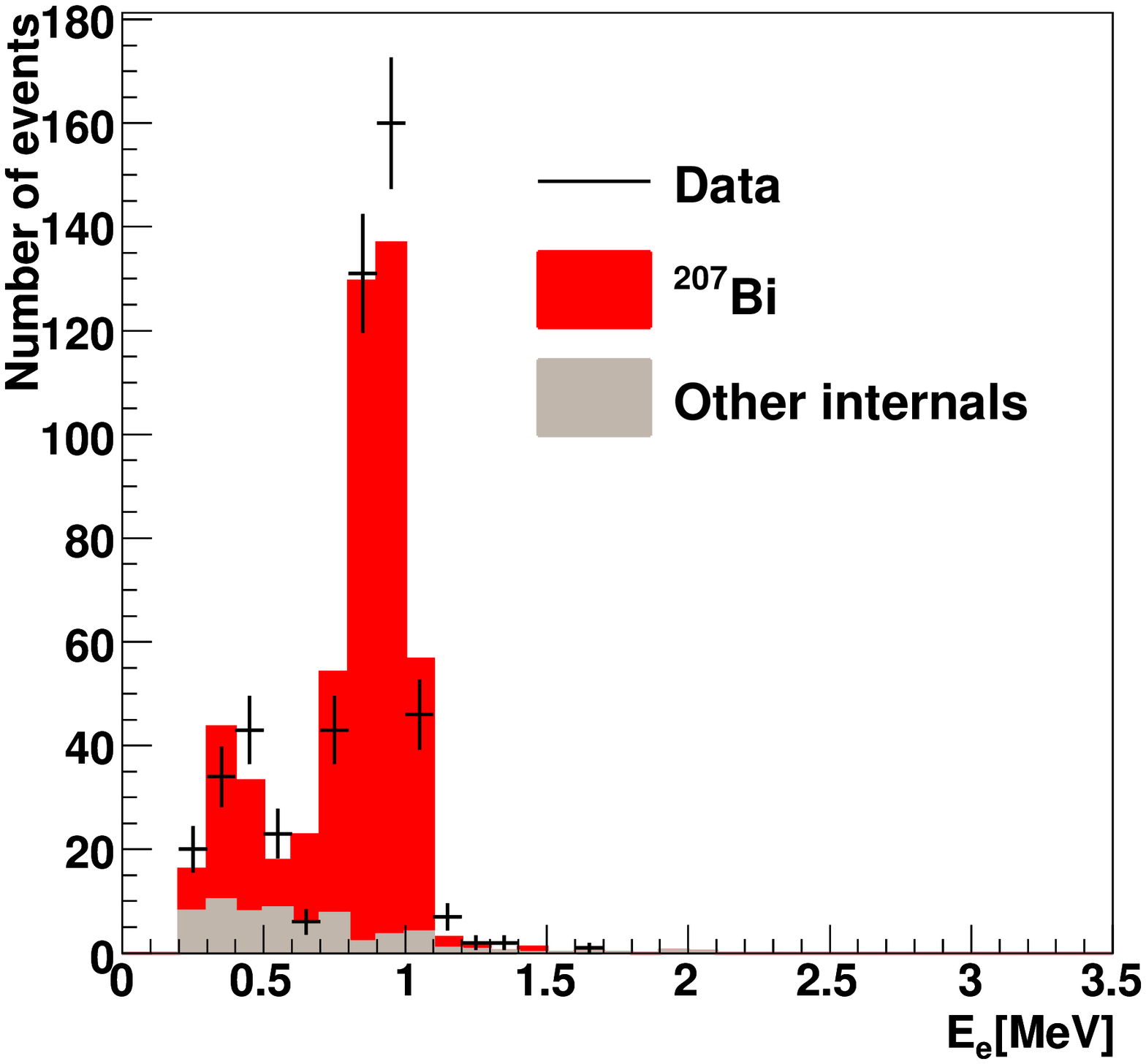}}
\subfigure{
b)\includegraphics[width=6.9cm]{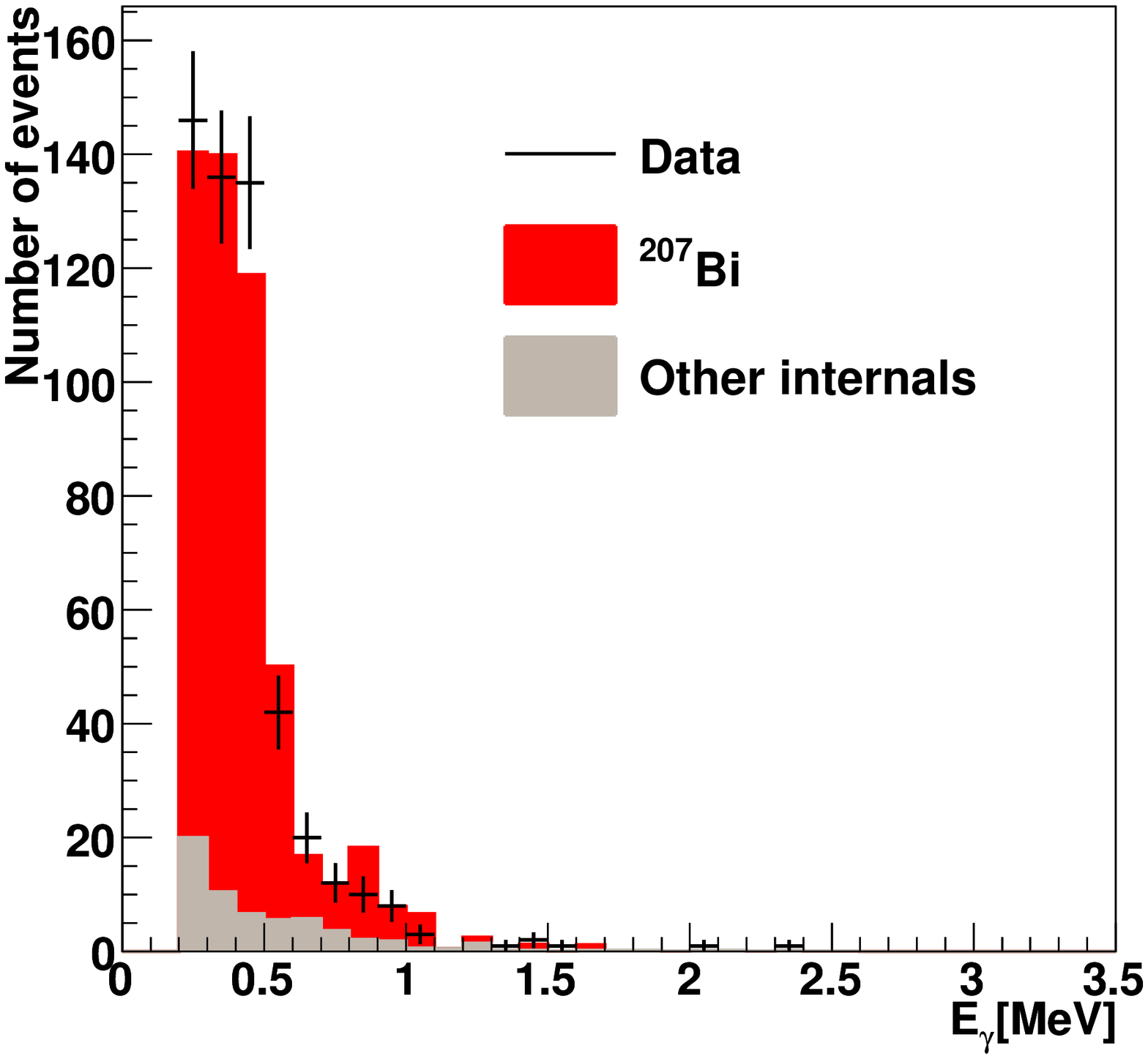}}
\caption[The energy of the electron and the  photon for  $e\gamma$ events originating from the hot-spot region]{The energy of a) the electron and b) photon for  $e\gamma$ events originating from the hot-spot region. The components of total background to $^{207}$Bi (shown in grey) are normalised to activities found from $e\gamma$ and other decay channels. The statistical uncertainties on the data points are shown with error bars. }
\label{fig-insidehs}
\end{figure}

To select electron-photon ($e\gamma$) events originating from the foil the following event selection criteria are applied:
\begin{itemize}
\item Only one electron track associated to a scintillator hit with deposited energy greater than 0.2~MeV.
\item The vertex of the $e\gamma$ event, which is found by extrapolating the electron track to the foil, is reconstructed inside the $^{150}$Nd foil (defined in Section~\ref{sec-ndg}). 
\item The track has at least one hit in one of the first two layers of drift cells.
\item Only one scintillator block is found which is not associated with a track, with energy greater than 0.2~MeV.
\item The internal TOF hypothesis  is greater than $4\%$, and the external TOF hypothesis is less than $1\%$.
\item The energy sum of all other scintillator deposits is  less than 0.15~MeV\@. This requirement  rejects events with more than one photon.   
\item The TOF calculation is less precise for particles with short trajectories. Thus the events with  photon and electron trajectory lengths less than 50~cm are removed.  The photons and electrons with   trajectory length  less than 50~cm are detected by the scintillators situated on top and bottom of the detector.
  Figures~\ref{fig-eglength}a and~\ref{fig-eglength}b show the electron track length and photon trajectory length before applying this requirement, respectively.  Photons can have longer trajectory lengths  than electrons because they  can cross  boundaries of  sector~5 and be detected by the scintillators in the neighbouring sectors. 
\end{itemize}
Figure~\ref{fig-hsbi207} shows the reconstructed $z$ and $\phi$ components of the vertex in the $^{150}$Nd foil for the $e\gamma$ events that pass the event selection criteria. Several hot-spots are seen in this figure. The two main hot-spots are in  the regions:
\begin{eqnarray}
1.82<\phi<1.827~{\rm rad}~~~{\rm and}~~~34<z<42~{\rm cm}\nonumber,\\
1.836<\phi<1.842~{\rm rad}~~~{\rm and}~~~6<z<12~{\rm cm}.~\label{eq-bi207hs}
\end{eqnarray}
Figures~\ref{fig-insidehs}a and~\ref{fig-insidehs}b  show  the electron and photon energies inside the hot-spot region. The normalisations of the $^{207}$Bi energy distributions are found by fitting simulated $^{207}$Bi events to data after background subtraction. 
 The  $^{207}$Bi activity is found to be ten times higher in these hot-spots and therefore this
   leads to the conclusion that these hot-spot areas are contaminated with $^{207}$Bi and  they are therefore removed from  the subsequent analysis. 

\begin{figure}
\centering
\includegraphics[width=10.0cm]{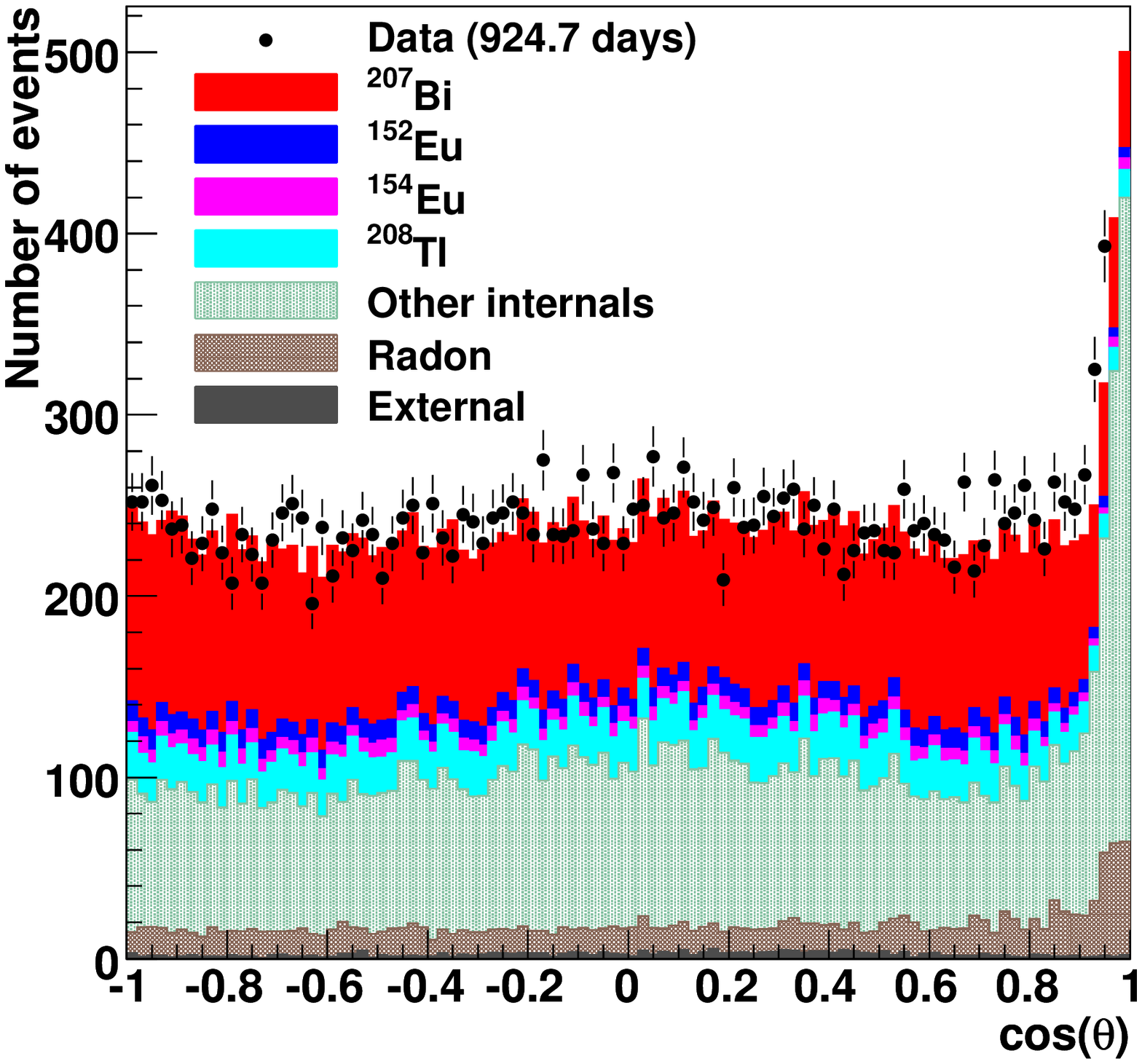}
\caption[he cosine of the angle between the electron and the photon in the $e\gamma$ channel]{The cosine of the angle between the electron and the photon in the  $e\gamma$ channel. The peak in  cosine of the angle more than 0.9 is mainly due to the   scattered single-electron events which  are shown in dashed  green.  The statistical uncertainties on the data points are shown with error bars.}
\label{fig-coseg-eg}
\end{figure}

\begin{figure}
\centering
\includegraphics[width=11cm]{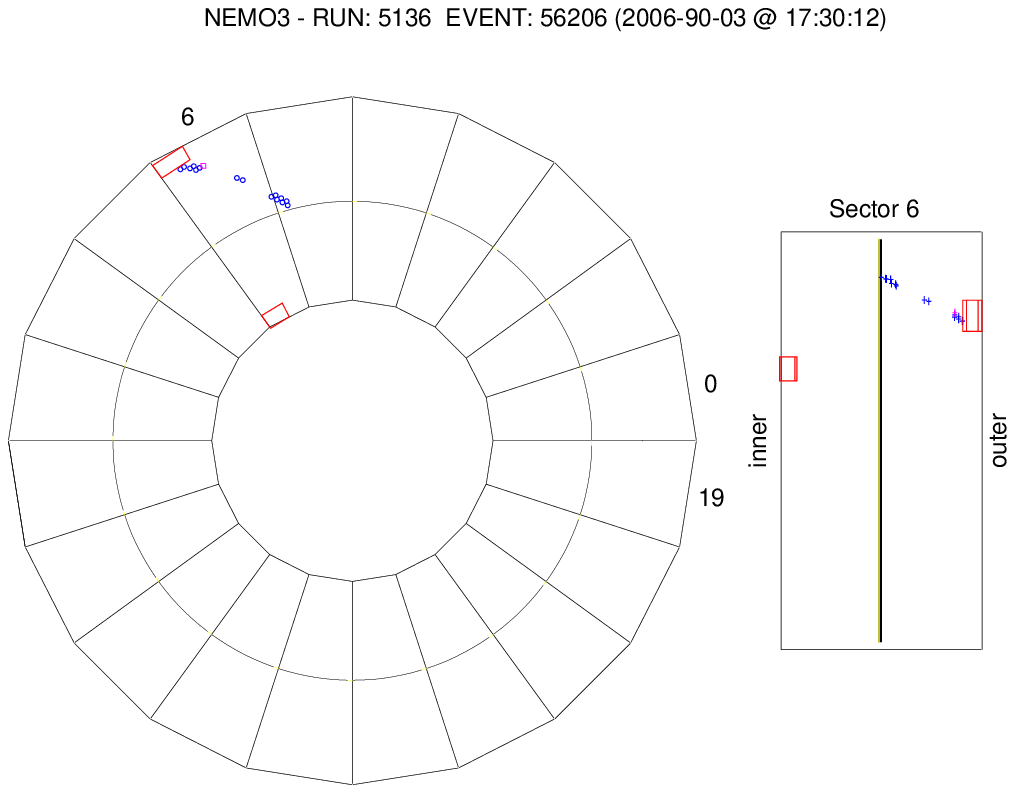}
\caption[A top and side view of an  $e\gamma$ event  originating from $^{150}$Nd foil in Sector~5]{A top and side view of an  $e\gamma$ event  originating from $^{150}$Nd foil in Sector~5.   This event is from data taken in  August 2006.} 
\label{fig-displayeg}
\end{figure}

Figure~\ref{fig-coseg-eg} shows the cosine of the angle between the electron and the photon. A single electron from a $\beta$ decay can be scattered by a scintillator block and create a photon which is detected by the neighbouring scintillator.  Thus  events are rejected if the cosine of the  angle between the photon and the electron  is more than 0.9. 
An $e\gamma$ event originating from the $^{150}$Nd foil is displayed in Figure~\ref{fig-displayeg}. The intermediate circle illustrates the source foil and the blue hits illustrate the electron track. The scintillator hits are shown in red.
\subsubsection{\boldmath{$^{208}$}Tl activity measurement}
\begin{figure}
\centering
\includegraphics[width=8.0cm]{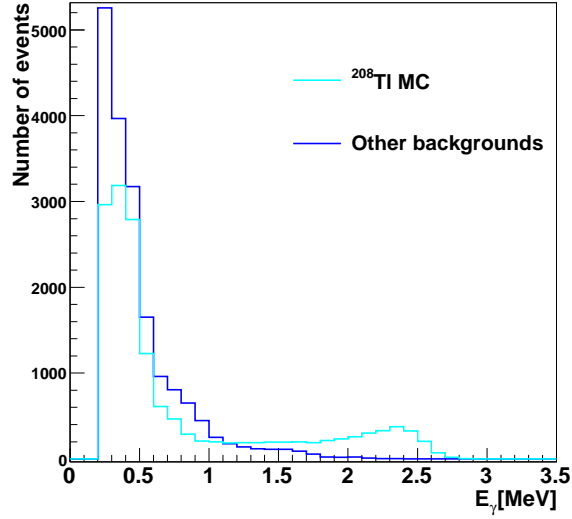}
\caption[The energy of the photon originating from $^{208}$Tl]{The energy of the photon originating from $^{208}$Tl is shown in cyan and the sum of all other background photon energies are shown in dark blue.  }
\label{fig-tl208optcut}
\end{figure}

The activity of $^{208}$Tl is  found in the energy region in which it dominates over all other background. Figure~\ref{fig-tl208optcut} shows  the photon energy for simulated $^{208}$Tl (normalised to an arbitrary number) and the sum of all other internal contaminants.  This plot  demonstrates that in order to measure the activity of $^{208}$Tl in the  electron and photon channel, the energy of the photon is required to  be more than 2.0~MeV to increase signal over background. 
As the normalisation factors of other internal backgrounds  at this stage of the analysis are not known this  cut is applied in order to reduce internal background events from other contaminants.

\begin{figure}
\centering
\subfigure{
a)\includegraphics[width=6.9cm]{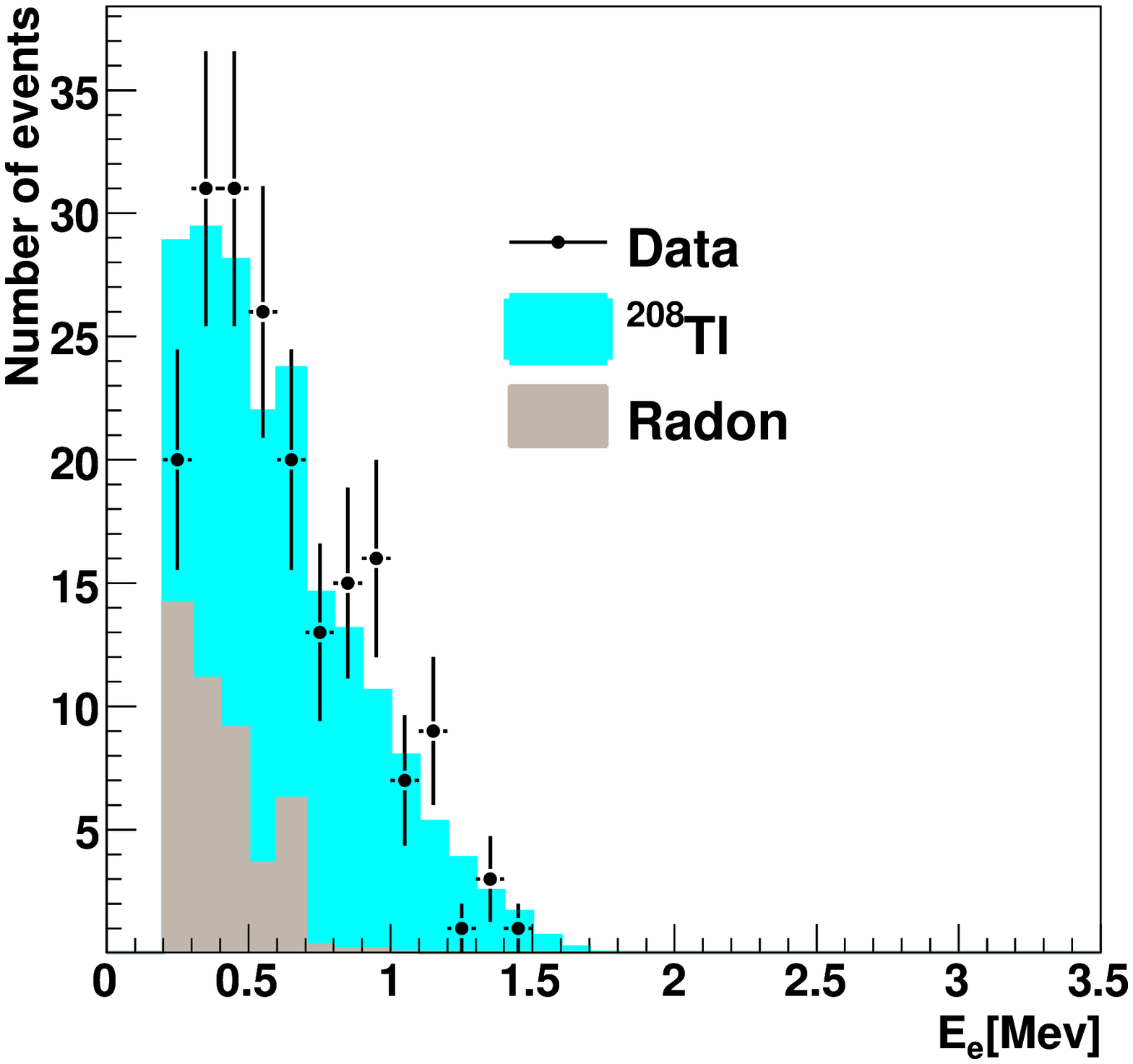}}
\subfigure{
b)\includegraphics[width=6.9cm]{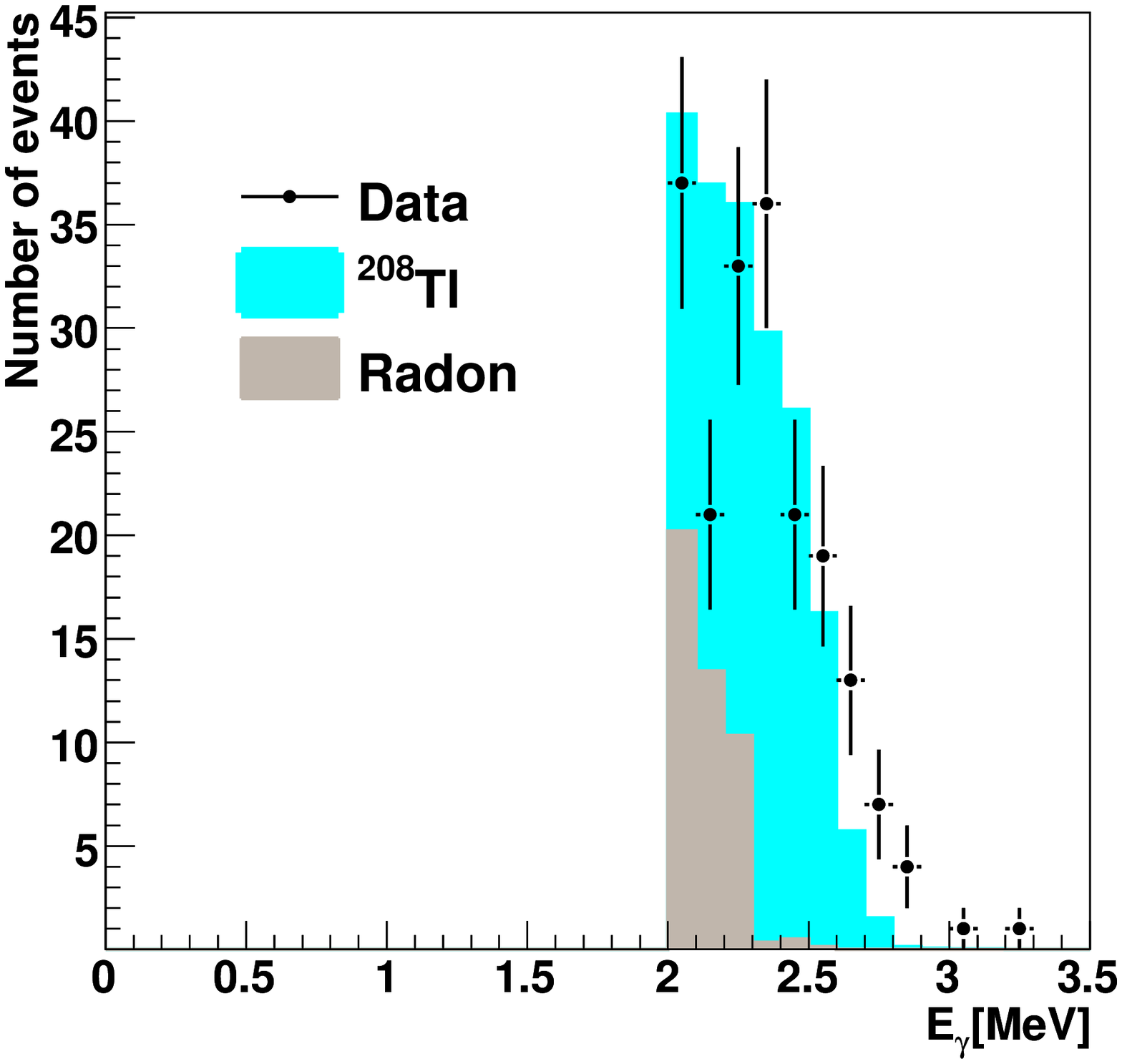}}
\caption[he distributions of electron and photon energies for the events with photon energy more than 2~MeV ]{ The distributions of the a) electron energy and  b) photon energy for the events with photon energy more than 2~MeV. The error bars show the statistical uncertainties on the data points. }
\label{fig-tl208energy}
\end{figure}

After applying the photon energy requirement, the only background to $^{208}$Tl which remains in the data sample is the radon from the tracker (described in Section~\ref{sec-radon}). The average activity of this background, which was measured by studying the electron-alpha channel~\cite{externalbkg}, is $0.45\pm0.07$~mBq.
Figures~\ref{fig-tl208energy}a and~\ref{fig-tl208energy}b show  the photon and the electron energies for data and  the simulated events  with photon energy more than 2~MeV\@.  In these figures, the distributions for simulated $^{208}$Tl and for data   minus radon background are normalised to the same number of events.  The event selection efficiency for $^{208}$Tl MC is $(0.390\pm0.004~{\rm(stat)})$\%. The number of data events that  pass the event selection criteria is $193$ of which $45.1\pm9.4$ events are estimated to be radon background.  This gives  a $^{208}$Tl activity of 
\begin{equation}
\label{eq-tl208eg}
A({\rm^{208}Tl})=0.47\pm0.05~{\rm(stat)~mBq}, 
\end{equation} 
Since the $^{150}$Nd source foil has a mass 50.7~gr, this value corresponds to 
\begin{equation}
 A^{\prime}=A/m=9.27\pm 1.02~{\rm(stat)~mBq/kg}.
\end{equation}
The isotopes~$^{208}$Tl, $^{212}$Bi and $^{228}$Ac are all in the decay chain of $^{232}$Th.  $^{212}$Bi can emit an $\alpha$ particle   and decay to $^{208}$Tl with a  branching ratio of 36\%~\cite{datasheetbi212}. Thus   the activities of $^{228}$Ac and $^{212}$Bi are 36\% of the  $^{208}$Tl activity.
The systematic uncertainty on this measurement and the final results of $^{208}$Tl activity will be given in Section~\ref{sec-systback}.
\subsubsection{\boldmath{$^{207}$}Bi activity measurement}
\begin{figure}
\centering
\subfigure{
a)\includegraphics[width=6.9cm]{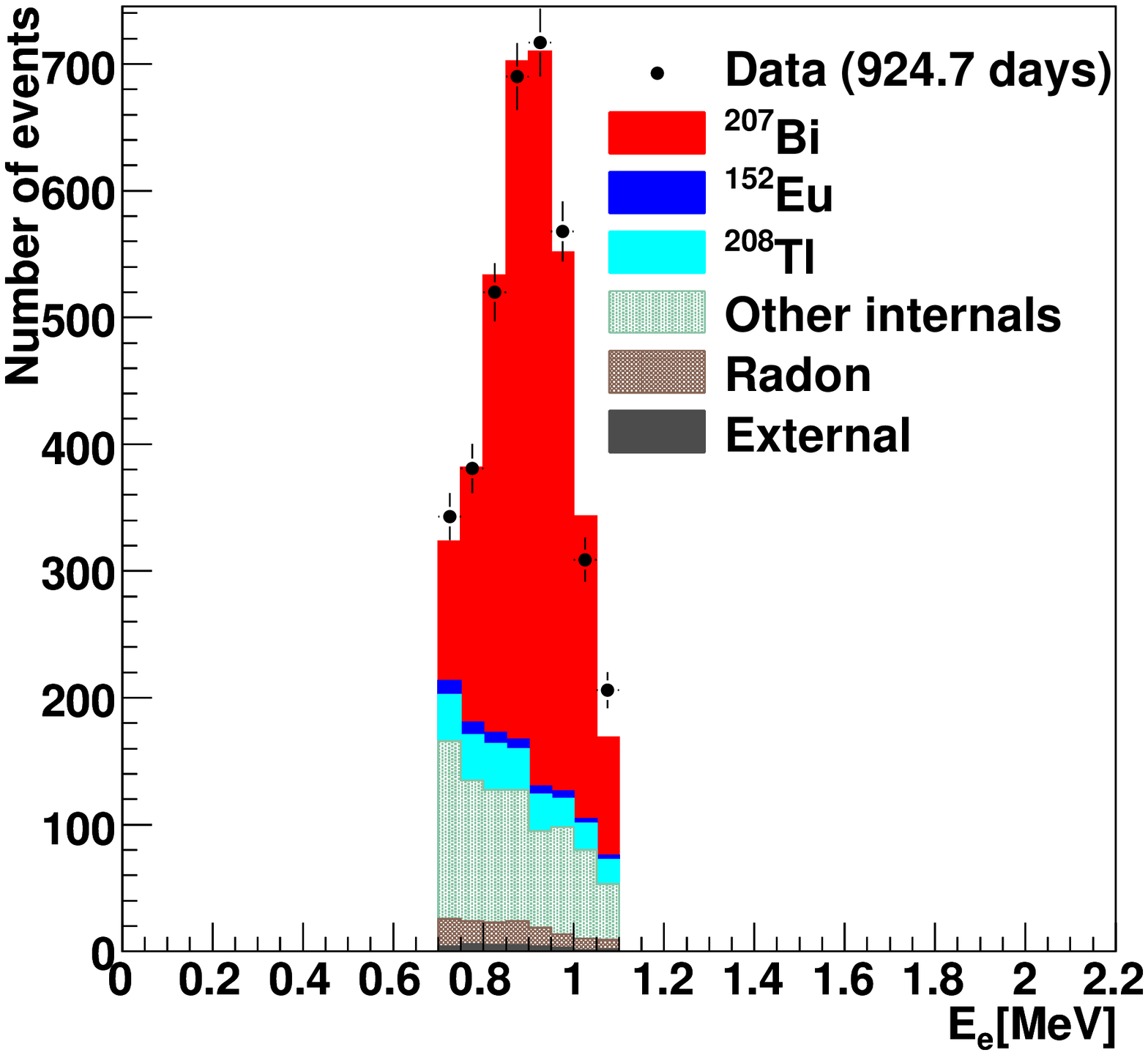}}
\subfigure{
b)\includegraphics[width=6.9cm]{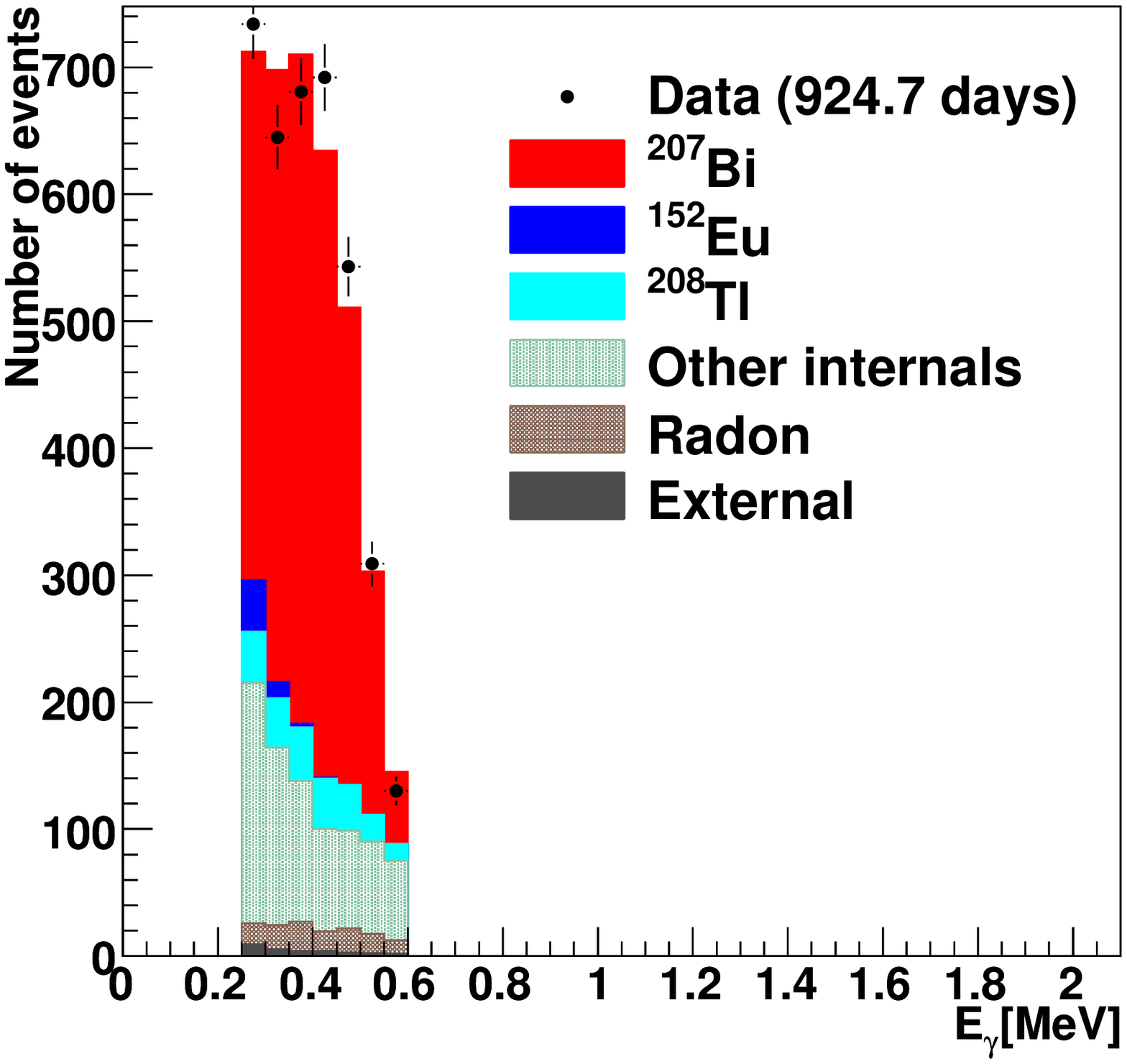}}
\caption[Electron and photon energy distributions for  energies  $0.7<E_{e}<1.1$~MeV and  $0.7<E_{e}<1.1$~MeV ]{Distribution of the energy of the a) electron and  b) photon energy for energies  $0.7<E_{e}<1.1$~MeV and  $0.7<E_{e}<1.1$~MeV.   The error bars show the statistical uncertainties on the data points.}
\label{fig-energy-bi207}
\end{figure}

To find  the $^{207}$Bi activity, the following energy selections  are applied:
\begin{eqnarray}
0.7<E_{e}<1.1~{\rm MeV},\\
0.25<E_{\gamma}<0.6~{\rm MeV}.
\end{eqnarray}
In these energy regions,  $^{207}$Bi dominates over all other background.
Figures~\ref{fig-energy-bi207}a and~\ref{fig-energy-bi207}b show the electron and photon energies in these energy regions. The simulated $^{207}$Bi is normalised to data minus the total number of other background events.  These backgrounds to $^{207}$Bi  are normalised to the  activities found by  studying different energy regions ($^{208}$Tl) or other decay channels~(Sections~\ref{sec-1e} and~\ref{sec-external}).  The number of candidate events is $3734$. The number of estimated background events is $1217\pm31$~(stat). The efficiency is equal to $(0.271\pm0.001~{\rm (stat)})$\%. Thus the activity of $^{207}$Bi is estimated to be 
\begin{eqnarray}
A({\rm^{207}Bi})=11.7\pm0.3~{\rm(stat)~mBq}, \\
 A^{\prime}=230.8\pm 6.2~{\rm(stat)~mBq/kg}.
\label{eq-bi207}
\end{eqnarray}
\begin{figure}[h]
\centering
\subfigure{
a)\includegraphics[width=6.9cm]{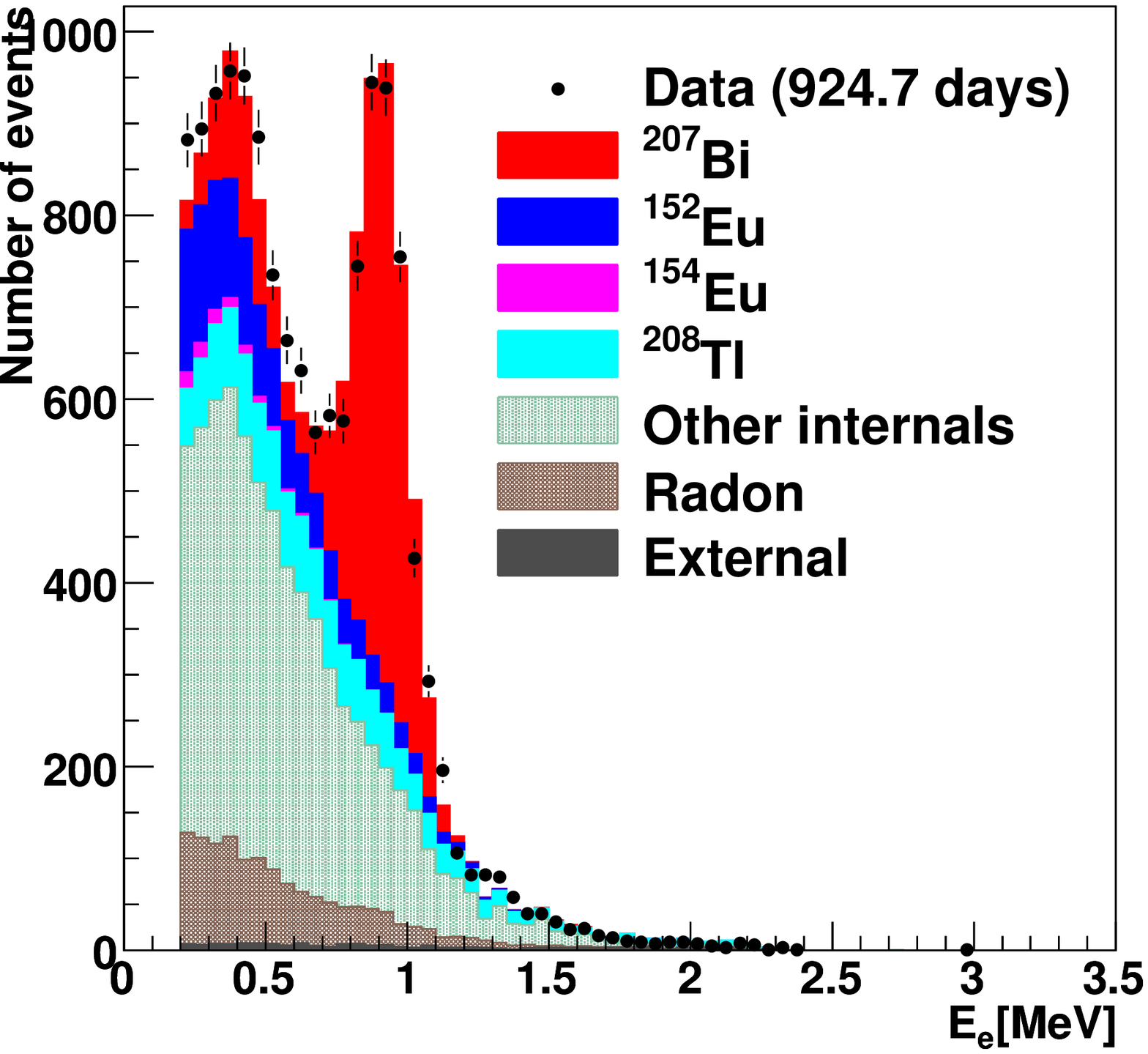}}
\subfigure{
b)\includegraphics[width=6.9cm]{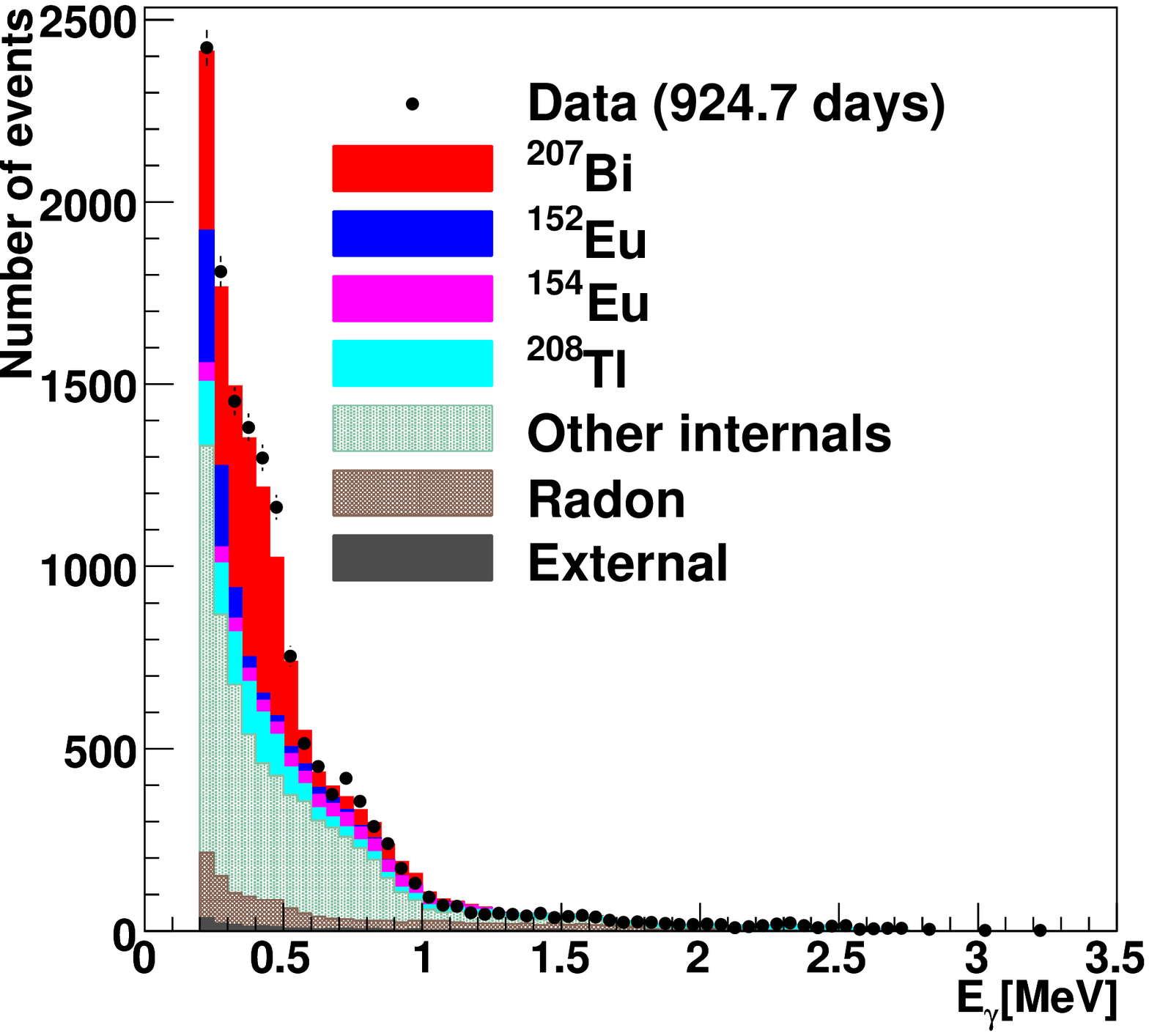}}
\subfigure{
c)\includegraphics[width=6.9cm]{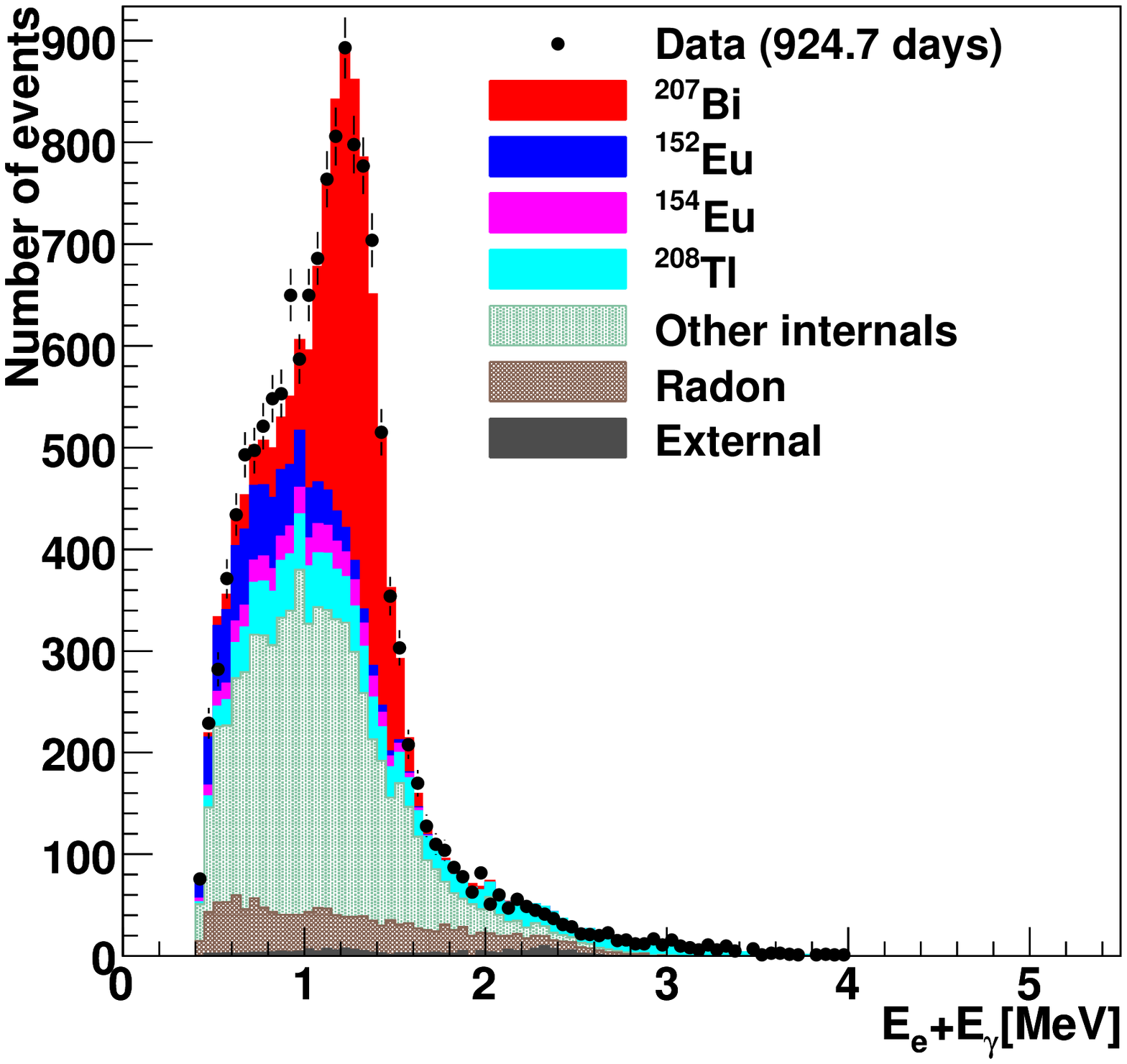}}
\caption[Electron and photon energies for the events passed the event selection criteria]{Distribution of the energy of the a) photon,  b) electron and c) electron plus photon for the events that pass the event selection criteria. The statistical uncertainties on the data points are shown with  error bars.}
\label{fig-eg-total}
\end{figure}
\subsubsection{\boldmath{$^{152}$}Eu and \boldmath{$^{154}$}Eu activity measurements}
The activities of  $^{152}$Eu and $^{154}$Eu are found by fitting their simulated photon energy spectrum to data minus all other background events simultaneously. A binned maximum likelihood  fit as  described in Section~\ref{sec-fit} is applied.  The fit is applied to the full range of the photon energy distribution. The total number of  $e\gamma$ data events is $14230$. The $^{207}$Bi and $^{208}$Tl activities are normalised to the activities given in Equations~\ref{eq-tl208eg} and ~\ref{eq-bi207}. The total number of  background events is $12752\pm 69$~(stat).  By applying the fit, the number of $e\gamma$ events from $^{152}$Eu and $^{154}$Eu are found to be $834\pm 37$~(stat) and $643\pm35$~(stat)  and  the  efficiencies for selecting $e\gamma$ events from these isotopes' decays are ($0.380\pm0.005~{\rm(stat)})$\% and $(0.630\pm0.005~{\rm(stat)})$\%, respectively. The errors on the number of $^{152}$Eu and $^{154}$Eu events are determined by the fit. The activities of $^{152}$Eu and $^{154}$Eu are:
\begin{eqnarray}
A({\rm^{152}Eu})=2.71\pm0.12~{\rm(stat)~mBq} \\
A^{\prime}=53.5 \pm 2.4~{\rm(stat)~mBq/kg},\\
A({\rm^{154}Eu})=1.26\pm0.07~{\rm(stat)~mBq} \nonumber \\
A^{\prime}=24.8 \pm 1.4~{\rm(stat)~mBq/kg}.
\end{eqnarray}
Figure~\ref{fig-eg-total} shows the energy of the electron, photon and the sum of the energies of these two particles. The simulated $^{208}$Tl, $^{207}$Bi, $^{152}$Eu and $^{154}$Eu  are normalised to the activities found in this section. Other background's energy distributions are normalised to activities found from other decay channels. These plots demonstrate that background MC fits data well in the internal $e\gamma$ channel.

\subsection{Electron  plus two photon channel}
\label{sec-egg}
\begin{figure}[b]
\centering
\includegraphics[width=9.0cm]{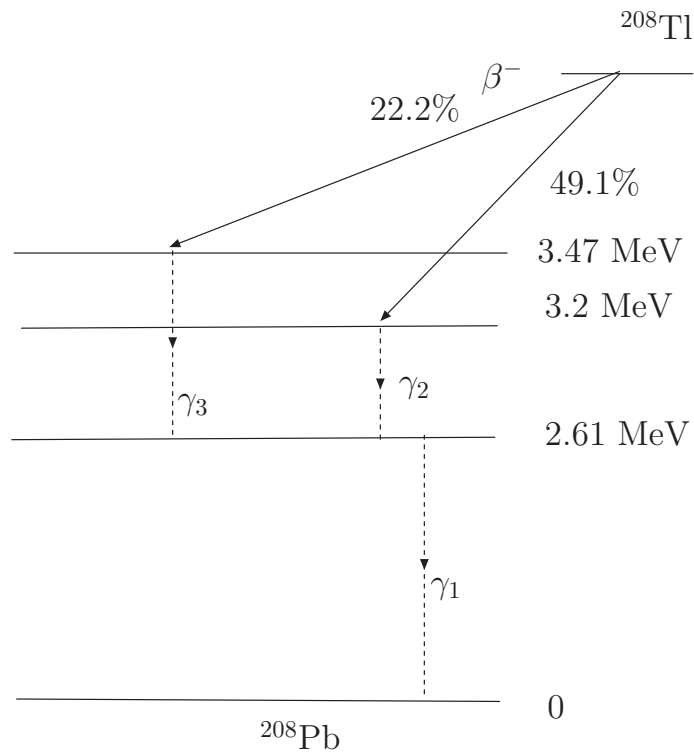}
\caption[A decay scheme of $^{208}$Tl]{A  decay scheme of $^{208}$Tl showing the relevant energy levels which can create $e\gamma\gamma$ events. }
\label{fig-tl208decay}
\end{figure}
\begin{figure}
\centering
\includegraphics[width=8.0cm]{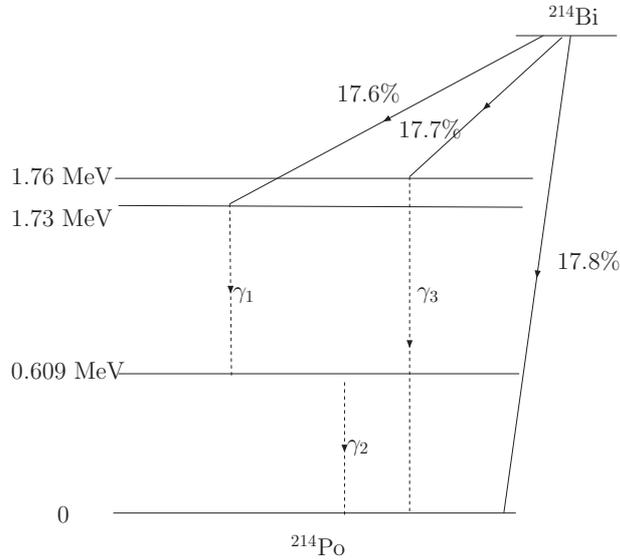}
\caption[A simplified decay scheme of $^{214}$Bi]{A simplified decay scheme of $^{214}$Bi showing the relevant energy levels which can create $e\gamma\gamma$ and $e\gamma$ events. }
\label{fig-bi214decay}
\end{figure}
\begin{figure}[h]
\centering
\includegraphics[width=8.0cm]{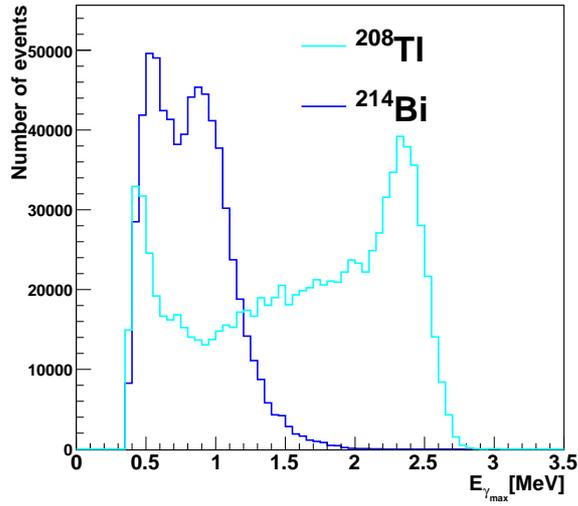}
\caption{Energy distribution of the  higher energy photon for simulated $^{208}$Tl and $^{214}$Bi events. }
\label{fig-bi214-tl208}
\end{figure}
\begin{figure}[h]
\centering
\includegraphics[width=10.0cm]{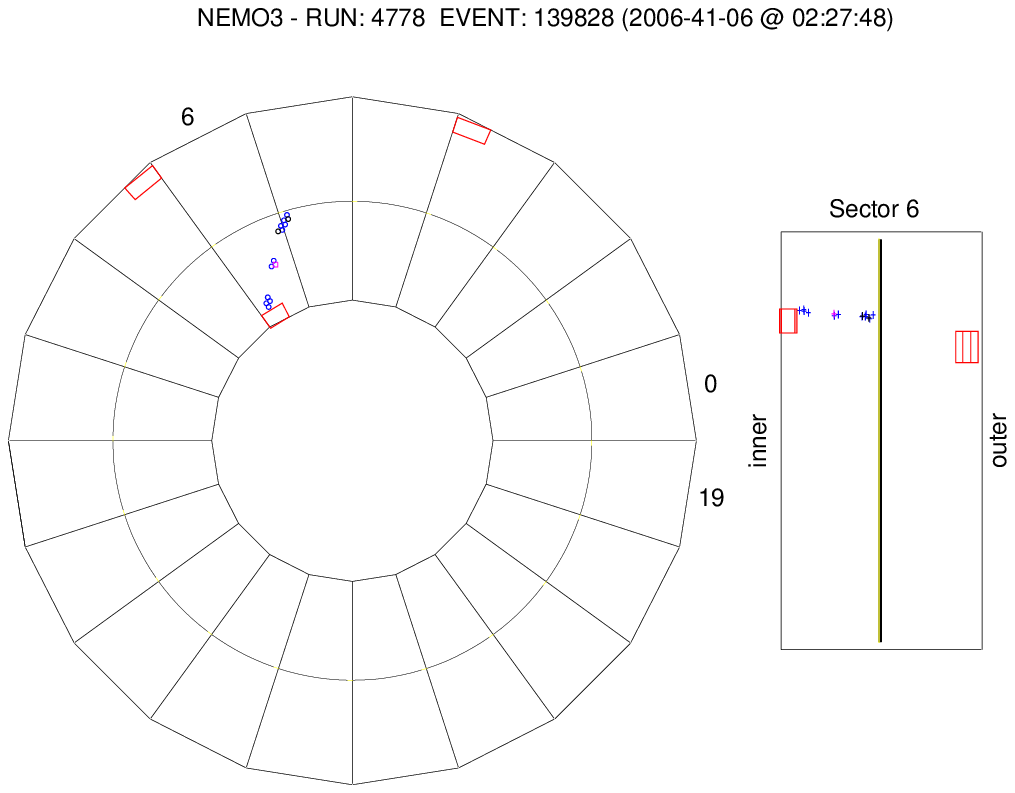}
\caption[An $e\gamma\gamma$ event display]{An $e\gamma\gamma$ event display. A top and side view of the event are shown. This event is from data taken in May 2006.}
\label{fig-eggdisplay}
\end{figure}
\begin{figure}
\centering
\subfigure{
a)\includegraphics[width=6.0cm]{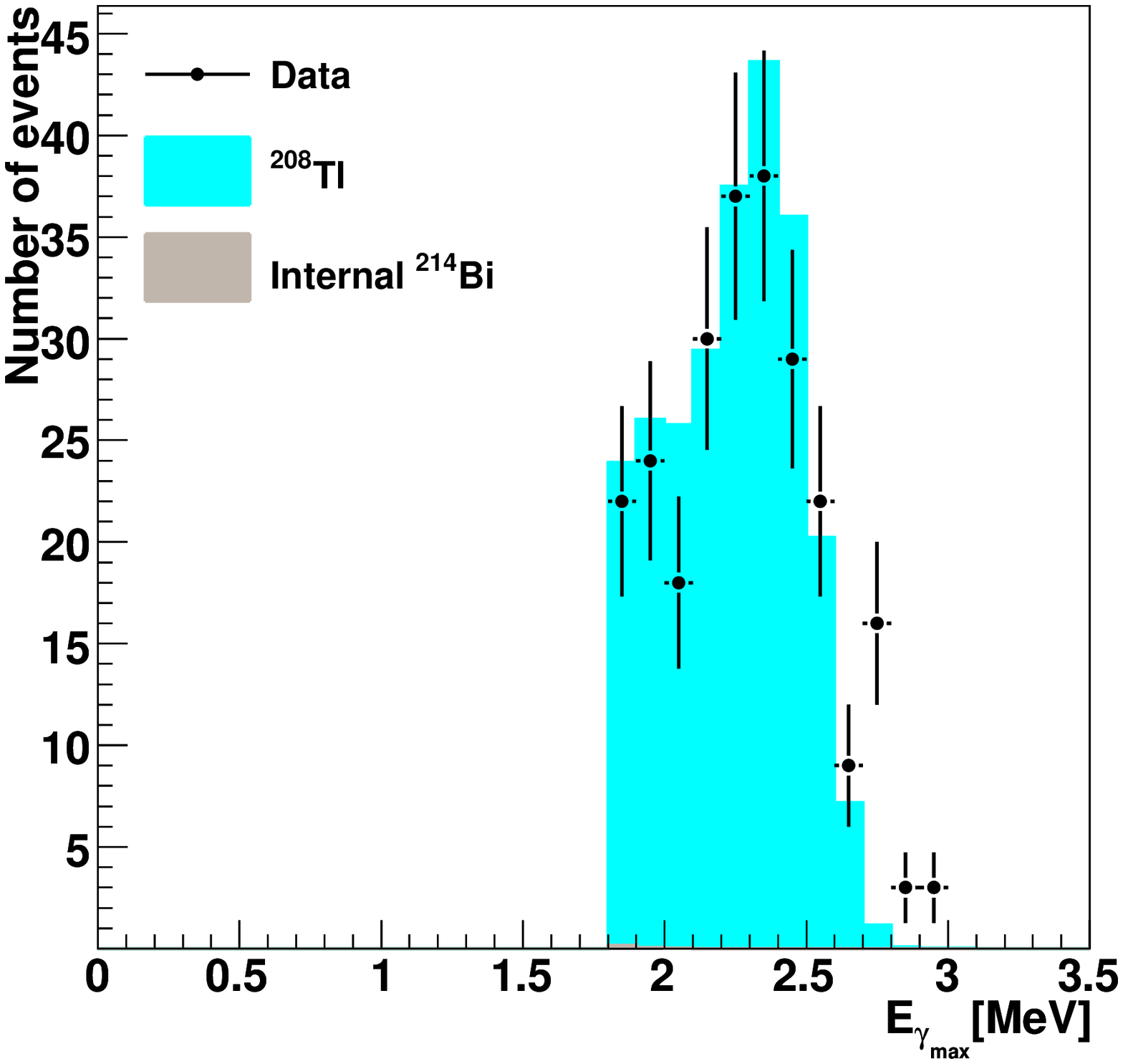}}
\subfigure{
b)\includegraphics[width=6.0cm]{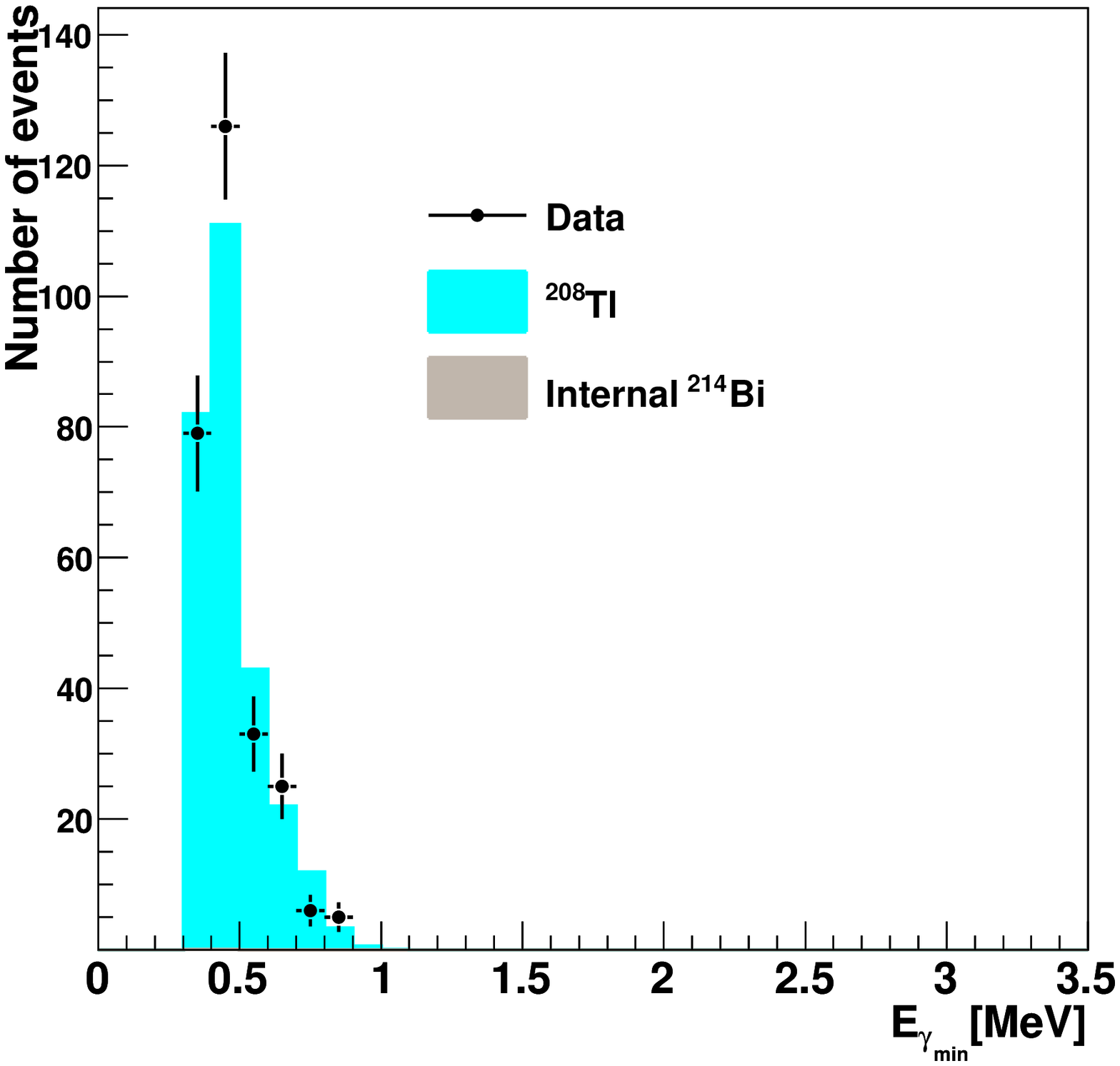}}
\subfigure{
c)\includegraphics[width=6.0cm]{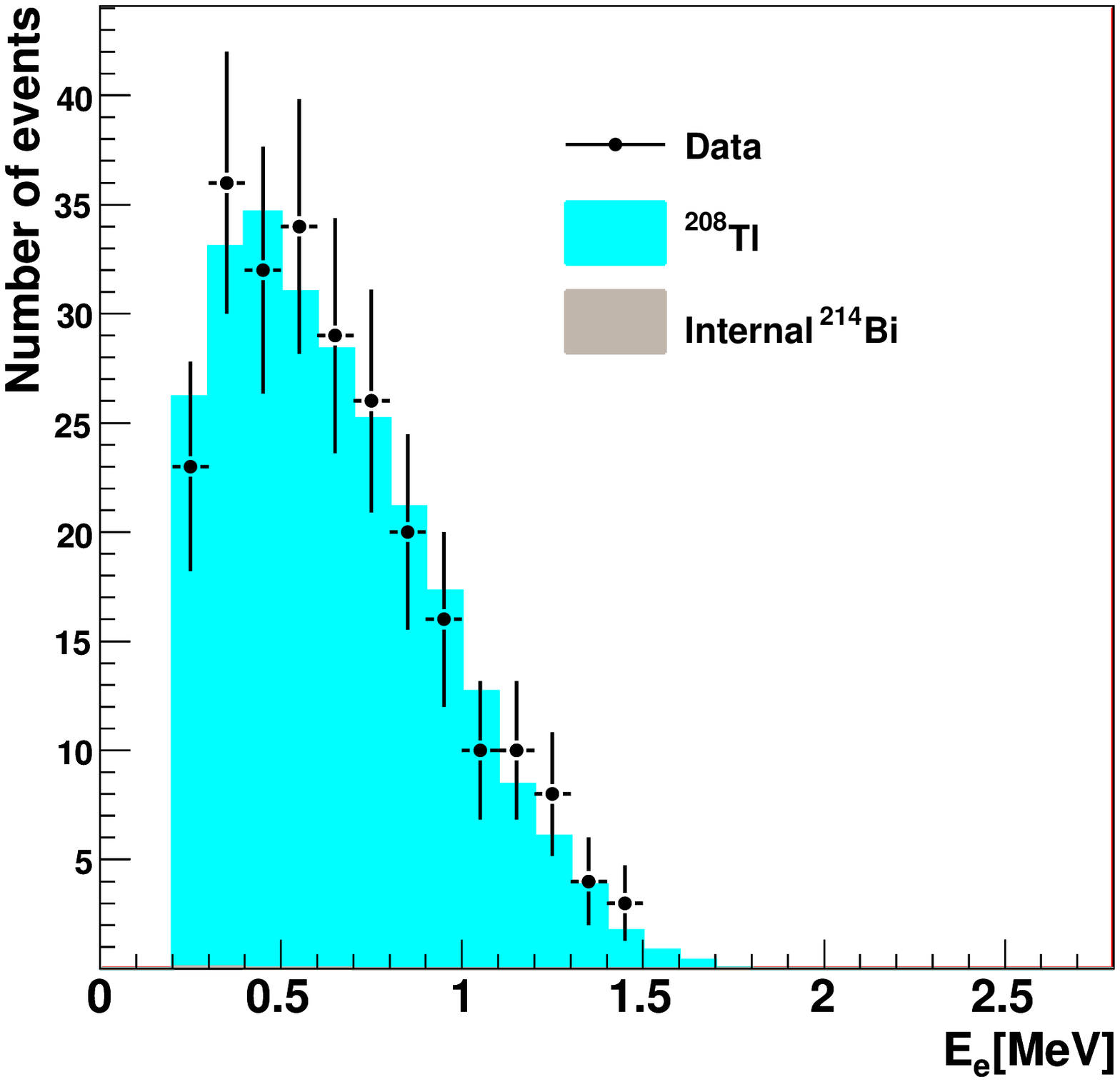}}
\subfigure{
d)\includegraphics[width=6.0cm]{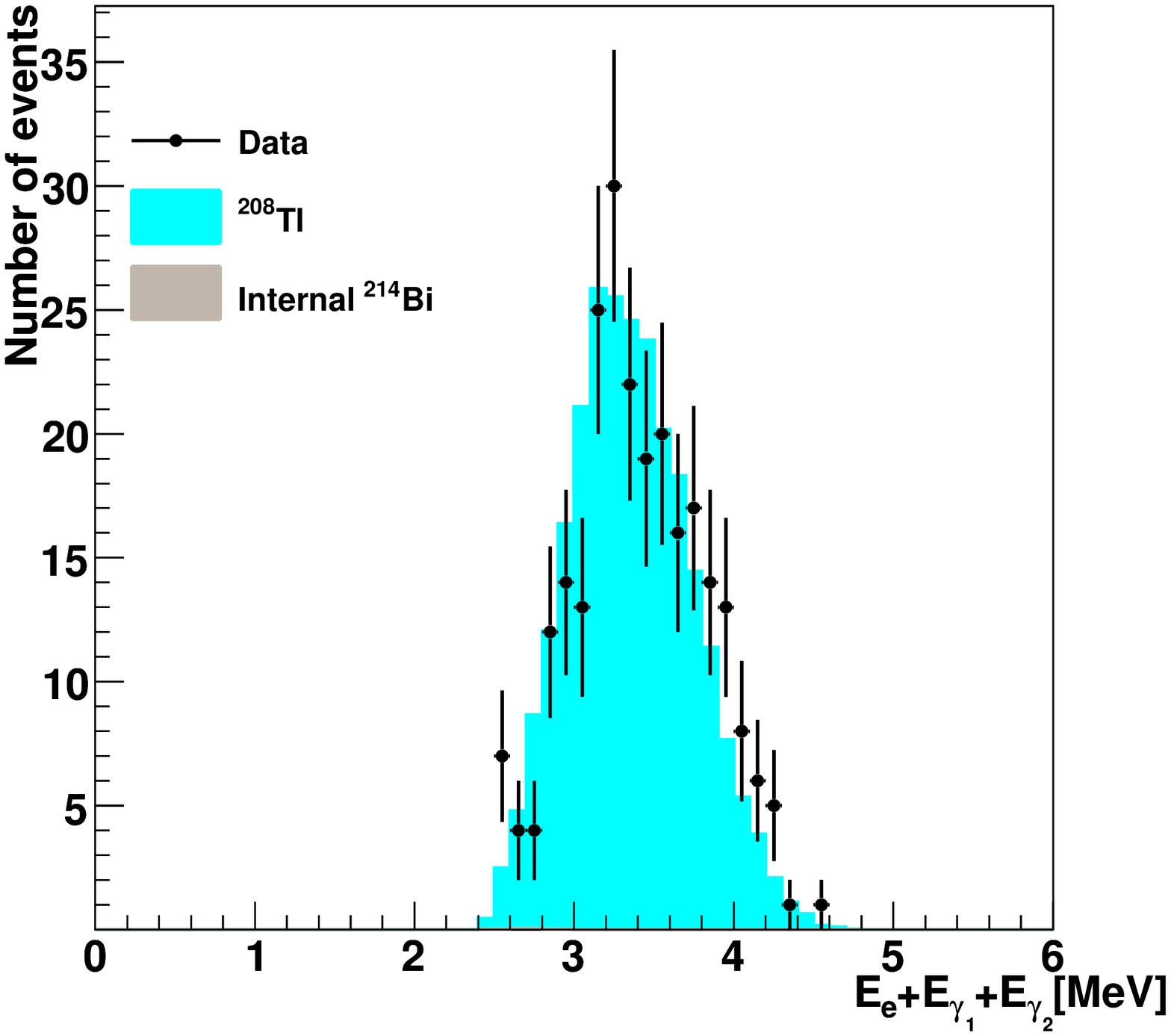}}
\caption[Energy distributions of high energy and low energy photons and electrons for $e\gamma\gamma$ events]{  The energy distribution of  the a) higher energy photon, b) lower energy photon, c) electron and d) sum of all particles for $e\gamma\gamma$ events. The error bars show the statistical uncertainties on the data points.}
\label{fig-energy1e2g}
\end{figure}
\begin{figure}
\centering
\subfigure{
a)\includegraphics[width=6.0cm]{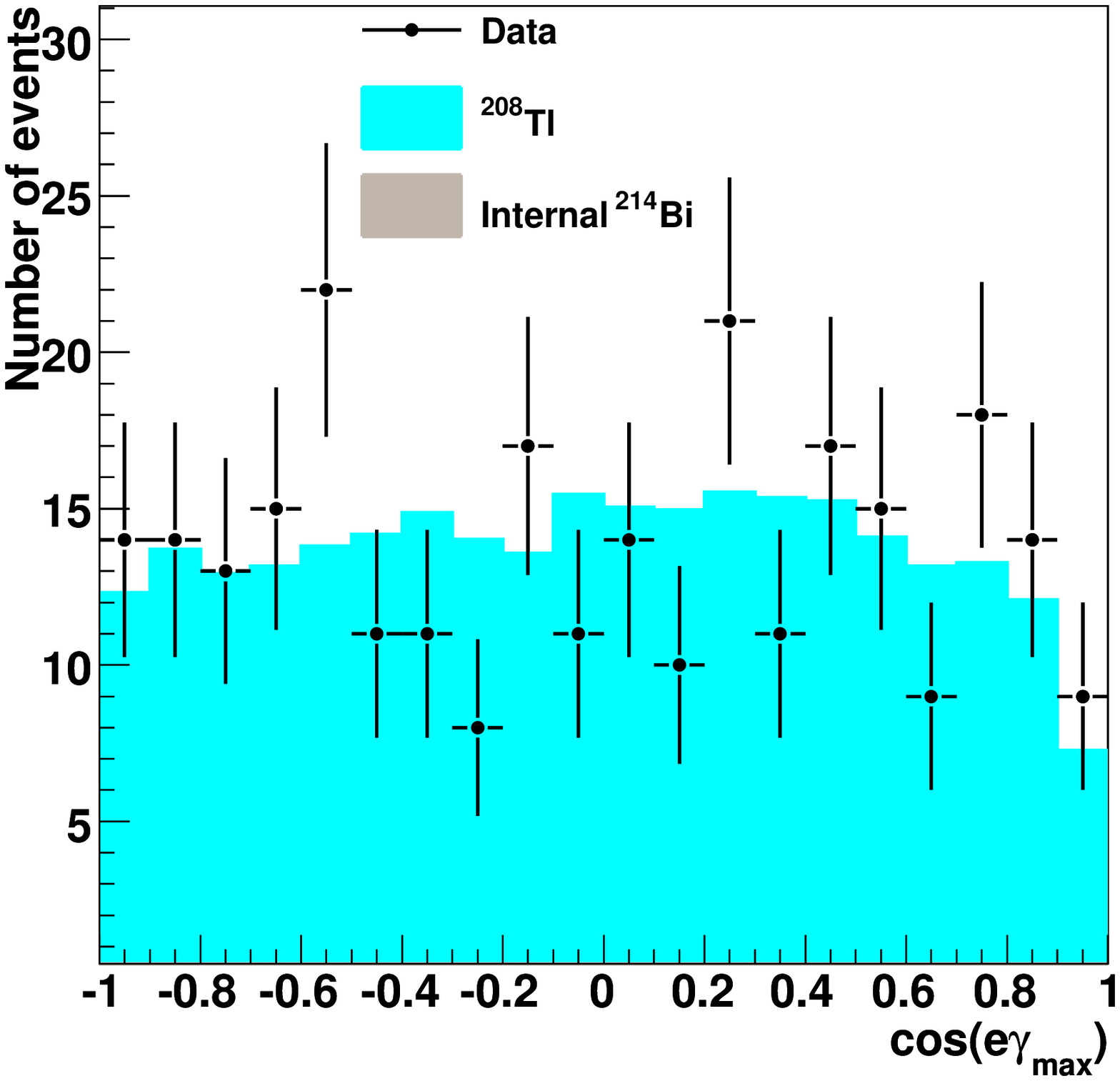}}
\subfigure{
b)\includegraphics[width=6.0cm]{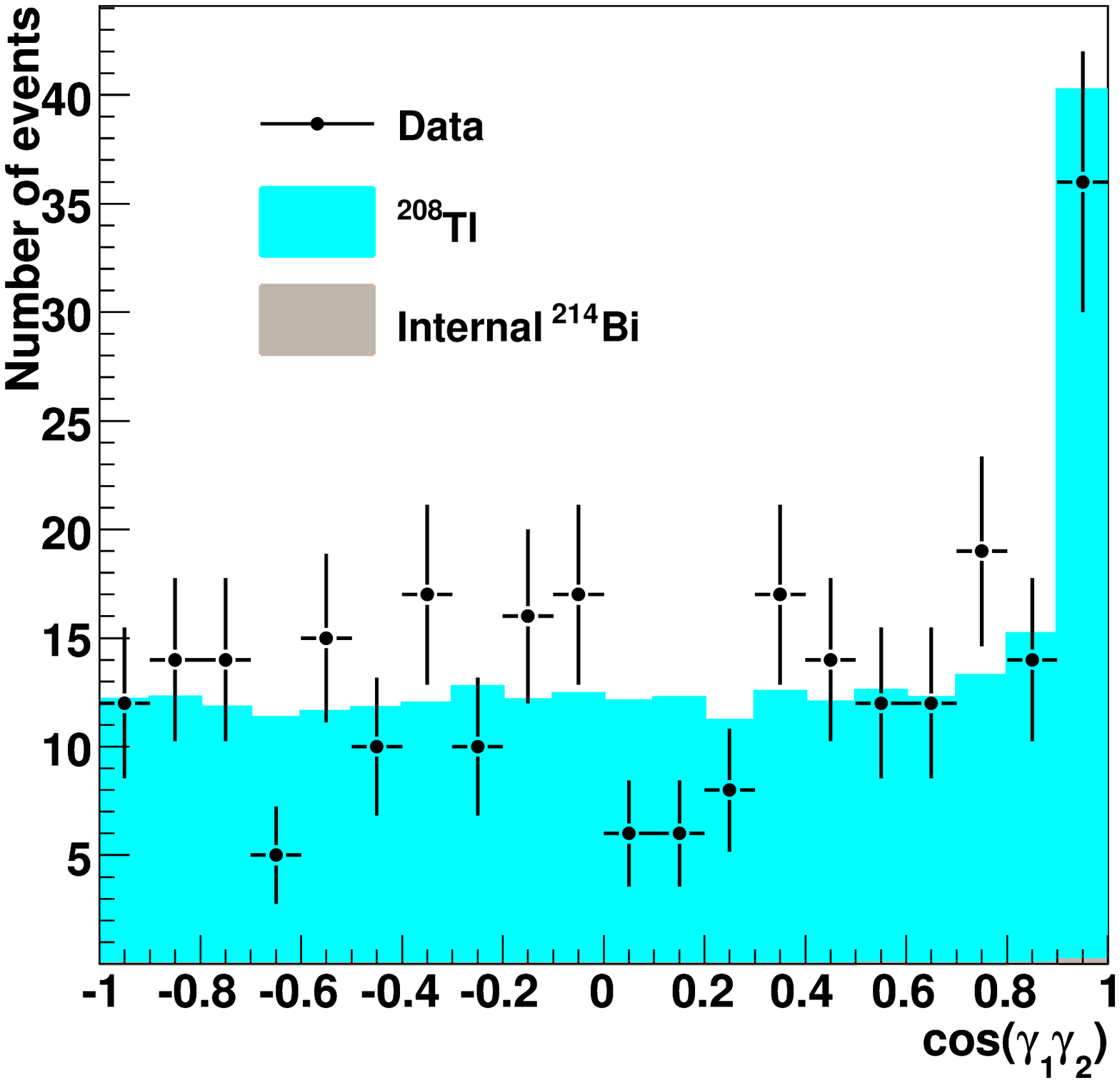}}
\caption[The cosine of the  angle between the higher energy photon and electron and two photons for $e\gamma\gamma$ events]{The cosine of the angle between a)  the higher energy photon and the electron  and b) the two photons for $e\gamma\gamma$ events. The error bars show the statistical uncertainties on the data points.}
\label{fig-coseg}
\end{figure}
The isotope~$^{208}$Tl can undergo a $\beta$ decay to  an excited state of $^{207}$Pb which de-excites  to the ground state  via two-photon emission. The first photon  has an  energy of 2.6~MeV and is accompanied by a second  photon with an energy of 0.58~MeV or 0.86~MeV. Figure~\ref{fig-tl208decay} shows the decay scheme of $^{208}$Tl. The main internal background to $^{208}$Tl in this channel is $^{214}$Bi which can undergo a $\beta$ decay and form  an excited state of $^{208}$Po which  de-excites either by emission of two photons with energies of 0.61~MeV and 1.12~MeV or a photon with energy  1.76~MeV (a  decay scheme is shown in Figure~\ref{fig-bi214decay}). 

To select one electron plus two photons originating from the foil, the following event selection criteria are applied:
\begin{itemize}
\item Only one electron track associated to a scintillator hit with deposited energy greater than 0.2~MeV.
\item The vertex of the $e\gamma\gamma$ event, which is found by extrapolating the electron track to the foil, is reconstructed inside the $^{150}$Nd foil (defined in Section~\ref{sec-ndg}).
\item The track has a hit in one of the first two layers of the tracker.
\item Two scintillator hits  are found with the  deposited energies greater than 1.8~MeV for the higher energy photon and greater than 0.35~MeV for the lower energy photon. These energy requirements which are based on the decay scheme of $^{208}$Tl are effective to reject internal $^{214}$Bi background. Figure~\ref{fig-bi214-tl208} compares the distribution of the higher energy photons for the simulated $^{214}$Bi and $^{208}$Tl.
\item The internal TOF hypothesis for each $e\gamma$ pair is more than $4\%$, and the external TOF hypothesis is less than $1\%$.
\item The energy sum of all other photon clusters is less than 0.15~MeV.
\item The length of the electron and photon trajectories is more than 50~cm.
\end{itemize} 
Figure~\ref{fig-eggdisplay} shows a typical internal $e\gamma\gamma$ event display. 
Figure~\ref{fig-energy1e2g}   shows the energy distributions of the photons and the electron for  $e\gamma\gamma$ events that pass the event selection criteria. The energy sum of all the decay particles is also shown. Figure~\ref{fig-coseg} shows the cosine of the angle between  particles. The simulated $^{208}$Tl is normalised to the same number of events as data minus $^{214}$Bi background. These plots show that the background MC describes data well in this channel. The expected number of $^{214}$Bi events is $0.22\pm0.02$~(stat). This value is found by normalising the simulated $^{214}$Bi events to the activity of $0.169$~mBq, which  was measured by studying  $e\alpha$ events originating from the foil~\cite{externalbkg}. 
  The number of data events is found to be $251$. The $^{208}$Tl event selection efficiency for this channel is $(0.55\pm0.01{\rm(stat)})\%$. The activity of $^{208}$Tl is  thus  found to be:
\begin{eqnarray}
\label{eq-tl208egg}
&A=0.57 \pm 0.04~{\rm(stat)}~{\rm mBq}& \\
&A^{\prime}= 11.04\pm 0.79~{\rm(stat)}~{\rm mBq/kg}.&
\end{eqnarray}

\subsubsection{Measurement of the  background uncertainties by comparing the e$\gamma$ and $e\gamma\gamma$ results }
\label{sec-systback}
To estimate  the systematic uncertainty on the measurement of the $^{208}$Tl activity,   results from the $e\gamma$ and $e\gamma\gamma$ channels are compared.   The mean value of the two activities  given in Equations~\ref{eq-tl208eg}  and~\ref{eq-tl208egg} is obtained. The systematic uncertainty is found by calculating the deviation of each activity from the mean of the two extracted activities. The final $^{208}$Tl activity is thus measured to be:
\begin{eqnarray}
A({\rm^{208}Tl})=0.51 \pm 0.03~{\rm (stat)} \pm 0.04~{\rm (syst)}~{\rm mBq},\\ 
A^{\prime}=10.06 \pm 0.59~{\rm (stat)} \pm 0.79~{\rm (syst)}~{\rm mBq/kg}.
\label{eq-tl208final}
\end{eqnarray}
By normalising  $^{208}$Tl background to this value and refitting, the $^{207}$Bi, $^{152}$Eu and $^{154}$Eu  activities are remeasured to be:
\begin{eqnarray}
A({\rm^{207}Bi})& = &11.3 \pm 0.3~{\rm (stat)},  \\
A({\rm^{152}Eu})& = &2.6\pm0.1~{\rm(stat)}, \\
A({\rm^{154}Eu})& = &1.24\pm 0.07~{\rm(stat)}. \label{eq-eu154final}
\end{eqnarray}

As with $^{208}$Tl, the uncertainties on the $^{207}$Bi, $^{154}$Eu and $^{152}$Eu activities are defined by the difference from the mean values. Thus the final activity results for these isotopes are 
\begin{eqnarray}
A({\rm^{207}Bi})& = &11.5 \pm 0.2~{\rm (stat)} \pm 0.2~{\rm (syst)~mBq}, \nonumber \\
&A^{\prime}= & 226.8\pm 3.9~{\rm(stat)} \pm 3.9~{\rm(syst)~mBq/kg},\\
A({\rm^{152}Eu})& = &2.6\pm0.1~{\rm(stat)} \pm 0.1~{\rm(syst)~mBq} \nonumber\\
&A^{\prime}=& 51.3 \pm 1.4~{\rm(stat)} \pm 2.0~{\rm(syst)~mBq/kg},\\
A({\rm^{154}Eu})& = &1.25\pm 0.05~{\rm(stat)}\pm 0.02{\rm(syst)~mBq} \nonumber \\
&A^{\prime}= & 24.6 \pm 1.0~{\rm(stat)} \pm 0.4~{\rm(syst)~mBq/kg}.
\end{eqnarray}

 The reconstruction of an additional photon in the $e\gamma\gamma$ channel gives the  systematic uncertainty on the photon detection efficiency. Other systematic uncertainties,  including the uncertainties on the electron detection efficiency (section~\ref{sec-systematic2n2b}), are negligible  for the purpose of these  results.
\subsection{Single-electron decay channel}
\label{sec-1e}
\begin{figure}[p]
\centering
\includegraphics[width=9.0cm]{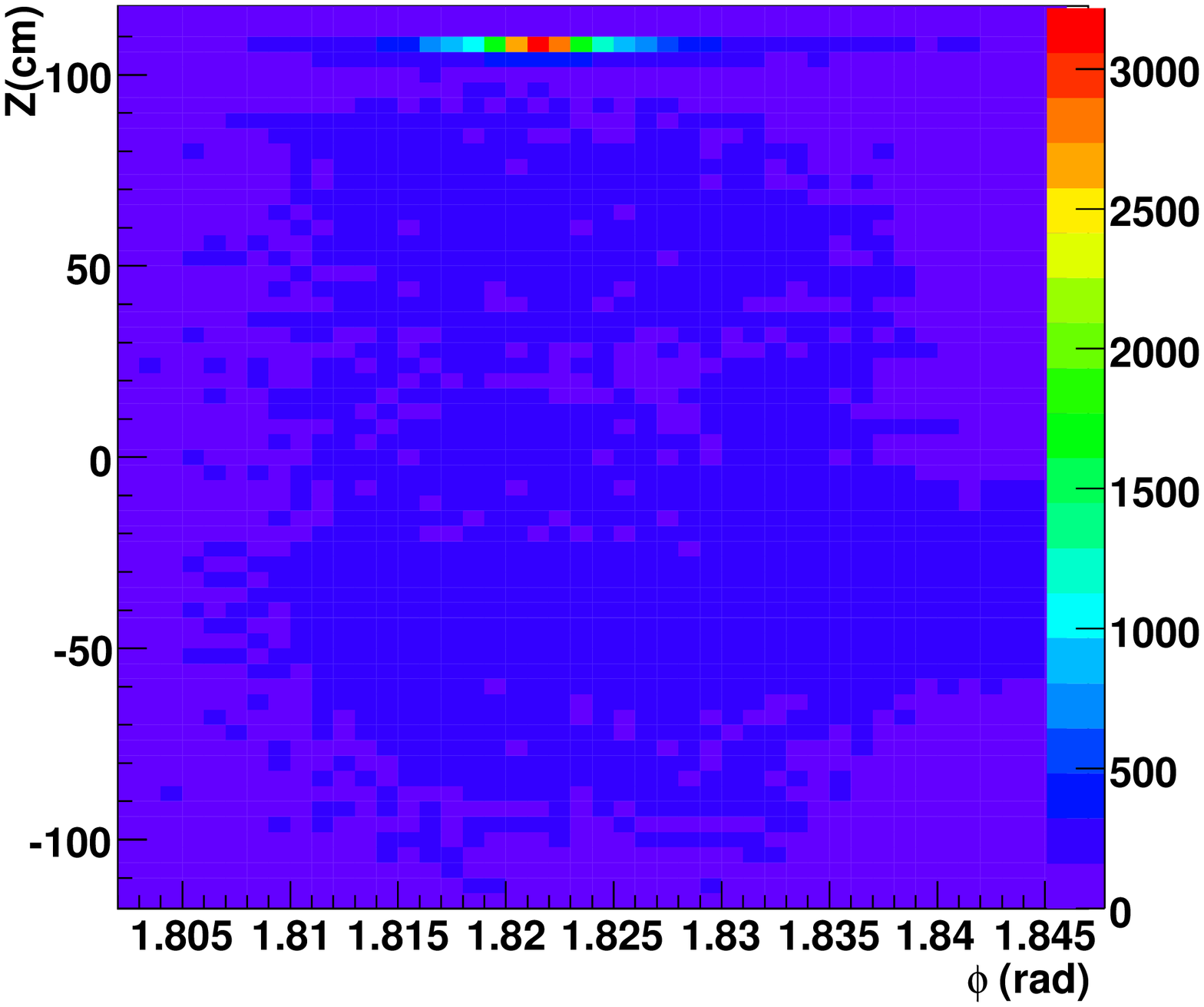}
\caption{The reconstructed $\phi$ component versus $z$  component for single-electron events.}
\label{fig-hotspotpa234m}
\end{figure}

\begin{figure}
\centering
\includegraphics[width=9.0cm]{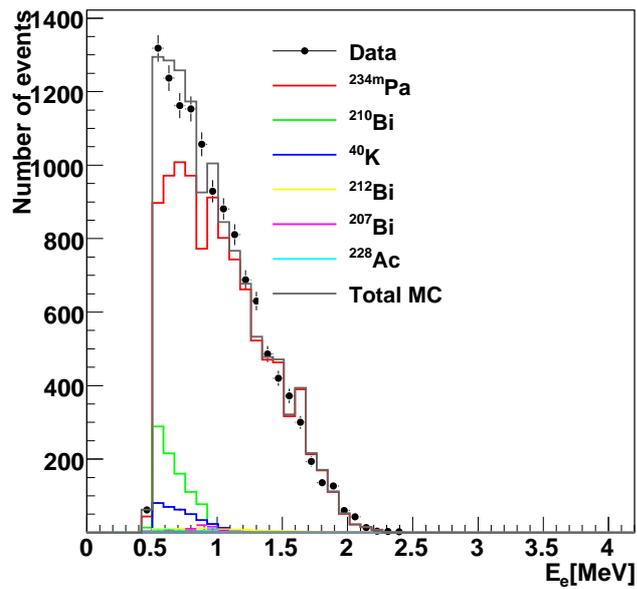}
\caption[The distribution of the electron energy in the region $1.815<\phi<1.827~{\rm and}~104<z<110$~cm for Phase~2 data]{The distribution of the electron energy in the region $1.815<\phi<1.827~{\rm and}~104<z<110$~cm for Phase~2 data. The statistical uncertainties on the data points are shown with error bars. This plot shows that the hot-spot region is highly polluted with $^{234m}$Pa.}
\label{fig-energyhotspot}
\end{figure}

\begin{figure}[p]
\centering
\includegraphics[width=10.0cm]{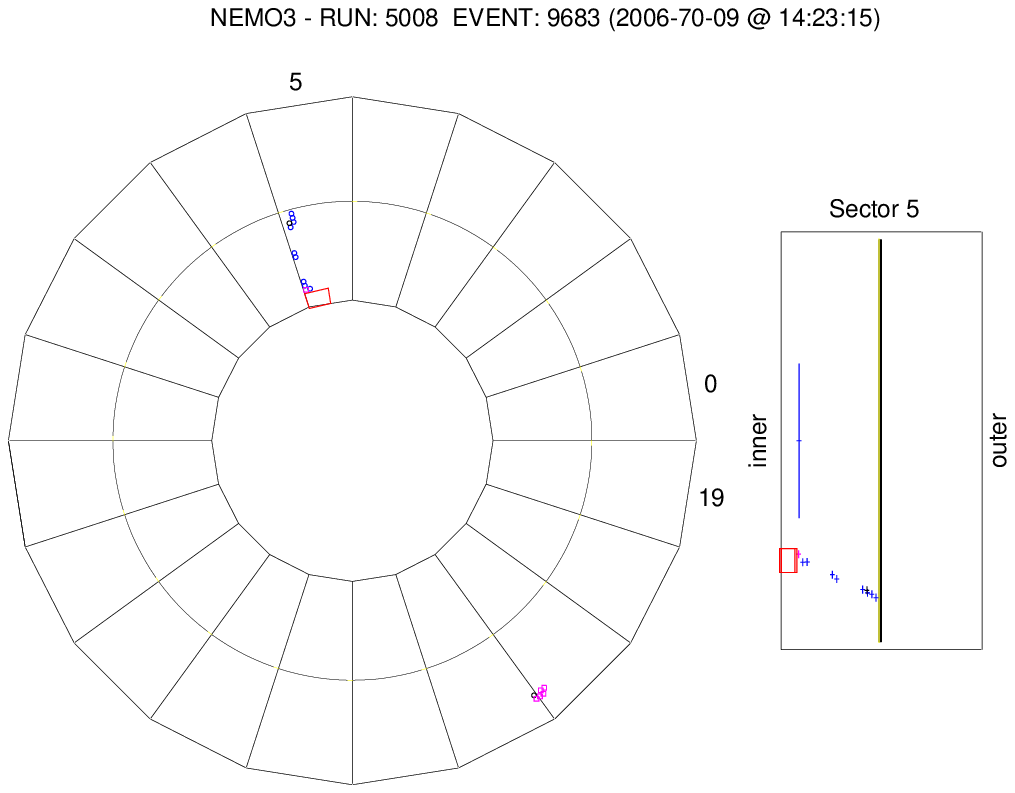}
\caption[A $\beta$ decay event display]{A $\beta$ decay event display. A top and side view of the event are shown. This event is from data taken in July 2006.}
\label{fig-1edisplay}
\end{figure}

The isotopes $^{234m}$Pa and $^{40}$K  predominantly undergo $\beta$ decay. Thus the activities of these internal contaminants are measured by studying  the single-electron~($1e$) channel. The following requirements are applied to select $1e$ events:
\begin{itemize}
\item Only one electron is found in the event with energy greater than 0.5~MeV.  This energy requirement eliminates a large amount of  low energy events which are not relevant for this analysis.
\item The electron track must originate from the $^{150}$Nd foil. 
\item The length of the track is greater than 50~cm.
\item The track has at least a hit in one of the first two layers of the tracker.
\end{itemize} 
Figure~\ref{fig-hotspotpa234m} shows the reconstructed $\phi$ component versus the  $z$ component for single-electron events.  A hot-spot is  observed in the region:
\begin{equation}
1.815<\phi<1.827~~~{\rm and}~~~104<z<110~{\rm cm},
\label{eq-pahs}
\end{equation}
which is highly contaminated with $^{234m}$Pa (Figure~\ref{fig-energyhotspot}) and has been subsequently removed from the analysis. Figure~\ref{fig-1edisplay} shows a $\beta$ decay event display. 
\begin{figure}
\centering
\subfigure{
a)\includegraphics[width=6.9cm]{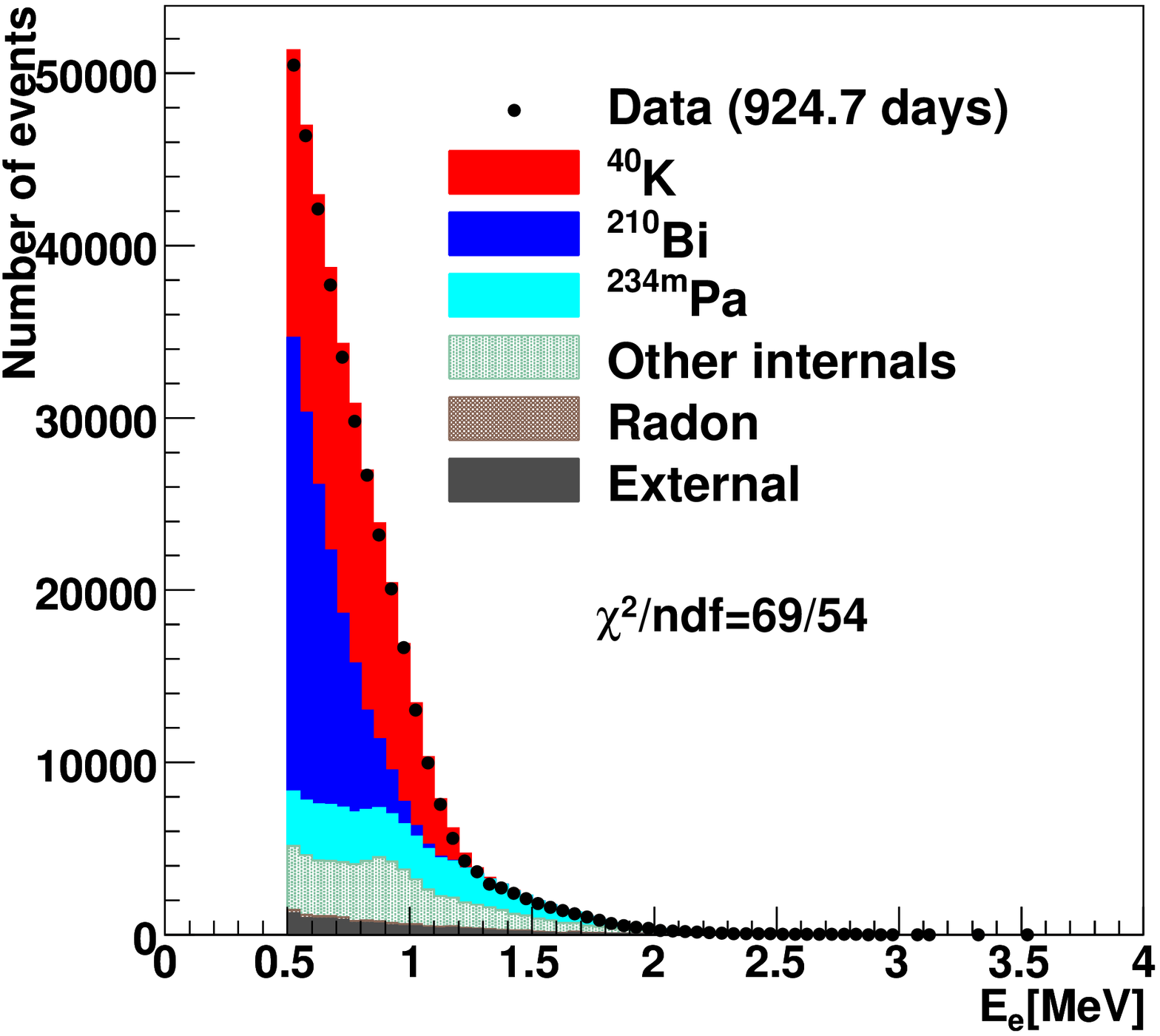}}
\subfigure{
b)\includegraphics[width=6.9cm,height=6.7cm]{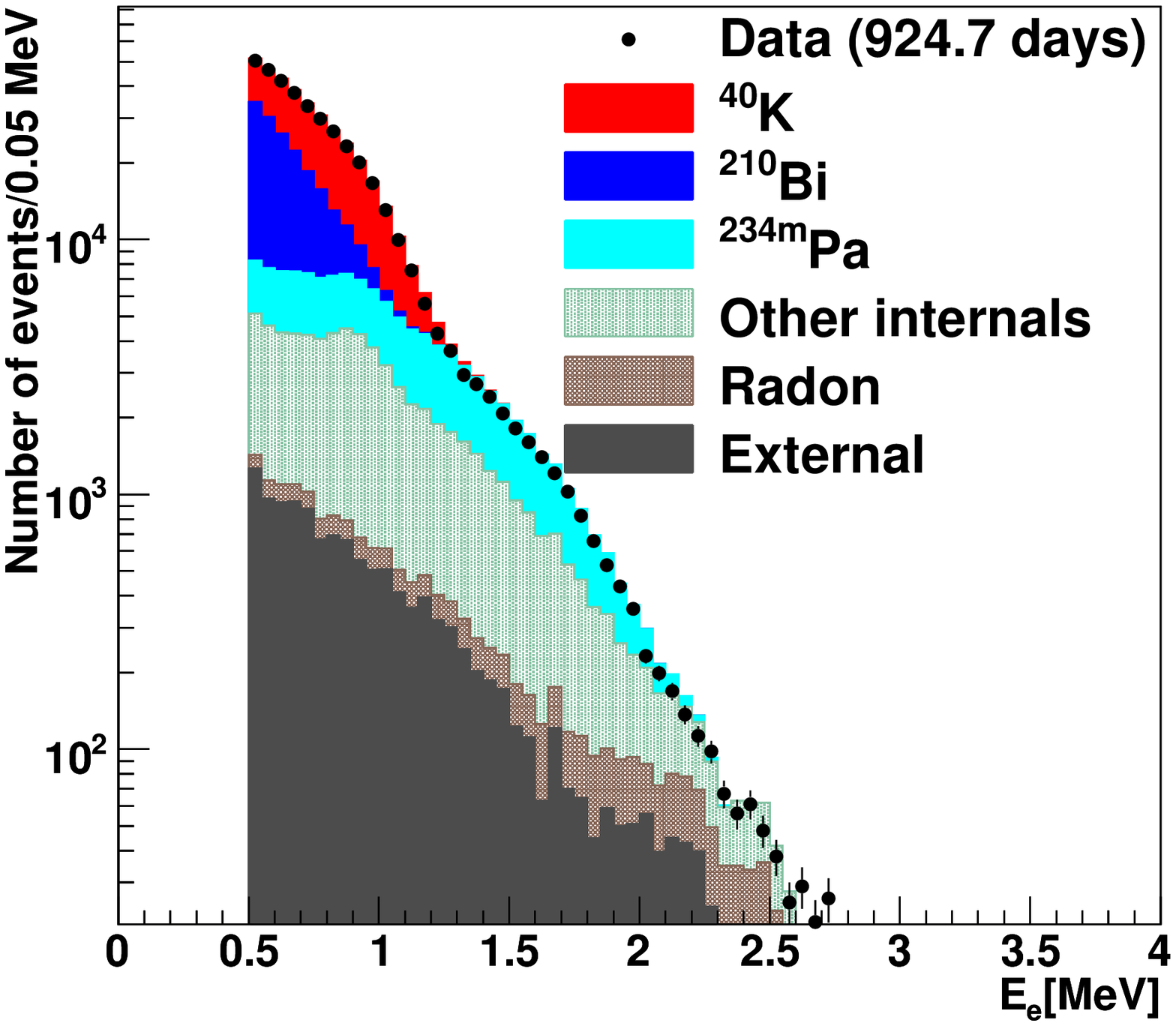}}
\caption[Distribution of the energy of the electron for the single-electron events that pass the event selection criteria]{Distribution of the energy of the electron for the single-electron events that pass the event selection criteria shown in  a) linear scale and b) log scale. A simultaneous data and MC fit is performed for $^{40}$K and $^{234m}$Pa. The error bars show the statistical uncertainties on the data points.}
\label{fig-1eenergy}
\end{figure}
$^{210}$Bi  from the radon  decay chain can settle on the tracker wires and undergo a $\beta$ decay which can mimic $1e$ events originating from the foil. The average activity of $^{210}$Bi on the surface of the wires is $12.61$~mBq~\cite{externalbkg} (more information about this background is given in Section~\ref{sec-radon}). Other internal backgrounds are normalised by the activities found in Section~\ref{sec-eg}.  The activities of $^{40}$K and $^{234m}$Pa are found by fitting the data simultaneously with the MC contributions of these two contaminants. 

Figure~\ref{fig-1eenergy} shows the energy distribution of the single-electron events that pass  the event selections.  This figure demonstrates that the background MC describes data well in this channel. The activities given from Equation~\ref{eq-tl208final} to  Equation~\ref{eq-eu154final} are used to normalise  internal backgrounds to the  $1e$ channel.  As before, the radon activity is normalised to average value of $0.45\pm0.07$~mBq (the external background components  and their activities are given in Section~\ref{sec-external}).
\begin{table}[b]
\centering
\begin{tabular}{|c|c|c|c|c|}
\hline
Contaminant & $N_{exp}$ & $N_{bgr}$ &  $N_{^{40}\rm K}$,  $N_{^{234m}\rm Pa}$ &Efficiency\\
\hline
$^{40}$K   &  $392123$  & $179922\pm698$ &  $161675\pm569$   & $18.7\%$\\     
\hline
$^{234m}$Pa & $392123$    &   $179922\pm698$ &  $50525\pm 431$  &   $26.1\%$\\
\hline
\end{tabular}
\caption[The internal contaminates of the $^{150}$Nd foil that mainly decay to an electron, the number of expected events]{The internal contaminants of the $^{150}$Nd foil that mainly decay to an electron, the number of expected events, the total number of background events and the efficiency of the event selection criteria. The errors on the number of $^{40}$K and $^{234m}$Pa events are determined by the fit.}
\label{tab-1e}
\end{table}
Table~\ref{tab-1e} shows the number of data events ($N_{exp}$), number of background events remaining in the data found by MC simulations ($N_{bgr}$), number of events originating from $^{40}$K and $^{234m}$Pa found from the fit ($N_{^{40}\rm K}$ and $N_{^{234m}\rm Pa}$) and  the event selection efficiency.
From this table, the activities of $^{40}$K and $^{234m}$Pa are found to be:
\begin{eqnarray}
A(^{40}{\rm K})=    10.8\pm 0.1~{\rm(stat)}\pm0.6~{\rm(syst)~mBq}\\
A^{\prime}=213\pm 2~{\rm(stat)}\pm12~{\rm(syst)~mBq/kg},\\
A(^{234m}{\rm Pa})= 2.42\pm 0.04~{\rm (stat)}\pm0.14~{\rm(syst)~mBq}\\
A^{\prime}=47.7\pm 0.8~{\rm(stat)}\pm 2.9~{\rm(syst)~mBq/kg}.
\end{eqnarray}
The systematic uncertainties on $^{40}$K and $^{234m}$Pa activities are found by  varying the energy selections  between $0.5$~MeV and $1.5$~MeV and refitting $^{234}$Pa and $^{40}$K MC distributions to data minus other backgrounds after each energy selection.
 The systematic uncertainties are found to be  $5\%$ for  $^{40}$K and $6\%$ for $^{234m}$Pa.

\section{The external background}
\label{sec-external}
\begin{figure}[tb]
\centering
\includegraphics[width=9.0cm]{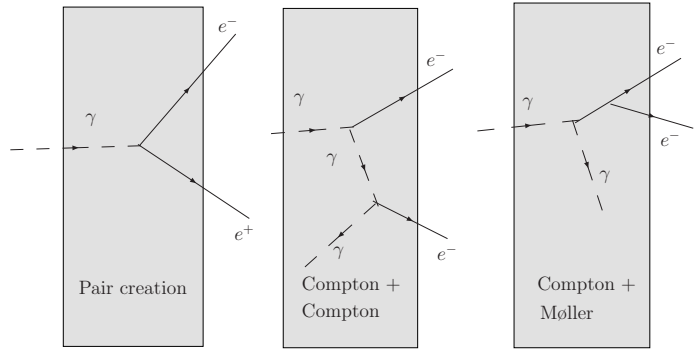}
\caption[Three  ways that the external background can mimic double beta decay.]{Three  ways that the external background can mimic double beta decay: the electron-positron pair production due to the photon interaction with the foil (the diagram on the left); double Compton scattering (the diagram in the centre); and  Compton followed by  M$\o$ller scattering (the diagram on the right). } 
\label{fig-external}
\end{figure}

The external background in the  NEMO~3 experiment is caused by electrons and photons generated outside the source foil. These particles can interact with the foil and mimic two electron tracks.  An external photon's interaction with  the foil can produce an electron-positron pair, which can be misidentified as  two electrons. A photon can  also  undergo Compton scattering and produce an electron and another photon. The second electron  can be produced by either M$\o$ller scattering of the electron or  Compton scattering of the second photon. Figure~\ref{fig-external} shows these three main ways that an external photon can mimic two -electron events originating from the foil. 

The second source of external background are   crossing electrons which are produced by Compton scattering of the external photons within the detector components. These electrons can be scattered by the foil and thus  detected by two different scintillators. As described in Section~\ref{sec-tof}, these events can be reduced by the TOF requirements.     

The external background is divided into two main categories based on the origin of the events: the radon background inside the tracker 
 and the background originating from outside the tracker (i.e. the calorimetry, the shielding and the air surrounding the detector). 

\subsection{The radon background}
\label{sec-radon}
Radon ($^{222}$Rn)  is out-gassed into the air from the surrounding rocks and can enter the detector either through leaks between the sectors, or through gas piping joins. Bismuth-214 ($^{214}$Bi) is one of the descendants of $^{222}$Rn  which can settle on the surface of the tracker's wires.  The $^{214}$Bi undergoes a $\beta$ decay  to  $^{214}$Po. The latter   has  a half-life of 161~$\mu$s and  decays to $^{212}$Pb via emission of an $\alpha$ particle. Thus the activity of $^{214}$Bi on the wires  was measured by studying   electron-alpha  ($e\alpha$) events~\cite{externalbkg}. 

Another source of  background which is part of the radon decay chain originating from the surface of the wires is $^{210}$Bi. This isotope is a $\beta$ emitter with  $Q$ value of 1.1~MeV. Therefore few $^{210}$Bi events can mimic $2\nu\beta\beta$ decay. However, since it contributes to the internal single-electron decay channel, it is important to estimate its activity. The activity of $^{210}$Bi was measured by studying single-electron events originating from the surface of the wires~\cite{externalbkg}.
Table~\ref{table-rad}  gives the activity of the $^{214}$Bi and $^{210}$Bi originating from the first layer of the drift cells situated in  sector~5.
\begin{table}
\centering
\begin{tabular}{|c|c|c|c|c|}
\hline
Contaminant & \multicolumn{4}{c|}{Activity~(mBq)}\\
\cline{2-5}
   & \multicolumn{2}{c|}{inner foil side} & \multicolumn{2}{c|}{outer foil side}\\
\cline{2-5}
   & Phase~1 & Phase~2 &Phase~1 &Phase~2 \\
\hline
$^{214}$Bi & $768\pm 10$&$177\pm 4$& $635\pm 8$ & $145\pm 3$\\
$^{210}$Bi & $ 15.5\pm 1.2$ &  $14.7\pm 1.0$ &   $13.4\pm 0.9$& $12 \pm 1$ \\
\hline
\end{tabular}
\caption[The activities of $^{214}$Bi and $^{210}$Bi originating from the closest layers to the foil  in sector~5]{The activities of $^{214}$Bi and $^{210}$Bi originating from the closest layers to the foil  in sector~5. Due to the installation of the anti-radon facility, the level of the $^{214}$Bi on the wires is reduced dramatically for the Phase~2 period~\cite{externalbkg}. The errors given are due to the statistical uncertainties.}
\label{table-rad}
\end{table}

The other main source of external background is the natural radioactivity of the detector components. The  PMTs represent an important source of background. The HPGe measurements~\cite{technical} show that the glass of the PMTs are contaminated with $^{214}$Bi, $^{40}$K, $^{208}$Tl and $^{228}$Ac.  A study of the external $e\gamma$ and crossing electron events~\cite{externalbkg} reveals that the iron shield and iron in the sector walls are also contaminated with these isotopes. It is also important to include $^{60}$Co and $^{234m}$Pa to the external background model~\cite{externalbkg} in order  to reproduce the crossing electron data. The magnetic shield of the PMTs and the wall of the sectors are contaminated with these isotopes. In addition,
before the installation of the anti-radon facility, a fraction of external $^{214}$Bi and $^{208}$Tl background originated from the air surrounding the detector.  Table~\ref{table-external} summarises the external background sources and the values measured in~\cite{externalbkg}.

\begin{table}
\begin{tabular}{|c|c|c|c|c|c|c|}
\hline
\multicolumn{2}{|c|}{External background} &\multicolumn{5}{c|}{Activity~(Bq)}\\
\cline{3-7}
\multicolumn{2}{|c|}{}& $^{214}$Bi & $^{228}$Ac & $^{208}$Tl & $^{40}$K & $^{60}$Co\\
\hline
\multicolumn{2}{|c|}{Glass of the PMTs} & 324          &     72        &    27       &    1078     &    --   \\
\hline
\multicolumn{2}{|c|}{Plastic scintillators}& --   &     --        &      --      &       21.5  & --\\
\hline
\multicolumn{2}{|c|}{Magnetic shield of the PMTs}& --  &    --        &   --        &  --       & 14.6 \\
\hline
\multicolumn{2}{|c|}{Sector iron walls}  &     9.1   &     8.5       &    3.1   &  100 &       6.1 \\
\hline 
\multicolumn{2}{|c|}{Internal tower} &     --       &        --       &  --       &    --       &18.4 \\
\hline
\multicolumn{2}{|c|}{Iron shield}   &      7359     &    1345               &   484     &    --        &-- \\

\hline
Air between the detector&P1:&566.5 &-- &11.5 &-- &--\\
 and the iron shield &P2: &0 &-- &0 &--&--\\
\hline
\end{tabular}
\caption[Components of the external background model]{Components of the external background model~\cite{externalbkg}.} 
\label{table-external}
\end{table}


%

\section{Validation of  the external background model}
The results presented in Section~\ref{sec-external} have been  obtained by studying foil crossing events throughout the detector. In order to validate the external background model for the $^{150}$Nd foil, only particles  interacting with this foil are selected. Two decay channels are used to check the external background model:  the crossing electron channel and the external $e\gamma$ channel. In both channels the simulated external backgrounds are normalised to the activities given in Table~\ref{table-external}. The distribution of the sum of  external background is then compared with data. 

\begin{figure}
\centering
\subfigure{
a)\includegraphics[width=6.4cm]{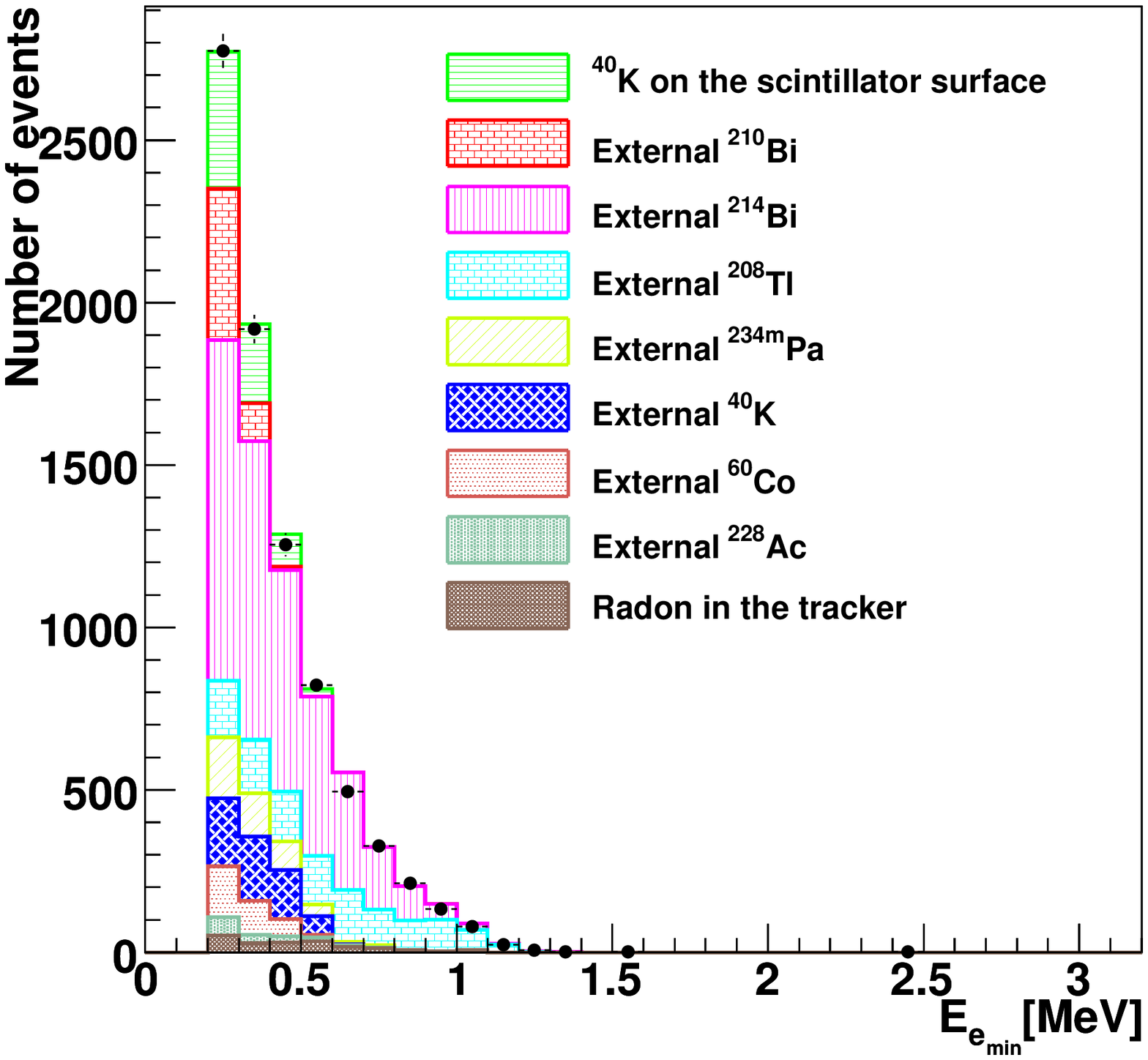}}
\subfigure{
b)\includegraphics[width=6.4cm]{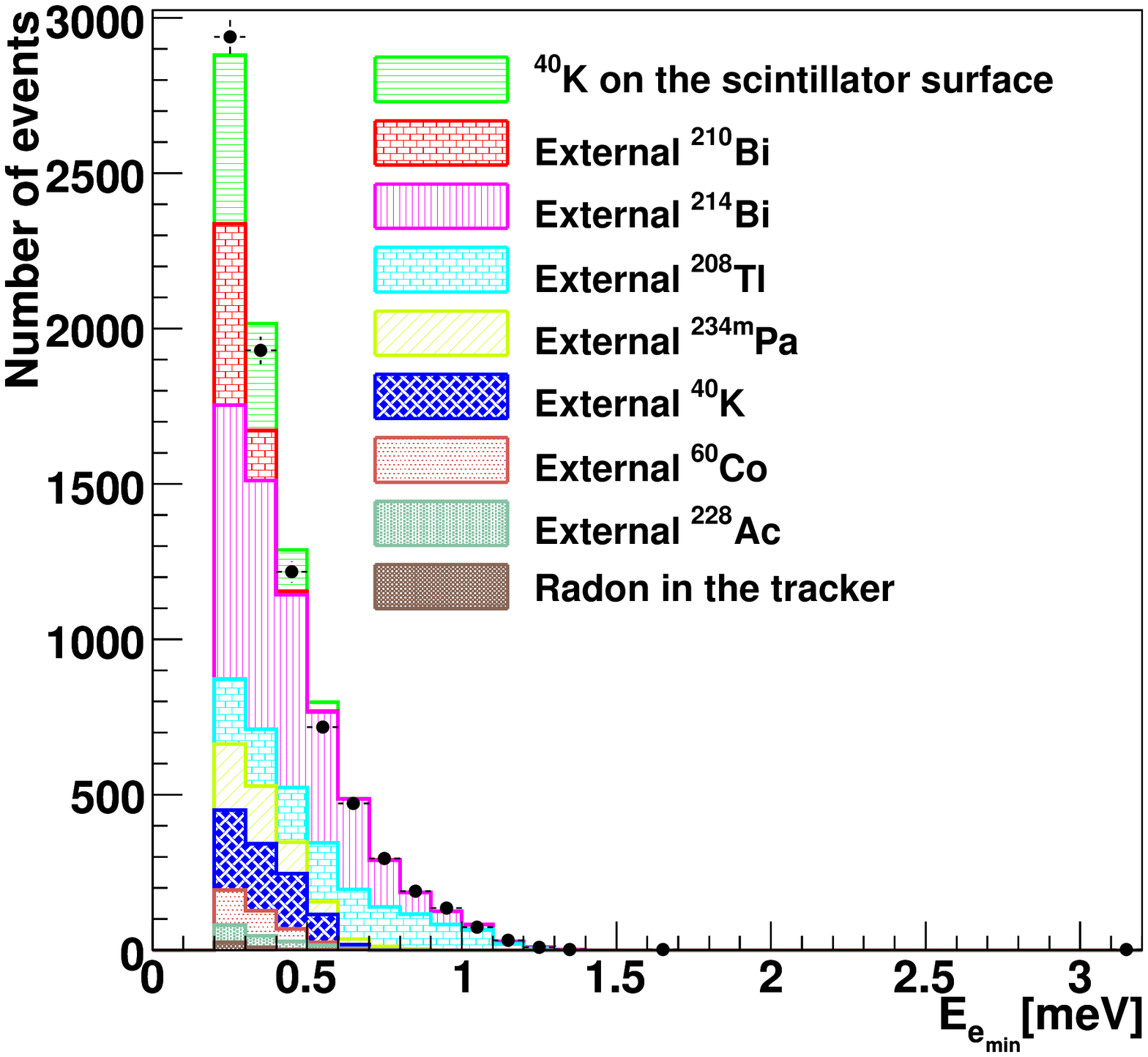}}
\subfigure{
c)\includegraphics[width=6.4cm]{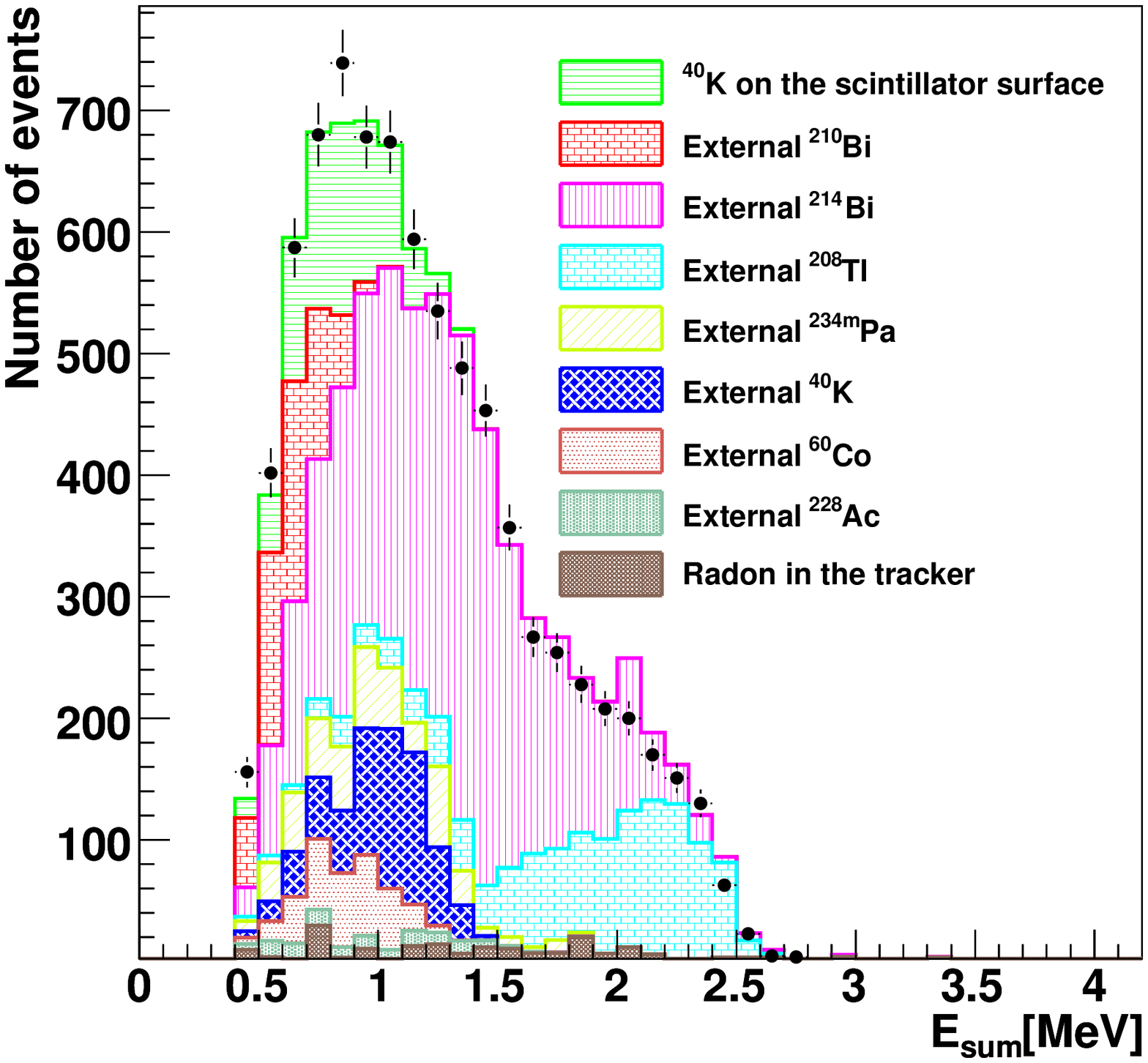}}
\subfigure{
d)\includegraphics[width=6.4cm]{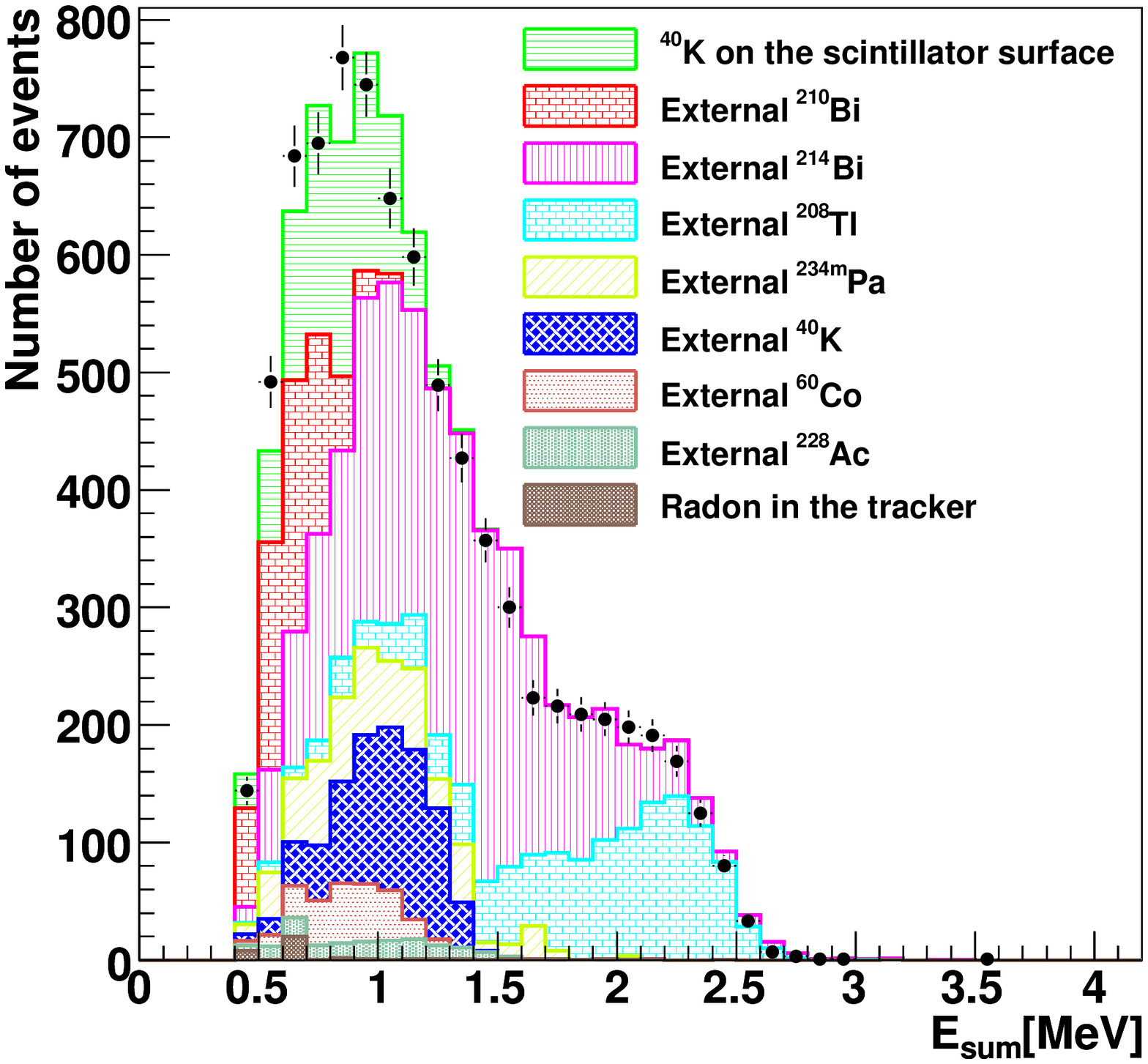}}
\subfigure{
e)\includegraphics[width=6.4cm]{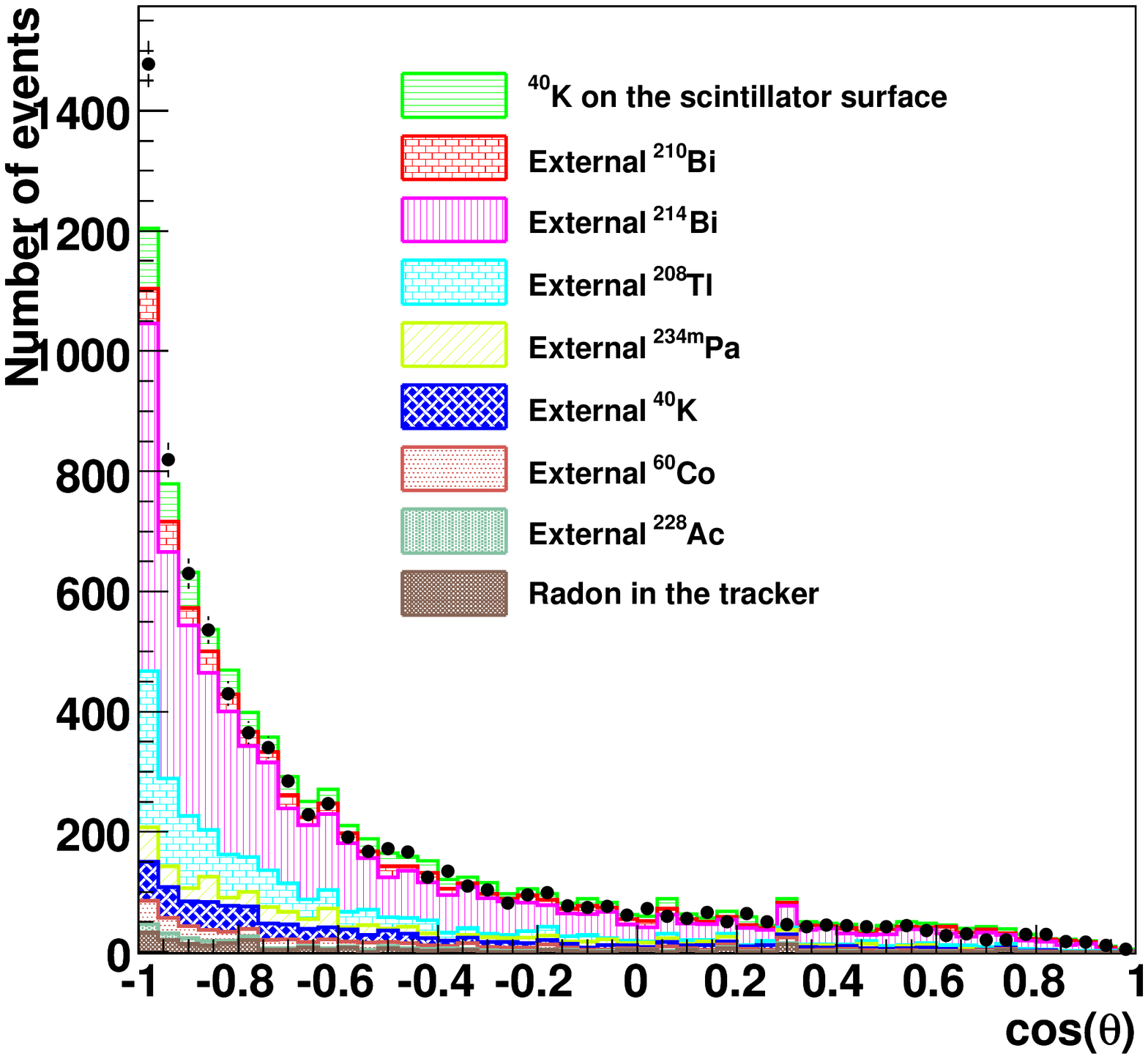}}
\subfigure{
f)\includegraphics[width=6.4cm]{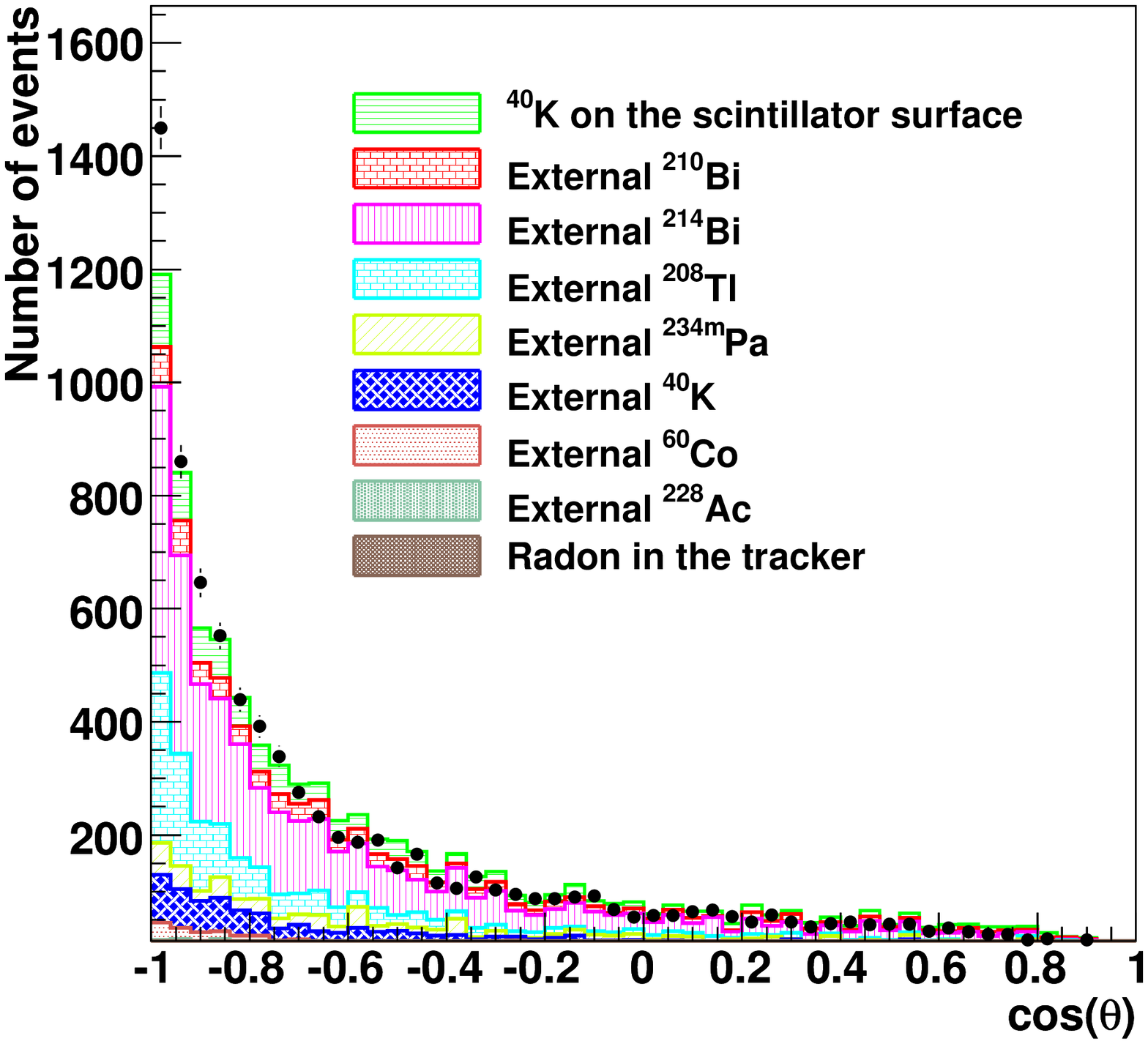}}
\caption[Energy  and cosine of the scattering angle for crossing electrons]{a,  b) energy of the crossing electron after scattering by the foil;  c, d) sum of the energies deposited in two scintillators and  e, f)  cosine of the angle between the two electron tracks. a, c, e) Phase~1 data; b, d, f) Phase~2 data. The statistical uncertainties on the data points are shown with error bars.}
\label{fig-ecrossing}
\end{figure}

The one crossing electron events create two tracks in the tracking chamber and two scintillator hits in the calorimeter. The external TOF hypothesis for these events must be more than $4\%$ and the internal TOF hypothesis must be less than $1\%$. 
Figures~\ref{fig-ecrossing} shows the  energy of the crossing electron after   scattering by the foil and detection by the second scintillator,  the sum of the energies deposited in the two scintillators and the cosine of the angle between the two electron tracks. Phase~1 and Phase~2 data are shown separately.  The total number of background events estimated by MC simulations is $8177\pm 103$~(stat)  for Phase~1 and $8197\pm 101$~(stat) for Phase~2. The number of data events are  $8050$    and  $8012$.  The difference between the number of  expected background events  and data events  gives $2.2\%$  uncertainty on   external background in this channel. 
\begin{figure}
\centering
\subfigure{
a)\includegraphics[width=6.4cm]{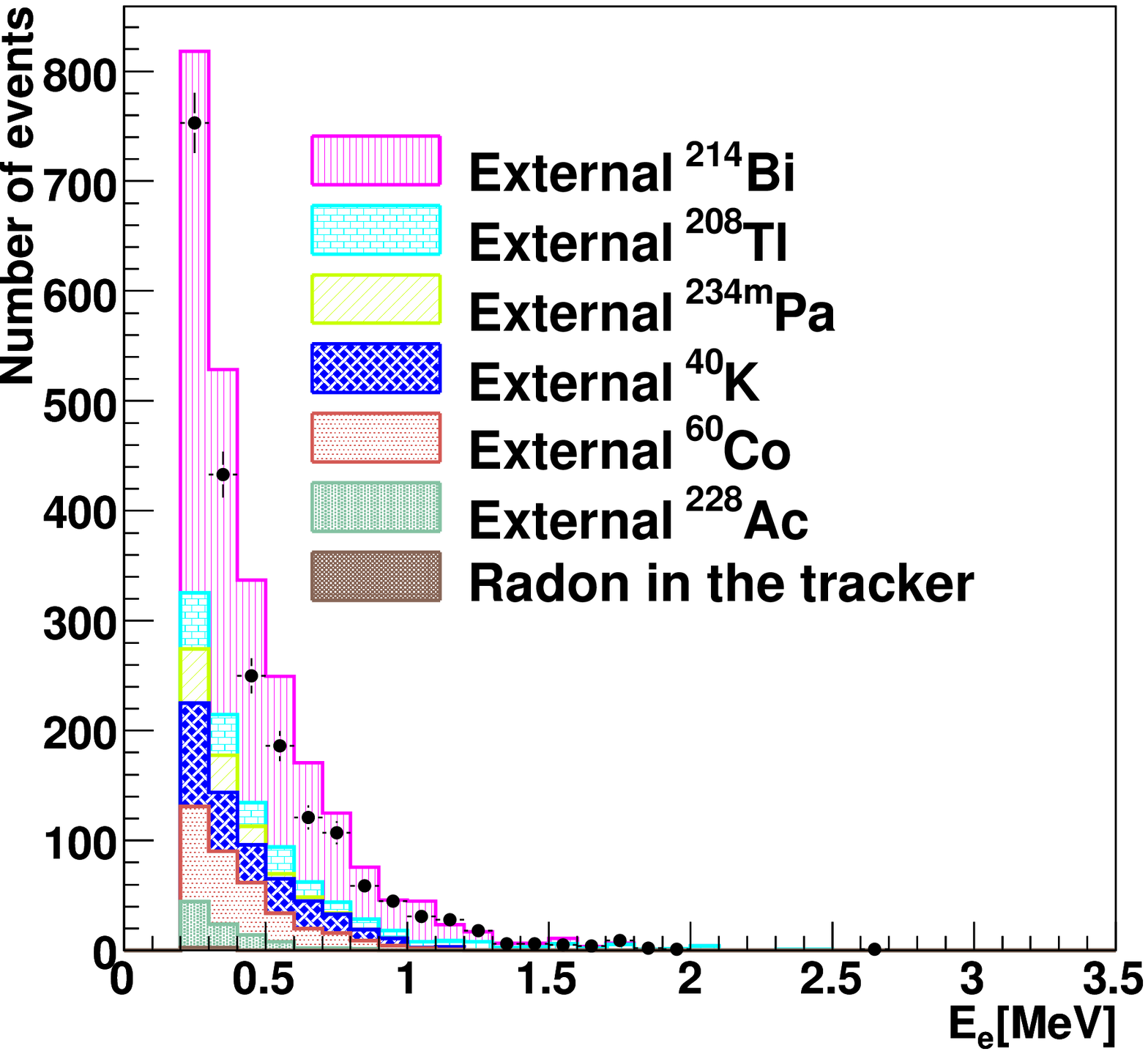}}
\subfigure{
b)\includegraphics[width=6.4cm]{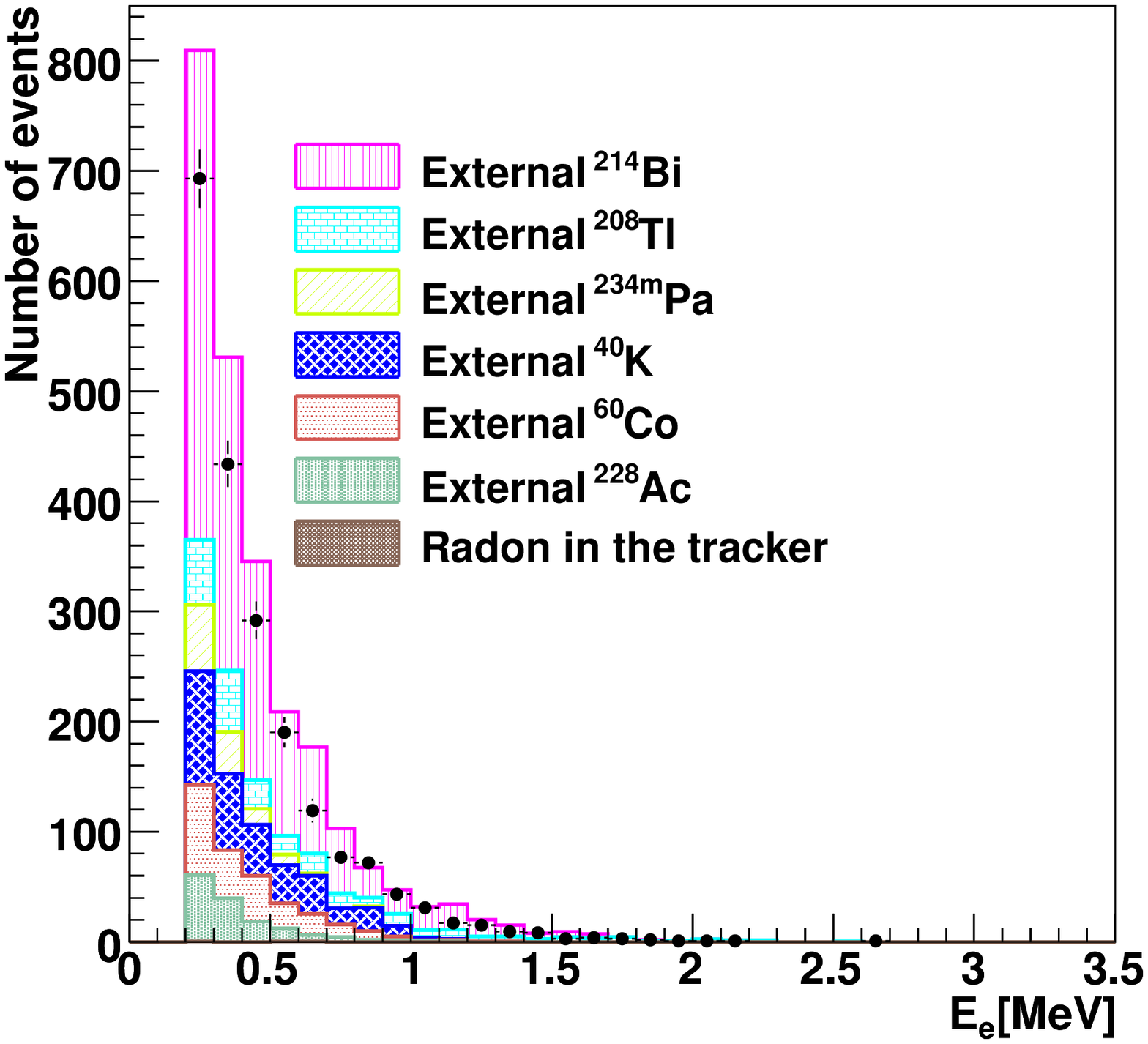}}
\subfigure{
c)\includegraphics[width=6.4cm]{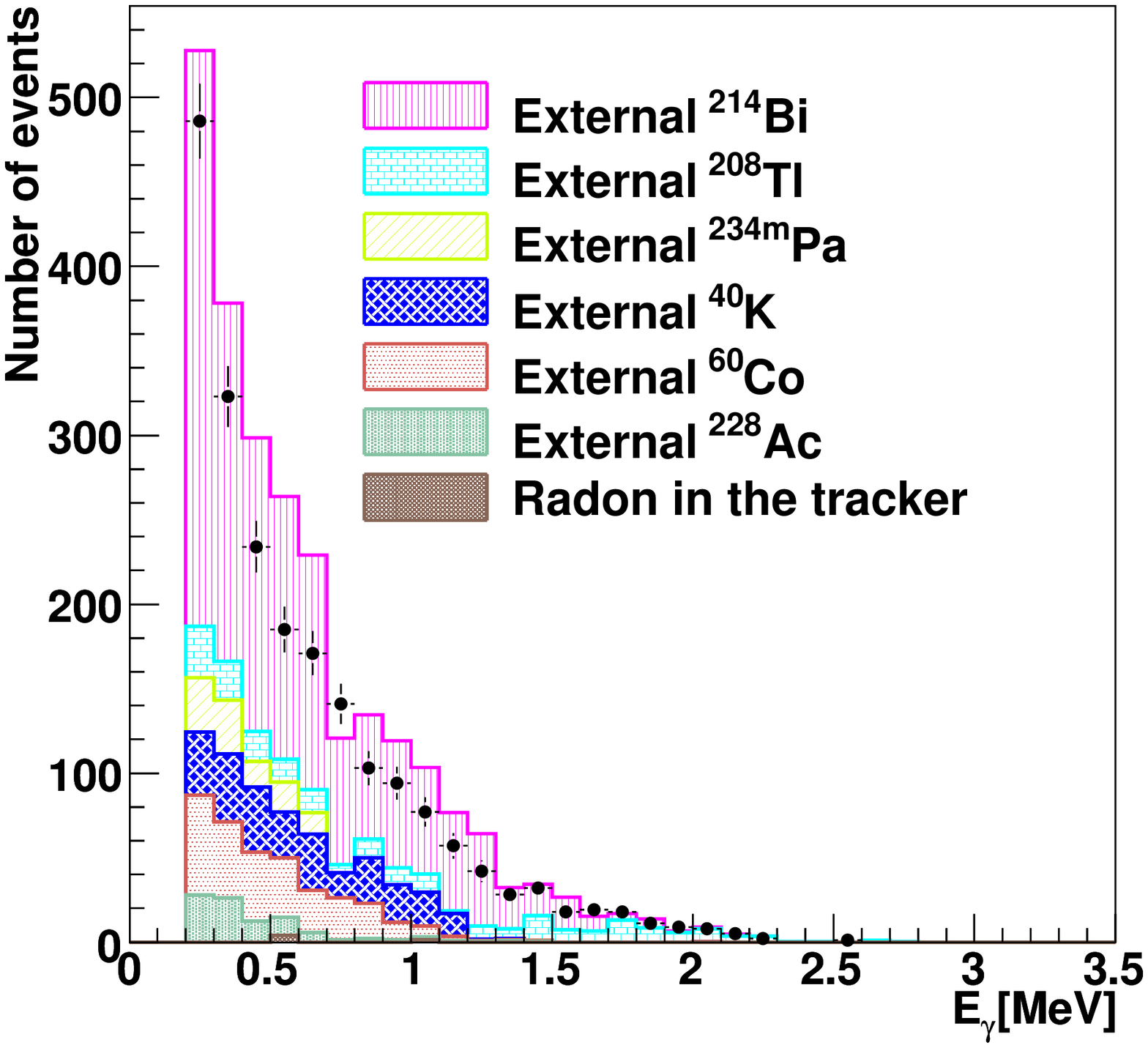}}
\subfigure{
d)\includegraphics[width=6.4cm]{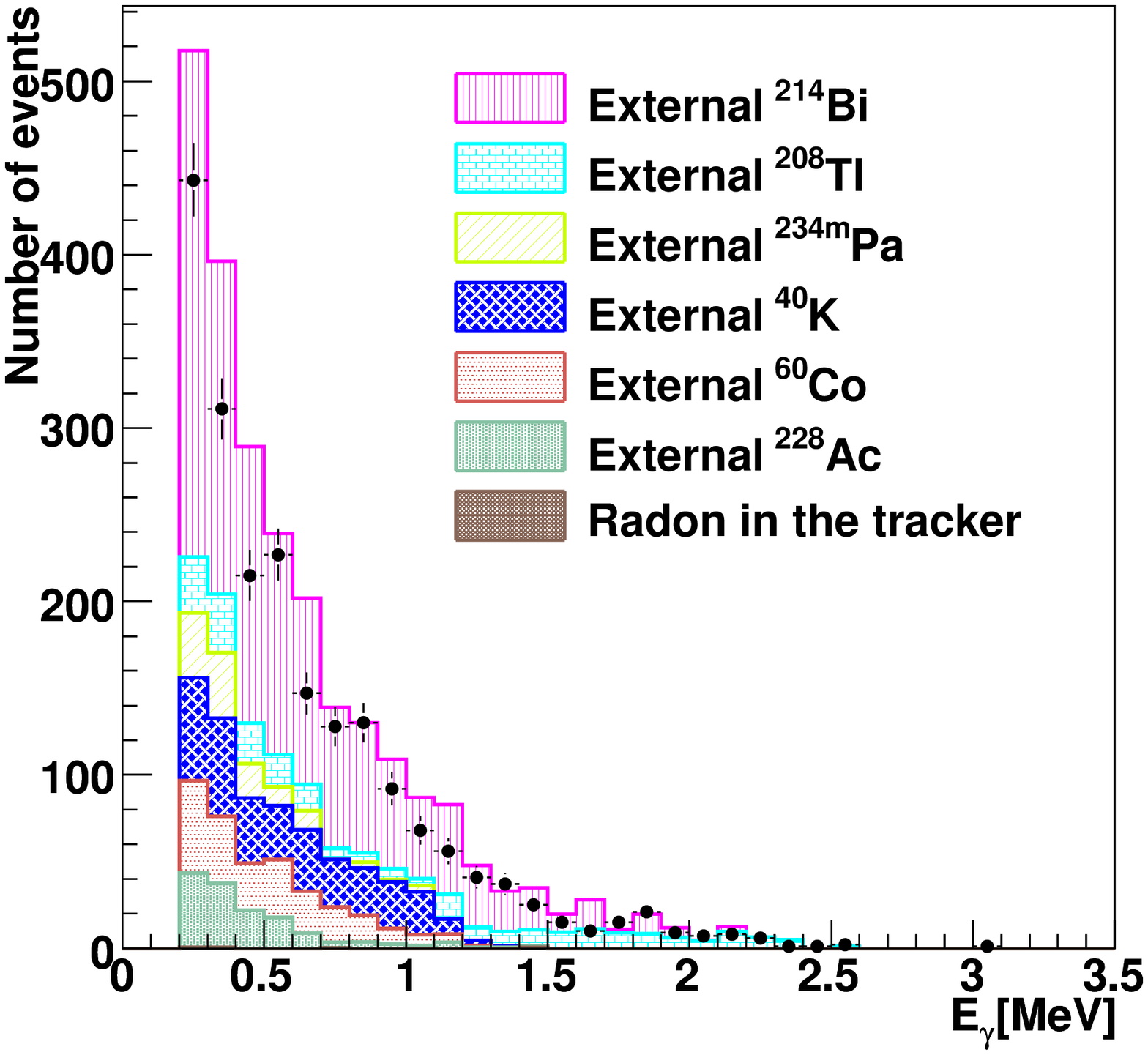}}
\subfigure{
e)\includegraphics[width=6.4cm]{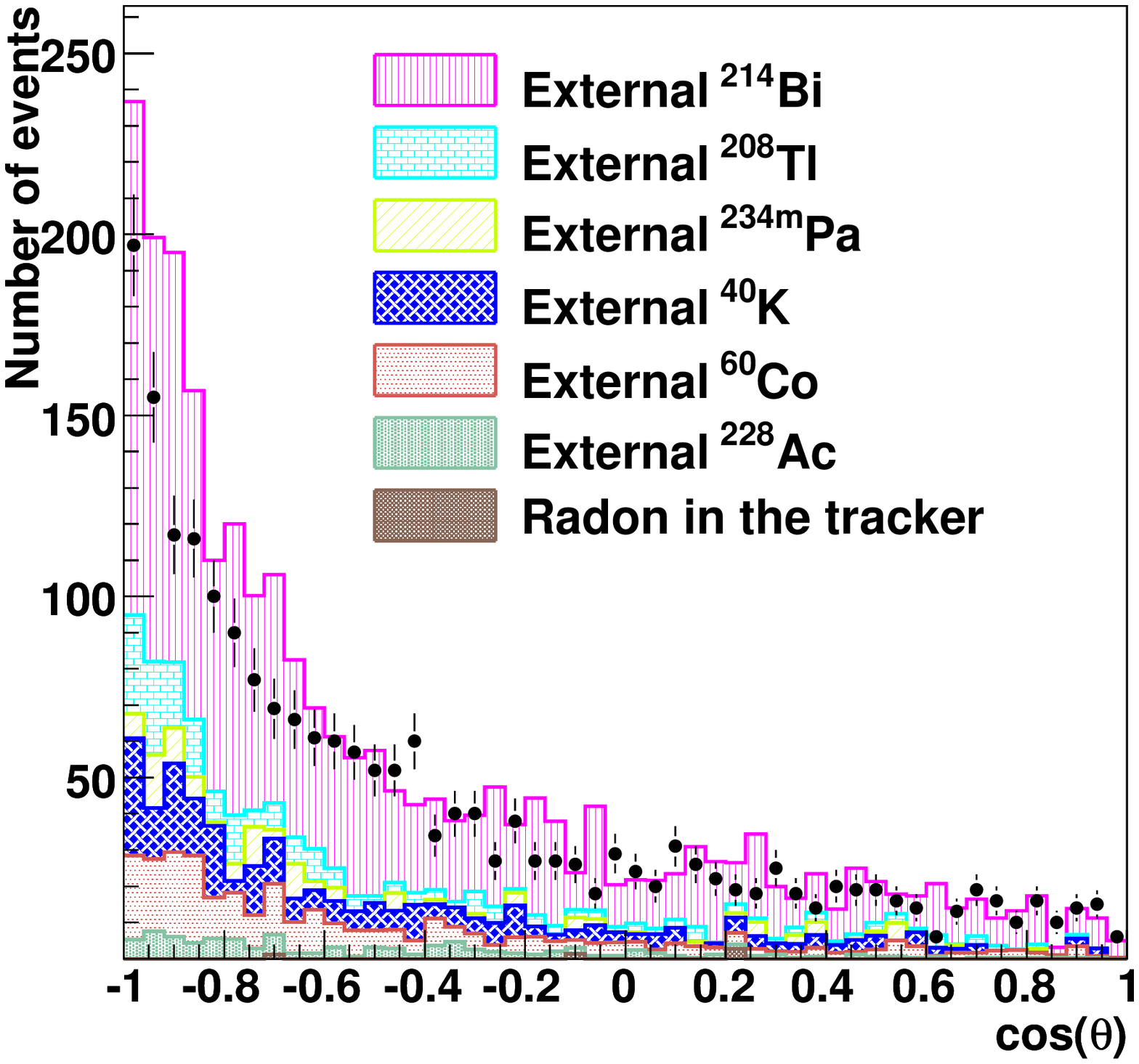}}
\subfigure{
f)\includegraphics[width=6.4cm]{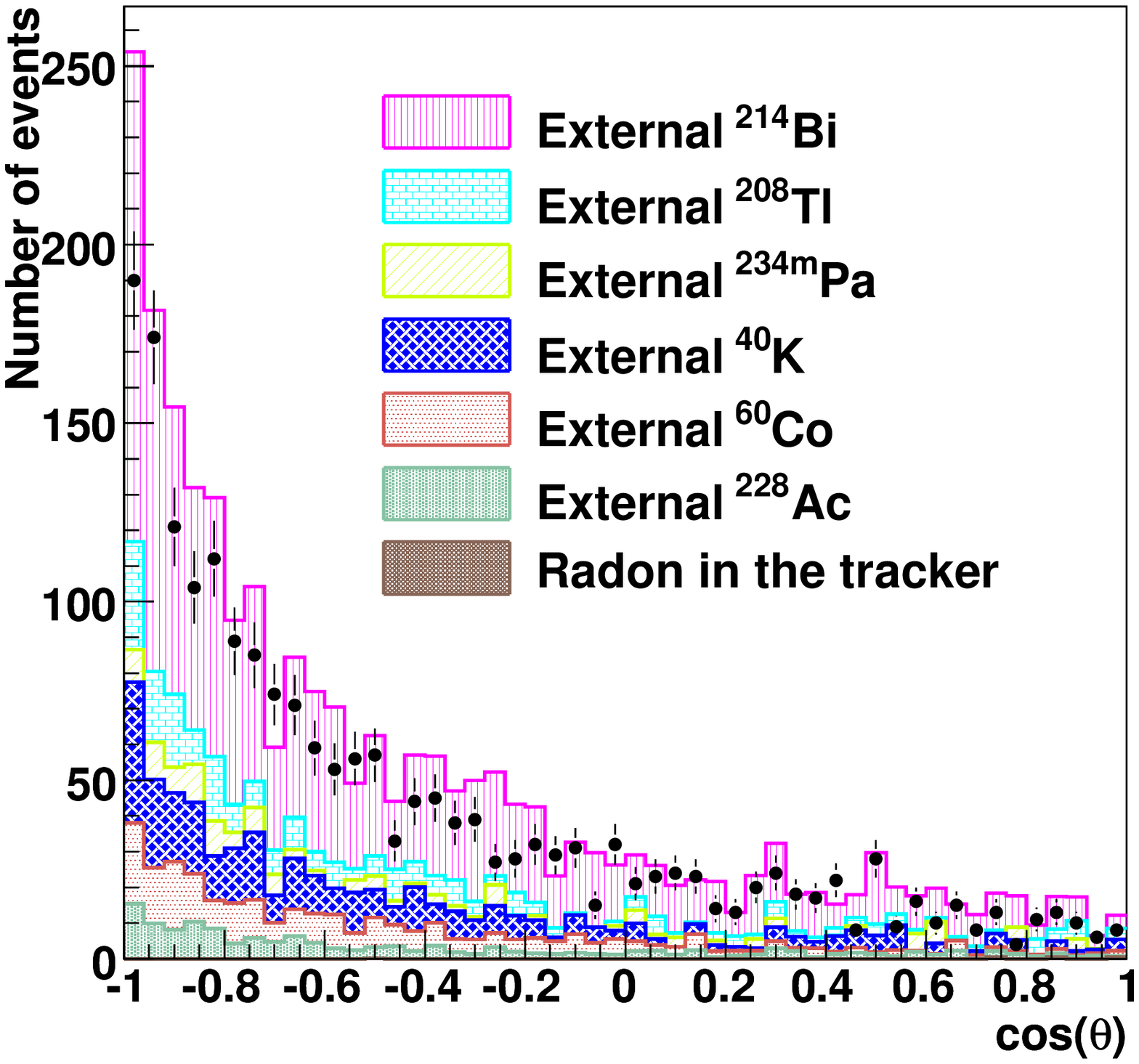}}
\caption[The electron and photon energy for external e$\gamma$ channel and the cosine of the angle between the two particles]{a, b) electron energy for external e$\gamma$ events; c, d) photon energy and  e, f) the cosine of the angle between the  electron and the photon. a, c, e) Phase~1 data; b, d, f) Phase~2 data. The statistical uncertainties on the data points are shown with error bars. }
\label{fig-exteg}
\end{figure}

To select external $e\gamma$ events, the same selections are applied as described in Section~\ref{sec-eg} except that the external TOF hypothesis for an $e\gamma$ event  has to be greater than $4\%$ and the internal TOF hypothesis has to be less than $1\%$. 
Figure~\ref{fig-exteg}  shows the electron energy, photon energy and cosine of the angle between the electron and the photon  for Phase~1 and Phase~2 data. The number of external $e\gamma$ events predicted by simulation are   $2480\pm55$~(stat) and $2399\pm55$~(stat),  respectively. The number of data events are $2065$ and $2016$. The difference between the number of expected background  events and data events  gives $16.7\%$  uncertainty on the external background in this channel.  

\section{Summary and discussion}
The $e\gamma$, $e\gamma\gamma$, $1e$ and $e\alpha$~\cite{externalbkg} channels were studied in order to  measure  the internal background activities with NEMO~3 data.  Table~\ref{tab-activitiesall} summarises the internal backgrounds, the  channels studied  and the activities measured.
\begin{table}
\centering
\begin{tabular}{|c|c|c|}
\hline
  Internal background & Channel &  Activity/mass~(mBq/kg) \\
\hline
\hline
$^{208}$Tl   &  $e\gamma$ and $e\gamma\gamma$     &  $10.1\pm1.0$      \\
\hline
$^{212}$Bi   &   from $^{208}$Tl activity       & $28.0\pm2.8$        \\
\hline
$^{228}$Ac  &   from $^{208}$Tl activity       &   $28.0\pm2.8$       \\
\hline
$^{207}$Bi    & $e\gamma$    & $226.8\pm5.5$  \\
\hline
$^{152}$Eu   &  $e\gamma$   & $51.3\pm2.4$ \\
\hline
$^{154}$Eu   &  $e\gamma$   & $24.6\pm1.1$ \\
\hline
$^{234m}$Pa   &      $1e$         &$47.7\pm3.0$ \\
\hline
$^{40}$K    &       $1e$        &   $213\pm12$ \\
\hline
$^{214}$Bi &     $e\alpha$ & $3.35\pm0.79$~\cite{externalbkg} \\
\hline
\end{tabular}
\caption{The  internal backgrounds activities  found from NEMO~3 data.}
\label{tab-activitiesall}
\end{table} 
The activities of $^{207}$Bi and $^{40}$K measured using NEMO~3 data are inconsistent  with the HPGe detector result. The $^{207}$Bi activity is found to be $1.8$ times more than  HPGe measurement.  As is observed in Figure~\ref{fig-hsbi207}, the foil has regions with  higher contamination of $^{207}$Bi than others  and it is possible  that HPGe  measured a sample of $^{150}$Nd foil with lower contamination~\cite{victort}. The activity of $^{40}$K is four times higher than the HPGe measurement. The HPGe detector is  a photon detector,  whereas $^{40}$K predominantly decays to an electron. The difference  can  therefore be due to the HPGe  detector's inability to detect electrons.

 In this thesis  a $17\%$ uncertainty on the external background model is used from the validation of the external background model with the $e\gamma$ channel. As the number of events from external background sources that mimic double beta decay is expected to be low, due to the small surface area of $^{150}$Nd foil (see Table~\ref{tab-act-ext}), this uncertainty will not contribute significantly to the uncertainty on the measurement of  $2\nu\beta\beta$ decay half-life.

%% file: twoneutrino.tex
\renewcommand{\baselinestretch}{1.6}
\normalsize
\chapter{Two-neutrino double beta decay of \boldmath{$^{150}$}Nd}
\label{chap-2nbb}
This chapter describes the measurement of the $2\nu\beta\beta$ half-life of $^{150}$Nd. The analysis is performed on data described in Section~\ref{sec-dataset}. The signature inside  the NEMO~3 detector  for $2\nu\beta\beta$ of $^{150}$Nd is two electrons emanating  from this foil. The event selection criteria for two-electron events are given in Section~\ref{sec-2e} followed by the presentation of the half-life result in Section~\ref{sec-2nbbresult}.
\section{Half-life definition} 
The number of expected events, derived from the radioactive decay law and  the event selection efficiency, $\epsilon$, of the considered channel, can be expressed as
\begin{equation}
N(t)=  \epsilon N_{at}(1- e ^{-\ln{2}(\frac{t}{T_{1/2}})}),
\label{eq-eventrate}
\end{equation}
where $N_{at}$ is the number of atoms in a sample. For the $^{150}$Nd sample in NEMO~3 with a mass of $36.55$~g, $N_{at}$ is equal to $1.462\times10^{23}$. The time $t$ is the data taking time, and $T_{1/2}$ is the half-life of the studied decay mode. Because of the large value of   $T_{1/2}$ for $2\nu\beta\beta$ decay compared to $t$, Equation~\ref{eq-eventrate} can be rewritten as
\begin{equation}
T_{1/2}= \epsilon N_{at} \ln{2} \frac{t}{N(t)}=
\epsilon N_{at} \ln{2} \frac{t}{N_{exp}-N_{bgr}^{tot}},
\label{eq-halflife2}
\end{equation}
where  $N_{exp}$ is the number of expected $2\nu\beta\beta$ events found from data and $N_{bgr}^{tot}$ is the total number of background events remaining in the data after applying the event selection criteria. 
\section{Two-electron event selection}
\label{sec-2e}
\begin{figure}
\centering
\subfigure{
a)\includegraphics[width=6.9cm]{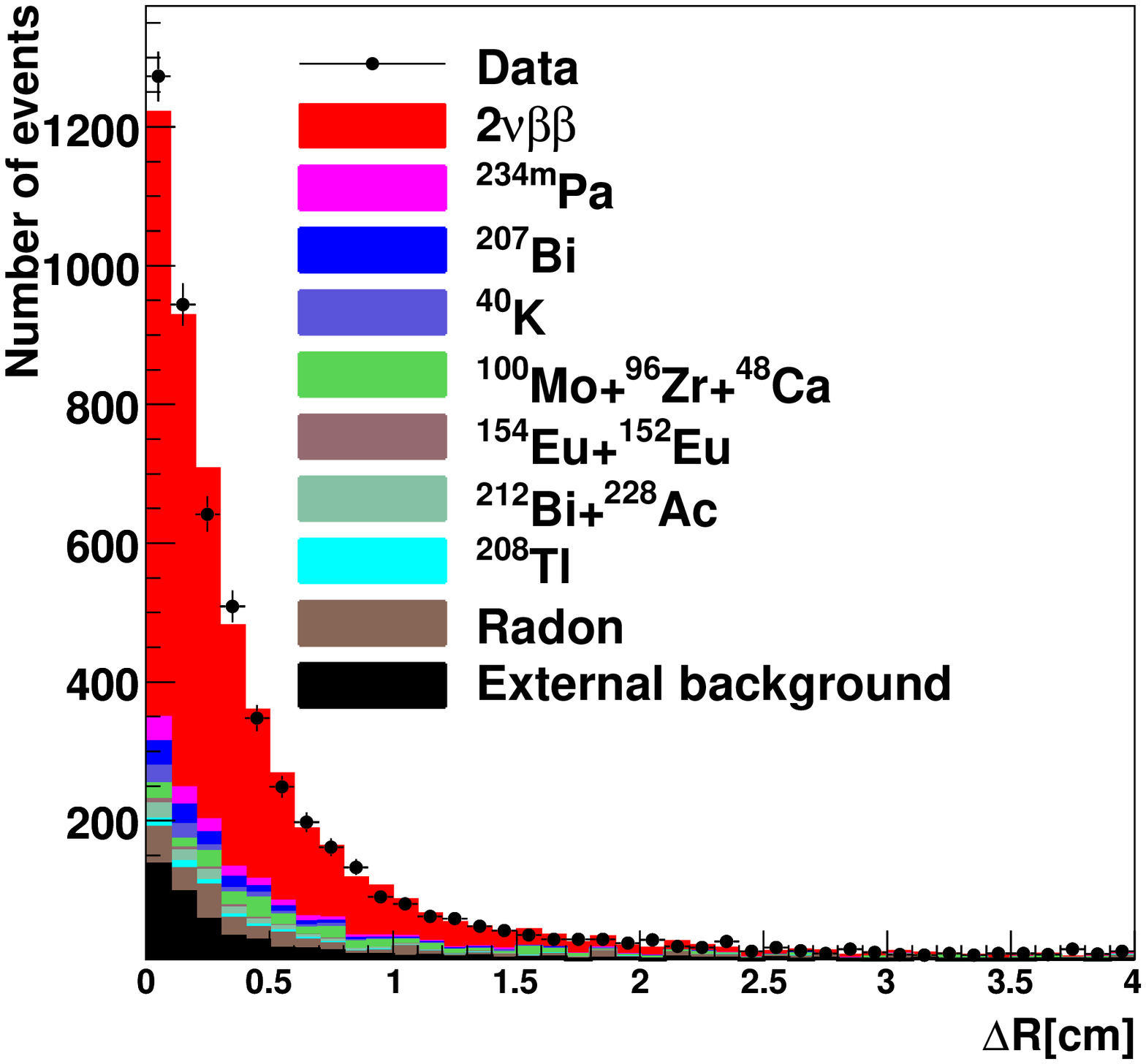}}
\subfigure{
b)\includegraphics[width=6.9cm]{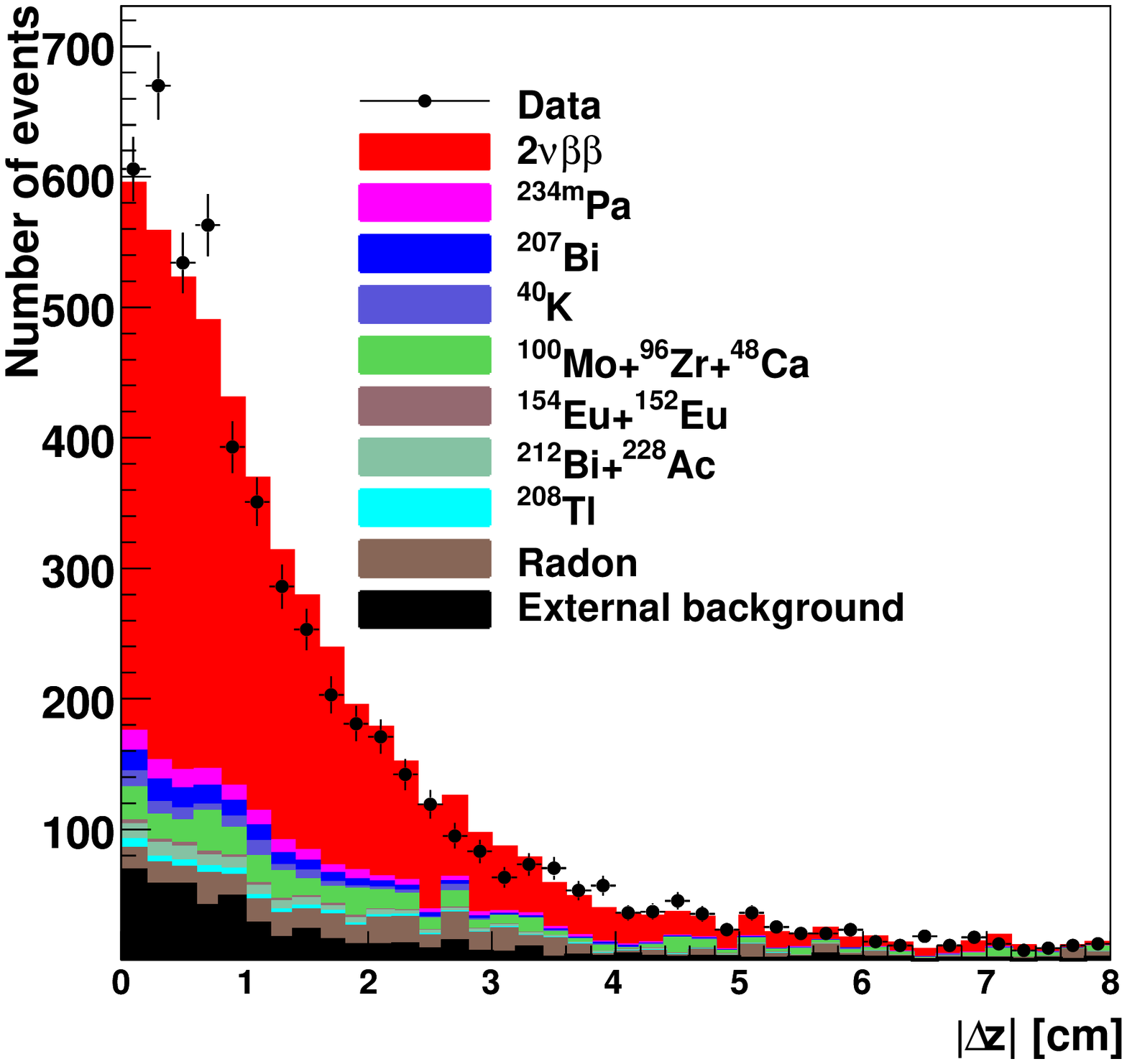}}
\caption[The vertex distribution in the $x-y$ plane and the vertical coordinate for two electron events]{The vertex distribution in the  $x-y$ plane ($\Delta R$ ) and the  vertical coordinate ($|\Delta z|$) for two electron events before the selections $|\Delta R|=\sqrt{(\Delta x)^{2}+(\Delta y)^{2}} < 2$~cm and $|\Delta z| < 4$~cm   are applied. The statistical uncertainties on the data points are shown with error bars. The $2\nu\beta\beta$ signal MC (red) is scaled to the  number of  background subtracted data events. }
\label{fig-dvert}
\end{figure}
\begin{figure}
\centering
\includegraphics[width=7.9cm]{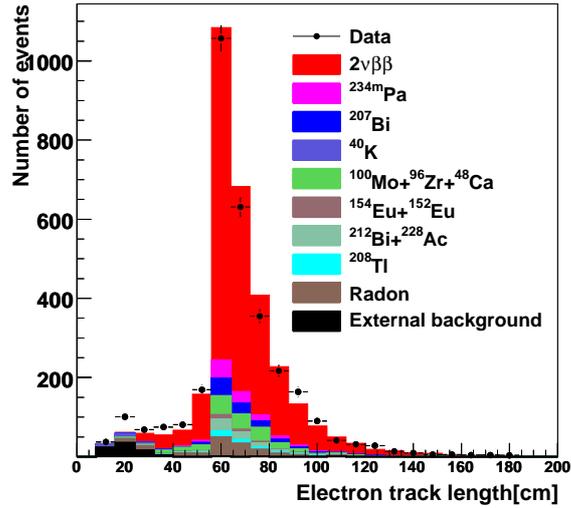}
\caption[The  track length distribution of one of the electrons of two-electron events]{The  track length distribution of one of the electrons of two-electron events before applying the track length $>30$~cm requirement. The error bars show the statistical uncertainties on the data points. }
\label{fig-trackl} 
\end{figure}

\label{sec-2e}
\begin{figure}
\centering
\subfigure{
a)\includegraphics[width=6.9cm]{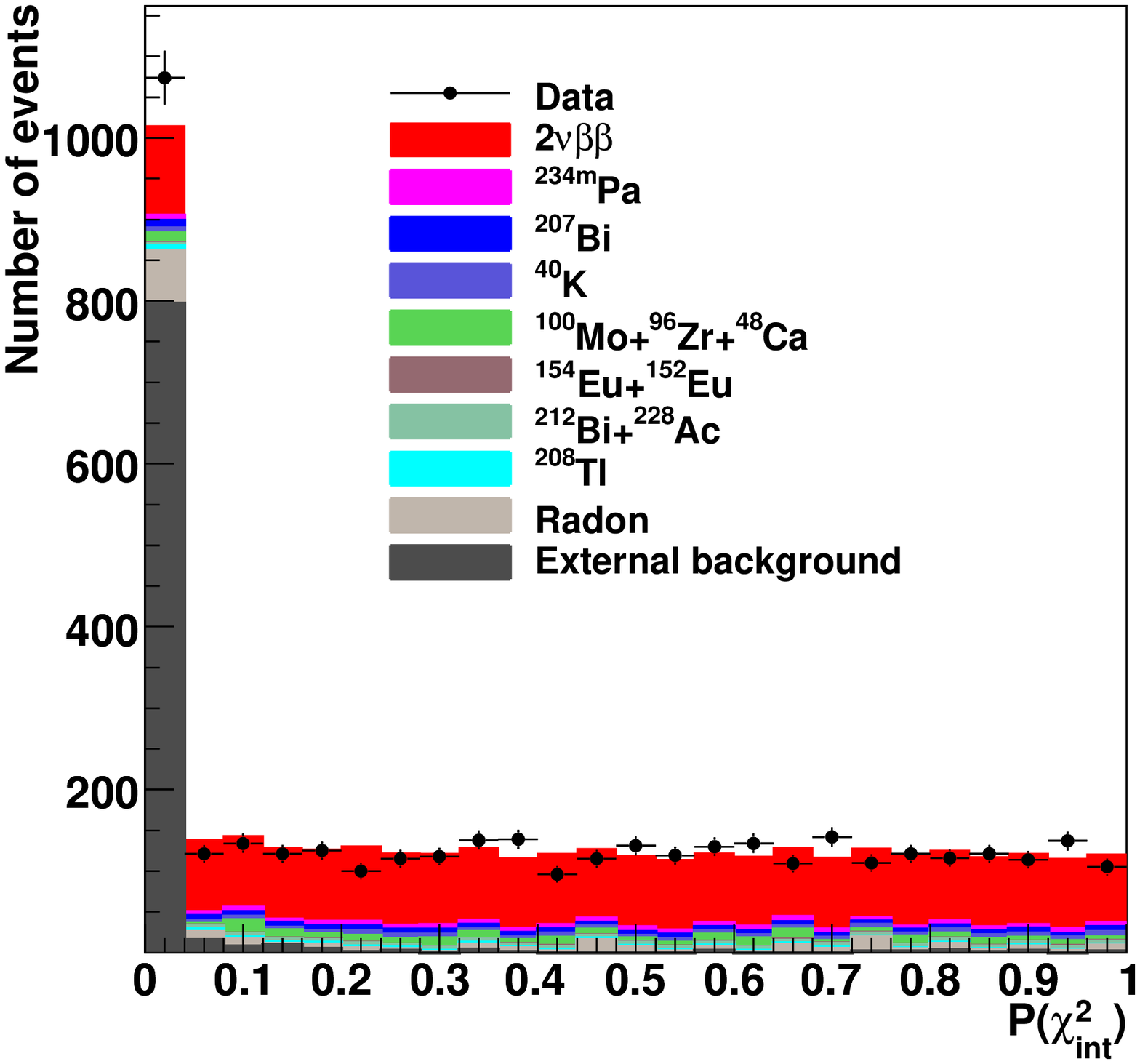}}
\subfigure{
b)\includegraphics[width=6.9cm]{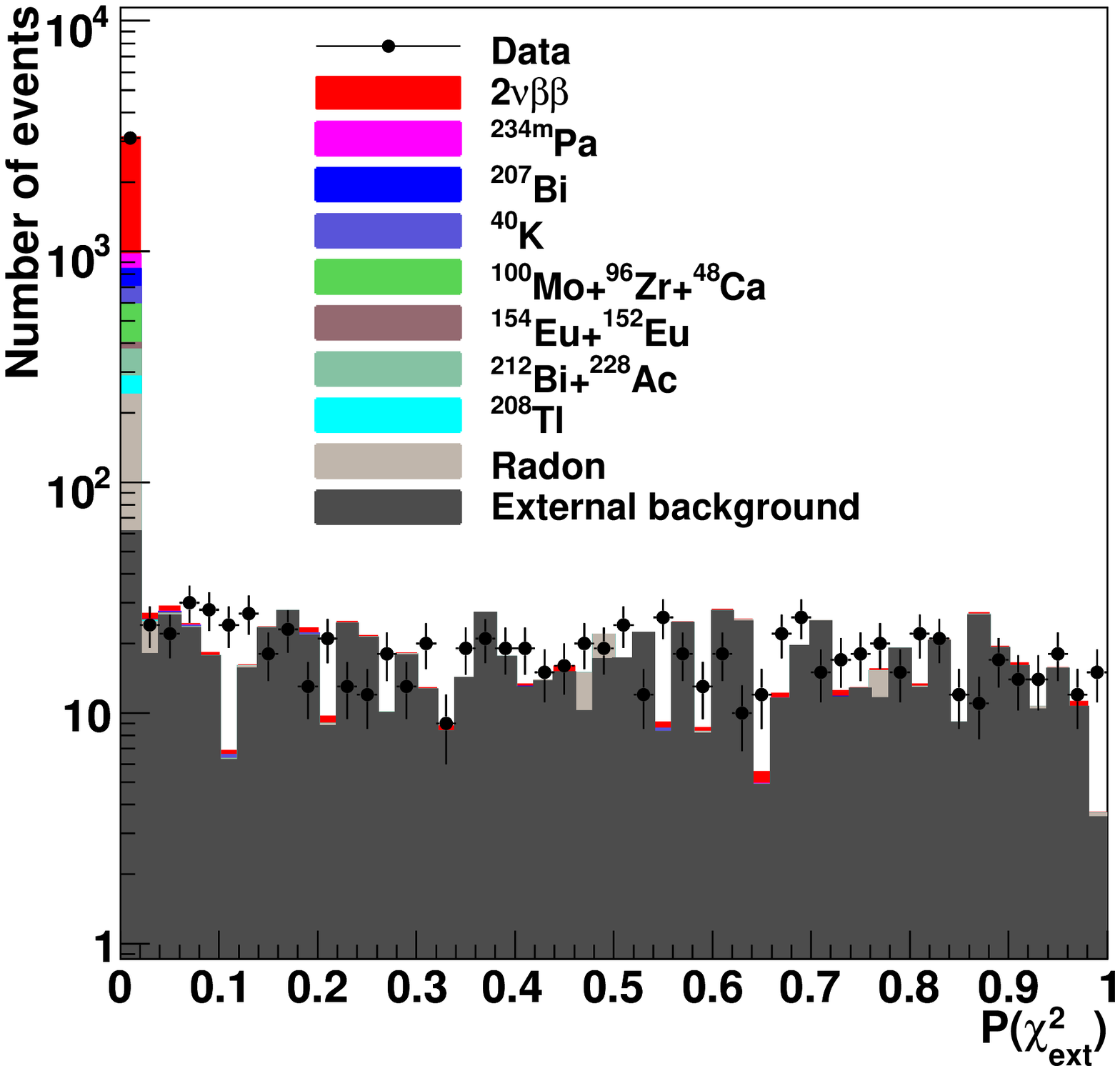}}
\caption[The distribution of the internal and external TOF hypothesis for two-electron events]{ The distribution of the  a) internal TOF hypothesis and b) external TOF hypothesis before applying the requirement of $P(\chi_{int}^{2})>0.04$ and $P(\chi_{ext}^{2})<0.01$.  The external background events (gray)  are normalised to the number of data events  for $P(\chi_{int}^{2})<0.04$ and  $P(\chi_{ext}^{2})>0.01$.  The error bars show the statistical uncertainties on the data points. }
\label{fig-tof2e}
\end{figure}
The following selections are applied to the data in order to reduce background to double beta decay:
\begin{itemize}
\item Two negatively charged particles are required. The tracks must be  associated with two scintillator hits.
\item Both tracks are required to originate from the $^{150}$Nd foil.
\item  To remove  the events from  hot-spot regions, events are rejected if they   originate from $^{207}$Bi and $^{234m}$Pa contaminated regions defined in Equations~\ref{eq-bi207hs} and~\ref{eq-pahs}.
\item The energy of each electron is required to be greater than 0.2~MeV.
\item Both electrons originate from a common vertex and have: 
\begin{itemize}
\item $|\Delta R|=\sqrt{(\Delta x)^{2}+(\Delta y)^{2}} < 2$~cm,
\item $|\Delta z| < 4$~cm,
\end{itemize}
where $\Delta R$ and $\Delta z$ are the  distances between the two tracks' intersection points with the foil in the horizontal plane, $x-y$, and in the $z$ coordinate.  Figures~\ref{fig-dvert}a and Figures~\ref{fig-dvert}b show the distribution of these variables before applying the vertex cuts and after applying all other cuts described above.  
\item In order to reduce the external background, the internal TOF hypothesis for two-electron events is required to be  greater than $4\%$, and the external TOF hypothesis to be  less than $1\%$. Figure~\ref{fig-tof2e} shows the distributions of the internal and external TOF hypotheses before applying the TOF hypothesis requirement and after applying all  other selections described above.  In these figures the number of external background events is normalised to the number of data events for $P(\chi_{int}^{2}<0.04)$ and $P(\chi_{ext}^{2}>0.01)$. 
\item The track length for each electron is greater than $30$~cm. This cut value is lower than the one applied to channels involving one electron track only,  as the TOF  measurement for two track events is more precise~\cite{technical}. Figure~\ref{fig-trackl} shows the  track length distribution for one of  the electrons before applying the track length requirement and after applying all other selections.  It is observed  that signal over background ratio is small for electrons with track length less than 30~cm and that the rate of events with track length $<$30~cm is not well simulated by the MC.
\item The tracks pass through one of  the first two layers of the tracker.
\end{itemize}
Table~\ref{tab-selectiondata} gives the   number of data events remaining  after  each event selection criteria is applied and Table~\ref{tab-selectionmc} gives the fraction of MC events remaining after each cut. This shows the effect of cuts on the $2\nu\beta\beta$ signal and the internal and external backgrounds with major contributions in the   $2\nu\beta\beta$ signal region. 

\begin{figure}
\centering
\includegraphics[width=11cm]{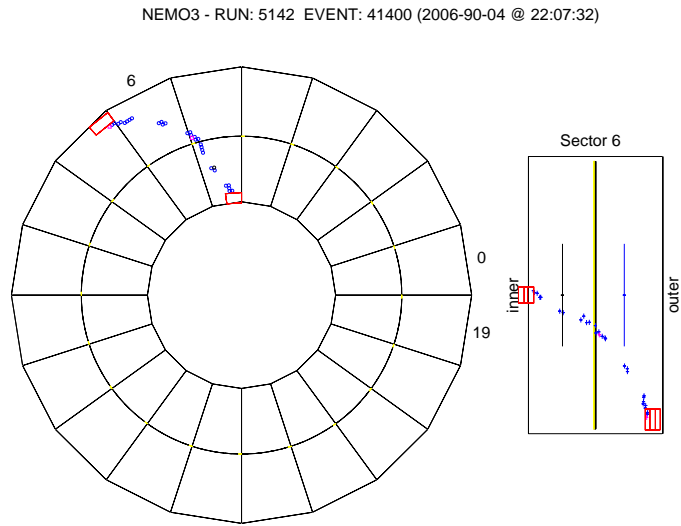}
\caption[A display of a two-electron event originating from $^{150}$Nd foil]{A display of a two-electron event originating from $^{150}$Nd foil. The top ($x-y$) and side  views ($z$ coordinate)  of the event are shown.}
\label{fig-eventdisplay2e}
\end{figure}

 In order to illustrate the effect of each cut on external backgrounds and radon in the tracker, the number of remaining simulated events after each cut is   shown for $^{214}$Bi in the glass of the PMTs and in the tracker. Figure~\ref{fig-eventdisplay2e} shows the display of a selected two-electron event  originating from the $^{150}$Nd foil. The electron tracks and scintillator hits  are shown.

\begin{table}
\centering
\footnotesize
\begin{tabular}{|c|c|c|}
 \hline
Cut&Cut & Number of events ($N$)\\
\hline
\hline
(1)&two tracks & $157575017$\\
\hline
(2) &tracks associated with scintillators & $14309481$\\
\hline
(3) & particles have negative charge & $1213245$\\
\hline
(4) &tracks originated from $^{150}$Nd foil & $9249$\\
\hline
(5) &Not originated from hot-spot regions & $8975$\\
\hline
(6) &Energy of each electron $>0.2$~MeV & $5657$\\
\hline
(7)&$\Delta R<2.0$~cm and $\Delta z<4.0$~cm  & $3983$\\
\hline
(8) &$P(\chi^{2}_{int})>0.04$ and $P(\chi^{2}_{ext})<0.01$ & $2891$\\
\hline
(9) &Length of each track $>30$~cm    & $2852$\\
\hline
(10) &Tracks pass through one of the first two layers & $2789$\\
\hline
\end{tabular}
\caption{Details of the two-electron event selection criteria. The number of events remaining in the data sample are presented after applying each selection.}
\label{tab-selectiondata}
\centering
\footnotesize
\vspace{1.0cm}
\begin{tabular}{|c||c||c|c|c|c|c||c|c|c}
\hline
                  Cut     & $2\nu\beta\beta$& $^{234m}$Pa& $^{40}$K &$^{208}$Tl&$^{228}$Ac & $^{207}$Bi& $^{214}$Bi tracker&$^{214}$Bi PMT\\
\hline
\hline
(1)                   &   $31.0\%$ &     $0.6\%$     &      $1.9\%$           & $0.55\%$     &      $0.30\%$         &  $0.07\%$ & $0.0027\%$   & $(3.5\times10^{-4})\%$\\  
\hline
(2)                  &   $15.6\%$  &  $0.3\%$       &    $0.08\%$          &     $0.27\%$     &       $0.15\%$          &  $0.035\%$&  $0.0013\%$  &  $(1.8\times10^{-4})\%$\\
\hline
(3)                     & $13.6\%$ &   $0.19\%$      &   $0.036\%$           &   $0.21\%$        &         $0.10\%$         & $0.026\%$  & $0.001\%$    & $(1.7\times10^{-4})\%$\\
\hline
(4)                 &   $10.4\%$  &     $0.14\%$      &   $0.03\%$            &   $0.17\%$      &       $0.083\%$           &$0.021\%$   & $(3.37\times10^{-4})\%$  & $(3.9\times10^{-5})\%$\\
\hline
(5)                 &     $10.1\%$  &  $0.12\%$       &    $0.028\%$           &    $0.16\%$    &      $0.079\%$            & $0.020\%$  & $(3.30\times10^{-4})\%$     & $(3.8\times10^{-5})\%$\\
\hline
(6)                  &  $8.7\%$   &    $0.1\%$       &    $0.012\%$          &     $0.13\%$   &      $0.057\%$        & $0.018\%$      &  $(3.37\times10^{-4})\%$  & $(3.8\times10^{-5})\%$\\
\hline
(7)                   & $7.8\%$    &    $0.08\%$      &     $0.0083\%$         &      $0.11\%$   &    $0.049\%$       & $0.016\%$  &   $(1.23\times10^{-4})\%$  & $(2.5\times10^{-5})\%$\\
\hline
(8)                   &  $7.4\%$   &     $0.078\%$     &       $0.0079\%$      &    $0.10\%$       &     $0.047\%$       & $0.0152\%$  &  $(6.3\times10^{-5})\%$ & $(1.1\times10^{-6})\%$\\
\hline
(9)                   &  $7.3\%$ &   $0.075\%$      &  $0.0078\%$           &    $0.10\%$       &      $0.046\%$       &  $0.015\%$  &  $(5.6\times10^{-5})\%$  &$(1\times10^{-6})\%$\\
\hline
(10)                   &   $7.2\%$ &  $0.074\%$      &  $0.0077\%$          &     $0.10\%$      &    $0.0457\%$         &  $0.015\%$ & $(4.9\times10^{-5})\%$ &$(1\times10^{-6})\%$\\
\hline
\end{tabular}
\caption{The effect of cuts on $2\nu\beta\beta$ signal and  internal and external background MC samples. The fraction of  events remaining after each cut  is given for the internal backgrounds with major contribution to the $2\nu\beta\beta$ signal  and  for $^{214}$Bi from the tracker and the PMTs.}
\label{tab-selectionmc}
\end{table}
\section{Double beta decay half-life}
\label{sec-2nbbresult}
After applying  all two-electron event selections, 2789 events remain. The number of  
Monte Carlo  events are normalised to the activities found in Chapter~\ref{chap-bgr} to estimate background 
remaining in the data sample. 
 Due to the resolution of the tracking detector, events from neighbouring foils ($^{100}$Mo, $^{48}$Ca and $^{96}$Zr) can be reconstructed as originating from the $^{150}$Nd foil. The $^{48}$Ca foil is heavily contaminated with $^{90}$Y  which itself is a source of background to $2\nu\beta\beta$~\cite{shivaking}.  The activities of these isotopes are given in  Table~\ref{tab-act-int}. 
   The total number of background events from neighbouring foils is estimated to be $168.1\pm13~({\rm stat})\pm14~({\rm syst})$ from MC simulations, where the systematic uncertainty is determined by the uncertainty on the half-life of these isotopes.

Table~\ref{tab-act-int}  summarises the internal background originating from $^{150}$Nd and  neighbouring foils. The number of total internal background events is  found to be $693\pm28$. The uncertainty on the number of background events is due to  the statistical errors and the systematic uncertainties on the activity measurements.  The systematic uncertainties on  $^{208}$Tl, $^{228}$Ac and $^{212}$Bi activities are  fully correlated. The same  is true for $^{214}$Pb and $^{214}$Bi. 
Table~\ref{tab-act-ext} summarises the  external background that mimic $2\nu\beta\beta$ events. The total number of external background events is estimated to be $53\pm 11$. The $20\%$ uncertainty includes statistical uncertainties and   a $17\%$ systematic uncertainty on the external background model. By adding up the number of internal and external background events, the total number of background events  is found to be $746\pm30$.            
\begin{table}
 \begin{center}
  \footnotesize
    \begin{tabular}{|c|c|c|c|}   
      \hline
     Contaminant & $A$~(mBq) &  Selection efficiency &  $N_{bgr}$ \\     
      \hline
       $^{234m}$Pa&    $2.42\pm0.14$    &   $0.074\%$          & $144\pm 13~({\rm stat})\pm9~({\rm syst})$   \\
       $^{40}$K &      $10.8\pm0.6$   &  $0.0077\%$        & $66\pm4~({\rm stat})\pm4~({\rm syst})$    \\
       $^{208}$Tl &     $ 0.51\pm 0.05$ &      $0.10\%$           & $46\pm 1~({\rm stat})\pm4~({\rm syst})$   \\
       $^{228}$Ac &  $14.2\pm1.4$      &$0.046\%$ & $52\pm2~({\rm stat})\pm 5~({\rm syst})$\\
       $^{212}$Bi &  $14.2\pm1.4$  &$0.029\%$ & $32\pm 2~({\rm stat})\pm3~({\rm syst})$\\    
       $^{207}$Bi &    $11.5\pm0.3$   &     $0.015\%$     & $138\pm3~{\rm(stat)}\pm 5~{\rm(syst)}$ \\
       $^{214}$Bi & $0.17\pm0.04$~\cite{externalbkg}     &   $0.097\%$    &  $13\pm 1~({\rm stat})\pm3~({\rm syst})$ \\
       $^{214}$Pb &   $0.17\pm0.04$~\cite{externalbkg}   &   $0.042\%$    &    $6\pm1~({\rm syst})$    \\
$^{152}$Eu        &  $2.6\pm0.1$        &       $0.0094\%$ &   $19\pm1~({\rm stat})\pm 1~({\rm syst})$       \\
$^{154}$Eu        &   $1.25\pm0.05$       &     $0.0087\%$    &  $9\pm 1~({\rm stat})\pm 1~({\rm syst})$   \\        
    \hline  
     $^{48}$Ca   &     $(4.3\pm0.4)\times 10^{-2}$~\cite{shivaking}         &$0.062\%$                  & $2.1\pm0.2~({\rm syst})$       \\
     $^{90}$Y    &  $30.6\pm3.1$~\cite{shivaking}        &   $9.1\times10^{-4}\%$               &   $22\pm3~({\rm stat})\pm 3~({\rm syst})$  \\
    \hline 
    $^{96}$Zr    & $(7.2\pm0.7)\times 10^{-2}$~\cite{mattnote}        & $0.082\%$                   &  $5\pm1~({\rm syst})$  \\
    \hline
    $^{100}$Mo  &      $0.13\pm0.01$~\cite{mo100paper}        & $1.34\times 10^{-3}\%$                & $139.0\pm 13~({\rm stat})\pm14~({\rm syst})$  \\
   \hline
        Sum      &                        &                 &   $693\pm28$         \\      
\hline               
   \end{tabular}
\end{center}
    \caption[Summary of the internal background to the $2\nu\beta\beta$ signal of $^{150}$Nd]{Summary of the internal background to the $2\nu\beta\beta$ signal of $^{150}$Nd. Each background's activity,  the two-electron selection efficiency and the number of events which contribute to two-electron events in 924.7 days are given. The statistical uncertainty due to the finite size  of MC samples are added in quadrature to the errors found in Chapter~\ref{chap-bgr}. }
     \label{tab-act-int}
\end{table}
\begin{table}
\begin{center}
\begin{tabular}{|c|c|}
\hline
 Background &    $N_{bgr}$ \\
\hline
Total external $^{214}$Bi  & $10.0\pm 4$~(stat) \\
Total external $^{60}$Co & $5\pm 3$~(stat) \\ 
$^{214}$Bi/$^{214}$Pb in the tracker & $31 \pm 3$~(stat) \\
$^{210}$Bi in the tracker &   $7\pm 1$~(stat)           \\
\hline
Sum     &         $53\pm11$\\ 
\hline
\end{tabular}
\caption[Summary of the external background remaining  in the two-electron sample after the event selection criteria in 924.7 days of data taking period]{Summary of the external background remaining  in the two-electron sample after the event selection criteria in 924.7 days of data taking. A systematic uncertainty of 17\% is added to the statistical uncertainties.}
\label{tab-act-ext}
\end{center}
\end{table}
\begin{figure}[h]
\centering
\subfigure{
a)\includegraphics[width=6.9cm]{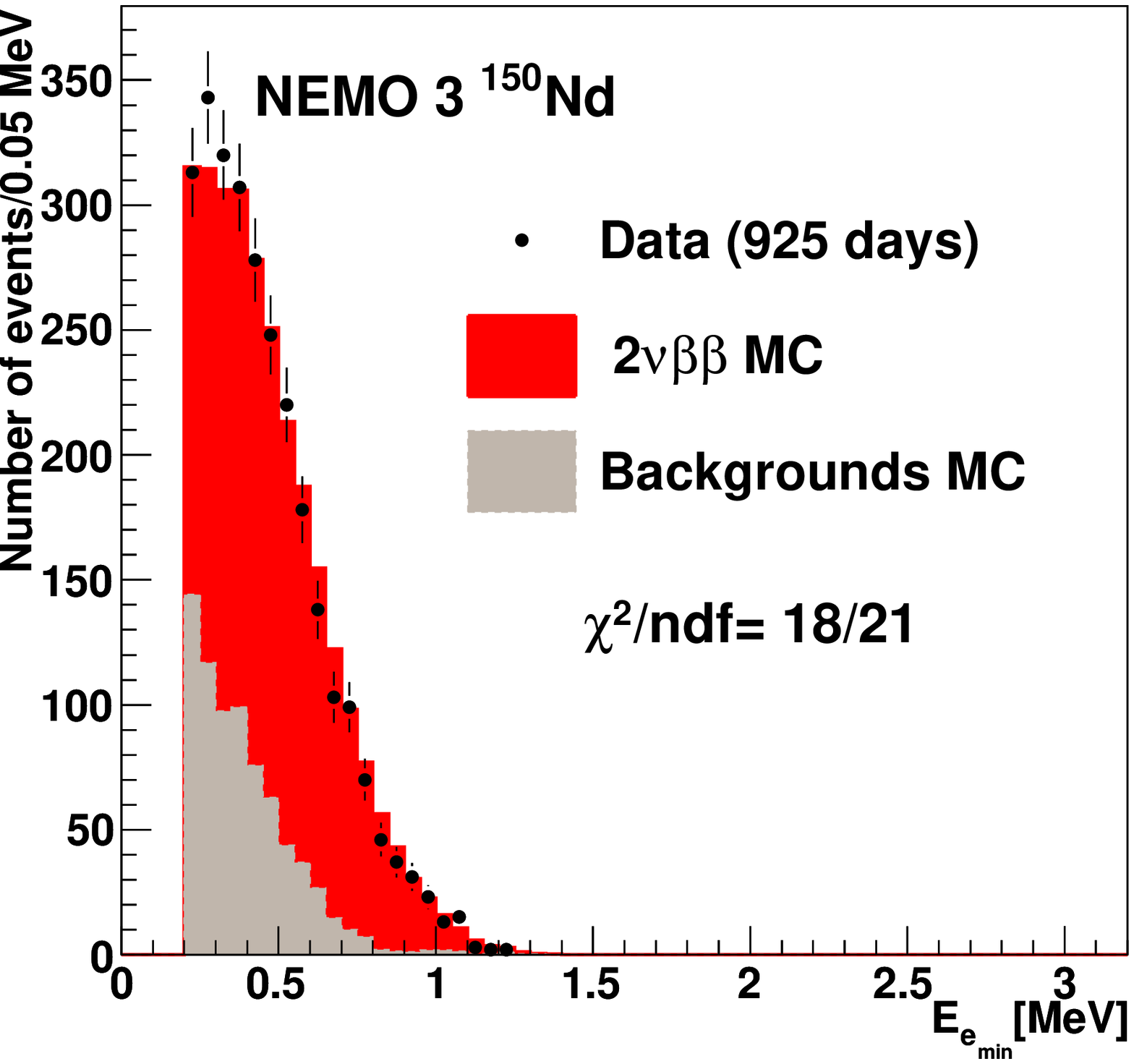}}
\subfigure{
b)\includegraphics[width=6.9cm]{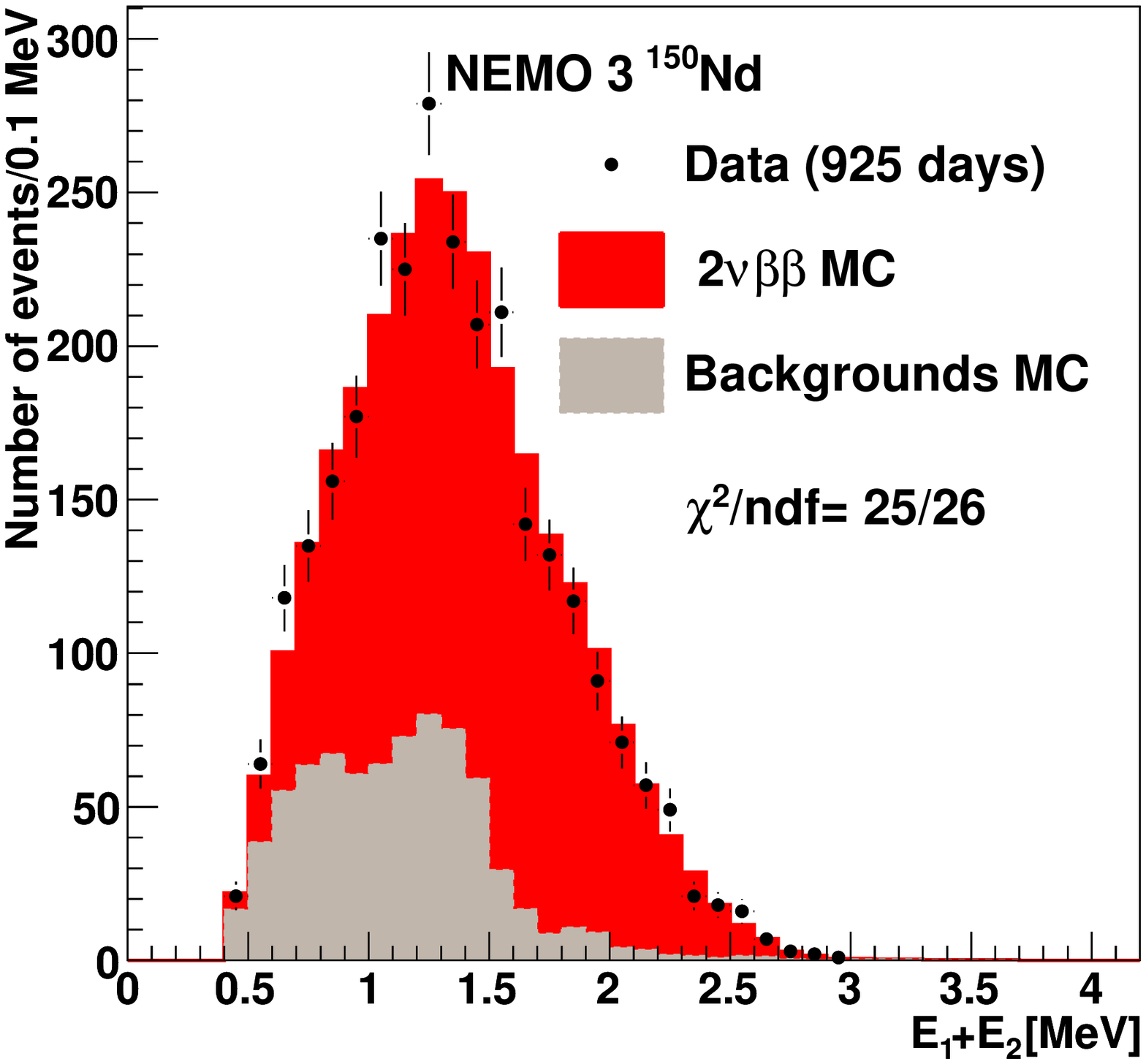}}
\subfigure{
c)\includegraphics[width=6.9cm]{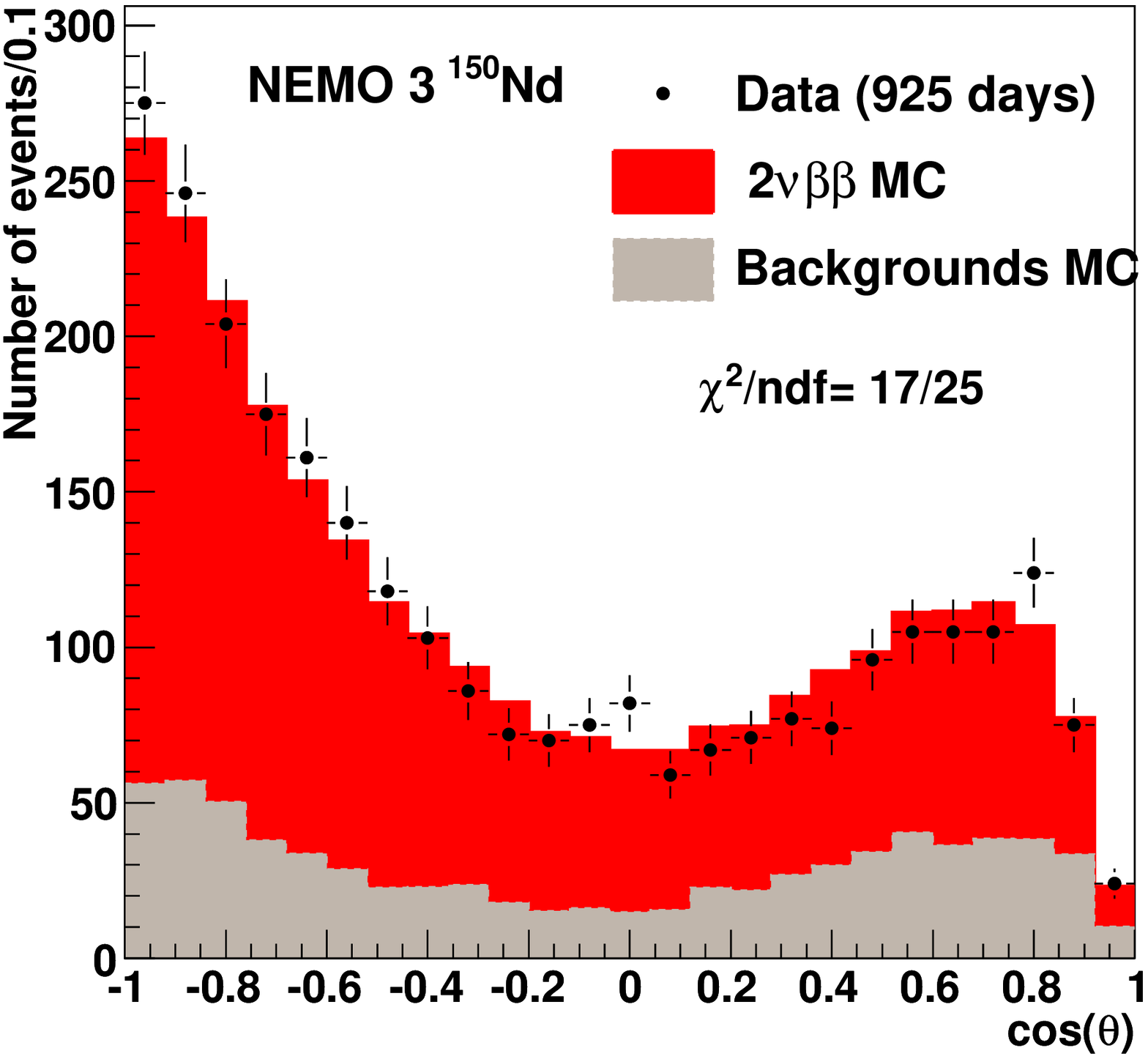}}
\caption[The distributions of energy of the electron with lower energy, energy sum of the two electrons and the cosine of the angle between the two electrons]{The distributions of a) energy of the electron with lower energy; b) energy sum of the two electrons  and c) the cosine of the angle between the two electrons. The sum of the  total radioactive background (gray) and  the $2\nu\beta\beta$ signal (red) is in  good agreement with data. The statistical uncertainties on data points are shown with error bars. }
\label{fig-2nbbplots}
\end{figure}

The energy distribution of electrons with the lower  energy, the energy sum of the two electrons and the opening angle between them are shown in Figure~\ref{fig-2nbbplots}. The normalisation of the $2\nu\beta\beta$  signal is found by scaling the  simulated MC events to the same number of data events minus the radioactive background events. The data are in good agreement with the sum of the background and the $2\nu\beta\beta$ signal distribution. The efficiency of the $2\nu\beta\beta$ event selection is $7.2\%$. After background subtraction  the half-life of the $2\nu\beta\beta$  decay is measured to be:
\begin{equation}
T^{2\nu}_{1/2}=(9.11^{+0.25}_{-0.22}~{\rm(stat.)}\pm0.62~{\rm(syst.)})\times 10^{18}~{\rm y}.
\end{equation}
\subsubsection{Systematic uncertainties}
\label{sec-systematic2n2b}
\begin{figure}
\centering
\includegraphics[totalheight=9cm, width=10cm]{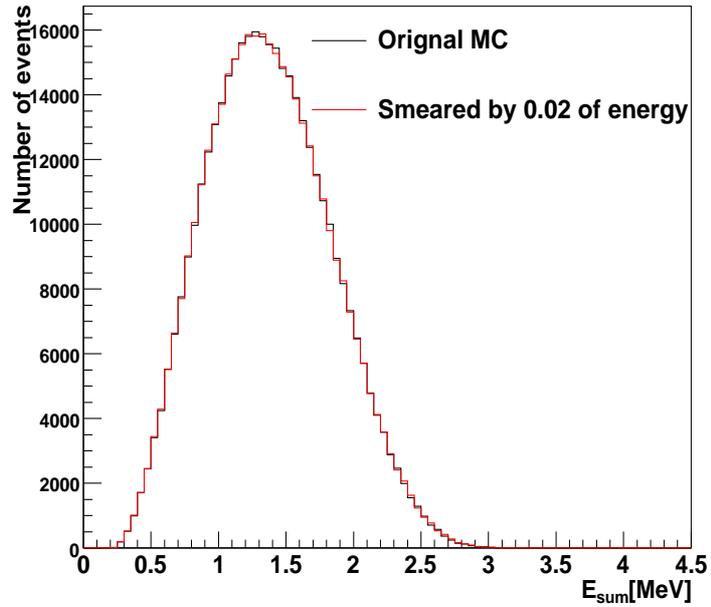}
\caption[The distribution of energy sum of the two electrons before and after smearing each individual electron energy by 2\%]{The distribution of energy sum of the two electrons for $2\nu\beta\beta$  before and after smearing each individual electron energy by 2\%.}
\label{fig-2nbbsmear}
\end{figure}
The breakdown of the systematic uncertainties on the $2\nu\beta\beta$ half-life is as  follows:
\begin{itemize}
\item The systematic uncertainty on the background estimate, obtained as the quadratic sum of the uncertainties on the estimates for the internal and external background, is 4\%,   which translates into an uncertainty on the $2\nu\beta\beta$ half-life of $1.5\%$.  
\item An uncertainty of $3\%$ is added  due to the uncertainty on the position of the $^{150}$Nd foil in the detector. This is estimated by comparing the half-life results before and after changing the position of the $^{150}$Nd  foil~(Section~\ref{sec-ndg}).
\item The TOF cut is changed from $4\%$ to $1\%$ in order to estimate the uncertainty on this cut. This yields $1\%$ uncertainty on the half-life results. 
\item There is a $5\%$ uncertainty on the efficiency calculation due to the inaccuracy of the GEANT simulation and tracking program. This was found by measuring the activity of the calibration sources~\cite{mo-excited}.
\item There is  $2\%$ uncertainty in the energy calibration coefficient~\cite{mo-excited} since the laser energy calibration was not used. The uncertainty on the shape of the $2\nu\beta\beta$ energy sum distribution  is studied by smearing the individual electron energies for $2\nu\beta\beta$ simulated events by $2\%$.  In this case the smeared energy  is defined as
\begin{equation}
E^{\prime}_{i}=E_{i}(1+0.02g),
\end{equation}
where $E_{i}$ is the energy of the $i^{th}$ electron and $g$ is a random number taken from a Gaussian distribution with a mean of zero and width of one. This yields a systematic uncertainty of $1.0\%$ on the $2\nu\beta\beta$ half-life. Figure~\ref{fig-2nbbsmear} shows the  distribution of energy sum of the two electrons  for simulated $2\nu\beta\beta$ events before and after smearing the electron energies by $2\%$.
\end{itemize}
\section{Comparison with previous results}
The $2\nu\beta\beta$ decay of $^{150}$Nd  was first observed in 1993 by a TPC experiment constructed by the Irvine group in California~\cite{nddiscovery}. 
The discovery  was confirmed by the  ITEP Moscow experiment later that year~\cite{ndconfirm}.  Both of these groups developed their experiments in order to measure the half-life of this process. The TPC experiment in ITEP Moscow used 51.5~g of ${\rm Nd_{2}O_{3}}$ and  
 presented a half-life  of $T_{1/2}=1.88^{+0.66}_{-0.39}~{\rm(stat)}\pm0.19~{\rm(syst)}\times 10^{19}$~y~\cite{moscowtpc}  in 1995. This half-life result had a large statistical uncertainty due to the short running time of the experiment (53~days).
\begin{figure}[!]
\centering
\includegraphics[width=11.0cm]{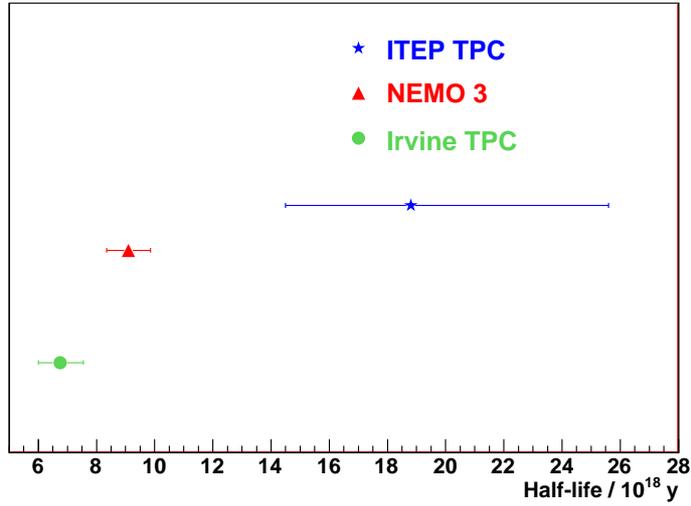}
\caption{Comparison of  NEMO~3 $2\nu\beta\beta$ half-life result (the result presented in this thesis) with ITEP and Irvine TPC experiments. The error bars present the quadratic sum of the statistical and systematic uncertainties.}
\label{fig-compare}
\end{figure}

The Irvine group's TPC experiment was situated  in an underground valve house in a canyon wall at the Hoover Dam in the USA~\cite{irvine}. The TPC was a rectangular polycarbonate box with internal dimension of $88\times 88\times 23~{\rm cm^{3}}$. The double beta decay source was in the centre and served as the central drift-field cathode. Anode and cathode wires  were set perpendicular to each other and  located near the walls of the TPC. A 1200~G magnetic field was applied perpendicular to the source plane. The helix of the particles provided the measurement of the momentum and thus the energy. A  Nd$_{2}$${\rm O_{3}}$ source with a mass $19.0$~g enriched to 91\% of $^{150}$Nd was used. After running the experiment for $262$ effective days, $476$ double beta decay events were observed. The efficiency of the detector for detecting two-electron events was $13\%$, resulting in a measured half-life of $6.75_{-0.42}^{+0.37}~({\rm stat})\pm 0.68~({\rm syst})\times 10^{18}$~y~\cite{irvine}.
Figure~\ref{fig-compare} compares the $2\nu\beta\beta$ half-life result presented in this thesis with the results of the ITEP Moscow  and Irvine  TPC experiments. This  measurement is slightly more than  two standard deviations higher than the Irvine group results and two standard deviations  lower than the ITEP Moscow result.

%% file: zeroneutrino.tex
\renewcommand{\baselinestretch}{1.6}
\normalsize
\chapter{Limits on different modes of neutrinoless double beta decay}
\label{chap-0nbb}
This Chapter presents searches for beyond the Standard Model  double beta decay processes as described in Sections~\ref{sec-0nbbth} to \ref{sec-majoronth}. It starts by introducing the  signature for  $0\nu\beta\beta$  and $0\nu\beta\beta\chi^{0}(\chi^{0})$ signals in the NEMO~3 detector and continues 
with setting limits on the  half-life of  each individual mode using the method described in Section~\ref{sec-colie}.  Limits are also set on neutrinoless double beta decay of $^{150}$Nd to excited states of $^{150}$Sm.
\section{Signal event selection}
The same two-electron event selections as  described in Section~\ref{sec-2e} are applied to simulated events from:
\begin{itemize}
\item $0\nu\beta\beta$  decay to the ground state of $^{150}$Sm via the  mass mechanism ($0^{+}_{\rm gs}$ $\langle m_{\nu}\rangle$).
 \item  $0\nu\beta\beta$  decay to the ground state of $^{150}$Sm  via right-handed currents ($0^{+}_{\rm gs}$~RC).
\item  $0\nu\beta\beta$ decay with emission of  Majoron(s). 
\item  $0\nu\beta\beta$  decay to the $2^{+}_{1}$ excited state  ($2^{+}_{1}$~RC). The mass mechanism of this decay is highly suppressed due to the nuclear angular momentum and only the right-handed current is enhanced~\cite{excite2prc}. 
 \item  $0\nu\beta\beta$ decay to the $0^{+}_{1}$ excited state ($0^{+}_{1}~\langle m_{\nu}\rangle$). The right-handed current of this decay is highly  suppressed due to the nuclear angular momentum and only the mass mechanism is enhanced~\cite{excited0pmm}.
\end{itemize}
The first three decays have  the same event topology as $2\nu\beta\beta$. The event topologies of the last two decays  are two electrons with emission of a photon ($2^{+}_{1}$~RC) or two photons   ($0^{+}_{1}~\langle m_{\nu}\rangle$). In this thesis it has not been attempted to reconstruct the additional photon(s). 

\begin{figure}[htp]
\centering
\subfigure{
a)\includegraphics[width=6.9cm]{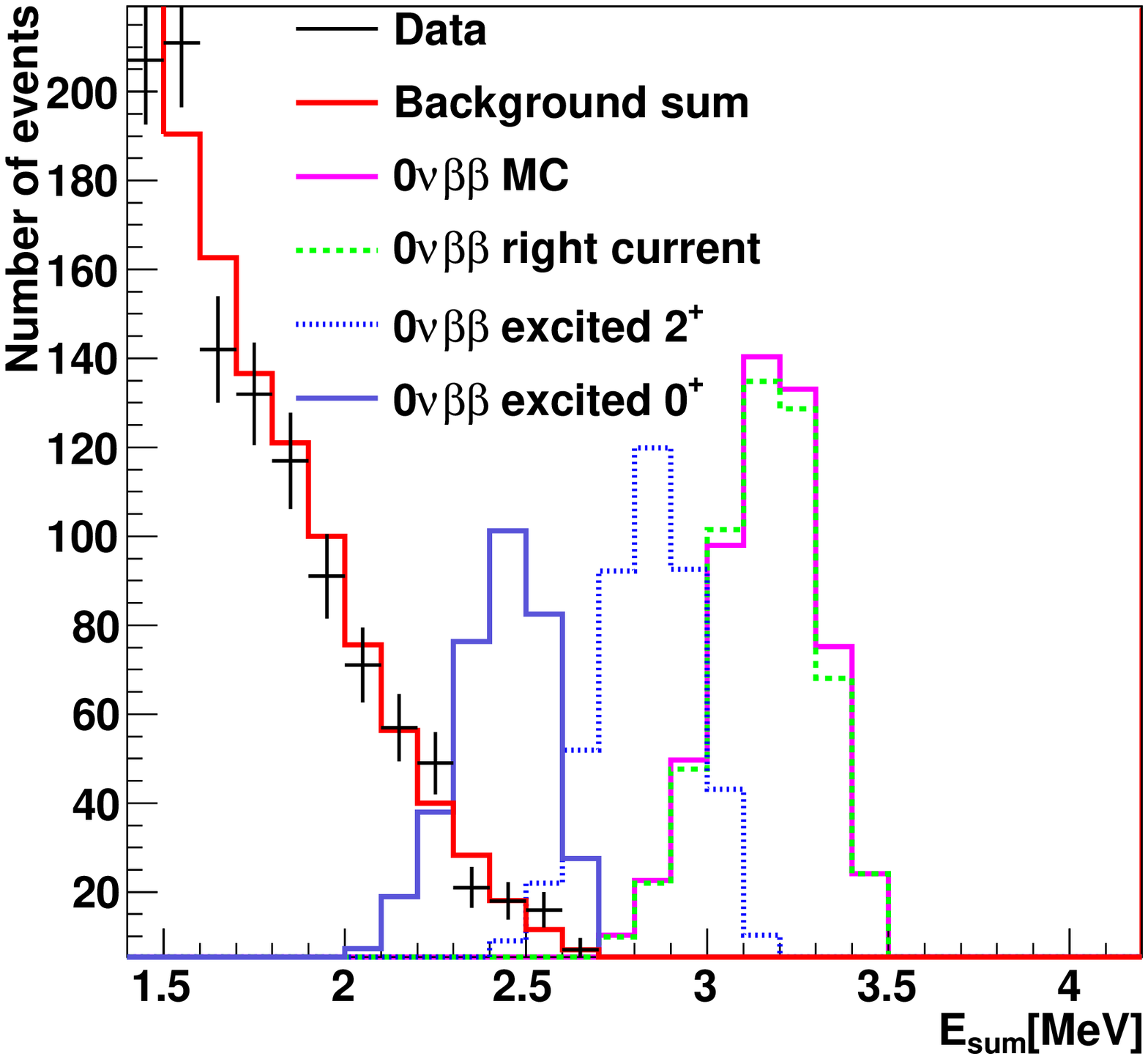}}
\subfigure{
b)\includegraphics[width=6.9cm]{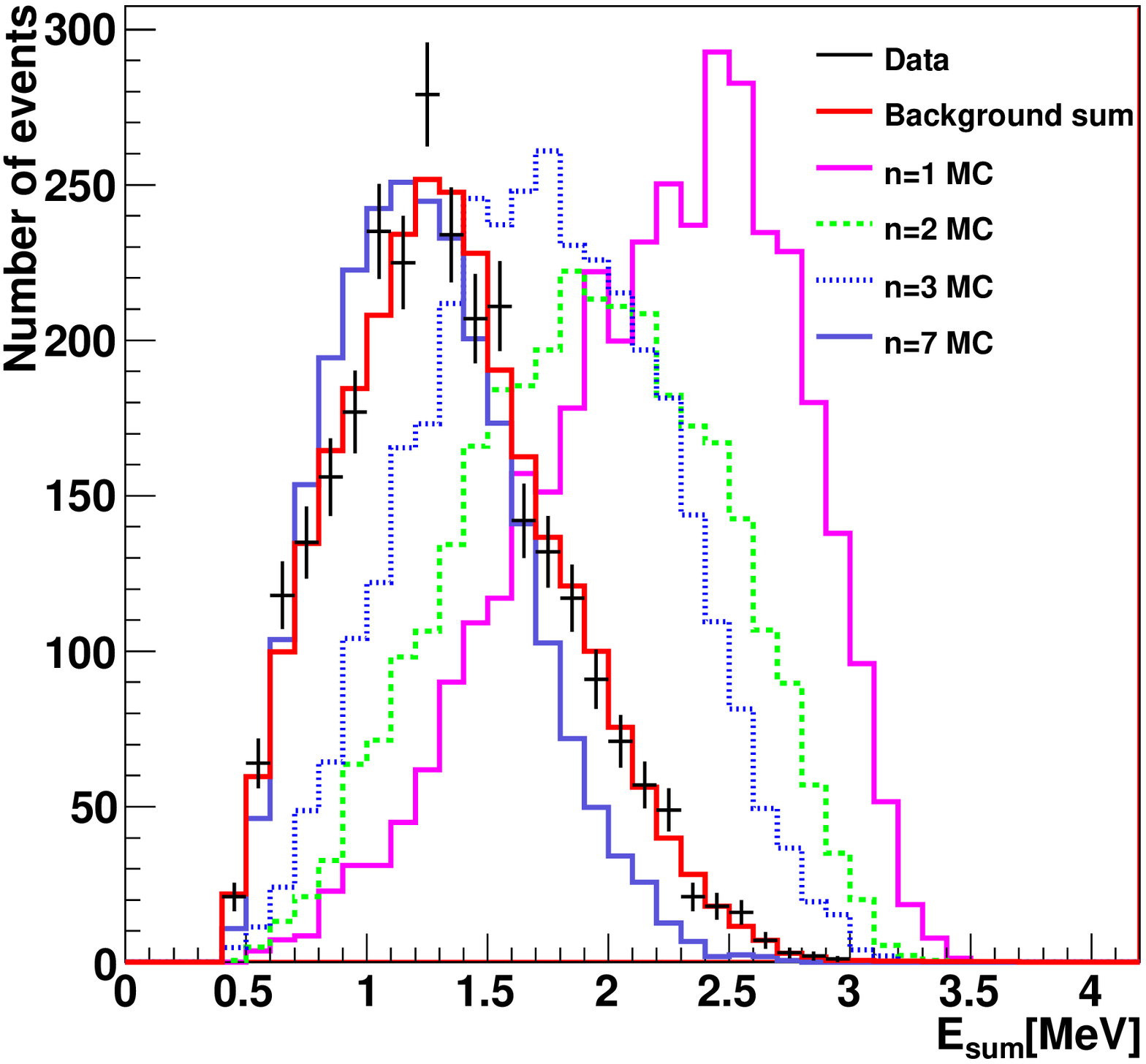}}
\caption[Electrons energy sum distributions for $^{150}$Nd decay to different modes of double beta decay ]{Energy sum distribution of electrons for a)  mass mechanism, right-handed current, decays to $2^{+}$ and $0^{+}$ excited states  and b) Majoron mode one (n=1), mode two (n=2), mode three (n=3) and mode seven (n=7). Data and $2\nu\beta\beta$ MC are also shown. The statistical uncertainties on the data points are shown with error bars.}  
\label{fig-signalesum}
\end{figure}

Table~\ref{tab-0nbbeff} present the event selection efficiency for each of these modes.
\begin{table}
\centering
\begin{tabular}{|c|c|c|c|c|c|c|c|c|}
   \hline
 & \multicolumn{4}{c|}{$0\nu\beta\beta$} &
   \multicolumn{4}{c|} {Majorons}           \\
\cline{2-9}
 & $0^{+}_{\rm gs}$ ($\langle m_{\nu}\rangle$)
 & $0^{+}_{\rm gs}$~RC
 & $2^{+}_{1}$~RC
 & $0^{+}_{1}$~($\langle m_{\nu}\rangle$)
 & n=1 &  n=2 &n=3 &n=7   \\
\cline{1-9}
efficiency (\%)     & $19.0$  &$10.9$ & $5.9$  & $2.1$  & $14.4$  & $12.2$ & $10.3$  &$4.9$\\
\cline{1-9}
\hline
\end{tabular}
\caption{Two-electron event selection efficiency for different  $0\nu\beta\beta$ decay modes.}
\label{tab-0nbbeff}
\end{table}
To set limits on the half-life of these modes, the whole energy sum distributions of the two electrons are investigated and no energy cuts are applied. Figure~\ref{fig-signalesum} shows the energy sum ($E_{sum}$) distribution of different simulated  modes, normalised to arbitrary numbers. In these figures the data and the sum of the backgrounds are also shown.
\section{Systematic considerations}
In all the above cases, there are energy regions where  the $2\nu\beta\beta$ background overlaps with the $0\nu\beta\beta$ signal. This is especially true for the Majoron modes three and  seven  which have the largest overlapping energy regions with the $2\nu\beta\beta$ background.  The $2\nu\beta\beta$ background's normalisation factor (half-life) in Section~\ref{sec-2nbbresult} was found by fitting it to data minus radioactive background, assuming that there was  no new physics in data. Thus, in order to have an unbiased  signal search,  the $2\nu\beta\beta$ statistical uncertainty on the $2\nu\beta\beta$  normalisation, which is found by fitting $2\nu\beta\beta$ background   to data minus radioactive background MC, was ``floated'' (see Section~\ref{sec-float}). This makes the $2\nu\beta\beta$ background normalisation factor  a free parameter.

\begin{figure}[htp]
\centering
\subfigure{
a)\includegraphics[width=6.0cm]{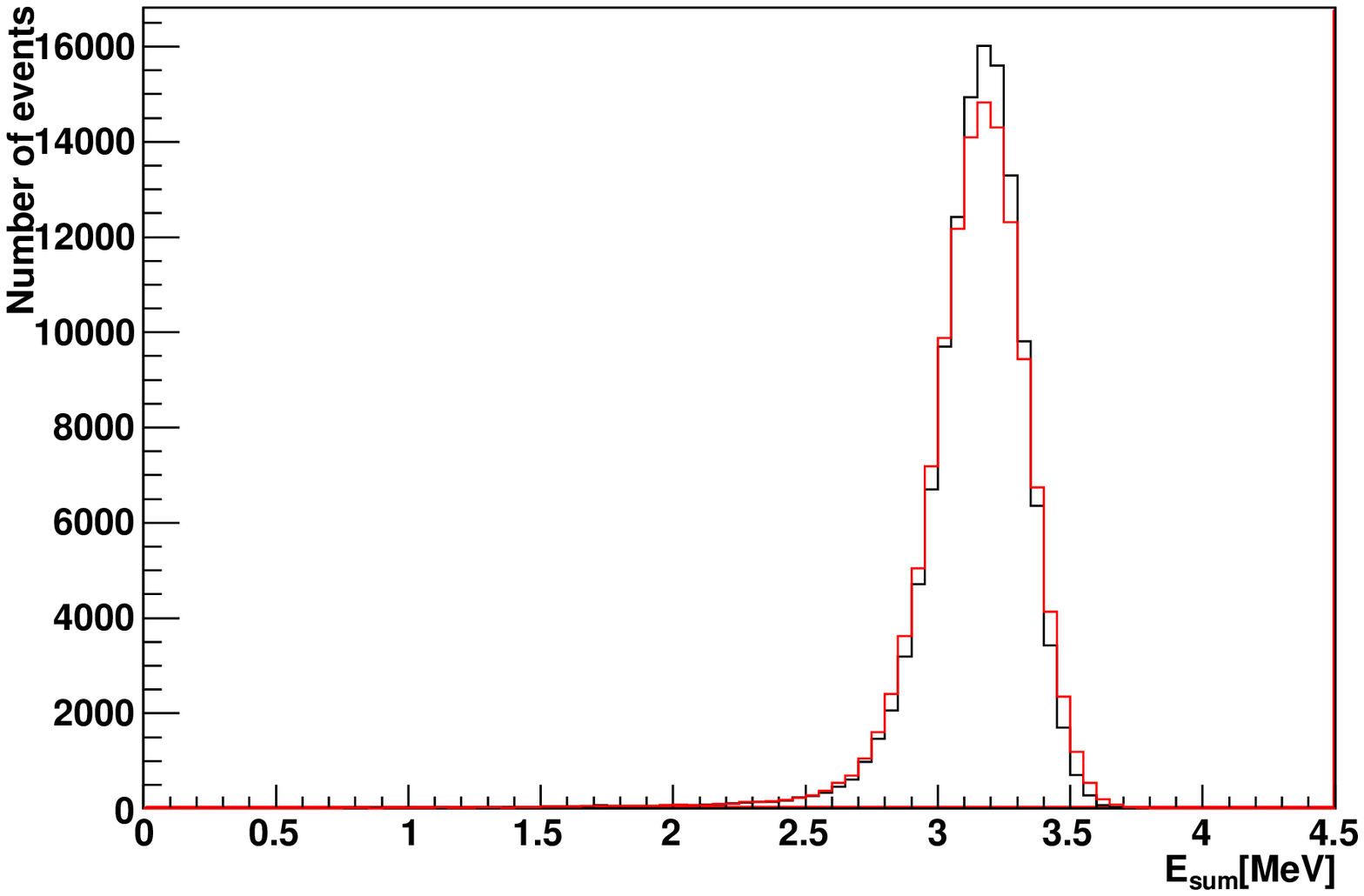}}
\subfigure{
b)\includegraphics[width=6.0cm]{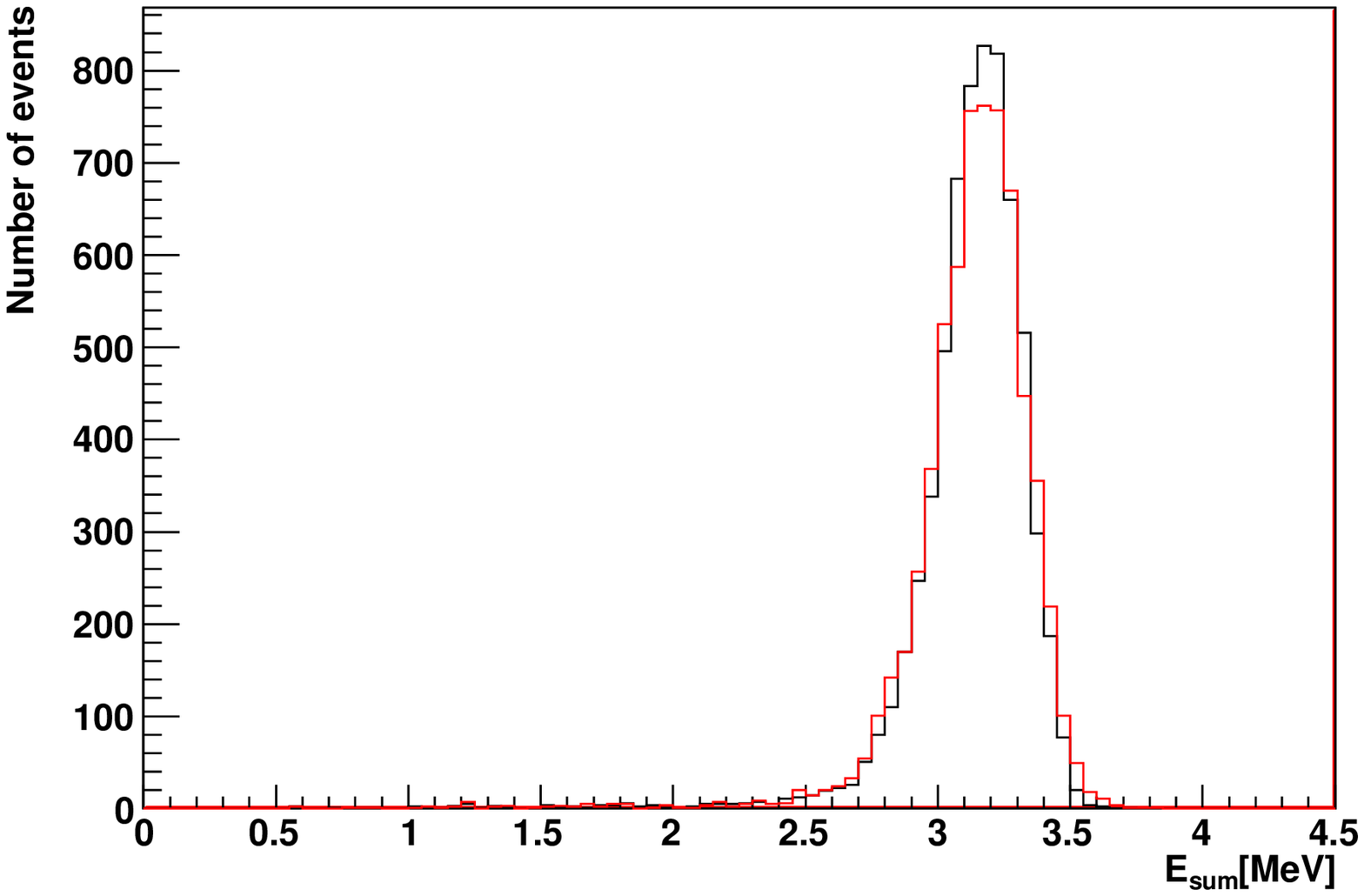}}
\subfigure{
c)\includegraphics[width=6.0cm]{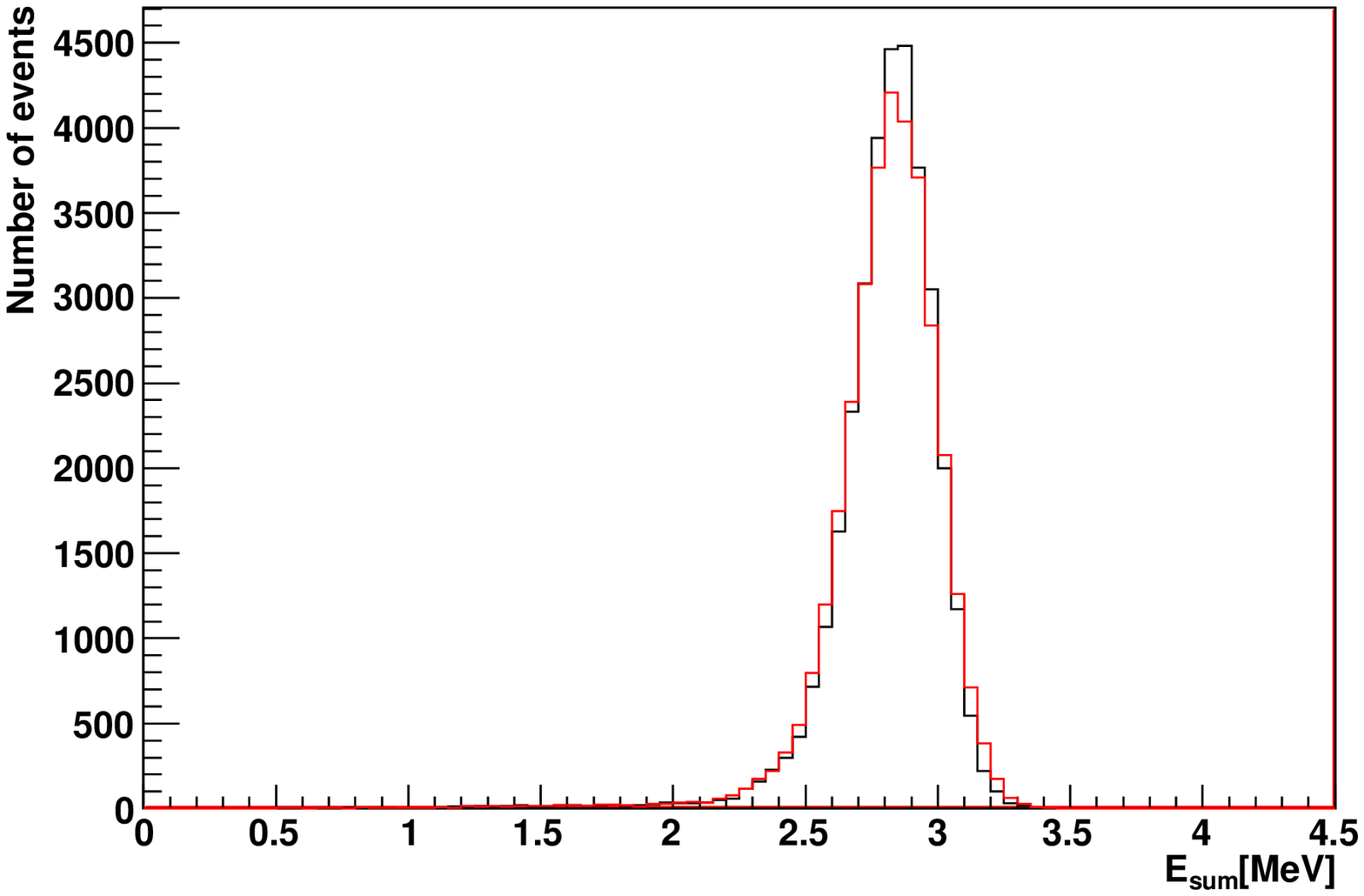}}
\subfigure{
d)\includegraphics[width=6.0cm]{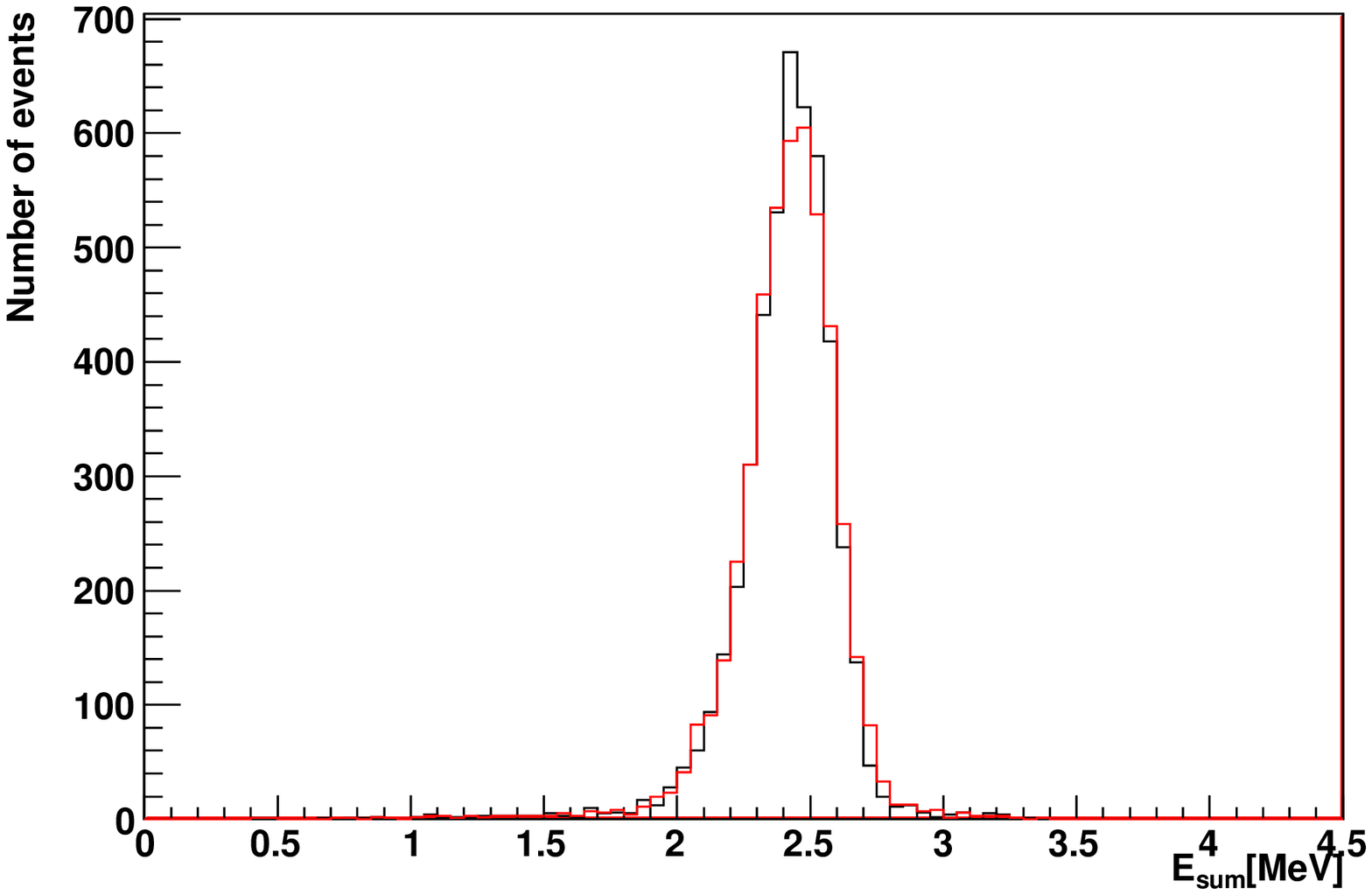}}
\subfigure{
e)\includegraphics[width=6.0cm]{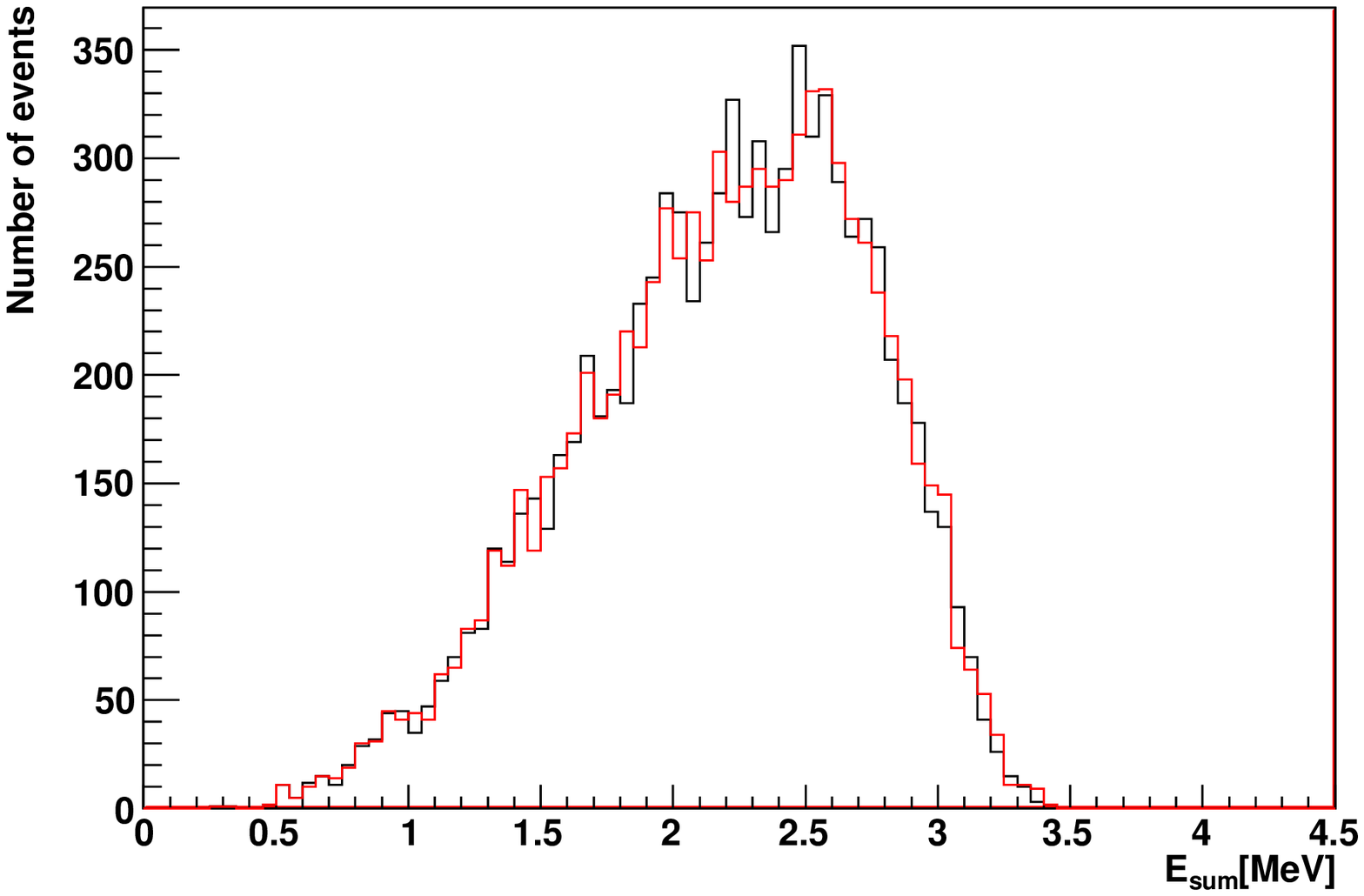}}
\subfigure{
f)\includegraphics[width=6.0cm]{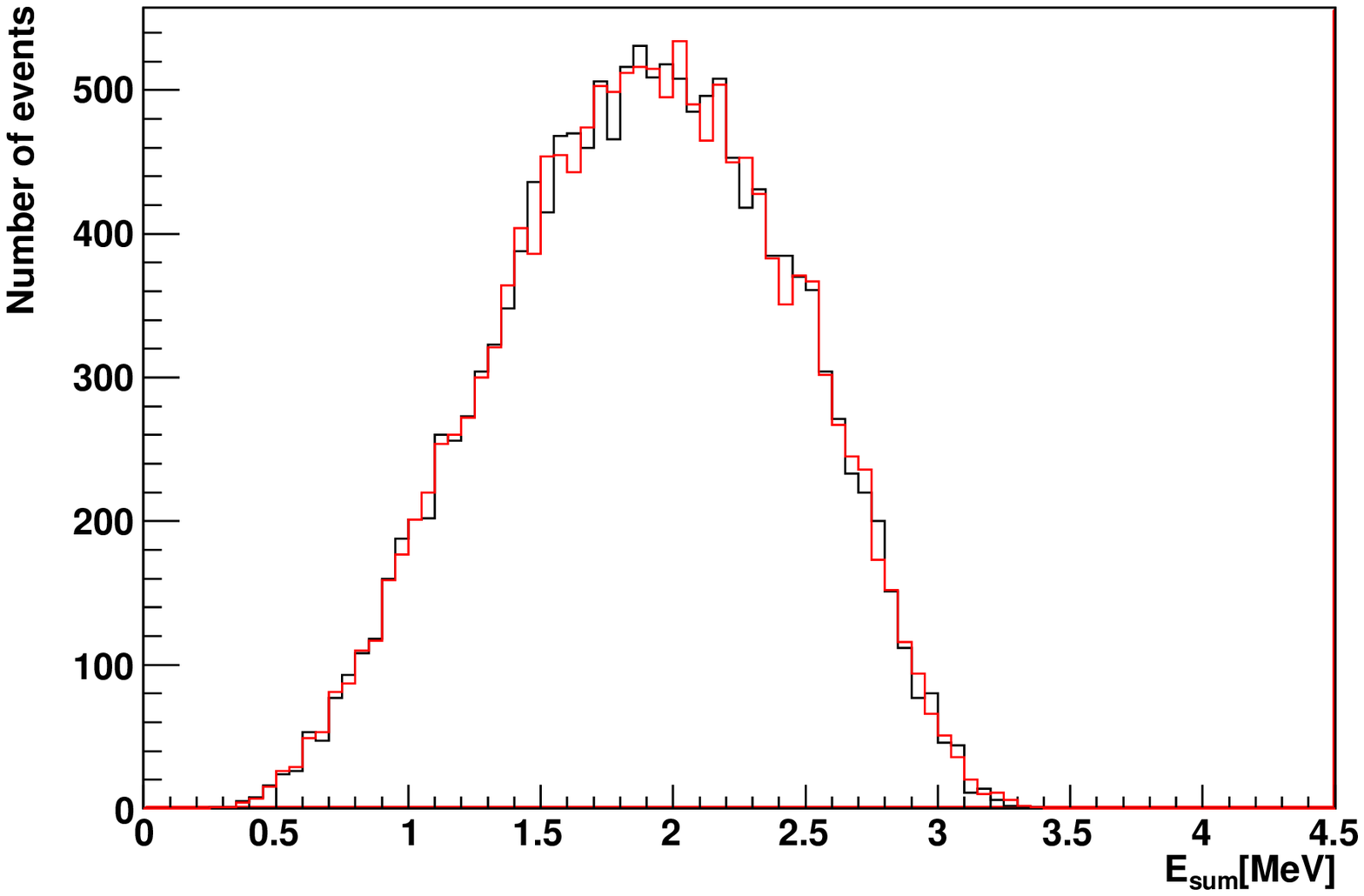}}
\subfigure{
g)\includegraphics[width=6.0cm]{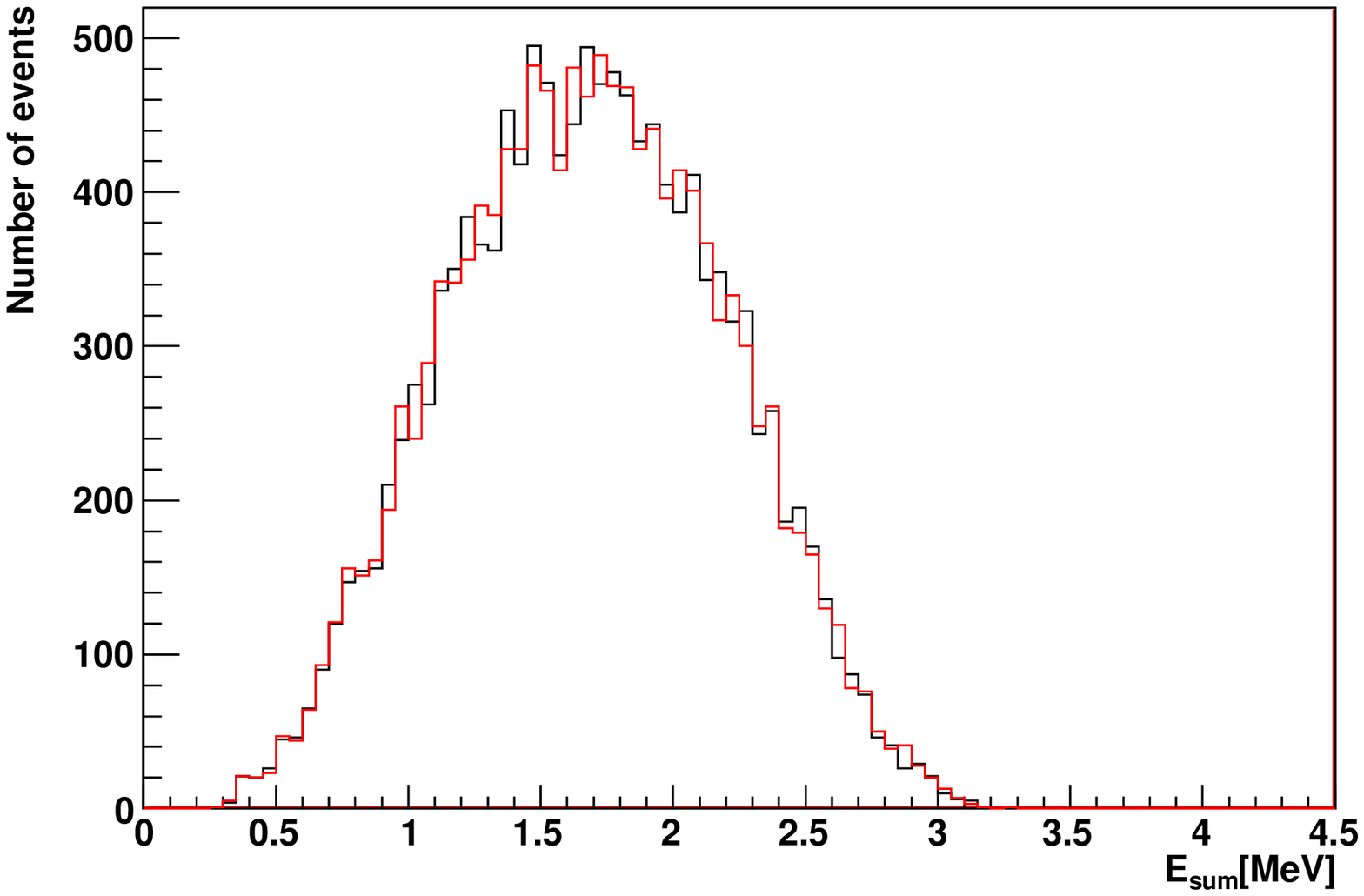}}
\subfigure{
h)\includegraphics[width=6.0cm]{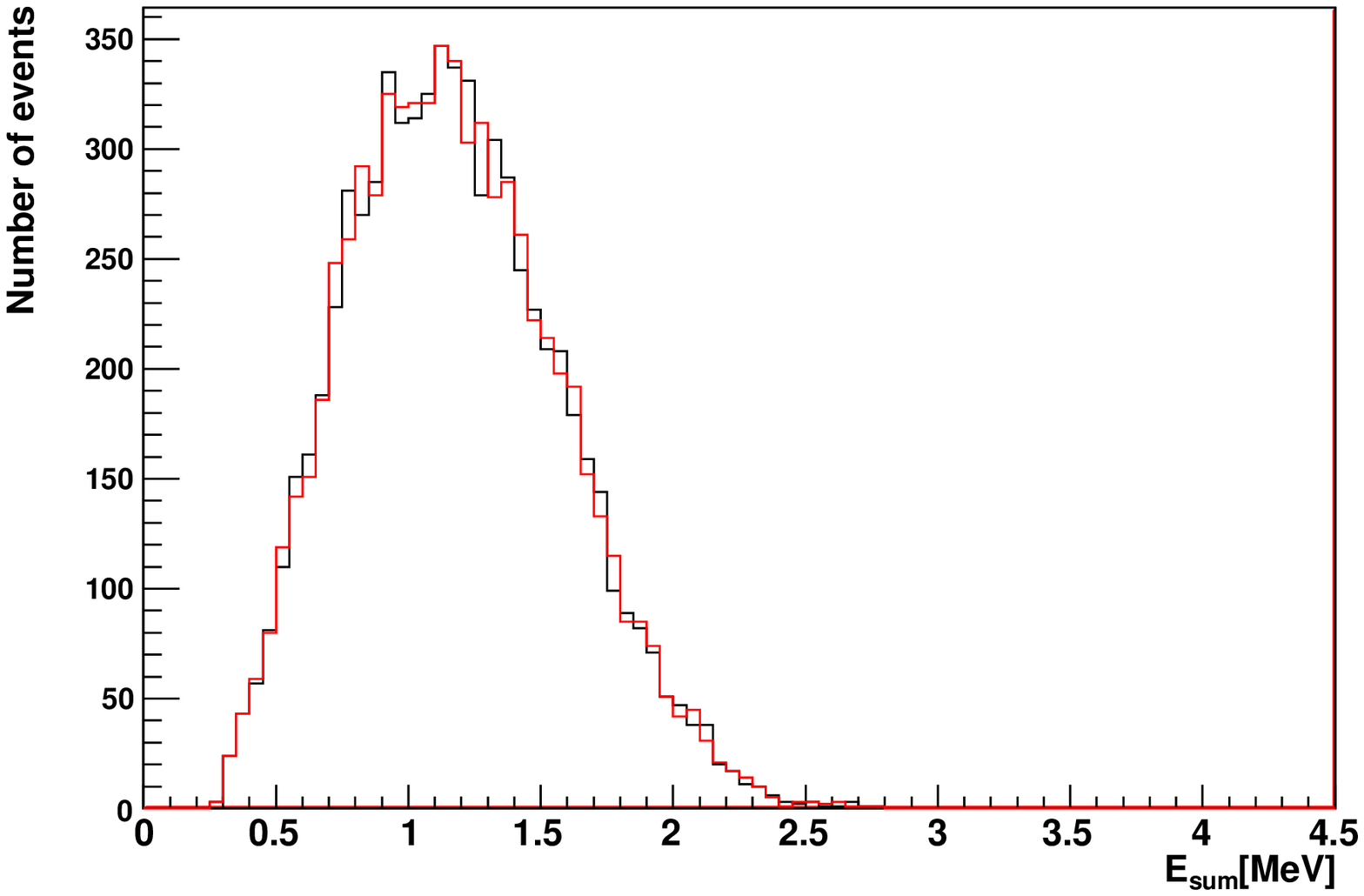}}
\caption[The energy sum distributions of different $0\nu\beta\beta$ decay modes before and after smearing the energy of each electron by 2\%]{The distribution of energy sum of the two electrons for a)~$0\nu\beta\beta$ mass mechanism, b)~$0\nu\beta\beta$ right-handed current, c) $0\nu\beta\beta$ decay to  $2^{+}$ excited state, d) $0\nu\beta\beta$ decay to  $0^{+}$ excited state, e)~Majoron mode one, f)~Majoron mode two, g)~Majoron mode three and h)~Majoron mode seven. The black lines show the energy distributions before smearing  the energy of each electron and  the red lines  shows the energy distribution of the smeared MCs.}
\label{fig-smear0nbbsignal}
\end{figure}

\begin{figure}[htp]
\centering
\subfigure{
a)\includegraphics[width=6.0cm]{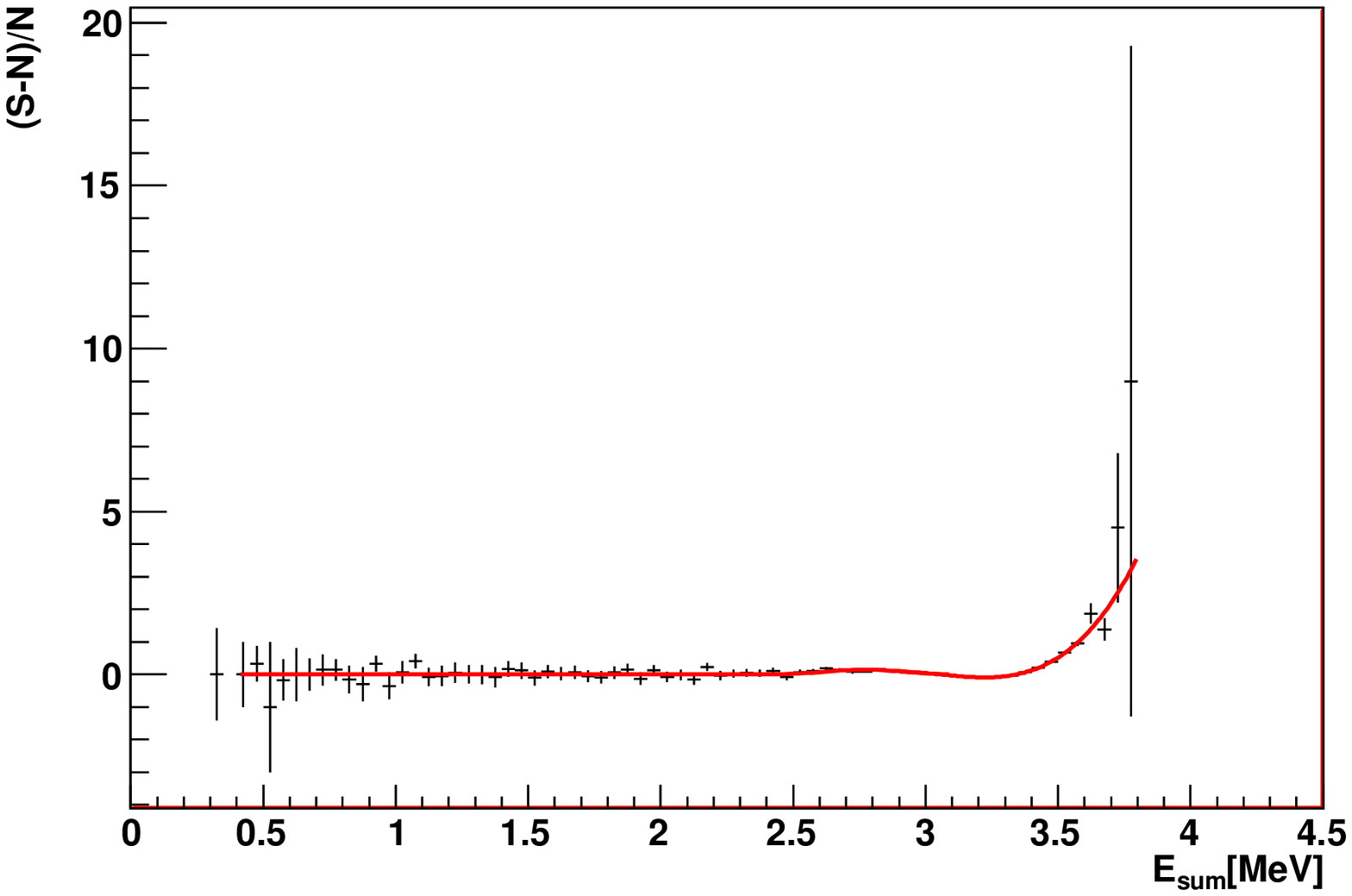}}
\subfigure{
b)\includegraphics[width=6.0cm]{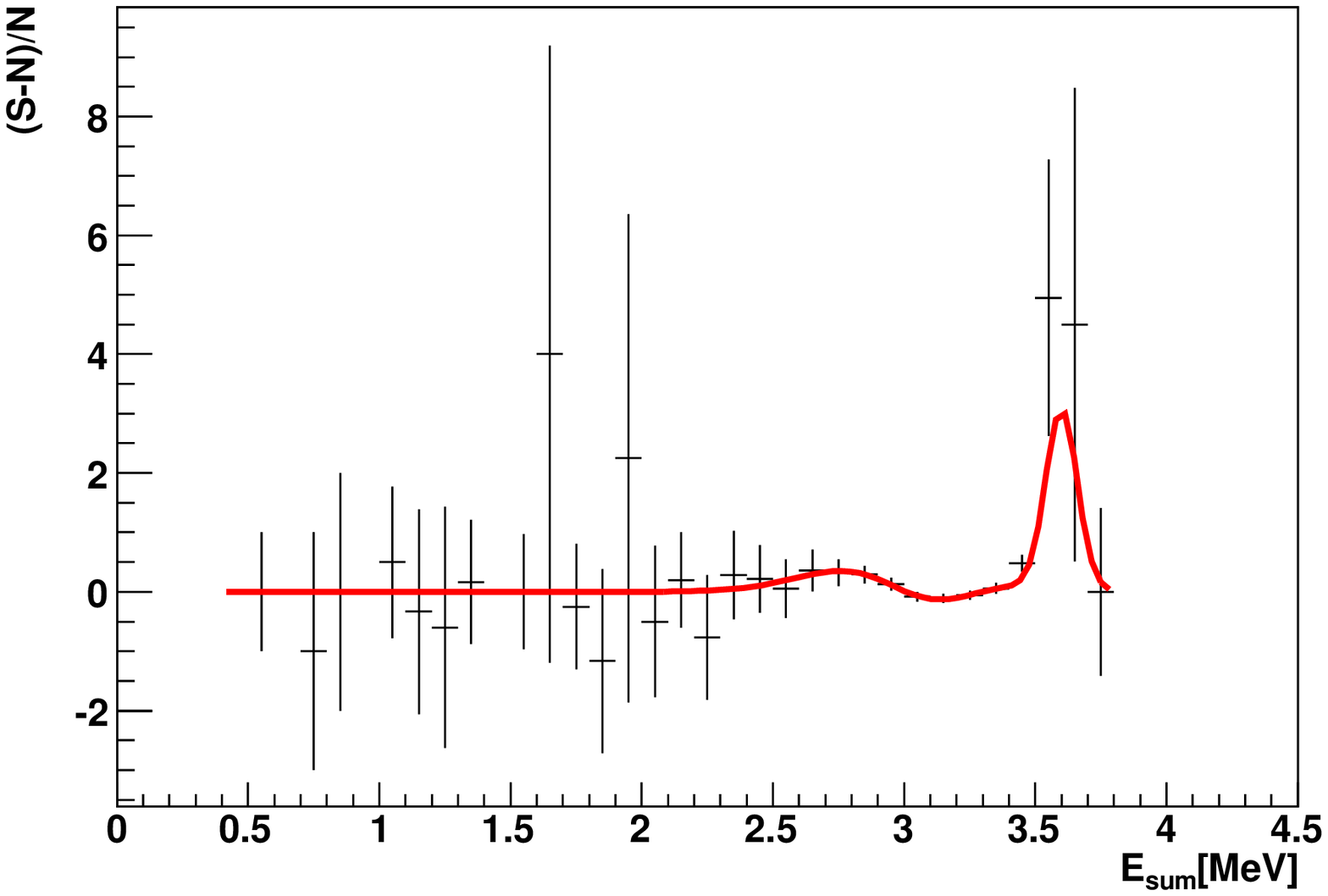}}
\subfigure{
c)\includegraphics[width=6.0cm]{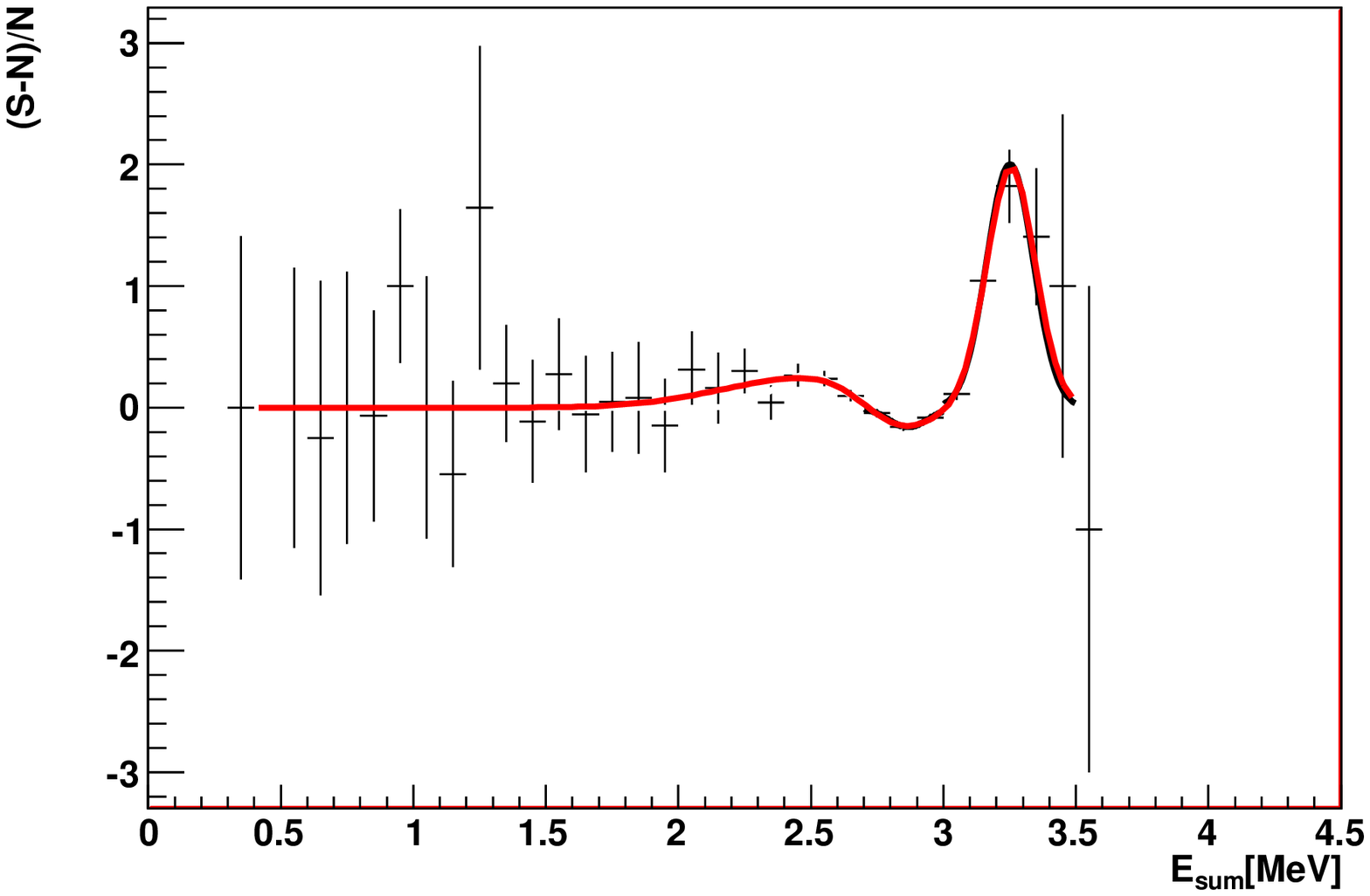}}
\subfigure{
d)\includegraphics[width=6.0cm]{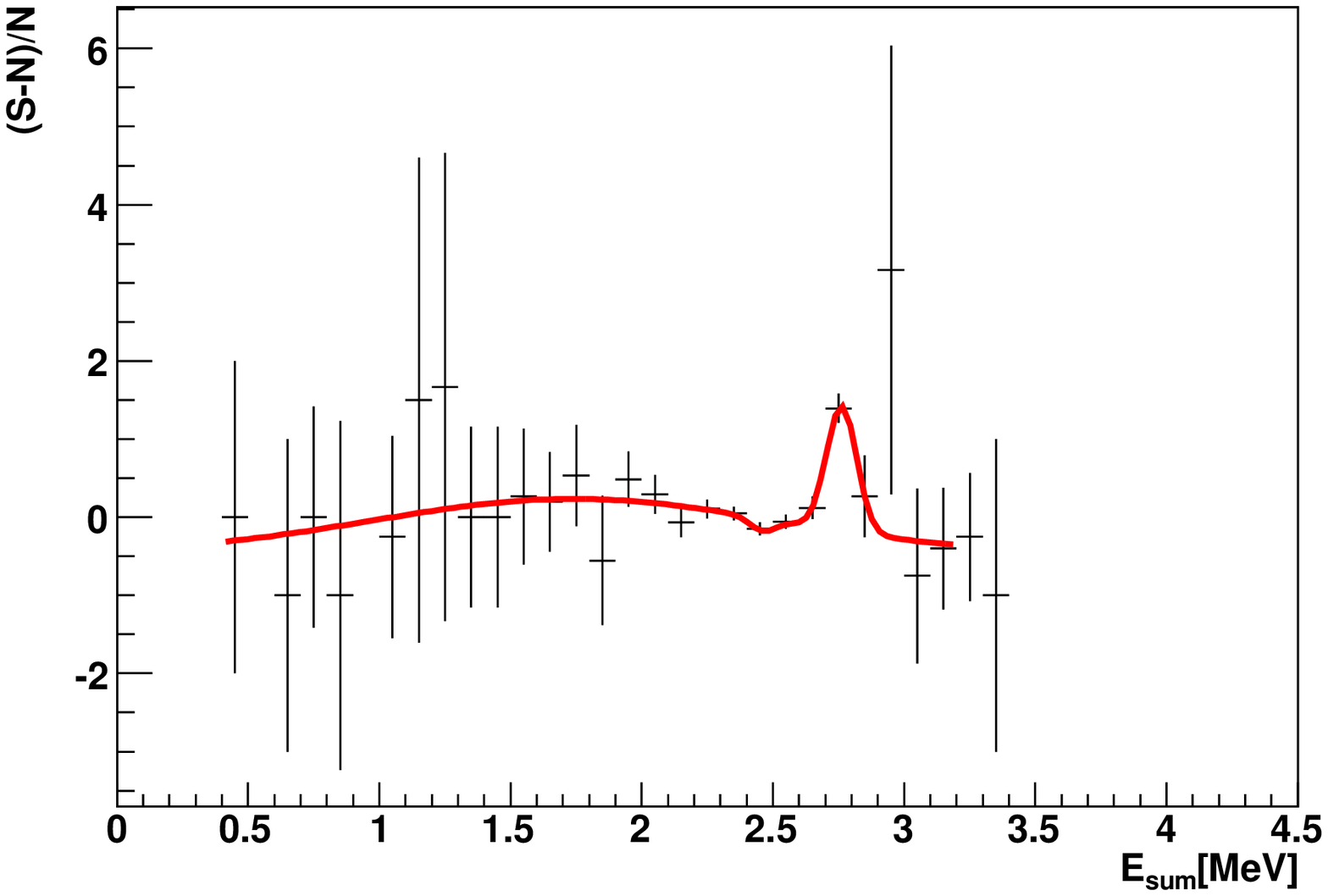}}
\subfigure{
e)\includegraphics[width=6.0cm]{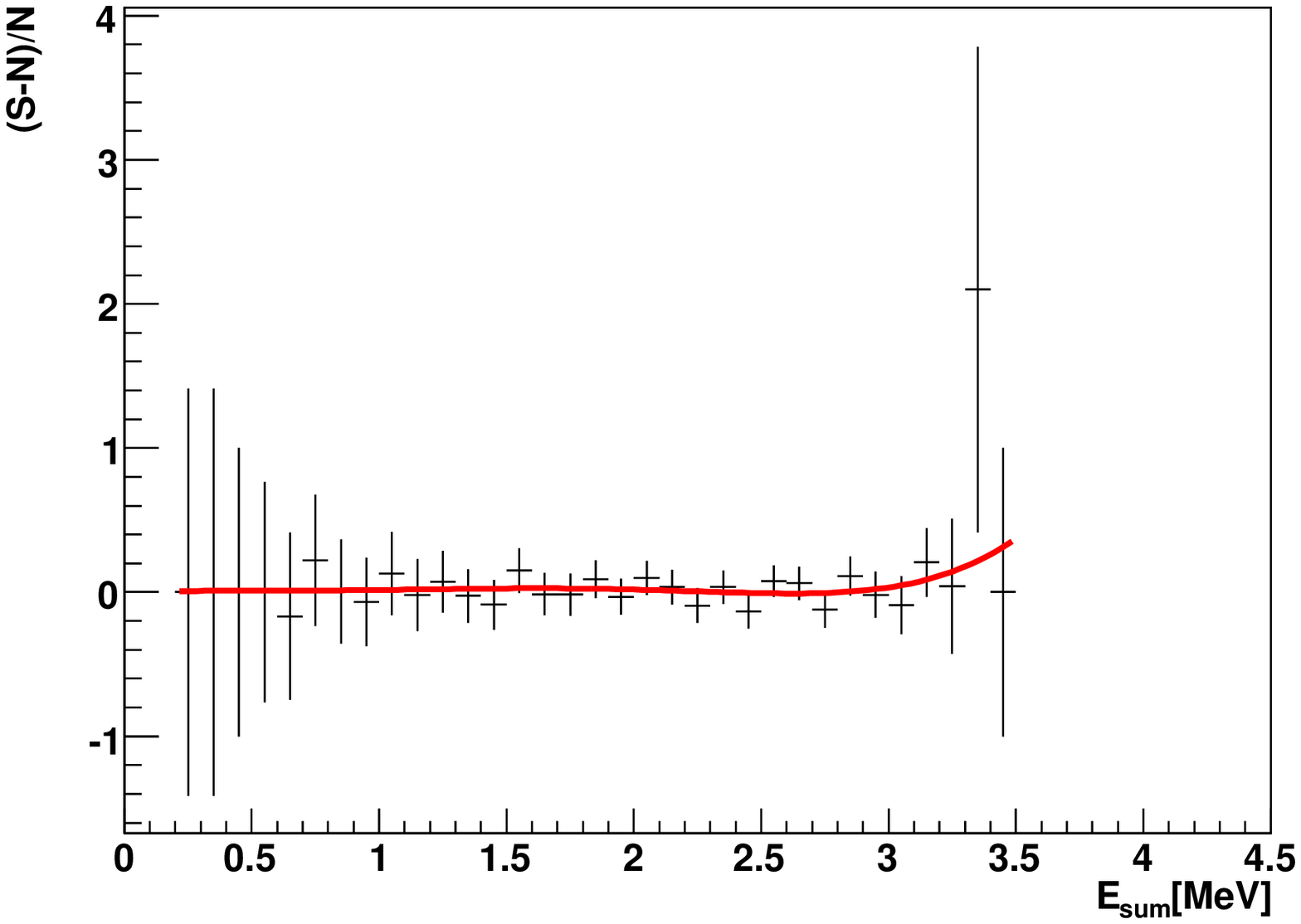}}
\subfigure{
f)\includegraphics[width=6.0cm]{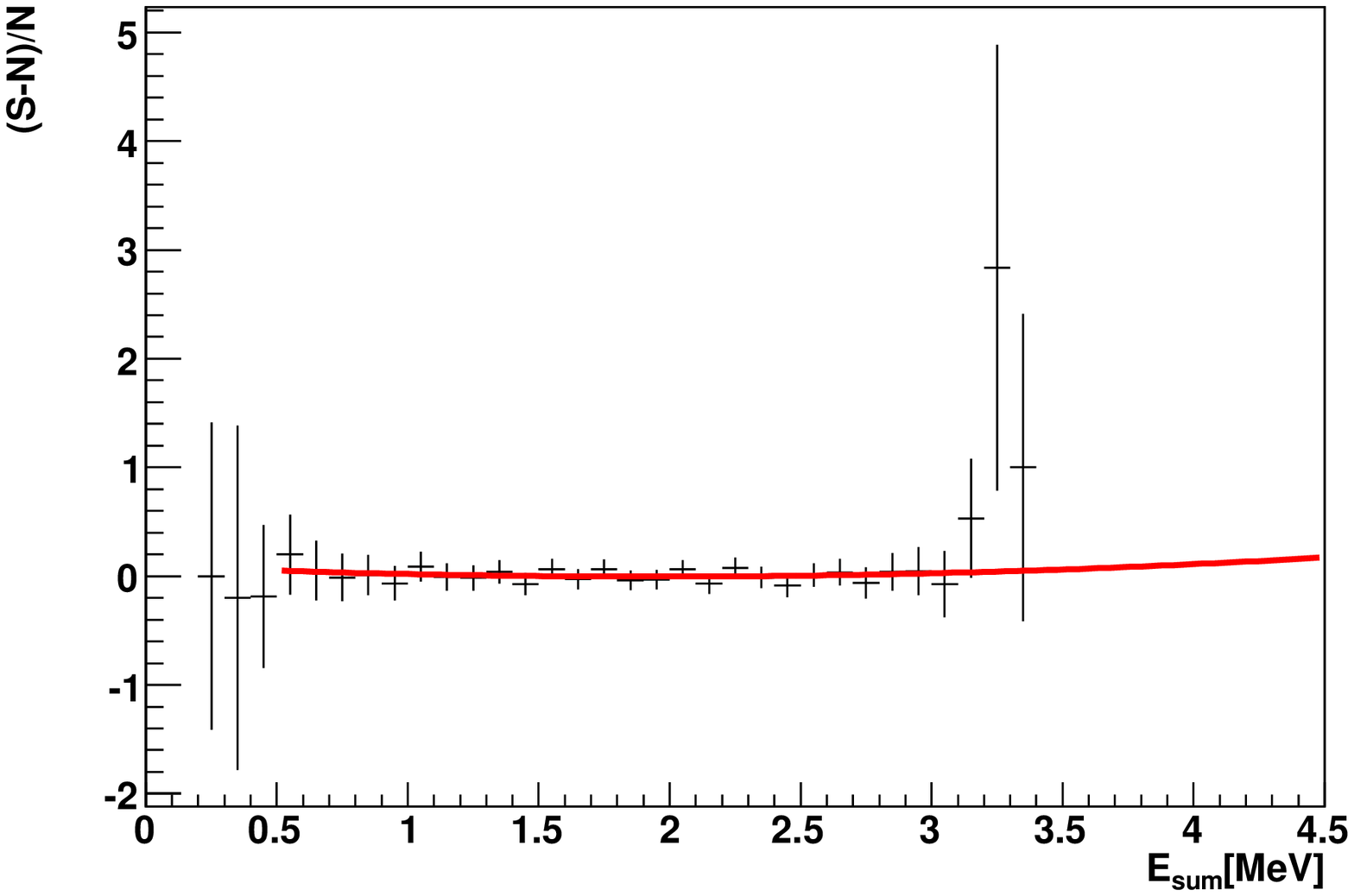}}
\subfigure{
g)\includegraphics[width=6.0cm]{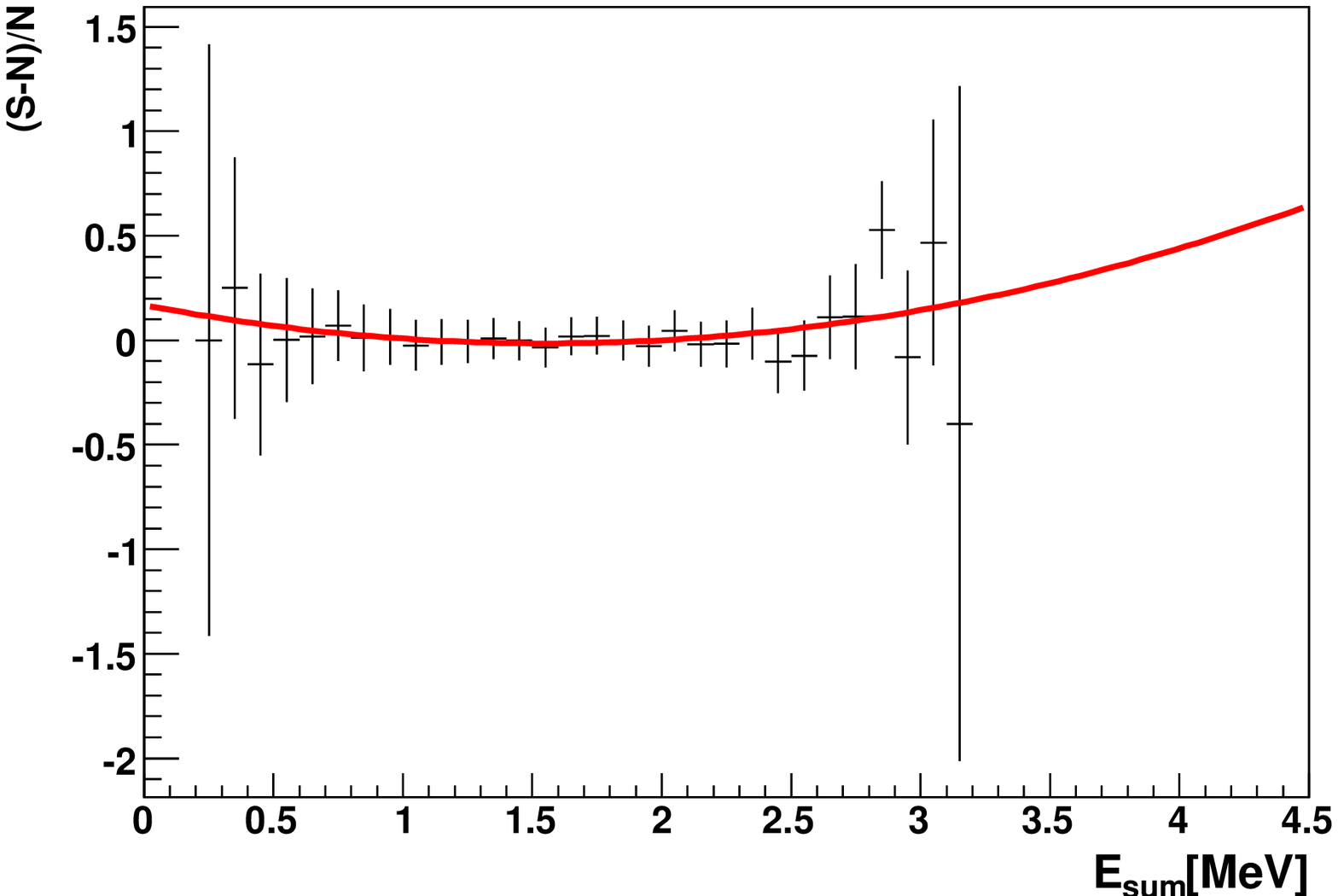}}
\subfigure{
h)\includegraphics[width=6.0cm]{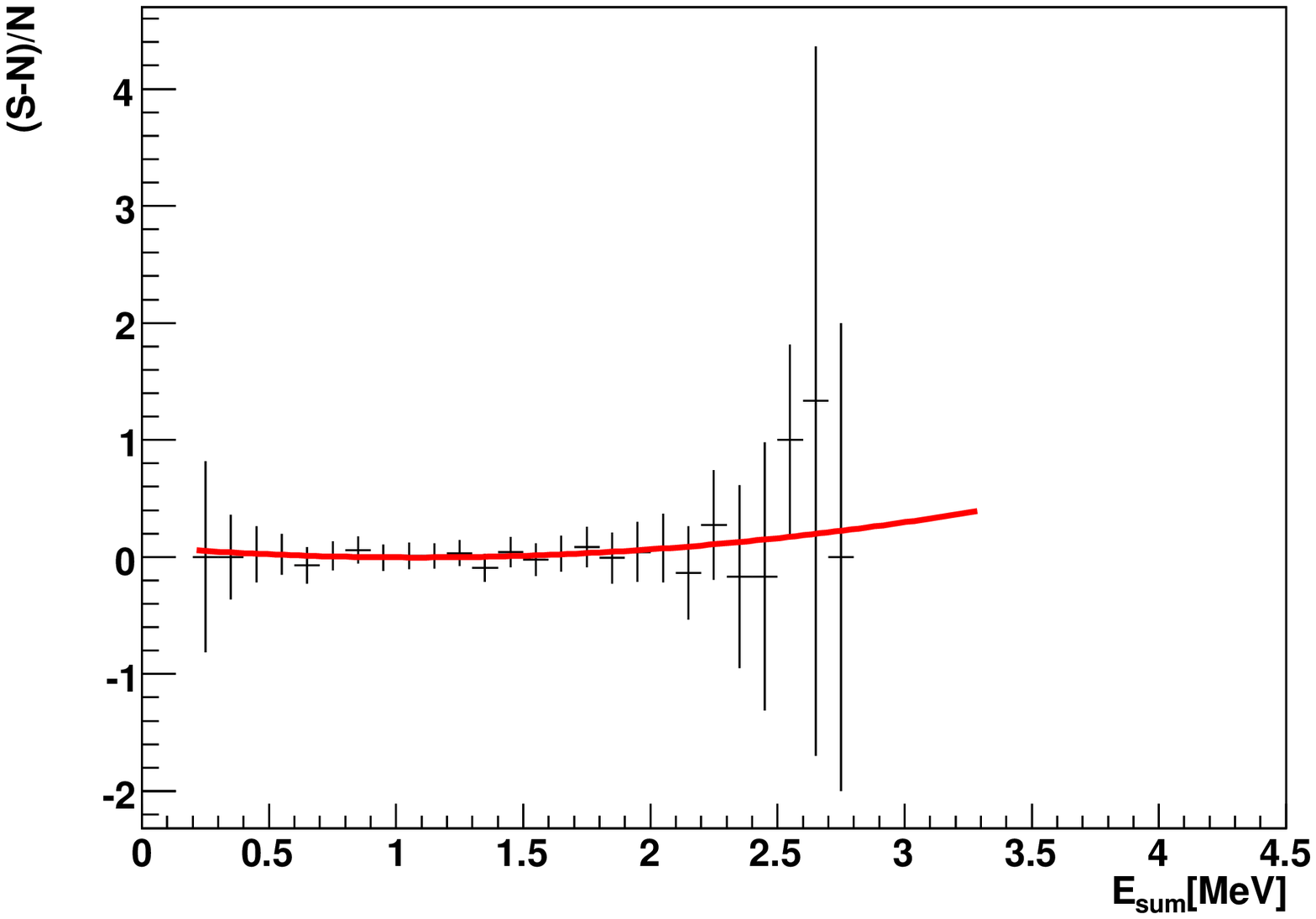}}
\caption[The distributions of (S-N)/N and fitted  functions]{The distributions of (S-N)/N and fitted  functions for  a) $0\nu\beta\beta$  mass mechanism, b). $0\nu\beta\beta$  right-handed current, c) $0\nu\beta\beta$ decay to $2^{+}$ excited state, d)  $0\nu\beta\beta$ decay to $0^{+}$ excited state, e) Majoron mode one, f) Majoron mode two, g) Majoron mode three and h) Majoron mode seven. The error bars show the statistical uncertainties.}
\label{fig-0nbbsn}
\end{figure}

\begin{figure}[htp]
\centering
\subfigure{
a)\includegraphics[width=6.0cm]{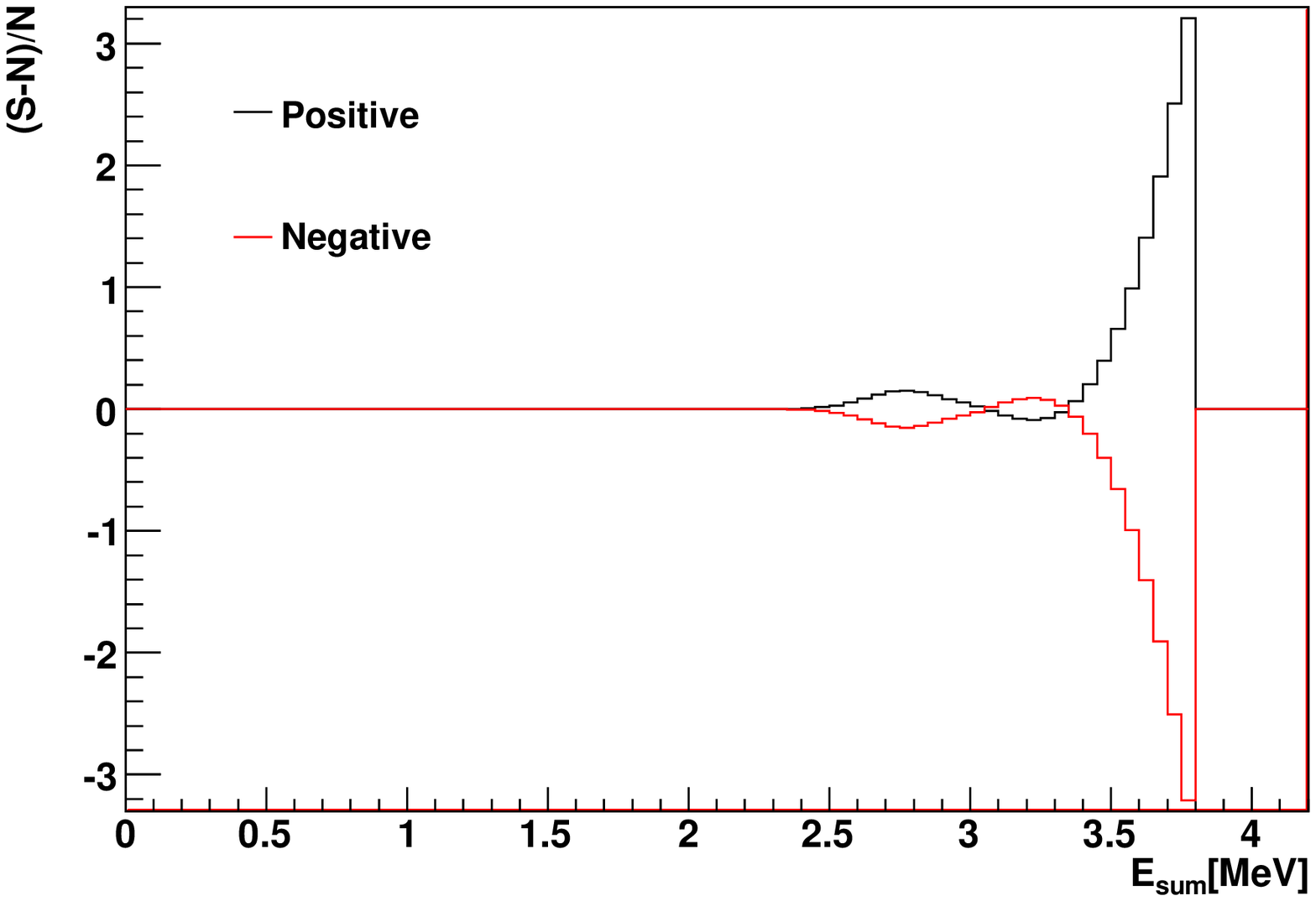}}
\subfigure{
b)\includegraphics[width=6.0cm]{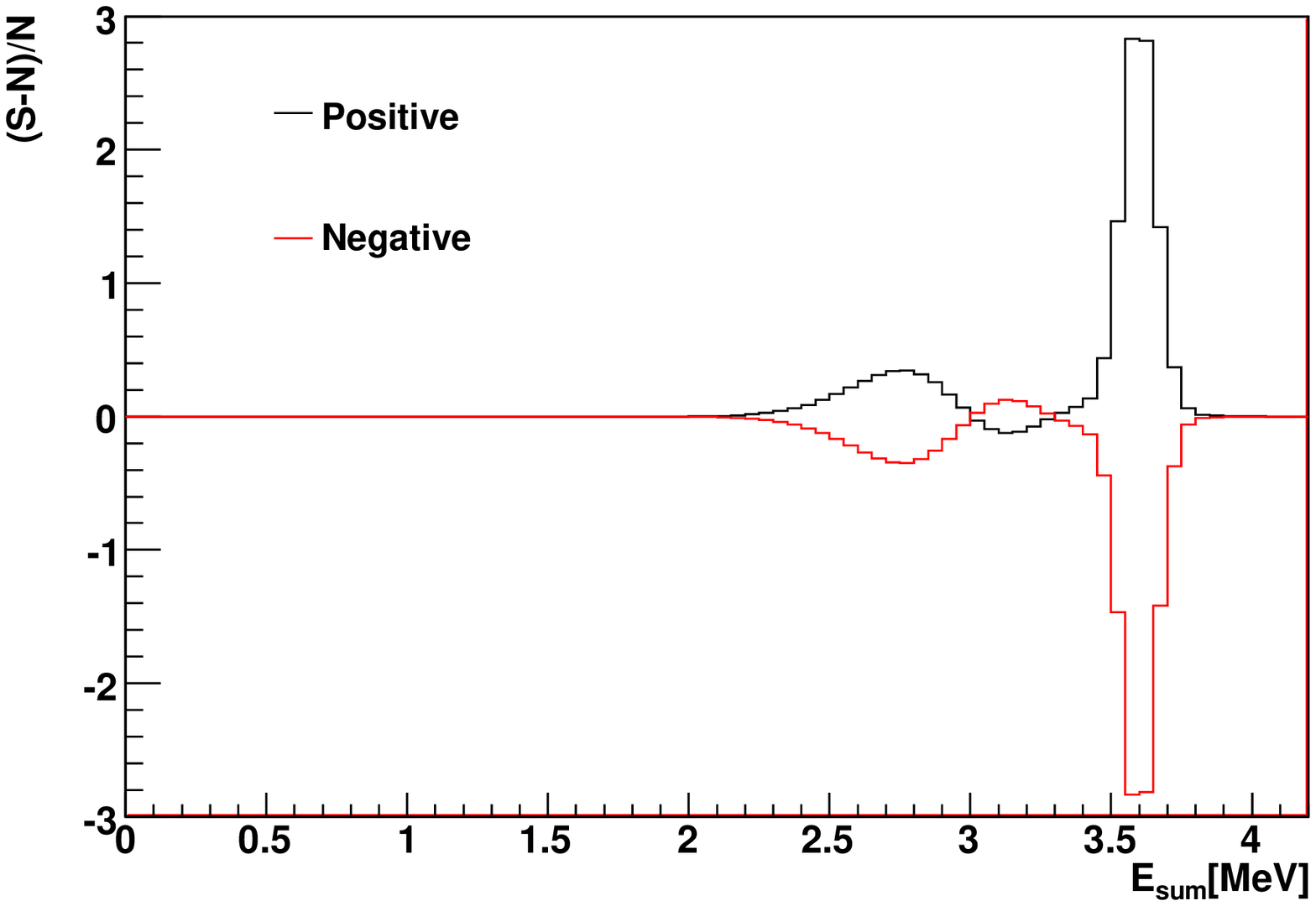}}
\subfigure{
c)\includegraphics[width=6.0cm]{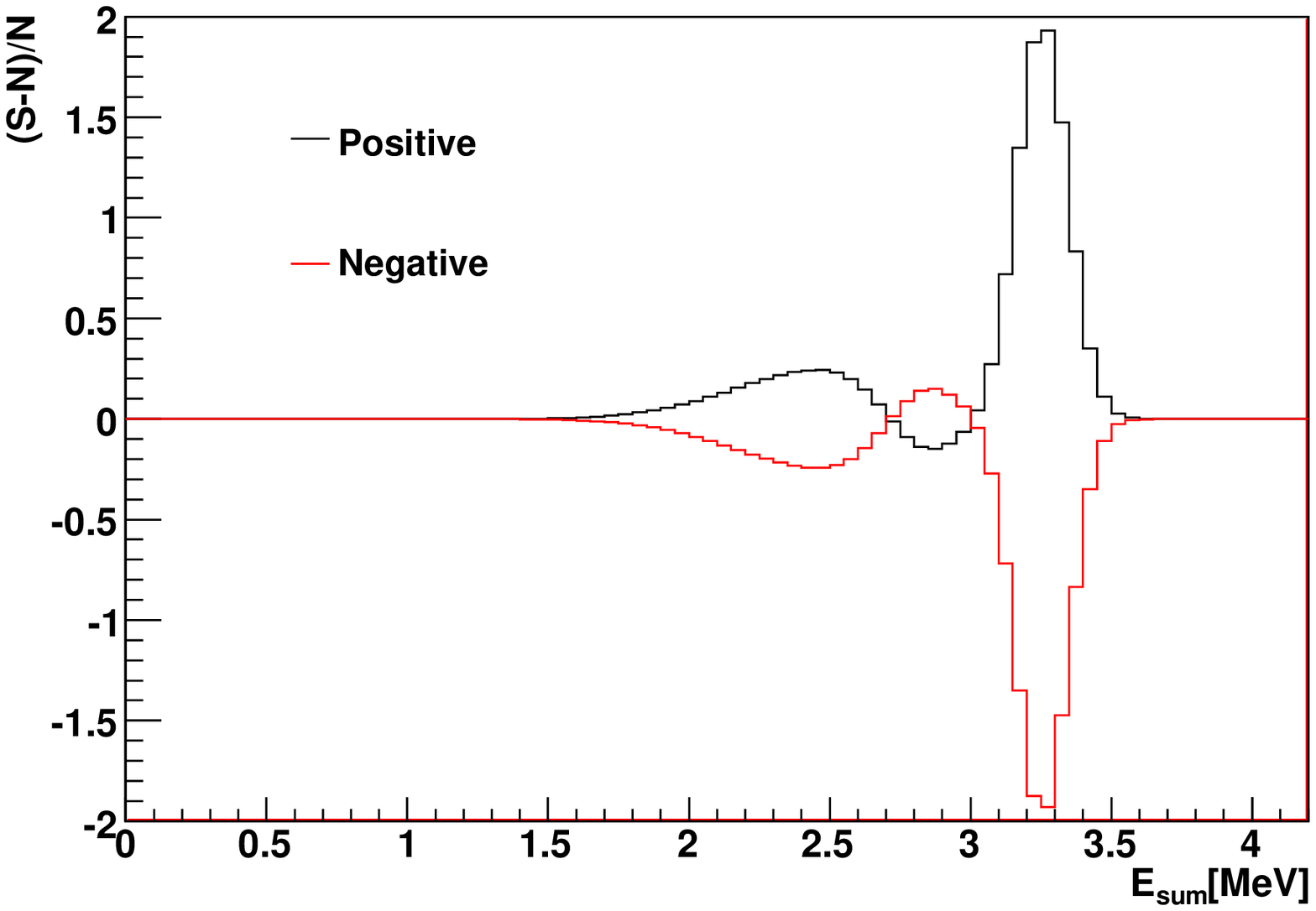}}
\subfigure{
d)\includegraphics[width=6.0cm]{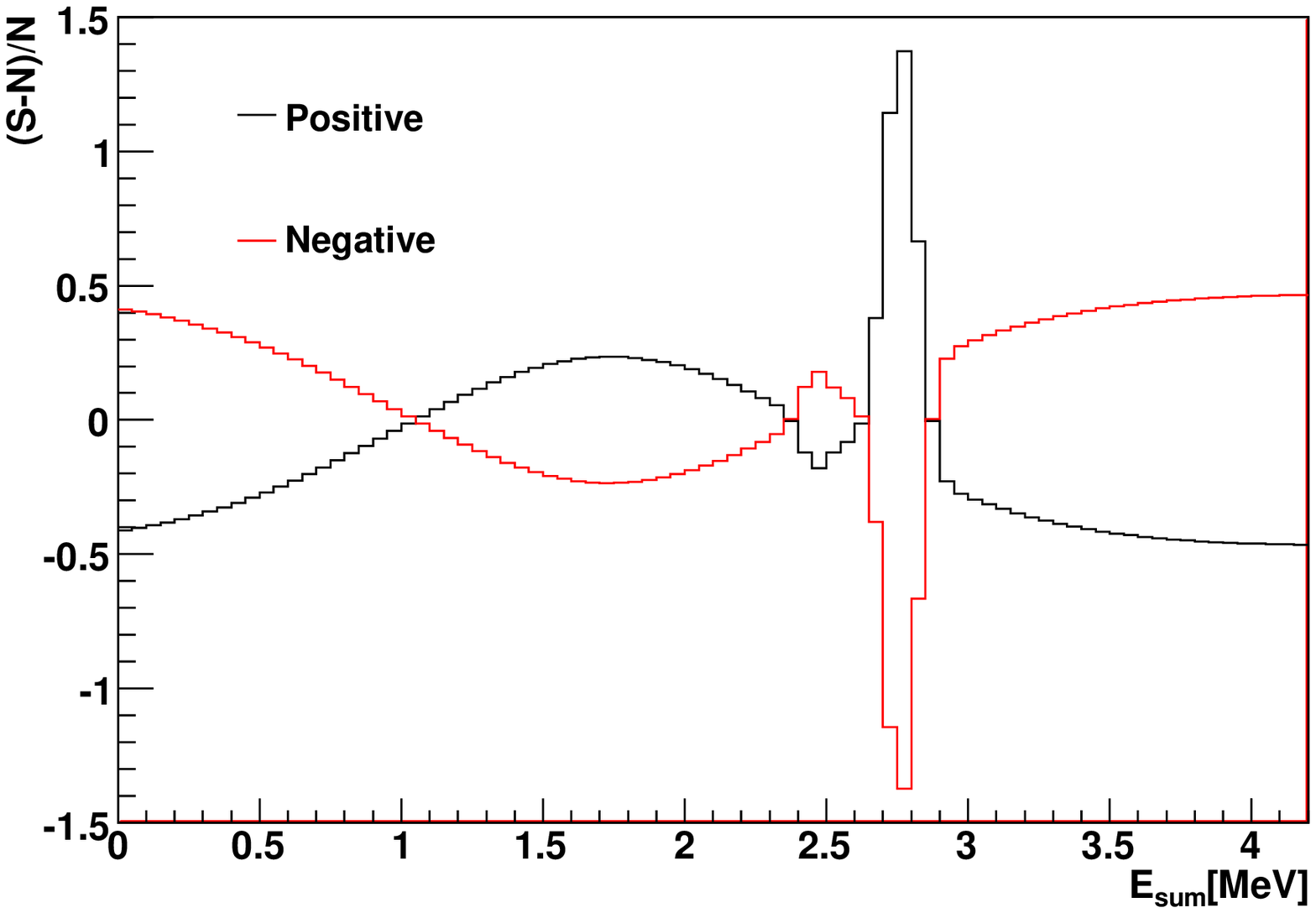}}
\subfigure{
e)\includegraphics[width=6.0cm]{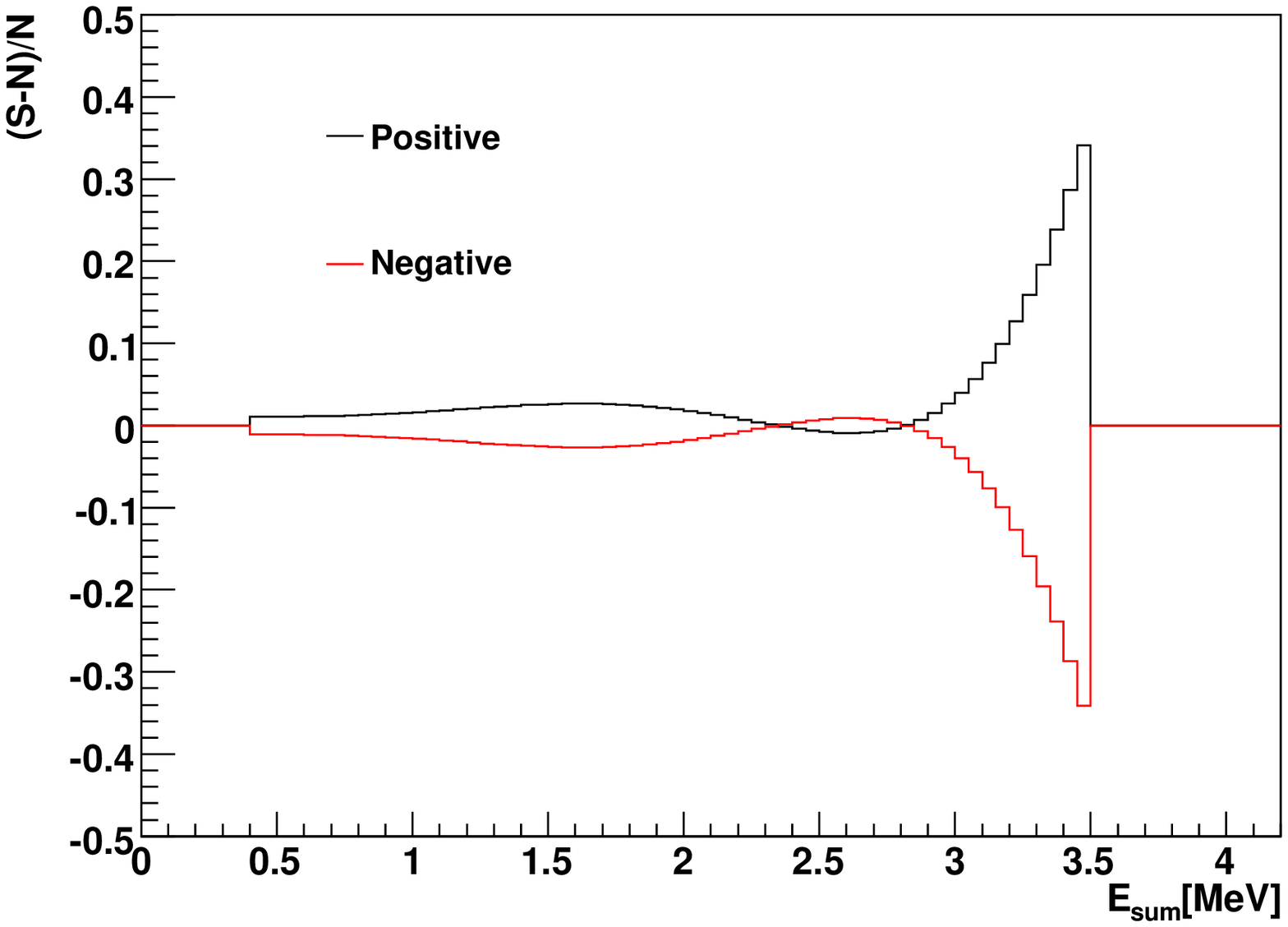}}
\subfigure{
f)\includegraphics[width=6.0cm]{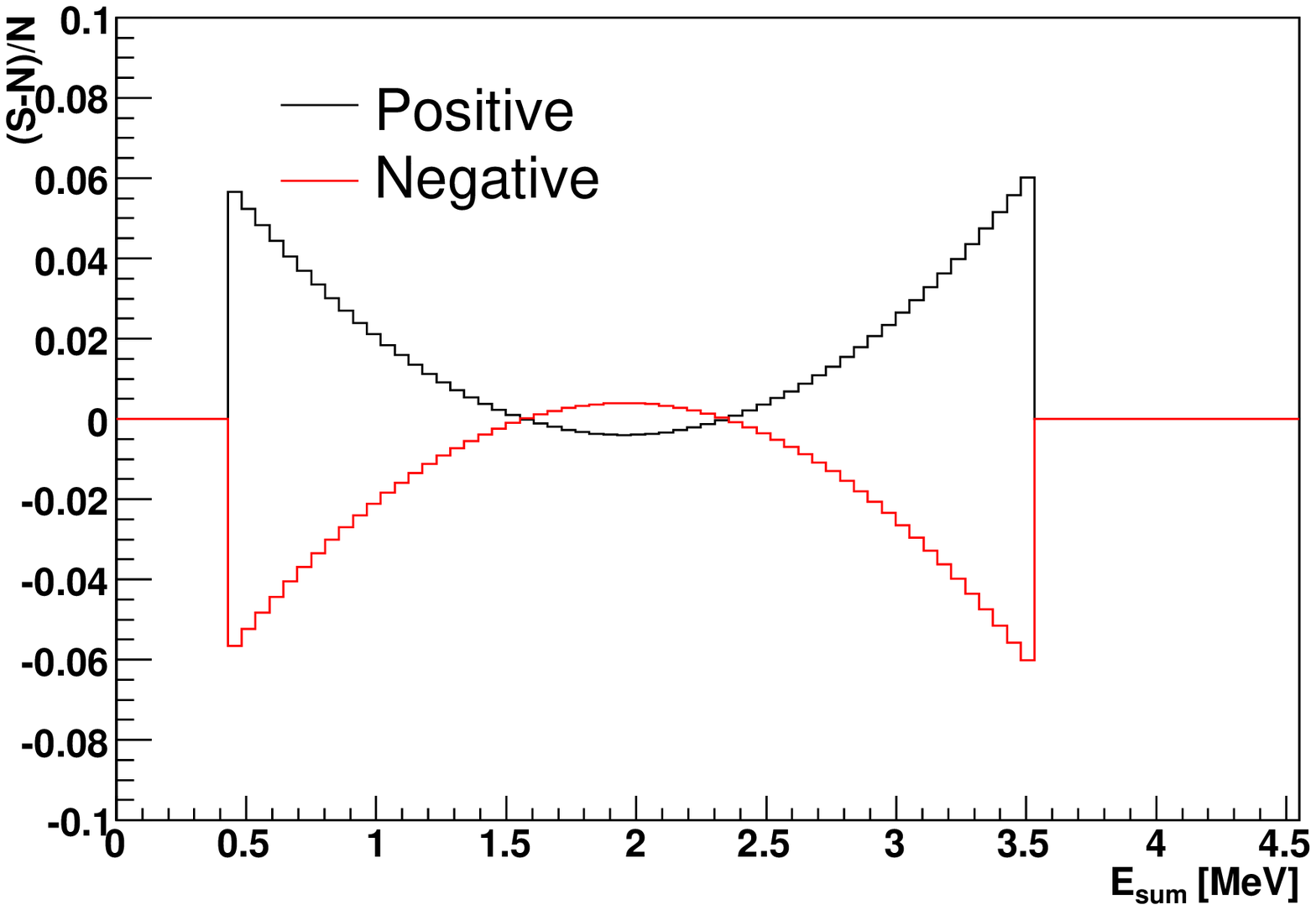}}
\subfigure{
g)\includegraphics[width=6.0cm]{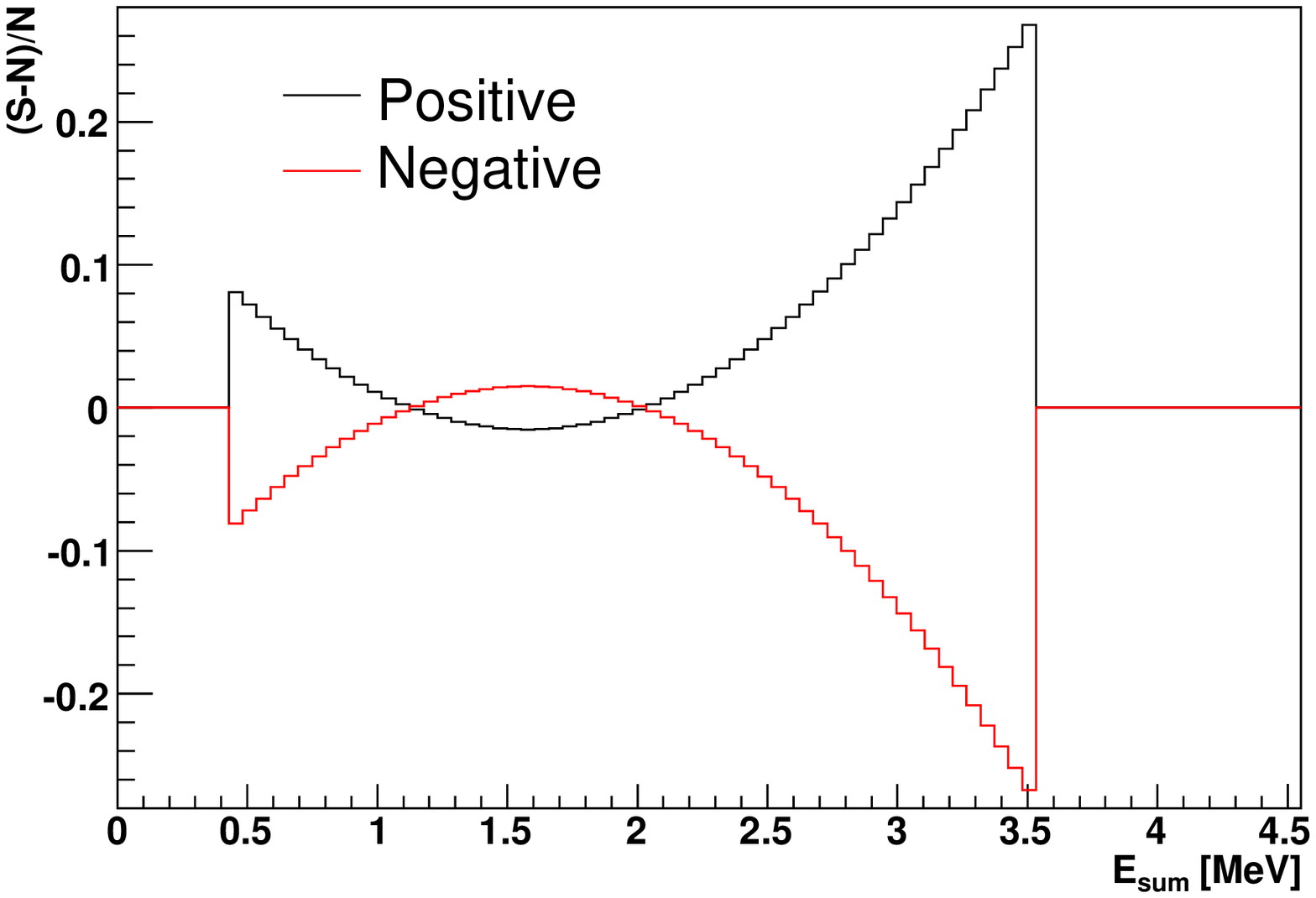}}
\subfigure{
h)\includegraphics[width=6.0cm]{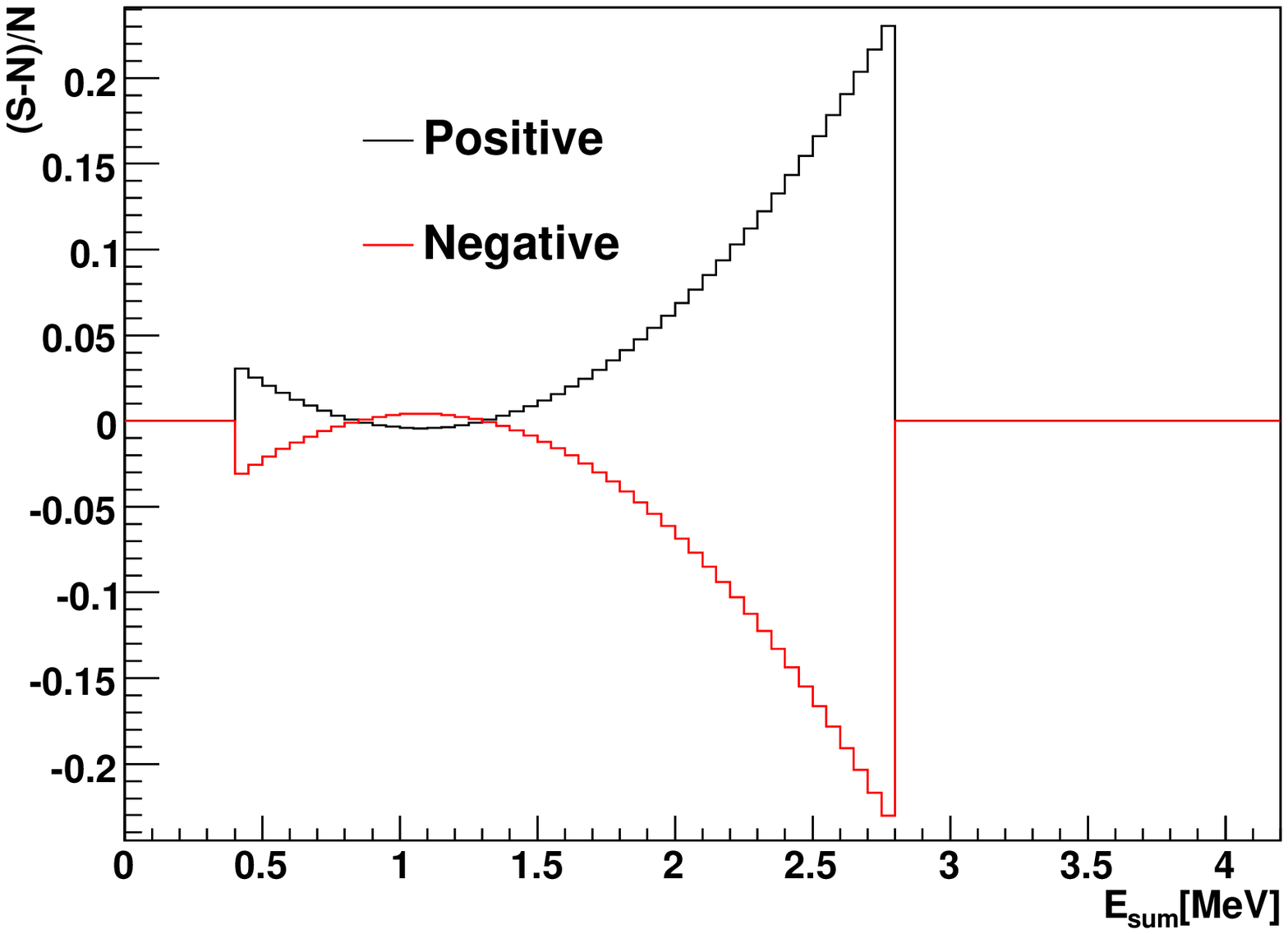}}
\caption[The estimated positive and negative shape systematics for energy distributions of different double beta decay modes.]{The estimated positive and negative shape systematics for energy distributions of  a) $0\nu\beta\beta$  mass mechanism, b) $0\nu\beta\beta$  right-handed current, c) $0\nu\beta\beta$ decay to $2^{+}$ excited state, d)  $0\nu\beta\beta$ decay to $0^{+}$ excited state, e) Majoron mode one, f) Majoron mode two, g) Majoron mode three and h) Majoron mode 7. }
\label{fig-snhisto}
\end{figure}

\begin{figure}
\centering
\subfigure{
a)\includegraphics[width=6.6cm]{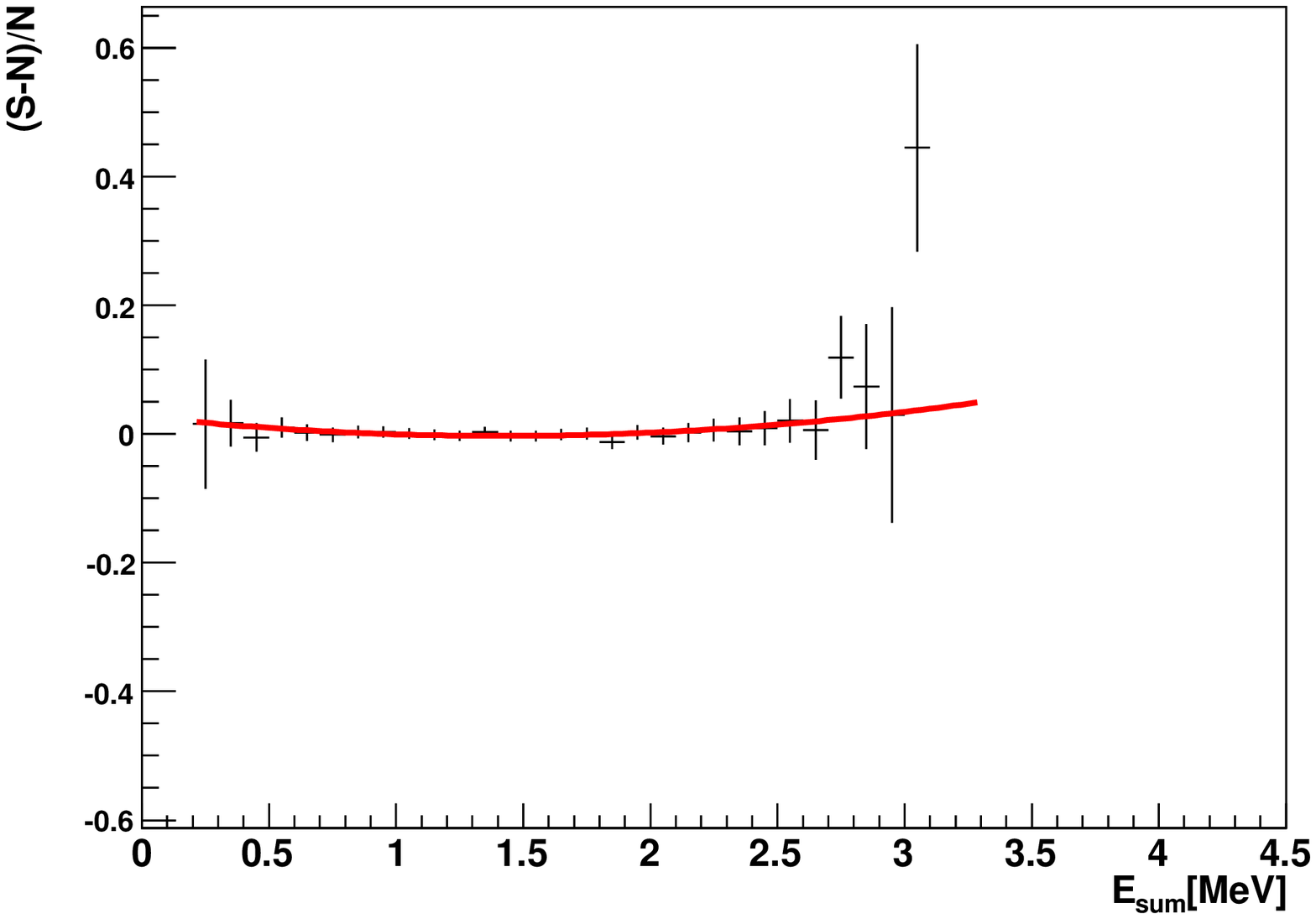}}
\subfigure{
b)\includegraphics[width=6.9cm]{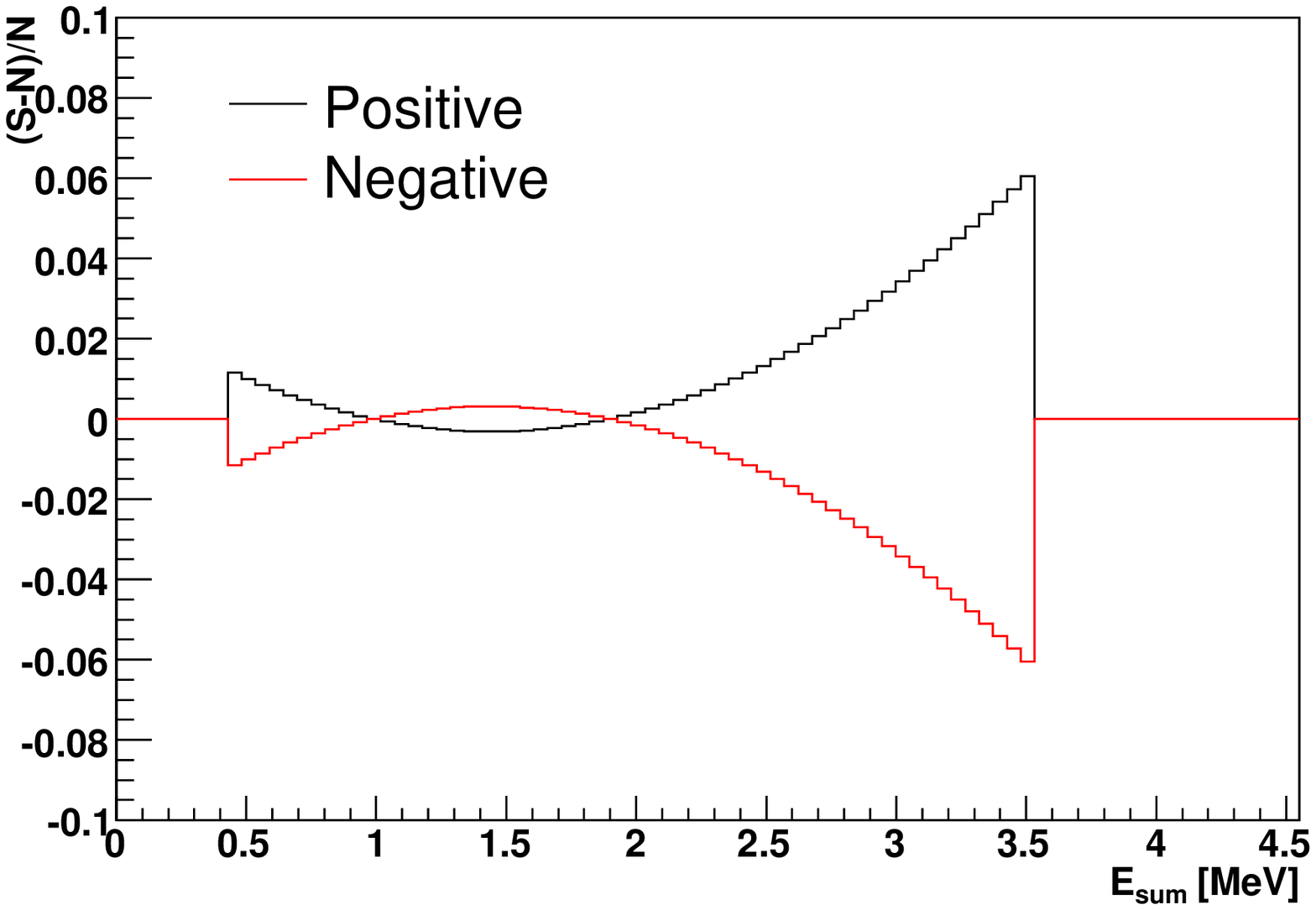}}
\caption{The distributions of a) (S-N)/N and fitted  second order polynomial function and b) the resulting positive and negative shape systematics for the energy distributions of  $2\nu\beta\beta$ decays. The error bars show the statistical uncertainties.}
\label{fig-2nbbsnfit}
\end{figure}

The uncertainty on the energy measurement affects the shape of the energy sum distribution of the signal and  is different in each bin. Figure~\ref{fig-smear0nbbsignal} shows the energy sum of the two electrons for different  simulated neutrinoless double beta decay modes before and after smearing the energy of each electron by 2\%~(Section~\ref{sec-systematic2n2b}). To find the shape uncertainty  on the energy, the ratio  (S-N)/N is plotted, where S is the smeared MC distribution and N is the nominal (unsmeared) MC distribution. These ratios for different $0\nu\beta\beta$ modes are shown in Figure~\ref{fig-0nbbsn}.  The histograms show  some  statistical fluctuations.    By fitting  the  histograms  to  different functions  the statistical fluctuations  can be reduced. Multi-Gaussian fits are performed for $0\nu\beta\beta$ modes and second order polynomial fits are performed for   $0\nu\beta\beta$  with emission of Majoron(s).  Figure~\ref{fig-snhisto} shows the resulting histograms after the fit. The inverted histograms (shown in red) are also constructed in order to estimate the negative uncertainty.  Figure~\ref{fig-2nbbsnfit}  shows the shape uncertainty on the  $2\nu\beta\beta$ background. 
The uncertainties that are considered for setting limits on $0\nu\beta\beta$ decays and   $0\nu\beta\beta$ decays with the emission  of Majoron(s) are presented  in Table~\ref{tab-syst}. The shape systematic uncertainties on signal and $2\nu\beta\beta$ background and the efficiency systematic uncertainties on  signal and radioactive background are assumed to be fully correlated.   
The shape systematic was found to contribute to the limit results with less than $1\%$.
\begin{table}[h]
\begin{tabular}{|c||c|c|c|}
\hline
Uncertainty type   &   signal &  $2\nu\beta\beta$ background & other backgrounds\\
\hline
Efficiency         &      5\%          &      --      &   5\%\\
Activities         &      --             &      --       & 4.0\%          \\
$2\nu\beta\beta$ half-life      &    --                &   2.7\% (Float)         & --                   \\
Energy measurement    &  Shape         &      Shape  &    --   \\
\hline         
\end{tabular}
\caption{The correlated systematic uncertainties on the $0\nu\beta\beta$ signal, the $2\nu\beta\beta$ and radioactive backgrounds as  considered for the  limit setting. }
\label{tab-syst}
\end{table}

\section{Limit Results}
\subsection{\boldmath{$0\nu\beta\beta$} mass mechanism}
\label{sec-0nbbresultsmm}
 In order to search for  $0\nu\beta\beta$, the $E_{sum}$ distributions of the $0\nu\beta\beta$ signal, $2\nu\beta\beta$ and radioactive backgrounds are used as inputs to the limit calculation. The search is performed using the profile likelihood ratio test statistic described in Section~\ref{sec-llrprofile} (Equation~\ref{eq-llrchi}). Table~\ref{tab-llr} gives the observed log likelihood ratio  (LLR$_{\rm obs}$), the median expected log likelihood ratio  (LLR$_{\rm med}$)  and  the log likelihood ratio for the background  with one and two standard deviations (LLR$_{\rm med}$$\pm 1\sigma$ and LLR$_{\rm med}$$\pm 2\sigma$). The results in this table show that  LLR$_{\rm obs}=4.6$~is in agreement with  LLR$_{\rm med}$ within about one standard deviation  and therefore there is no sign of a $0\nu\beta\beta$ signal. The $CL_{B}$ value  is 0.14 and therefore also consistent with a background only observation. To set  limit on the observed and median expected  number of events at 90\%~CL, the $0\nu\beta\beta$ MC is re-scaled until $CL_{S+B}=0.014$ ($CL_{S}=0.1$).
\begin{table}[!]
\centering
\begin{tabular}{|c|c|c|c|c|c|}
\hline
LLR$_{\rm obs}$ & LLR$_{\rm med}$$-2\sigma$ &  LLR$_{\rm med}$$-1\sigma$ &  LLR$_{\rm med}$ &  LLR$_{\rm med}$$+1\sigma$ &  LLR$_{\rm med}$$+2\sigma$ \\
\hline
4.6 & 5.4 & 4.4 & 2.0 &  -0.4 & -3.6 \\
\hline
\end{tabular}
\caption{Values of the observed and expected log likelihood ratios.}
\label{tab-llr}
\end{table}

\begin{figure}
\centering
\includegraphics[width=10cm]{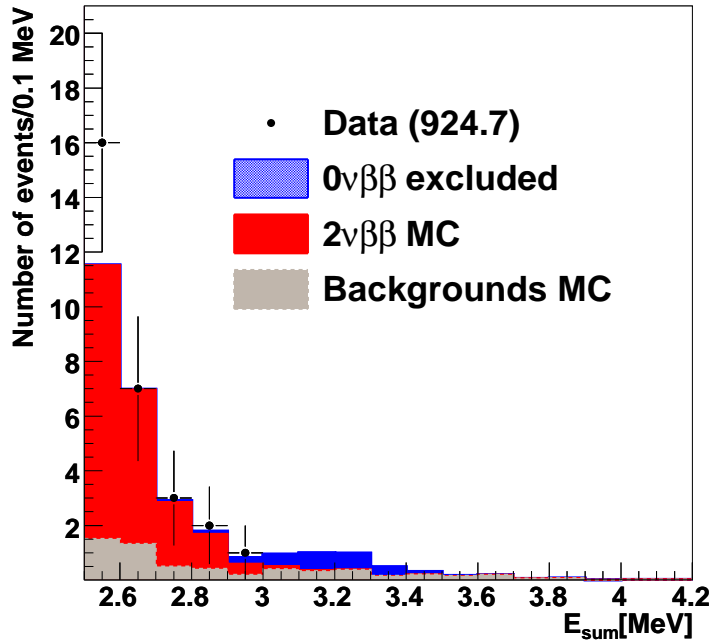}
\caption[Distribution of the energy sum of the two electrons, $E_{sum}$, for  $E_{sum}>2.5$~MeV]{Distribution of the energy sum of the two electrons, $E_{sum}$, for  $E_{sum}>2.5$~MeV. The data are compared to the total background, consisting of radioactive background and the $2\nu\beta\beta$ expectations. The error bars show the statistical uncertainties on the data points. A MC simulation of the signal with half-life of $1.8\times 10^{22}$~y is also shown.}
\label{fig-0nbblimit}
\end{figure}

The observed  upper limit on the number of events from the $0\nu\beta\beta$ mass mechanism   is found to be $2.7$ at 90\%~CL. With the  detector efficiency of $19\%$ and using Equation~\ref{eq-halflife2}, the lower bound on the $0\nu\beta\beta$ half-life is found to be:
\begin{equation}
T_{1/2}^{0\nu}>1.8\times10^{22}~{\rm y}~(90\%{\rm~CL}).
\end{equation}
This is consistent with the median  expected limit   at 90\% CL of
\begin{equation}
T_{1/2}^{0\nu}>1.33\times10^{22}~{\rm y}~(90\%{\rm~CL}).
\end{equation}
This limit on the half-life is converted into a limit on the effective Majorana  neutrino mass, $\langle m_{\nu}\rangle$, using an NME of  $3.14-4.04$~\cite{nmend} and $G^{0\nu}$ of $2.69\times 10^{-13}$~y$^{-1}$~\cite{Gnend}. The   experimental lower limit on the half-life of $^{150}$Nd translates into  an effective neutrino mass  limit of 
\begin{equation}
\langle m_{\nu}\rangle<1.5-2.5~{\rm eV}.
\end{equation}
The uncertainty on the $\langle m_{\nu}\rangle$ limit is due to the uncertainty in  NME calculations.
Taking into account the nuclear deformation will modify this conclusion. The suppression of the NME for $^{150}$Nd is estimated to be a factor $2.7$ ~\cite{nddeformation2}. This increases the upper limit to
\begin{equation}
 \langle m_{\nu}\rangle<4.0-6.3~{\rm eV}.
\end{equation}
Figures~\ref{fig-0nbblimit} shows the $E_{sum}$ distribution for $E_{sum}>$2.5~MeV. The radioactive backgrounds, shown in grey, consist  of $3.5\pm 0.9$~$^{208}$Tl events and  $0.6\pm0.2$~$^{214}$Bi events. Other backgrounds are found to be negligible in this energy region. A $0\nu\beta\beta$ distribution   with half-life of $1.8\times 10^{22}$~y, corresponding to $2.7$ events expected for the observed 90\% CL, is shown in blue.
\subsection{Other \boldmath{$0\nu\beta\beta$} modes } 
\begin{figure}[!]
\centering
\subfigure{
a)\includegraphics[width=10.cm]{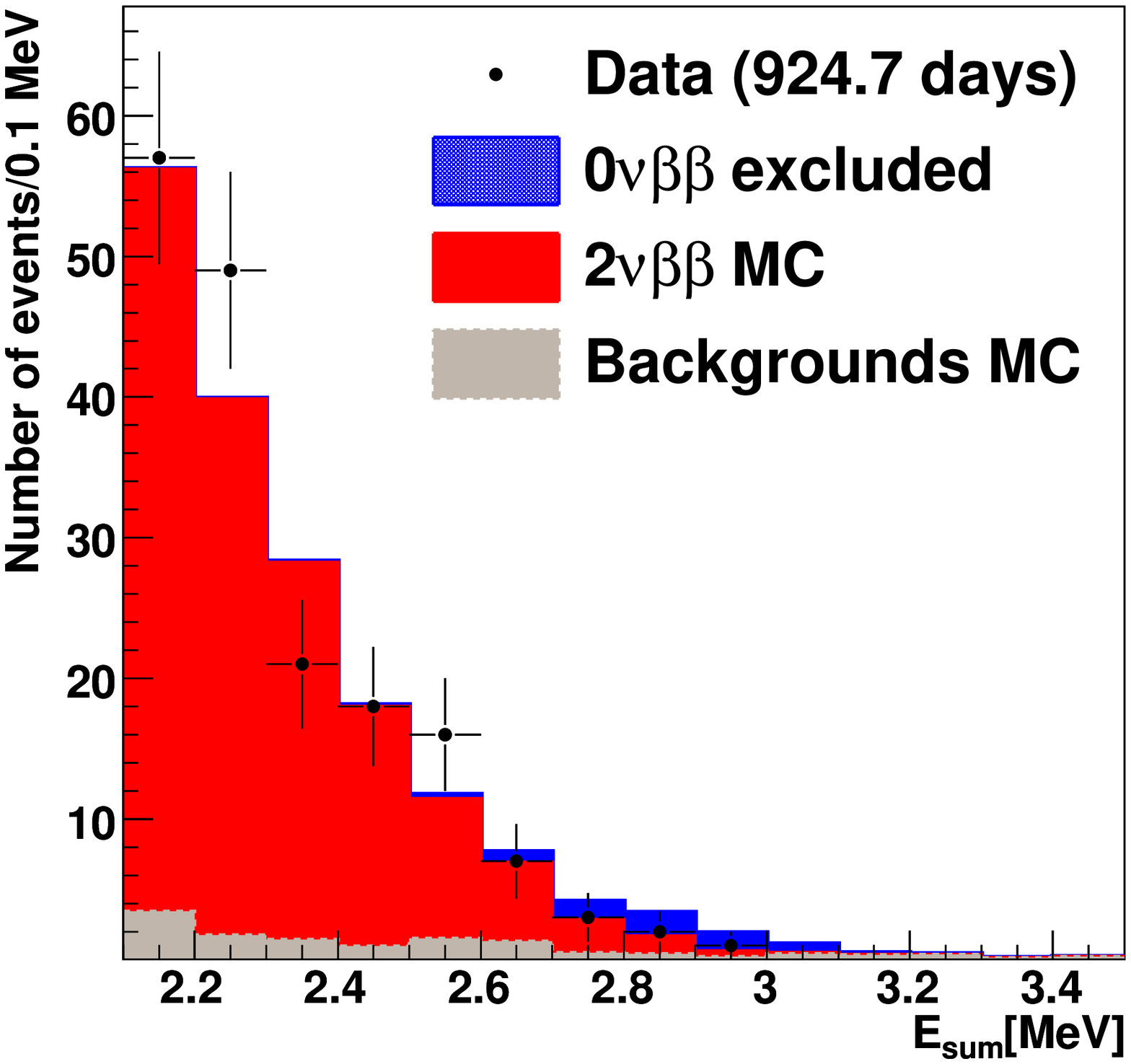}}
\subfigure{
b)\includegraphics[width=10.cm]{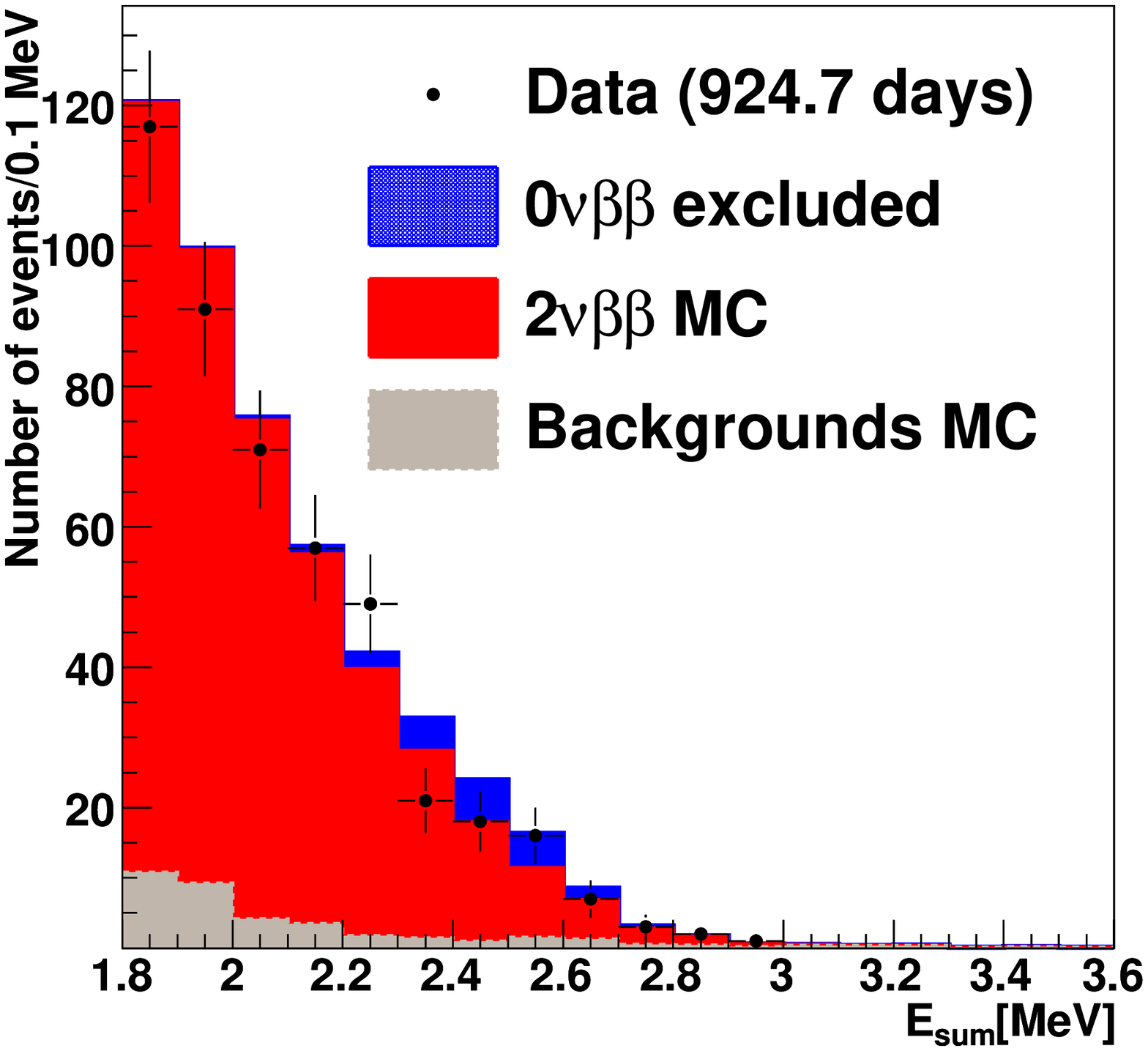}}
\caption[The $E_{sum}$ distribution of neutrinoless double beta decay to the excited states. The signal is normalised to the excluded number of events]{The $E_{sum}$ distribution of neutrinoless double beta decay to a) $2^{+}_{1}$ and b) $0^{+}_{1}$ excited states. The signal is normalised to the exclusion limit. The statistical uncertainties on the data points are shown with error bars.  }
\label{fig-excitedex}
\end{figure}

\begin{figure}
\centering
\subfigure{
a)\includegraphics[width=6.9cm]{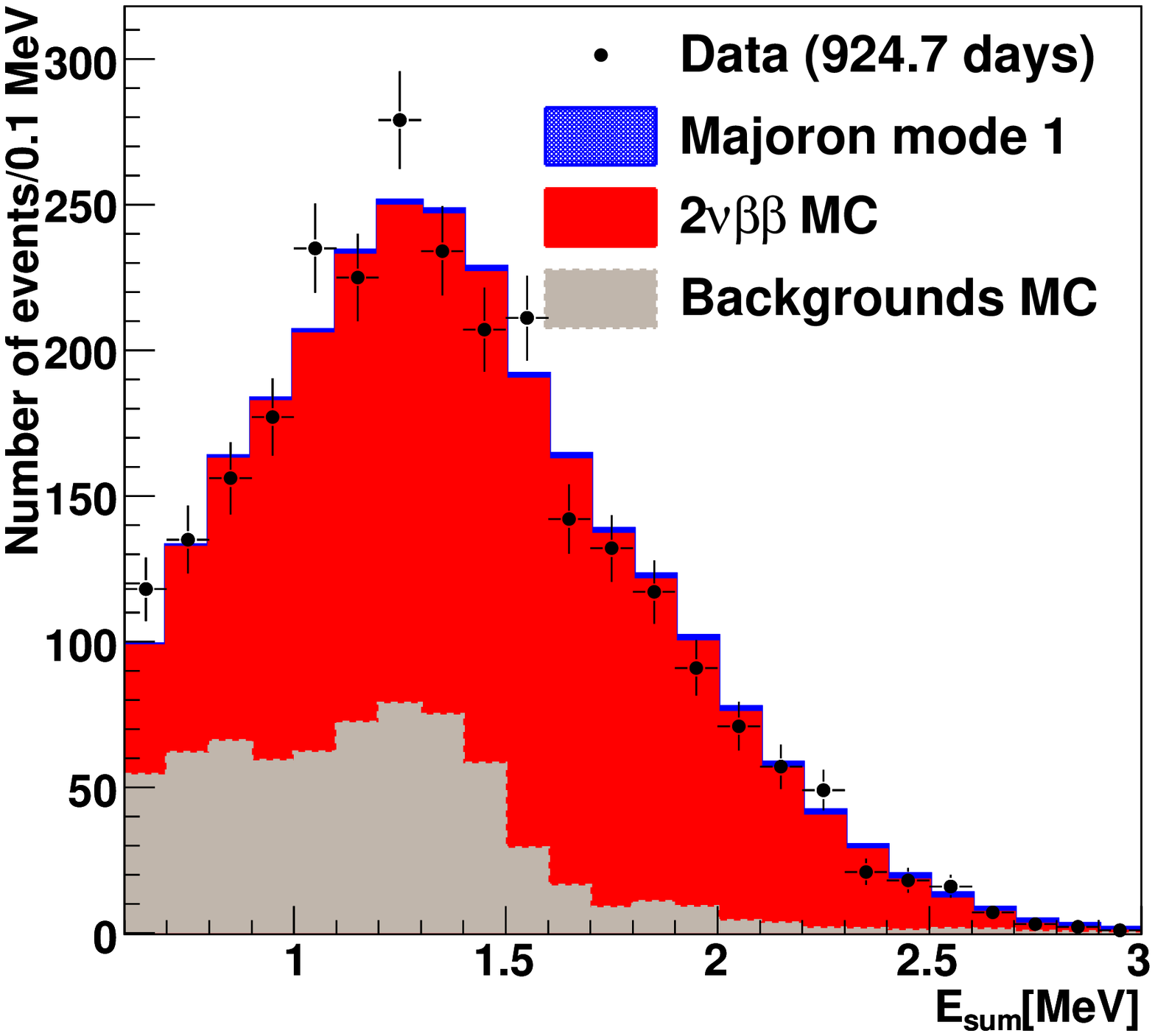}}
\subfigure{
b)\includegraphics[width=6.9cm]{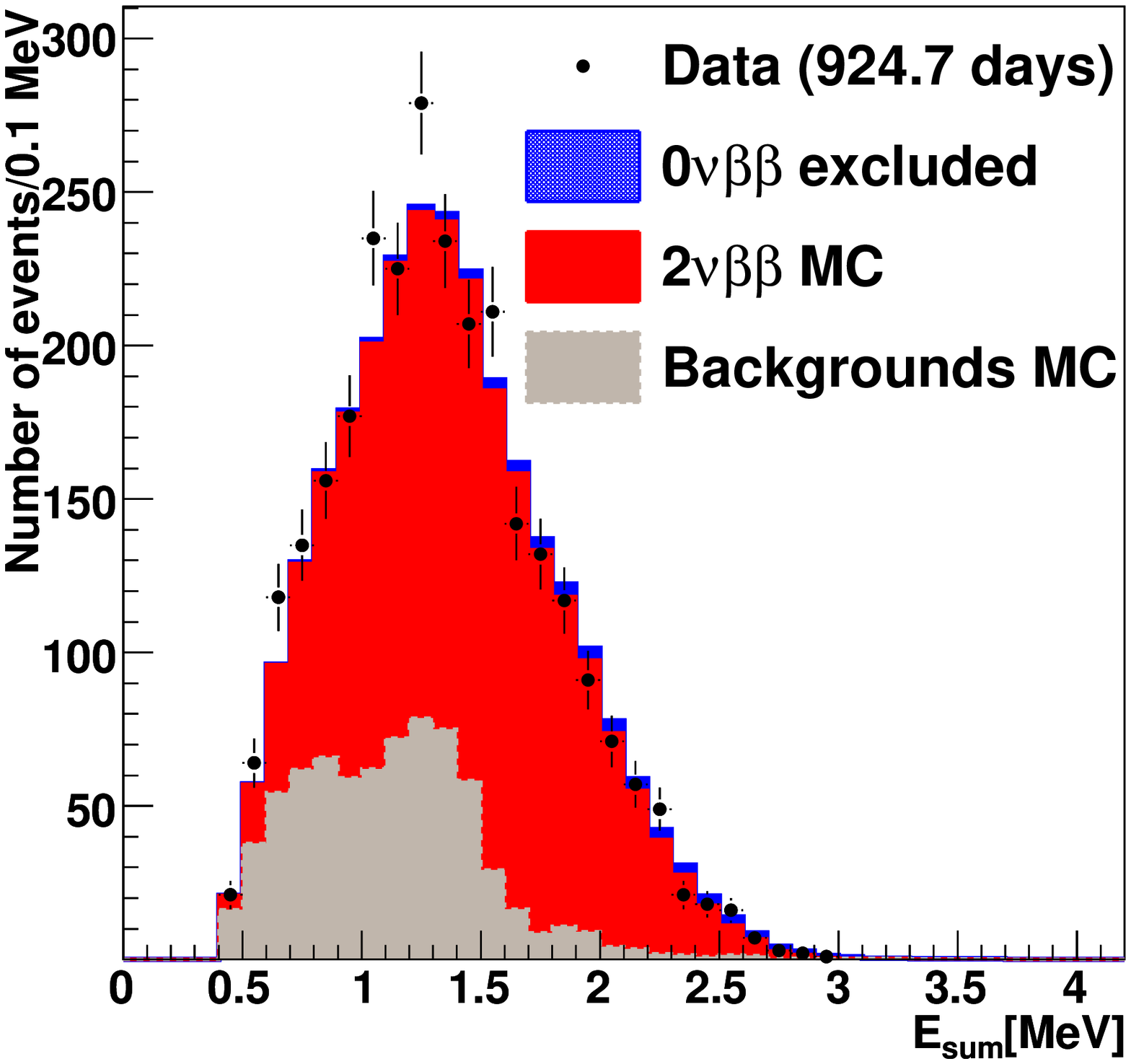}}
\subfigure{
c)\includegraphics[width=6.9cm]{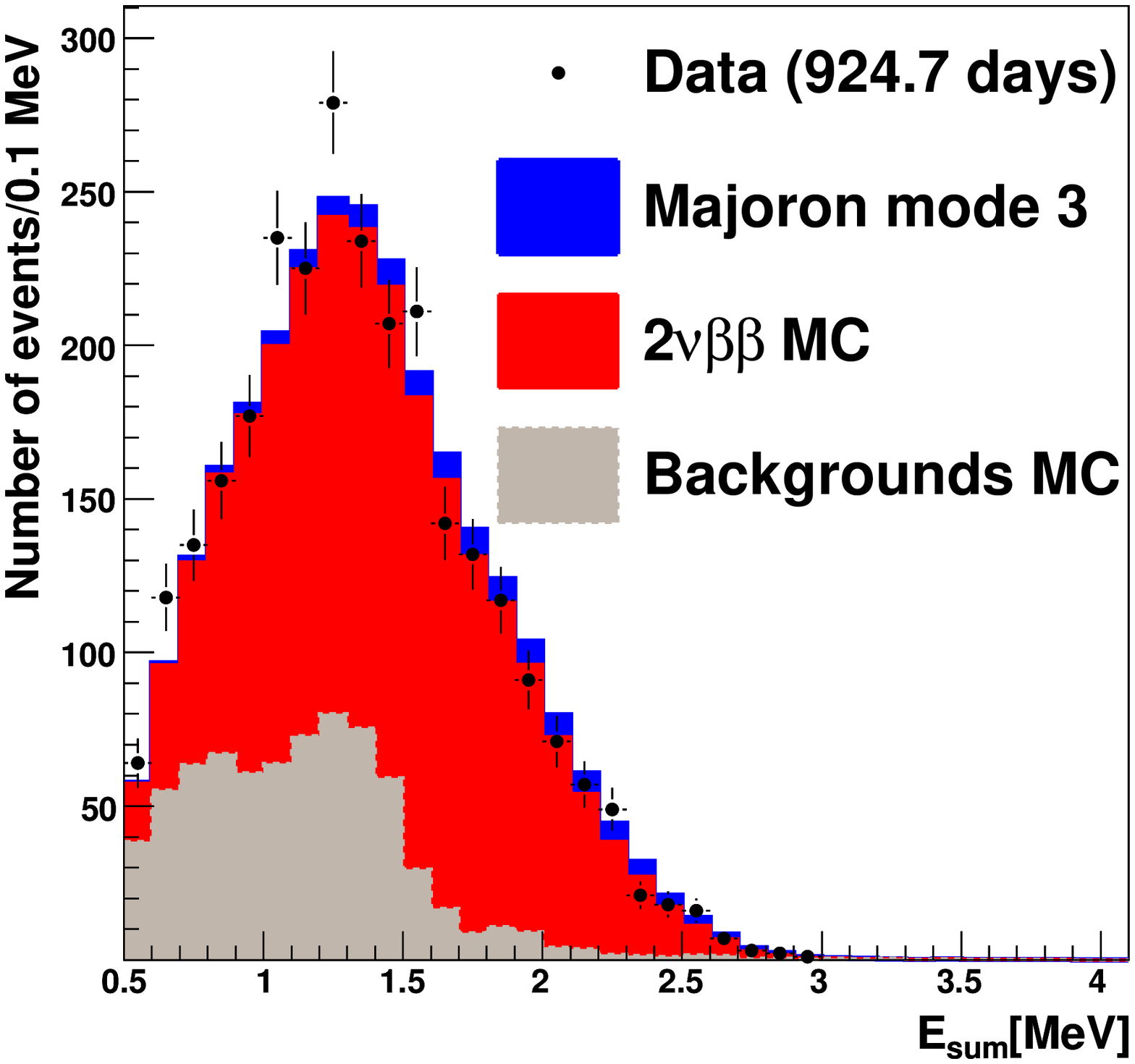}}
\subfigure{
d)\includegraphics[width=6.9cm]{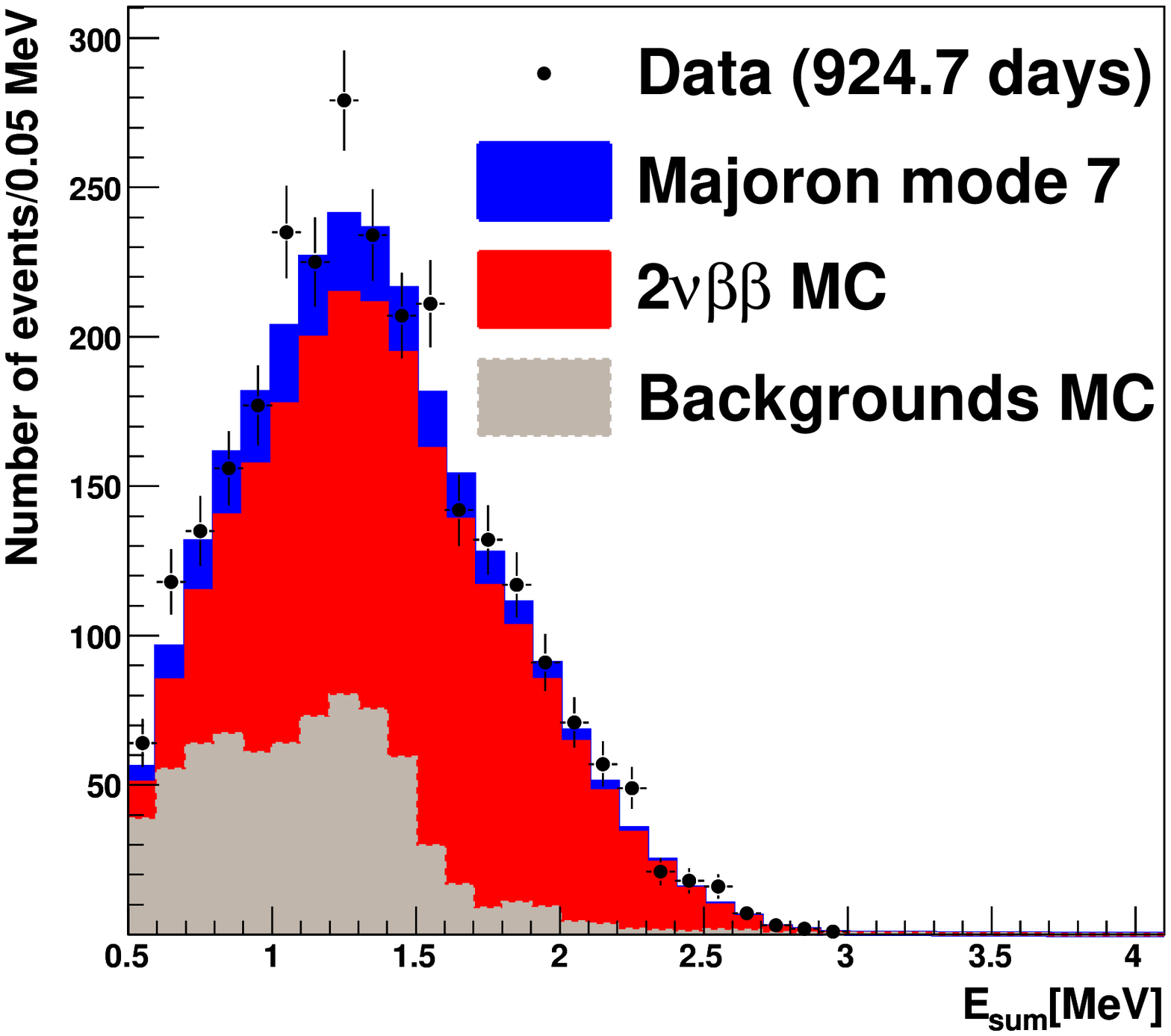}}
\caption[The $E_{sum}$ distribution of different Majoron modes]{The $E_{sum}$ distribution of a) Majoron mode one, b) Majoron mode two, c) Majoron mode three and d) Majoron mode seven.  The signal is normalised to the exclusion limit. The statistical uncertainties on the data points are shown with error bars. }
\label{fig-majoronex}
\end{figure}

Table~\ref{tab-othermodes} gives the number of events corresponding to the  median expected and observed  limits and the corresponding lower bounds on the  half-life of different  neutrinoless double beta decay processes. All limits are set  at 90\% CL.
\begin{table}[h]
\footnotesize
\centering
\begin{tabular}{|c|c|c|c|c|c|c|c|c|}
   \hline
 & \multicolumn{4}{c|}{$0\nu\beta\beta$} &
   \multicolumn{4}{c|} {Majorons}           \\
\cline{2-9}
 & $0^{+}_{\rm gs}$ ($\langle m_{\nu}\rangle$)
 & $0^{+}_{\rm gs}$~RC
 & $2^{+}_{1}$
 & $0^{+}_{1}$
 & n=1 &  n=2 &n=3 &n=7   \\
\cline{1-9}
med. exp. number  of events & $3.64$ & $3.64$  &  $5.82$  &$17.6$ & $15.5$ & $51.8$ &$114.7$& $200$\\
observed number of events& $2.7$  & $2.65$ &  $6.38$  & $22.0$ & $23.7$& $57.5$ & $115.1$ &$270$ \\
\hline
med. exp.  $T_{1/2}\times 10^{21}$~y     & $13.3$  &$7.7$ & $2.6$  & $0.3$  & $2.4$  & $0.6$ & $0.23$ &  $0.062$ \\
observed $T_{1/2}\times 10^{21}$~y   &$18.0$  &  $10.6$& $2.4$ & $0.24$ & $1.6$ &$0.54$ & $0.23$ & $0.046$ \\
\cline{1-9}
\hline
\end{tabular}
\caption{90\% median expected and observed upper limits  on the number of events  and  lower limits  on the half-life, $T_{1/2}$ for different modes of neutrinoless double beta decay. }
\label{tab-othermodes}
\end{table}
Figures~\ref{fig-excitedex} and \ref{fig-majoronex} show  the $E_{sum}$ distribution of different neutrinoless double beta decay. Each  signal is  normalised to the same number of events as   the corresponding exclusion limit. 
As the $2\nu\beta\beta$ background  was  left unconstrained, a different
  $2\nu\beta\beta$ background normalisation is found for each exclusion signal limit. 
The  lower limit on the half-life of the Majoron mode one leads to the upper limit on the  Majoron coupling with the  neutrino to be:
\begin{equation}
\langle g_{M1}\rangle<(0.64-1.05)\times 10^{-4}.
\end{equation}
To calculate this limit, the same NME as for the $0\nu\beta\beta$ mass mechanism and the phase space factor, $G^{\beta\beta}=6.40\times 10^{-15}$~$y^{-1}$~\cite{Gnend}  is used.  Considering $^{150}$Nd deformation changes the limit to:
\begin{equation}
\langle g_{M1}\rangle < (1.7-3.0)\times10^{-4}.
\end{equation}
The upper limit on the coupling of neutrino with right-handed currents and Majoron mode two and three  are not presented in this thesis as the NME calculations of these modes  are out of date~\cite{fedorchris}.
\section{Summary and discussion}
The observed limit on the half-life of neutrinoless double beta decay of $^{150}$Nd was found to be 
\begin{equation}
T_{1/2}^{0\nu}>1.8\times10^{22}~{\rm y}~(90\%{\rm CL}).\nonumber
\end{equation}
This half-life limit has improved the previous limit on this isotope by a  factor of ten.

 The best previous limit results were measured  by the Institute for Nuclear Research of the USSR in 1986~\cite{previouslimit} for $50.5$~g of enriched $^{150}$Nd.
 The experiment had a simple  set-up  using a central  foil surrounded by four scintillation counters providing calorimetry and event reconstruction.  This experiment gave a  lower bound on the $0\nu\beta\beta$ half-life of,  $T_{1/2}^{0\nu}>1.7\times10^{21}~{\rm y}~(95\%{\rm~CL})$. Despite  this significant improvement on the half-life limit, the  upper bound on effective neutrino mass has a large uncertainty due to the uncertainties on the NME calculations compared to other $0\nu\beta\beta$ isotopes. Therefore, further progress in the calculation of the NME for $0\nu\beta\beta$ decay of $^{150}$Nd  is required  to improve  the $\langle m_{\nu}\rangle$ limit  found from this isotope.  The NME  of other neutrinoless double beta decay modes  are also required to be  recalculated with improved calculating tools. This is especially true for  right-handed current $0\nu\beta\beta$ searches.

The lower limit on the half-life of the Majoron mode one decay of $^{150}$Nd was measured to be
\begin{equation}
T_{1/2}^{0\nu\beta\beta\chi}>1.6\times10^{21}~{\rm y}~(90\%{\rm CL}),\nonumber
\end{equation}
 which has  improved the previous limit significantly ($T_{1/2}^{0\nu\beta\beta\chi}>3.31\times10^{20}~{\rm y}~(90\%{\rm CL}$)~\cite{irvine}).
The corresponding limit on the Majoron-neutrino coupling is
\begin{equation}
 \langle g_{M1}\rangle < (1.7-3.0)\times10^{-4}.\nonumber
\end{equation}
 This limit is comparable with limits found with $^{100}$Mo ($\langle g_{M1}\rangle < (0.4-1.9)\times10^{-4})$)  and $^{82}$Se ($(0.66-1.7)\times10^{-4}$)~\cite{momajoron}. This result is obtained although only   $36.55$~g of  $^{150}$Nd is used, in comparison with $6.9$~kg of $^{100}$Mo and $0.93~$kg $^{82}$Se. This is due to the dependence of $G_{\alpha}^{\beta\beta}$ on $Q_{\beta\beta}^{7}$.

%% file: conclusion.tex
\renewcommand{\baselinestretch}{1.6}
\normalsize
\chapter{Conclusion}
\label{chap-summary}
In  this thesis a precise  measurement of the  $2\nu\beta\beta$ half-life of $^{150}$Nd and searches for different modes of neutrinoless double beta decay using NEMO~3 data have been presented.  These involve measurements of the  activities of backgrounds originating from the internal sources and validation of the external background model. The results given in this thesis have been submitted for publication~\cite{prlpaper}.

The internal background activities are measured by studying $e\gamma$, $e\gamma\gamma$ and $1e$ channels. Hot-spot regions  due to  $^{207}$Bi  and $^{234m}$Pa contamination are observed in  the $e\gamma$ and $1e$ channels, respectively. These  regions are subsequently removed from the analysis. The $^{207}$Bi and $^{40}$K activities found in this thesis have invalidated the HPGe results for these isotopes.
The $^{234m}$Pa, $^{207}$Bi, $^{40}$K,  $^{228}$Ac isotopes are found to be the major internal background sources to $2\nu\beta\beta$ and their activities  are measured with uncertainties less than $6\%$.  The $^{208}$Tl isotope is the major background to $0\nu\beta\beta$ and its activity is measured with an uncertainty  of $10\%$.   
 
The number of two-electron background events originating   from  the neighbouring  double beta decay sources and their contaminants are  found to  be $22\%$ of the total background events. The external background model used in NEMO~3 was validated by selecting external particles that are interacting with $^{150}$Nd foil. A $17\%$ uncertainty on the model  is used from the validation with the external $e\gamma$ channel.  The total external background contribution to $2\nu\beta\beta$ is found to be  $7\%$  of the total background events. After applying the  two-electron event selection and subtracting the background events, the half-life of $2\nu\beta\beta$ is measured to be: 
\begin{equation}
 T^{2\nu}_{1/2}=(9.11^{+0.25}_{-0.22}~{\rm(stat.)}\pm0.62~{\rm(syst.)})\times 10^{18}~{\rm y}.\nonumber
\end{equation} 
The dominant systematic uncertainties are due to the uncertainties on the  tracking efficiency~($5\%$), $^{150}$Nd foil position~($3\%$) and number of radioactive background events~($1.5\%$). 

A search is performed for different  $0\nu\beta\beta$  modes by studying the energy sum distributions of the two electrons in the final state. Limits on $0\nu\beta\beta$ signals are set  using the profile-likelihood ratio technique and by floating the $2\nu\beta\beta$ background.   This limit setting technique has been used for the first time in the NEMO experiment.
 The lower limits on  the neutrinoless double beta decay half-lives of $^{150}$Nd for the mass mechanism ($0\nu\beta\beta$) and the Majoron mode one decay ($0\nu\beta\beta\chi$)   are,
\begin{eqnarray}
T_{1/2}^{0\nu}&>&1.8\times10^{22}~{\rm y}~(90\%{\rm CL}),\nonumber\\
T_{1/2}^{0\nu\beta\beta\chi}&>&1.6\times10^{21}~{\rm y}~(90\%{\rm CL}).\nonumber
\end{eqnarray}
Both results  significantly improve the previous limits on neutrinoless double beta decay half-lives of this isotope and lead to the upper limits on the effective neutrino mass and Majoron-neutrino coupling of 
\begin{eqnarray}
&\langle m_{\nu}\rangle<&1.5-6.3~{\rm eV}, \nonumber\\
&\langle g_{M1}\rangle <&(0.64-3.0)\times10^{-4}.\nonumber
\end{eqnarray} 
The NME calculation of $0\nu\beta\beta$ must be improved  in order to reduce the uncertainty on these limits. 
Limits are also set for the first time on half-lives of  right-handed currents, excited state $0\nu\beta\beta$ decays and several other models leading to emissions of Majoron(s). 